%% file: main.tex
\documentclass[%
 aip,
 amsmath,amssymb,
 nofootinbib,
 reprint,%
]{revtex4-1}
\usepackage[english]{babel}
\usepackage{amsmath}
\usepackage{bbold}
\usepackage{mathtools}
\usepackage{nicefrac}
\usepackage{siunitx}
\usepackage[braket,qm]{qcircuit}
\usepackage{graphicx}
\usepackage[colorlinks=true,allcolors=blue]{hyperref}
\usepackage{listings}
\usepackage{tikz}
\usepackage{todonotes}
\usepackage{booktabs}
\usepackage{algorithm}   % pseudocode snippet
\usepackage{algpseudocode} % pseudocode snippet
\usepackage{ulem} % sout command
\usepackage[most]{tcolorbox}
\usepackage{xcolor}
\usepackage{adjustbox}
\usepackage{enumitem}

\newcommand{\Rx}{\(\texttt{R}_{\texttt{X}}\)}
\newcommand{\Ry}{\(\texttt{R}_{\texttt{Y}}\)}
\newcommand{\Rz}{\(\texttt{R}_{\texttt{Z}}\)}
\newcommand{\Rxx}{\(\texttt{R}_{\texttt{XX}}\)}
\newcommand{\Ryy}{\(\texttt{R}_{\texttt{YY}}\)}
\newcommand{\Rzz}{\(\texttt{R}_{\texttt{ZZ}}\)}
\newcommand{\Cx}{\(\texttt{C}_{\texttt{X}}\)}
\newcommand{\Cy}{\(\texttt{C}_{\texttt{Y}}\)}
\newcommand{\Cz}{\(\texttt{C}_{\texttt{Z}}\)}
\DeclareMathOperator{\Tr}{Tr}

\begin{document}
\title{A Practical Introduction to VQE: Methods, Challenges, and Applications in Physics and Chemistry}
\author{Jannis Ehrlich}
\affiliation{Fraunhofer-Institut für Werkstoffmechanik IWM, Wöhlerstraße 11, 79108 Freiburg, Germany}
\author{Felix Rupprecht}
\affiliation{Institute of Quantum Technologies, German Aerospace Center (DLR), Wilhelm-Runge-Str. 10, 89081 Ulm, Germany}
\author{Konstantin Lamp}
\author{Elias Walter}
\author{Alejandro D. Somoza}
\affiliation{Institute of Engineering Thermodynamics, German Aerospace Center (DLR), Wilhelm-Runge-Str. 10, 89081 Ulm, Germany}
\affiliation{Helmholtz Institute Ulm, Helmholtzstr. 11, 89081 Ulm, Germany}
\author{Manuel Enns}
\affiliation{Fraunhofer-Institut für Werkstoffmechanik IWM, Wöhlerstraße 11, 79108 Freiburg, Germany}
\author{Christian Elsässer}
\affiliation{Fraunhofer-Institut für Werkstoffmechanik IWM, Wöhlerstraße 11, 79108 Freiburg, Germany}
\affiliation{Freiburger Materialforschungszentrum, Universität Freiburg, Stefan-Meier-Straße 21, 70194 Freiburg, Germany}
\author{Birger Horstmann}
\affiliation{Institute of Engineering Thermodynamics, German Aerospace Center (DLR), Wilhelm-Runge-Str. 10, 89081 Ulm, Germany}
\affiliation{Helmholtz Institute Ulm, Helmholtzstr. 11, 89081 Ulm, Germany}
\affiliation{Department of Physics, Ulm University, Albert-Einstein-Allee 11, 89081 Ulm, Germany}
\author{Daniel F. Urban}
\affiliation{Fraunhofer-Institut für Werkstoffmechanik IWM, Wöhlerstraße 11, 79108 Freiburg, Germany}
\affiliation{Freiburger Materialforschungszentrum, Universität Freiburg, Stefan-Meier-Straße 21, 70194 Freiburg, Germany}

%% TODO list:
% Decide journal for submission
% AQS, JCP, Molecules, New Journal of Physics/ Journal of Physics: Cond Mat

\begin{abstract}

The Variational Quantum Eigensolver (VQE) has become a central hybrid quantum–classical method for addressing electronic structure and materials science problems on near‑term quantum hardware. As researchers across physics, chemistry, and quantum information engage with VQE, they encounter a wide range of algorithmic choices, including problem encoding, ansatz construction, classical optimization, and measurement strategies to choose from. This article provides a structured overview of these components and discusses practical factors that influence VQE performance in real settings, such as noise characteristics, available error‑mitigation techniques, and hardware‑dependent circuit considerations. Drawing partly on our early experiences, we highlight aspects that may be especially relevant for newcomers and outline common challenges observed in practical workflows. Our aim is to offer an accessible entry point for researchers approaching VQE‑based quantum simulation and to support informed decision‑making as the method and underlying hardware continue to evolve, which we support by discussion of examples from physics and quantum chemistry. 
\end{abstract}

\maketitle

\tableofcontents

% \newpage
\input{introduction}

\input{VQE}

\input{simulating}
\input{exampleO2}
\input{exampleHubbard}

\input{outlook}

\section*{Acknowledgements}
This project was made possible by the Competence Center Quantum Computing Baden-Württemberg (KQCBW) through projects QuESt, QuESt+, KQCBW24 and KQCBW25. Further appreciation is extended to the Center for Electrochemical Energy Storage Ulm \(\&\) Karlsruhe (CELEST). The authors acknowledge support by the state of Baden-Württemberg through bwHPC and the German Research Foundation (DFG) through grant no INST 40/575-1 FUGG (JUSTUS 2 cluster).
%This work has been funded by the Ministry of Economic Affairs, Crafts and Tourism Baden-Württemberg through the Competence Center Quantum Computing Baden-Württemberg (KQCBW).

%\bibliographystyle{unsrt} %unsrt abbrv aipnum4-2 % doesnt really work
% \bibliographystyle{apsrev4-2}
\bibliographystyle{aip_mod}
\bibliography{literature}

% \printbibliography

\end{document}

%% file: introduction.tex
\section{Introduction}

% \comment{\textbf{Journals:} PRX Quantum (tutorial); JCTC (review); Rev. Mod. Phys., JCP, JPCA (ACS) ...}\\

Understanding correlated electron systems is one of the grand challenges in quantum chemistry and materials science. Electronic correlations influence reaction dynamics and catalytic properties and give rise to various phases in solids like magnetic order and superconductivity. Strong electron-electron interactions pose a fundamental challenge to numerical simulations as they cannot be resolved by effective mean-field theories like Hartree-Fock or Density Functional Theory.  

While several computational methods running on classical computers like Configuration Interaction (CI) and Coupled Cluster (CC) for molecules or exact diagonalization (ED), density matrix renormalization group (DMRG) or Quantum Monte Carlo (QMC)  for solid state systems were developed, these methods always face the challenge of an exponentially growing Hilbert space, as correlation effects and active‑space sizes increase. In contrast,
quantum computers naturally encode many‑body quantum states, with the number of qubits scaling approximately linearly with the number of spin‑orbitals,  and thus they allow for a direct, efficient simulation of electronic quantum systems, like Feynman proposed in his famous talk \cite{Feynman1982}. 

The adiabatic time evolution of the system towards the ground state in combination with Quantum Phase Estimation (QPE) \cite{Kitaev1995} followed this idea and, thus, was one of the first proposed applications for quantum computers (QC). Unfortunately, QPE requires a large number of  controlled Hamiltonian operations, leading to deep and resource‑intensive quantum circuits on top of the already demanding adiabatic evolution. This combination of circuit depth, entangling‑gate count, and overall qubit requirements needed in order to achieve chemical precision goes far beyond the capabilities of current Noisy Intermediate‑Scale Quantum (NISQ) devices. 

To circumvent these limitations, the variational quantum eigensolver (VQE) was proposed \cite{Peruzzo2014} as a hybrid quantum–classical algorithm, which prepares a parametrized trial state on a quantum device and minimizes its energy using a classical optimization routine. In VQE, the long coherent evolutions are replaced by shorter parametrized circuits whose parameters are updated classically. Likewise, expectation‑value measurements replace phase estimation, enabling a ground‑state energy approximation at significantly reduced circuit depth and qubit overhead. 

Although the Variational Quantum Eigensolver is conceptually simple, its practical application involves several interconnected challenges. The choice of encoding and ansatz determines the circuit structure, accuracy, and resource requirements, thereby demanding a balance between expressivity, symmetry preservation, and circuit depth. Classical optimization must operate in the presence of sampling noise and rugged optimization landscapes, and can be further hindered by barren plateaus, where gradients become too small for efficient training. Expectation-value estimation often dominates the overall computational cost because realistic Hamiltonians decompose into many Pauli terms that must be measured repeatedly. In addition, hardware noise, limited coherence times, and compilation constraints necessitate error-mitigation and hardware-aware circuit design. Consequently, successful VQE implementations require carefully balancing trade-offs across all stages of the workflow.
 
In response to these practical considerations, significant research has focused on different aspects of VQE, ranging from more efficient protocols for calculating expectation-values\cite{Rubin2018,Arrasmith2020, Gokhale2019} and associated error mitigation techniques \cite{Temme2017, Giurgica2020, vandenBerg2023, vandenBerg2024}, the development of new optimizers tailored to noisy and stochastic objectives\cite{Nakanishi2020, Ostaszewski2021}, and the design of ansatz families that balance expressivity and hardware constraints\cite{Gard2020, Mondal2023, Halder2024, Fedorov2022, Haidar2025}. Moreover, VQE was used on actual quantum computing hardware for a wide variety of systems: molecular systems ranging from 2-6 qubits (H\(_2\) \cite{Peruzzo2014}, LiH and BeH\(_2\) \cite{Kandala2017}) up to 12 qubits (H\(_{12}\)) \cite{Arute2020} as well as solid-state model Hamiltonians, including the transverse‑field Ising model\cite{Kandala2017, Le2025}, Hubbard models\cite{Reiner2019,Cade2020} and the Anderson impurity model\cite{Rungger2019, Ehrlich2025} (4 qubits). Classical simulations of VQE have been demonstrated for up to 1000 qubits by employing matrix product states \cite{Shang2023}. Especially in strongly correlated regimes these are limited in their accuracy due to their finite bond dimension.

These studies highlight both the feasibility of near‑term demonstrations and the persistent bottlenecks in scaling to chemically accurate simulations of larger systems.
Despite steady progress in devices and methodology, increasing the size of simulatable systems, a clear quantum advantage for electronic‑structure tasks using VQE remains an open milestone. However, its proven applicability already led to the development of post‑VQE approaches that improve the calculated ground‑state energies and allow the calculation of excited states \cite{McClean2017, Ollitrault2020} and to VQE-extensions that directly target the determination of excited states\cite{Nakanishi2019, Higgott2019}%

There are already several reviews available which cover VQE. As part of broader reviews on variational quantum algorithms (VQA), Cerezo et al.\ \cite{Cerezo2021} offer an overview of the basic concepts with only limited VQE specific details, while Bharti et al. \cite{Bharti2022} provide more insight into certain VQA components without an in-depth discussion. The review of McArdle et al.\ \cite{mcardle_quantum_2020} takes a look at quantum computational algorithms, including VQE, from a quantum-chemistry viewpoint. While mentioning the important aspects and referring to many important papers, it does not provide a detailed discussion. The reviews of Fedorov et al.\ \cite{Fedorov2022} and Harville \cite{harville_recent_2026} introduce the whole VQE algorithm but mainly focus on the choice of ansatz design and initial state preparation. Similarly, Belaloui et al.\  focus on the ansatz but support their discussion with numerical calculations, without extending the analysis to optimization, measurement, or mitigation strategies. For other aspects such as error mitigation \cite{Cai2023}, there are method‑specific reviews outside the immediate VQE scope, which do not integrate these techniques into a full VQE workflow. The very comprehensive VQE review by Tilly et al.\ \cite{Tilly2022}, offers an extensive and technically detailed account of essentially all VQE components, however its breadth and density present a steep entry barrier for newcomers seeking a structured, guided introduction. 

Our tutorial is intended to offer scientists from chemistry, physics, material sciences, and quantum information a clear and accessible entry point to quantum computing and, in particular, to the application of VQE for electronic‑structure problems. Our goal is to provide a practical, workflow‑oriented introduction to VQE, focusing on its core components, the most relevant literature, and the key aspects to consider when performing a VQE calculation, based on our own experience. These considerations are summarized at the end of each section and illustrated with examples from both condensed‑matter model Hamiltonians and quantum‑chemistry applications at the end of the paper, which together aim to give newcomers a coherent and guided starting point.

This article is therefore structured as follows. In Sec.\ \ref{Sec:VQE} we introduce the VQE algorithm and outline its main workflow. We discuss the core components that shape a practical VQE implementation: the construction of a compact active space (Sec.\ \ref{Sec:activespace}), the mapping from fermionic to qubit operators (Sec.\ \ref{Sec:Mapping}), commonly used ansatz circuits for quantum‑chemistry and solid-state physics applications (Sec.\ \ref{Sec:Ansatz}), and the transpilation of these circuits to hardware‑specific gate sets (Sec.\ \ref{Sec:Transpilation}). The measurement of expectation-values is presented in Sec.\ \ref{Sec:MeasExpVal}, followed by a discussion of classical optimization strategies in Sec.\ \ref{Sec:Optimizer} and the issue of barren plateaus related to it in Sec.~\ref{Sec:barren_plateaus}. We conclude the presentation of the workflow with an overview of recent error‑mitigation techniques relevant for NISQ devices in Sec.\ \ref{Sec:Error_NISQ}. In Sec.\ \ref{Sec:simulating_qcircuits}, we describe how classical simulators can be used to emulate quantum‑circuit behavior. Finally, we illustrate the complete VQE procedure through two representative examples: molecular oxygen as an example from quantum chemistry in Sec.\ \ref{Sec:O2} and the 1-dimensional Hubbard model as a solid‑state physics example in Sec.\ \ref{Sec:Hubbard}. 

We begin in Sec.\ \ref{Sec:VQE} by formalizing the variational ground‑state objective, establishing notation, and outlining the closed‑loop quantum–classical workflow that underpins the design choices analyzed in the subsequent sections.

%% file: VQE.tex
\section{The Variational Quantum Eigensolver}\label{Sec:VQE}
%--> Flowchart  --> Jannis
The Variational Quantum Eigensolver (VQE) is an algorithm to determine the electronic ground state of physical systems like molecules or solid-state systems. The computation of this state is of great interest as it determines essential properties of the system, like molecular binding energies, the equilibrium geometry, vibrational and magnetic properties. By definition, the ground state \(\left|\Psi_0\right>\) of a Hamiltonian $H$ is the state with the lowest energy. 
Thus, the inequality
\begin{align}
 E_0 = \frac{\langle \Psi_0 | H | \Psi_0 \rangle}{\langle \Psi_0 |\Psi_0 \rangle} \leq \frac{\langle \Psi | H |\Psi \rangle}{\langle \Psi | \Psi \rangle} 
\end{align}
holds for any arbitrary state \(\left|\Psi\right>\). When the quantum state is parameterized, \(\left|\Psi\right> = \left|\Psi(\mathbf{\theta})\right> \), by a multidimensional parameter \(\mathbf{\theta}\) such that the whole Hilbert space is covered, there exists a set of parameters \(\mathbf{\theta}_0\) leading to the ground state. Thus, by employing the variational principle, \(\mathbf{\theta}\) can be varied in search for the ground state. 

The VQE transfers this general principle to a hybrid quantum-classical algorithm. A quantum computer prepares a parameterized quantum state and evaluates the expectation value of the Hamiltonian for this state. This value serves as the cost function of a classical optimizer, which adjusts the parameters in order to obtain the optimal parameter set referring to the lowest possible energy.

To make this procedure operational, a number of building blocks must be assembled. The quantum device first prepares a classically obtained initial state, which is then transformed by a parameterized quantum circuit implementing the unitary \(U(\mathbf{\theta})\), the so-called ansatz. Subsequent quantum measurements provide the data needed to calculate the expectation value of the Hamiltonian for this state. The classical optimizer must be chosen such that it performs reliably in the chosen setting of quantum noise and sampling statistics. Likewise, the selection and design of most components depend not only on the physical system under consideration but also on the specific quantum processor or emulator, on which the quantum calculation is executed.

\begin{figure*}[t]
    \centering
\includegraphics[width=\textwidth]{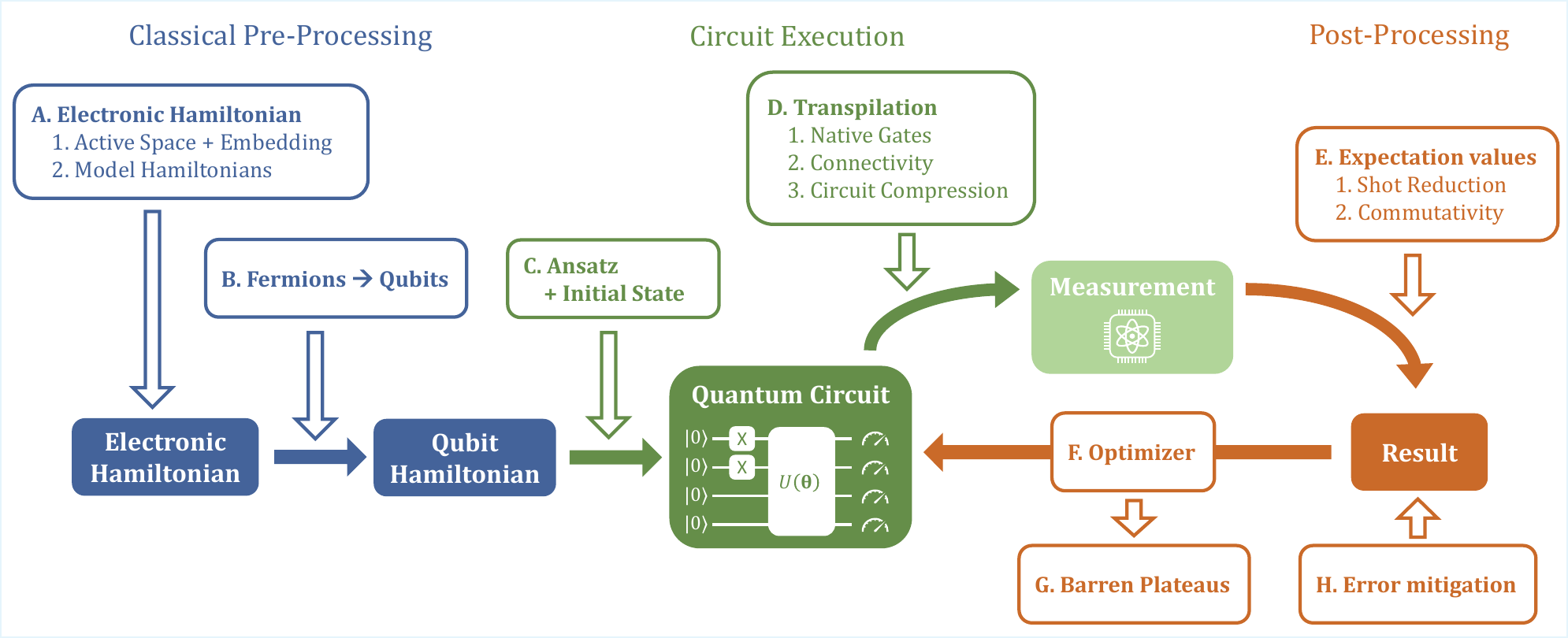}
    \caption{
    \textbf{Schematic overview of the Variational Quantum Eigensolver (VQE) workflow for electronic structure calculations.} The pipeline is organized into three main stages: Classical Pre-Processing (blue) encompasses the construction of the electronic Hamiltonian with active space selection and embedding techniques (Sec.~\ref{Sec:Hamiltonian}), fermion-to-qubit mappings such as Jordan--Wigner or Bravyi--Kitaev (Sec.~\ref{Sec:Mapping}), and the selection of a parametrized ansatz with appropriate initial state preparation, typically Hartree--Fock (Sec.~\ref{Sec:Ansatz}). 
    Circuit Execution (green) involves transpilation of the logical circuit to hardware-native gates, qubit connectivity constraints, and circuit compression optimizations (Sec.~\ref{Sec:Transpilation}), followed by measurement on the quantum processor. 
    Post-processing (orange) includes the estimation of expectation values with shot reduction and commuting group strategies (Sec.~\ref{Sec:MeasExpVal}), classical optimization of variational parameters (Sec.~\ref{Sec:Optimizer}), where key challenges such as barren plateaus in the optimization landscape are indicated (Sec.~\ref{Sec:barren_plateaus}). Finally, error mitigation techniques to correct for hardware noise are introduced (Sec.~\ref{Sec:Error_NISQ}). 
    The hybrid loop iterates until convergence to the ground-state energy estimate.}
\label{Fig:schematic}
\end{figure*}

Therefore, the following elements, as shown in Figure \ref{Fig:schematic}, constitute the essential building blocks of a VQE:
\begin{itemize}
 \item (Qubit) Hamiltonian of the system
 \item Parameterized quantum circuit (ansatz)
 \item A quantum processor or emulator 
 \item Measurement of the expectation values
 \item Mitigation of errors
 \item Classical optimizer
\end{itemize}
The subsequent sections describe the role of each component and provide some guidance on the considerations which are relevant for their practical implementation, starting with the physical formulation of the problem.

\subsection{Electronic Hamiltonian}\label{Sec:Hamiltonian}

The electronic structure problem in quantum chemistry and condensed matter physics targeted by VQE is typically formulated in second quantization in terms of one-body potentials and two-body (pairwise) interactions.
The non-relativistic electronic Hamiltonian for interacting electrons in the Born-Oppenheimer approximation reads
\begin{equation}
    \label{eq:H_second_q}
    H = \sum_{pq} h_{pq} a^\dagger_p a_q + \frac{1}{2} \sum_{pqrs} g_{pqrs} a^\dagger_p a^\dagger_q a_r a_s\,,
\end{equation}
with 
\begin{align*}
    h_{pq} &=  \int \mathrm d \mathbf {x_1} \ \phi^*_p (\mathbf {x_1}) \mathcal{O}_1 (\mathbf{x_1}) \phi_q (\mathbf {x_1}) \qquad
    \text{and  } \\
    g_{pqrs} &= \int \mathrm d \mathbf x_1 \mathrm d \mathbf x_2 \ \phi^*_p (\mathbf x_1)\phi^*_q (\mathbf x_2)  \mathcal{O}_2 (\mathbf{x_1},\mathbf{x_2}) \phi_r (\mathbf x_2)\phi_s (\mathbf x_1)
\end{align*}
being the integrals over the one-body operator $\mathcal{O}_1$ and the two-body operator $\mathcal{O}_2$ with respect to the set of one-electron basis functions (orbitals) $\{\phi_p\}$. In the Fock state \(f\) the occupation of each of these orbitals is indicated by a string of integers 0 (empty) and 1 (occupied) where the Fermionic creation and annihilation operators $a^\dagger_p$ and $a_p$ change the occupation for the basis function with index $p$.

The type of basis function used to evaluate the electronic integrals is highly dependent on the problem.
For molecules, atomic (Gaussian-type) orbitals are typically used. For solids and surfaces the most common basis sets are plane waves\cite{Kresse1996,Clark2005,Giannozzi2009}, mixed atom and plane wave basis \cite{Elsasser1990,Ho1992, Meyer1998}, (fully linearized) augmented plane waves (FLAPW)\cite{Singh2005,fleurCode,Blaha2020}, Wannier orbitals\cite{Pizzi2020}, 
or Gaussian orbitals\cite{Lippert1997}.
Important limitations are the basis set incompleteness error (BSIE) and basis set superposition error (BSSE) \cite{parrill_basis_2017}. Finding an appropriate basis set is a challenging problem and subject to ongoing research \cite{papajak_what_2012, lee_approaching_2021, mester_basis_2023}.

The second-quantized Hamiltonian \eqref{eq:H_second_q} can in principle be directly solved on a quantum computer by mapping fermions $a_p^\dagger$ to qubits\cite{mcardle_quantum_2020}. However, this typically requires at least one qubit per fermionic degree of freedom $p$ (cf.~Sec.~\ref{Sec:Mapping} below). Even for moderately-sized systems, the total number of required qubits can become intractably large for both current quantum hardware and classical simulation methods. On the other hand, often only few degrees of freedom contribute strongly to electronic correlation effects, while for the rest a mean-field description can be sufficiently accurate. 
In the next subsections, we discuss three complementary strategies for reducing the complexity of Eq.~\eqref{eq:H_second_q}. First, in \textit{ab initio} electronic-structure calculations, the Hamiltonian is restricted to a carefully chosen active space containing the most relevant orbitals. Second, in condensed-matter physics, correlated subspaces and effective low-energy model Hamiltonians are constructed from a restricted set of degrees of freedom describing the low-energy physics of interest. Third, embedding techniques are used to incorporate the effect of the remaining environment into the reduced Hamiltonian, thereby combining a compact correlated description with an approximate treatment of the surrounding system.

\subsubsection{Active Space Selection} \label{Sec:activespace}

\begin{figure}
    \centering
    \includegraphics[scale=0.5]{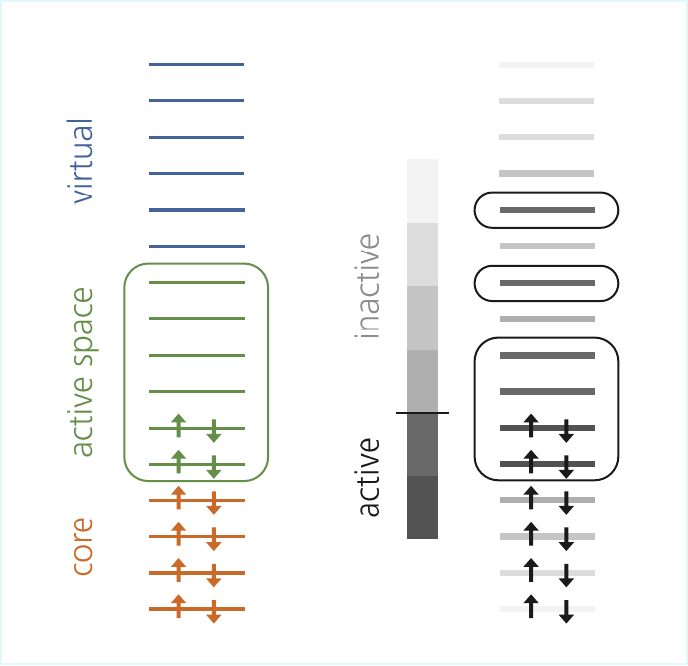}
    \caption{Molecular electronic orbitals can be separated into core (occupied), active, and virtual (empty) orbitals. Left: Selection based on orbital energies, i.e.\ orbitals around the Fermi energy are considered active. Right: Selection based on a different criterion, e.g.\ orbital entropies, where light (dark) gray indicates low (high) entropy.}
    \label{Fig:active_space}
\end{figure}

In molecular and solid-state systems, core (fully occupied) and virtual (empty) orbitals are often energetically decoupled from the valence region. Consequently, they contribute negligibly to strong correlation effects and can be excluded from the explicit treatment. The Hamiltonian \eqref{eq:H_second_q} can therefore be projected onto a reduced Hilbert subspace, termed the active space (AS) in quantum chemistry or the correlated subspace in solid-state theory.
Restricting the solution to a relevant subspace is standard in methods such as complete active space configuration interaction (CASCI) and complete active space self-consistent field (CASSCF) \cite{Roos1987}. While selection criteria generally depend on the target operator, this section focuses on the electronic Hamiltonian and orbital-based representations.
Canonical molecular or crystalline orbitals often delocalize electron density, which complicates truncation while retaining the capability to describe complex systems. Therefore, simply selecting orbitals around the Fermi energy is not optimal. Ideally, all orbitals contributing to static correlation must be included in the AS. However, static correlation lacks a unique rigorous definition, often necessitating manual selection based on chemical intuition. Although frequently successful, this approach lacks a systematic procedure, which hinders reproducibility. Fig.~\ref{Fig:active_space} illustrates the partitioning of orbitals into core, active, and virtual sets using different selection criteria.

A robust metric for automatic selection is using natural orbital occupation numbers (NOONs) \cite{Pulay1988,Khedkar2019}. Orbitals with fractional occupations ($0+\epsilon < n_i < 2-\epsilon$) indicate correlation and are included in the AS, while fully occupied or empty orbitals are excluded. Several advanced criteria for automatic active space selection have been proposed \cite{Bao2018,Sayfutyarova2017,Claudino2019,Kolodzeiski2023,stein_automated_2016,Stein2019,Ding2023}, often optimized for specific applications or embedding schemes. In the context of quantum computing, approaches inspired by quantum information theory are commonly used. These methods select orbitals based on one- and two-orbital entropies, which serve as quantitative measures for orbital correlation and entanglement \cite{stein_automated_2016,Stein2019,Ding2023}.

\subsubsection{Correlated-Subspace Construction}
In condensed matter systems, a closely related concept arises, where the relevant low-energy degrees of freedom are identified via the band structure in reciprocal space rather than in terms of individual molecular orbitals. The correlated subspace is typically defined by the relevant low-energy bands around the Fermi energy and is described by an effective tight-binding Hamiltonian with interactions. Based on a DFT-calculation or an experimental band-structure, these bands are first projected onto a localized orbital basis, such as atomic-like orbitals or Wannier functions\cite{Pizzi2020}, which define the subspace basis. In a formal downfolding approach, high energy degrees of freedom outside this subspace are subsequently integrated out by projection of the full Hamiltonian to this subspace, leading to renormalized effective parameters that explicitly incorporate virtual processes into the low-energy non-interacting Hamiltonian. Alternatively, in a fitting-based approach, these parameters are adjusted to reproduce the band dispersion of the selected subspace, such that these renormalization effects are implicitly contained. Finally, interaction parameters for the correlated orbitals are obtained from constrained DFT\cite{Kaduk2012}, constrained random-phase approximation, or\cite{Aryasetiawan2004} empirical estimates. These resulting effective models are of intrinsic interest, as they provide a qualitative understanding of complex physical phenomena. A prominent example is the single-band Hubbard model \cite{Hubbard1963}, the simplest lattice model incorporating only on-site interactions, which turns into the Heisenberg spin model\cite{Heisenberg1928} in the strong-interaction limit.

\subsubsection{Embedding}
Once a suitable active space or correlated subspace is found, the inactive environment needs to be taken into account on a mean-field level using a quantum embedding method \cite{Sun2016}. Embedding methods allow to construct an effective Hamiltonian for the active orbitals, which includes interactions from and between inactive orbitals calculated with Hartree--Fock (HF) or density functional theory (DFT). For this purpose, several different methods based on the electron density ((local) density functional embedding theory, DFET\cite{Libisch2014, Wesolowski2015, Jacob2014}), the density matrix (density matrix embedding theory, DMET \cite{Knizia2012,Knizia2013,Wouters2016}) or Green's functions (dynamical mean-field theory, DMFT\cite{Georges1996, Kotliar2006}, self-energy embedding theory, SEET\cite{Lan2017, Rusakov2019}, quantum defect embedding theory, QDET\cite{Ma2021}) have been developed, as discussed in Ref.\ \onlinecite{Vorwerk2022}. The remaining active space or correlated subspace model is significantly smaller and can be solved with more accurate methods.

\begin{tcolorbox}[colback=blue!5!white, colframe=blue!75!black, title=Active and Correlated Space]
Accuracy strongly depends on choice of active/ correlated subspace, careful selection is crucial
\begin{itemize}[leftmargin=*]
    \item Subspace size determines accuracy/cost trade-off
    \item Freeze chemically inert orbitals; include all (strongly) correlated orbitals until target observables converge
    \item Advanced active space selection criteria (natural-orbital occupations, entanglement/ mutual information) can improve accuracy
    \item Use embedding methods to incorporate the influence of the inactive environment
\end{itemize}
\end{tcolorbox}

\subsection{From Fermionic Hamiltonians to Qubits}\label{Sec:Mapping}

After we have defined the Hamiltonian of interest for which we want to determine the ground state, we need to represent it in terms of qubits on a quantum computer, i.e.\ we need to map a system of interacting fermions onto interacting spins. While a fermionic Fock state \(f\) describes the occupation of orbitals, the quantum register is described by a qubit state \(q\), which correspond to binary strings indicating spin-up or spin-down configurations in the \(S_z\)-basis. Therefore, in the first step, fermionic Fock states \(f\) are associated with qubit states $q$, and in the second step the creation and annihilation operators are expressed as combinations of Pauli spin operators acting on these qubits.
More precisely, the fermionic operators are rewritten as tensor products of Pauli spin operators $\mathbf{P}_\lambda$, called Pauli strings.
For $N$ qubits, we have:
\begin{align}
    \mathbf P_\lambda &= Q_N \otimes Q_{N-1} \otimes \cdots \otimes Q_2 \otimes Q_1,\\
    Q_i &\in \{ \mathbb 1, \sigma_x, \sigma_y, \sigma_z \} \equiv \{I,X,Y,Z\}\,.
\end{align}

Any fermionic operator, e.g.\ the electronic Hamiltonian, then becomes a linear combination of Pauli strings
\begin{equation}\label{Eq:PauliHamiltonian}
    H = \sum_\lambda c_\lambda \mathbf P_\lambda\,.
\end{equation}

In order to compare different fermion-to-qubit mappings, the \textit{Pauli weight} or locality of each Pauli string, i.e.\ the number of non-identity operators, and the total number of non-commuting Pauli strings in the desired expectation value are the most important quantities to look out for. 
These quantities are related, since smaller Pauli weight strings commute more easily. As we need to execute different circuits on the hardware only for non-commuting Pauli strings, we can save computation time with larger commuting groups (cf.\ Sec.\ \ref{Sec:MeasExpVal}). 

Moreover, the realization of a physically motivated ansatz (e.g.\ UCC- or Hamiltonian-based, cf.\ Sec.\ \ref{Sec:Ansatz}) as a quantum circuit depends on the chosen mapping, such that less circuit operations are required for terms with lower Pauli weight. However, if non-identity operations are on not-connected qubits and the quantum processor has a limited connectivity, swap operations are required as two-qubit operations can only be performed on neighboring qubits, which enlarges the circuit substantially. 

In the following part of this section we present a short list of the most commonly used mappings \cite{mcardle_quantum_2020} and discuss some of their advantages and disadvantages.
In order to facilitate the notation, we introduce the operators
\begin{align}
    \sigma^+_p = &\frac{1}{2} (X_p + \mathrm{i}Y_p) = \vert 0 \rangle \langle 1 \vert\\
    \left( \sigma^+_p \right )^\dagger = \sigma^-_p = &\frac{1}{2} (X_p - \mathrm{i}Y_p) = \vert 1 \rangle \langle 0 \vert \,.
\end{align}

\subsubsection{Jordan--Wigner Mapping} \label{Sec:JWMapping}
The Jordan-Wigner transformation maps each fermionic orbital to one qubit, which typically means that the index of the fermionic orbital coincides with the qubit index such that the state of qubit \(p\) corresponds to the occupation of the \(p\)-th fermionic orbital, i.e.\ $q_p = f_p$. 

In other words, if the fermionic orbital with index $p$ is occupied, the qubit with index $p$ will be in the $\vert 1 \rangle$ state. This makes the Jordan--Wigner mapping easy to work with, because we can directly read the corresponding Fock state from each bitstring.

When it comes to transforming the operators, we have to be aware of the parity factor arising from the exchange anti-symmetry of fermions. For any occupied orbital $i<p$ we obtain a prefactor of $(-1)$ when we apply an operator to orbital $p$. Therefore, we can get the correct parity by appending a chain of $Z$ operations for all orbitals $i<p$.
The mappings of operators and the encoding of orbital indices is therefore given by
\begin{align}
    a_p &\mapsto \sigma^+_p Z_{p - 1} \cdots Z_1\,,\\
    a^\dagger_p &\mapsto \sigma^-_p Z_{p - 1} \cdots Z_1\,,\\
    q_p &= f_p\,.
\end{align}

We note that the Jordan--Wigner mapping stores the occupation of the fermionic orbital locally in each qubit, but the parity is stored non-locally in all preceding qubits.
This is a big disadvantage of the method, because a fermionic operator acting on the last basis function will result in a Pauli operator acting on all qubits, so it has the largest possible Pauli weight. However, as fermionic operators often appear pairwise, parts of these Z-operator strings may cancel.

\subsubsection{Parity Mapping} \label{Sec:PMapping}

The idea of parity mapping is to locally store the parity instead of the orbital occupation. 
Using the encoding 
\begin{align}
q_p = \left ( \sum_{i=0}^p f_i \right ) \mod 2
\end{align}
where \(f_i\) corresponds to the fermionic occupation of orbital \(i\), the parity of the orbital $p$ is encoded in the qubit $p-1$. 
However, the information about orbital occupation now depends on the preceding state.
This becomes important, when we want to figure out how the creation and annihilation operators transform.
If we want to annihilate an electron in orbital $p$, then the parity up to state $p$  should become the same as the parity up to state $p-1$. Thus, if the preceding state is $\vert 1 \rangle$ we have to apply the creation operator $\sigma^-_p$, while we have to apply the annihilation $\sigma^+_p$ if $p-1$ is $\vert 0 \rangle$. This condition can be checked with the operators $\vert 1 \rangle \langle 1 \vert = I - Z$ and $\vert 0 \rangle \langle 0 \vert = I + Z$, respectively.
If the orbital is occupied, the parity changes, so we have to annihilate if the preceding state is occupied and we have to create if it is not.
We now run into a similar issue as with the Jordan--Wigner mapping. A change in the value of qubit $p$ also affects the value in $p-1$. Since we do not want to change state $p+1$, we also have to flip it and as a consequence all states $i>p$. In the end, we need to apply a chain of X operations.
This leads to the full transformation

\begin{align}
    a_p &\mapsto X_N \cdots X_{p+1} \left ( \sigma^+_p \left( I + Z \right)_{p-1} - \sigma^-_p \left( I - Z \right)_{p-1} \right )\\
    a^\dagger_p &\mapsto X_N \cdots X_{p+1} \left ( \sigma^-_p \left( I + Z \right)_{p-1} - \sigma^+_p \left( I - Z \right)_{p-1} \right )\\
    q_p &= \left ( \sum_{i=0}^p f_i \right ) \mod 2
\end{align}

Evidently, the Pauli weight of the parity mapping is similar to the one of the Jordan--Wigner mapping.

However, it has the great advantage that we usually know the number of particles in our system. We might even split the system into two spin sectors and conserve the number of particles in each sector. Since our qubit state $p$ is the sum of all occupations with modulo 2 up to the fermionic orbital with index $p$, we know that the last qubit in each spin sector will always remain in the same state, as the number of occupied orbitals is constant. This can be used to eliminate one qubit for each spin sector and, thus, reduce the computational effort for the problem at hand.

\subsubsection{Bravyi--Kitaev Mapping}

The JW mapping is local in occupation but non-local with respect to the parity, while it is vice-versa for the parity mapping, so that the Pauli weights can be up to \(\mathcal{O}(N)\).  
The Bravyi--Kitaev mapping tries to circumvent this problem by storing both the occupation and the parity semi-locally. With a clever grouping of qubits the Pauli weight is \(\mathcal{O}(\log_2 N)\) in the worst case for any fermionic operator, a significant improvement over the other encodings, especially for large qubit numbers \cite{tranter_comparison_2018}. 

However, this benefit requires more involved encoding, where, basically, qubits with an even index \(j\) encode the occupation of the \(j\)th orbital, while those with an odd index encode the parity of orbitals up to \(j\). Formally, this encoding is provided by the recursively defined matrix.
%
%\begin{align}
%    %q_p &= \left ( \sum_{j=0}^p \beta_{pj} f_j \right ) \mod 2 \\
%    \beta_1 &= [1] \\
%    \beta_{2^{k+1}} &= \begin{bmatrix}
%        \beta_{2^k} & 0 \\
%        \mathbf{A} & \beta_{2^k}\\
%    \end{bmatrix} \\
%    A_{ij} &= \begin{cases}
%        0\text{ if } i\in [1,2^k), j \in [1,2^k],\\
%        1\text{ if } i = 2^k, j \in [1,2^k],\\
%    \end{cases}
%\end{align}
%which increases in size in each iteration. 
%
As the explicit form of transformation rules for fermionic operators would require significant additional discussion, we leave them to the corresponding literature \cite{Tilly2022, Bravyi2002, Seeley2012, tranter_comparison_2018}. 

As pointed out in Ref.\ \onlinecite{Havlicek2017}, this encoding is equivalent to the classic Fenwick tree graph with $2^k$ nodes. This graph encoding can be utilized to map any number of qubits which may not be a power of 2 and is advantageous for lattice systems.
The scheme has been adapted recently to other tree encodings such as the ternary tree \cite{Jiang2020optimalfermionto}.

\subsubsection{Reducing Qubit Counts by Exploiting Symmetries}\label{Sec:symmetry_reduction}

Exploiting symmetries in the Hamiltonian allows for a reduction in the number of required qubits. For instance, the parity mapping inherently utilizes symmetries to reduce the qubit count by two compared to the Jordan--Wigner transformation. However, this primarily restricts the simulation to states with the correct electron parity.
A generalized framework for exploiting Abelian symmetries was introduced by Bravyi et al.\ \cite{bravyi_tapering_2017}, known as qubit tapering. This method identifies symmetry operators that commute with the Hamiltonian and applies a unitary transformation to isolate these symmetries on specific qubits. Once isolated, these qubits commute with the entire Hamiltonian and can be replaced by their corresponding eigenvalues (
±1). This process effectively removes the qubits from the simulation, dividing the problem into distinct symmetry sectors.

Provided the target symmetry sector is known, either by construction or by evaluating the initial state, this technique can significantly reduce the computational cost by lowering the qubit count and, consequently, circuit depth.

\subsubsection{Targeting Lower-Dimensional Hilbert Spaces} 
When non-Abelian symmetries are present, such as the SU(2) symmetry generated by the total spin operator $S^2$, conventional symmetry tapering based on Abelian symmetries may not achieve the minimal qubit count. For any fermionic system with a fixed number of electrons $N$ in $M$ spin orbitals, the number of possible electronic configurations is \( \binom{M}{N}\), or even only \(\binom{M/2}{N_\uparrow}\binom{M/2}{N_\downarrow}\) for $N_\uparrow$ electrons in the spin-up (alpha) sector and $N_\downarrow$ electrons in the spin-down (beta) sector. In contrast to the mappings discussed above, which encode the full \(2^M\)-dimensional Fock space and subsequently exploit symmetries to remove qubits, alternative mappings can be constructed directly for the physically relevant symmetry sector \cite{moll_optimizing_2016,shee_qubit-efficient_2022,gandon_quantum_2025}. By targeting this lower-dimensional Hilbert space from the outset, these encodings can achieve the minimal qubit count. However, the resulting mappings often require more complex operator representations (e.g.\ more Pauli strings) and larger classical preprocessing overheads \cite{cheng_optimal_2025}.

\subsubsection{Lattice Mappings}\label{Sec:lattice_mappings}
\begin{figure}
    \centering
    \input{2D_lattice}
    \caption{Regular square lattice with snake-like enumeration of sites.}
    \label{Fig:2D_Snake}
\end{figure}
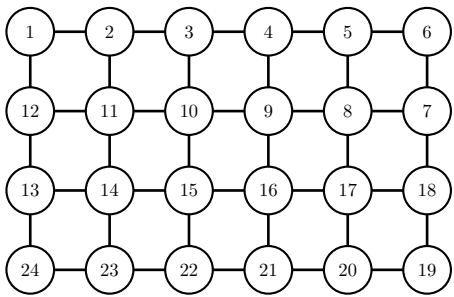
Although the fermion‑to‑qubit mappings discussed so far can be applied universally, lattice models allow us to exploit their regular geometric structure to obtain more efficient mappings. Consider a two-dimensional square lattice, which is the prototypical model-system for correlation effects in solids especially in the context of high-temperature superconductors. We enumerate orbitals of which \(n_l\) are in each line in a snake-like pattern (cf.\ Fig.\ \ref{Fig:2D_Snake}). Then the JW mapping would lead to local Pauli strings for hoppings and interactions with nearest neighbors on the same line while those for neighbors in the same column would become increasingly non-local when they are further away from the snakes bend, so that the maximal Pauli weight becomes \(\mathcal{O}(2n_l)\) where \(n_l\) is the length of each line~\cite{Verstraete2005}. This can be reduced by applying the Bravyi--Kitaev mapping to a Pauli weight of \(\mathcal{O}(\log n_l)\)~\cite{Havlicek2017}. Due to the relevance of this model, several other mappings were applied or specifically developed for this kind of system. The superfast Bravyi--Kitaev~\cite{Havlicek2017, Setia2018} and the generalized superfast encoding~\cite{Setia2019} reduce the Pauli weight compared to the BK mapping, but require additional qubits. Similarly, the auxiliary fermion scheme~\cite{Verstraete2005} as well as the compact encoding \cite{derby_compact_2021, Derby2021b} require \(2\) and \(1.5\) times the number of qubits of the physical system, but have a Pauli weight of 4 and 3, respectively, for a two-dimensional lattice. Similar mappings are subject to ongoing work with the aim to improve the balance between required qubits and locality of the Pauli string \cite{Chiew2023,Dyrenkova2026,Chiew2026}.

\begin{tcolorbox}[colback=blue!5!white, colframe=blue!75!black, title=Fermionic Encoding]
\begin{itemize}[leftmargin=*]
    \item While Jordan--Wigner mappings might be easy to understand, consider using more efficient mappings to reduce Pauli weight 
    \item Reorder/localize orbitals for a more compact representation (influence on two-qubit gates and measurement variance)
    \item Exploit symmetries (e.g.\ particle number, spin‐parity) to taper qubits or more efficient mappings
    \item For lattice models or other lower-dimensional Hilbert spaces: find balance between qubit count and Pauli weight
\end{itemize}
\end{tcolorbox}

\subsection{Choosing the Ansatz Circuits} \label{Sec:Ansatz}

After establishing a fermion‑to‑qubit mapping of the Hamiltonian, the next crucial step in a VQE calculation is the construction of a suitable ansatz circuit to explore the resulting qubit Hilbert space. This parameterized circuit is used to prepare the quantum state based on which the expectation value of the Hamiltonian is measured. In theory, the only requirement on this parameterized circuit is its \textit{expressibility}. This means, that by varying the parameters of the ansatz every quantum state of the relevant Hilbert space can be reached. A system of \(M\) spin-orbitals has a Hilbert space of dimension \(2^M\) and would require \(2^M-2\) parameters to be fully covered. The reduction by 2 is due to normalization and the arbitrarity of a global phase of the quantum state present in any quantum system. Fortunately, we have further knowledge about physical properties and symmetries of the system, by which we can restrict the Hilbert space significantly: First, if the number of electrons \(N\) is conserved, the dimension of the Hilbert space and thus the number of required complex parameters reduces to \(\binom{M}{N}\). Second, if the system is time-reversal invariant, the ground state wave function has to be real, such that the number of real parameters can be reduced by a factor of two. Third, if the system exhibits particle-hole symmetry, the dimension can be reduced by another factor of 2. Although the parameter-scaling for an expressive ansatz improved to a combinatorially smaller subspace, we still need  a large number of parameters, which can lead to the problem of barren plateaus in the optimization (cf.\ Sec.\ \ref{Sec:barren_plateaus}) and long circuits which acquire many errors on NISQ hardware (cf.\ Sec.\ \ref{Sec:Error_NISQ}). 

In practice, however, it is often possible to obtain good agreement with the exact state with significantly less parameters. Therefore, we need an ansatz circuit to cover at least the part of the Hilbert space that contains the ground state of the system, which can be regarded as minimal expressivity. Unfortunately, there is no generic recipe for constructing such a state, so physical intuition and trial and error are required. For small systems this is feasible, while it becomes difficult for large ones. This difficulty is further increased by the fact that weakly and strongly interacting regimes (e.g.\ at equilibrium vs.\ dissociation limit for molecules) have qualitatively different ground-states, so that an ansatz constructed for one setting might not work for another one. 

To circumvent these difficulties, generic circuit patterns were developed which typically consist of the preparation of an initial state and an ansatz. A reference state obtained from classical calculations like Hartree--Fock is typically chosen as initial state. The ansatz is a parameterized circuit motivated either by hardware considerations, such as limited circuit depth, gate availability, and connectivity, or by chemical and physical insights into the problem. Repeating ansatz layers can improve the expressibility, which, however, does not guarantee that the correct ground state can be reached. Such circuits are fundamentally constructed from a small set of elementary operations, most notably single-qubit gates (\texttt{X}, \texttt{Y}, \texttt{Z}, \texttt{H}, \texttt{S}) and parameterized rotations (\Rx, \Ry, \Rz), together with controlled two-qubit gates (\Cx, \Cy, \Cz) and parameterized entangling gates (\Rxx, \Ryy, \Rzz).  In the following we present initialization strategies and some prominent and recent ansatz patterns whose properties are summarized in Tab.\ \ref{Tab:ansatze}.

\begin{table*}
    \renewcommand{\arraystretch}{1.8}
    \begin{tabular}{c|c|c|c|c}
    Method & Depth & Parameters & 2Q-gates & Symmetry \\ \hline
    Hardware Efficient Ansatz & \(\mathcal{O}(L)\) & \(\mathcal{O}(NL)\) & \(\mathcal{O}((N-1)L)\) & - \\ \hline
    Efficient Symmetry Preserving & \(\mathcal{O}(2\binom{N-1}{M})\) & \(\mathcal{O}(2\binom{N}{M}-2)\) & \(\mathcal{O}(3\binom{N}{M})\) & X \\ \hline
    UCCSD & \(\mathcal{O}((N-M)^2M L)\) & \(\mathcal{O}((N-M)^2M^2 L)\) & \(\mathcal{O}((2q-1)N^4 L)\) & X \\ \hline
    Variational Hamiltonian Ansatz & \(\mathcal{O}(CL)\) & \(\mathcal{O}(CL)\) & \(\mathcal{O}(2(q-1)CL)\) & X \\ \hline
    \end{tabular}
    \caption{Summary of circuit depth, required number of parameters, number of two-qubit gates, and whether the ansatz is particle or \(S_z\) symmetry preserving. \(L\): Number of layers/repetition of the ansatz, \(N\): Number of orbitals/qubits, \(M\): Number of electrons, \(q\): average number of Pauli strings, \(C\): Number of commutative terms in Hamiltonian}\label{Tab:ansatze}
\end{table*}
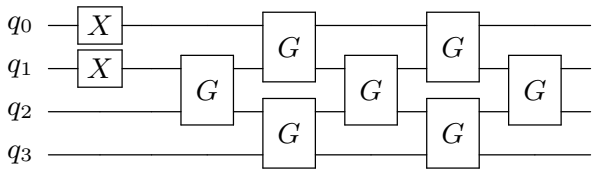
\begin{figure}
    \centering
    \input{Givens_prep}
    \caption{Schematic showing the preparation of an initial state based on Givens rotations for a system of four qubits at half filling.}
    \label{Fig:initial_state}
\end{figure}

\subsubsection{Initial State}\label{Sec:Ansatz_Init}
Initial states should be computationally easily accessible from classical pre-calculations, close to the expected exact result and simple to prepare on quantum hardware. Effective non-interacting systems where electron-electron interactions are considered only on a mean-field level, can be solved efficiently even in a self-consistent way on classical computers. Their Hamiltonian consists only of single excitations (i.e.\ terms with one creation and one annihilation operator), which is a particle-number preserving quadratic Hamiltonian. The solution of general quadratic Hamiltonians are called Fermionic Gaussian states (FGS), which is a linear combination of Slater determinants in this specific case. Efficient algorithms to prepare FGS with a maximal depth of \(O(N^2)\) (\(N\): number of qubits) were recently proposed \cite{Tubman2018, Jiang2018}. Those algorithms construct a circuit of Givens rotations, which correspond to parameterized two-qubit gates of the form
\begin{align}
    \mathcal{G}(\theta) = \begin{pmatrix}
                           1 &       0        &      0       & 0 \\
                           0 & \cos(\theta)   & \sin(\theta) & 0 \\
                           0 & \sin(\theta)   & \cos(\theta) & 0 \\
                           0 &       0        &      0       & 1
                      \end{pmatrix}.
\label{eq:Givens}
\end{align}
The Givens-rotations maintain the particle number and can be realized by a combination of a \(R_{xx}\) and a \(R_{yy}\) gate or three \(C_X\) gates in combination with single-qubit rotations.

A direct choice for a quadratic Hamiltonian is taking only the non-interacting part of the Hamiltonian. Alternatively, the explicit electron-electron interaction can be treated as a mean-field for a single electron, generating a self-consistent one-electron problem. These mean-field approaches, like e.g.\ Hartree--Fock (HF), can be solved efficiently. There are, in general, two ways to deal with the solutions of quadratic Hamiltonians: Either one prepares the corresponding solution state as Slater determinant on the quantum computer as discussed before, or one transforms the Hamiltonian to the one-particle solution state, which corresponds for HF to a transformation from the atomic basis to molecular orbitals. In this basis, the ground state is a product state, such that occupied states are initialized by \(X\)-gates on the respective qubits to the initial \(|0\rangle\) state. While this initial circuit is optimal in length, especially in solid state systems, the Hamiltonian in the atomic basis may turn out to be simpler than in the molecular basis \cite{Gratsea2024} leading to a trade-off between shorter and simpler circuits and a lower number of required measurements.
Other classical methods beyond HF, like configuration interaction single and double (CISD), coupled cluster singles and doubles (CCSD), complete active space configuration interaction (CASCI), or selective configuration interaction (SCI), provide ground states that are expressed as superpositions of Slater determinants. These states can serve as initial states for quantum algorithms and may be prepared using dedicated multi-determinant state-preparation techniques. Those methods are numerically more costly but still feasible with proper approximations. Alternatively, the density matrix renormalization group (DMRG) provides an approximate ground state in form of a matrix product state (MPS) \cite{schollwockDMRG2011,verstraeteDMRG2023,provazza_fast_2024}. There are already several approaches to create MPS states on a quantum computer \cite{Rudolph2022, iqbal2022preentanglingquantumalgorithms, Malz2024, Dov2024, jaderberg2025variationalpreparationnormalmatrix}, however, there is still a lot of ongoing research aiming for improvement. 

In order to start the optimization from this classically obtained state, the initial parameters of the ansatz circuits have to be chosen such that it corresponds to the identity map. For ansätze of UCC-  or VHA-type this is already achieved by initializing all parameters to zero. For the more complicated case of hardware-efficient ansätze an identity block initialization strategy was suggested in Ref.\ \onlinecite{Grant2019} with a simplified variation in Ref.\ \onlinecite{Gratsea2024}. However, as the classical solution is often a local minimum or saddle point in the energy landscape, some optimizers may get stuck. Starting from a completely random set of initial parameters on the other hand has a high probability to start the optimization in a barren plateau region \cite{McClean2018} where all gradients vanish (see Sec.\ \ref{Sec:barren_plateaus}). An approach to overcome these two limitations is to add a small perturbation to the identity initialization. Alternatively, for ansätze of UCCSD type, one can perform a classical UCCSD pre-optimization and use the resulting parameters as initial state.

A different approach exploits the efficient simulatability of Clifford circuits. Those circuits consist only of gates, that map the Pauli group to itself, i.e.\ unparametrized one-qubit (\texttt{X}, \texttt{Y}, \texttt{Z}, \texttt{H}, \texttt{S}) and two-qubit (\Cx, \Cy, \Cz) gates. In the pre-optimization stage, the variational parameters are restricted, such that the full ansatz remains in the class of Clifford circuits, which allows an efficient classical determination of the optimal parameter configuration  \cite{PhysRevA.111.062413}. 
\begin{tcolorbox}[colback=blue!5!white, colframe=blue!75!black, title=Choosing an Initial State]
\begin{itemize}[leftmargin=*]
\item Use physically motivated states from classical calculations with large overlap to the target state.
\item Either prepare the reference state as a quantum circuit or transform the Hamiltonian into a basis where this state is simple.
\item Examples: Hartree-Fock state in molecular orbitals; CCSD amplitudes as initial parameters for a UCCSD ansatz.
\end{itemize}
\end{tcolorbox}

\subsubsection{Hardware Efficient Ansatz}\label{sec:HEA}

\begin{figure}
    \centering
    \input{EffSU2_CX}
    \input{EffSU2_CZ}
    \caption{Schematic representation of hardware-efficient ansätze with $L=2$  and different choice of rotation and entangling layers, without (top) and with (bottom) final rotation layer.}
    \label{Fig:HEff_ansatz}
\end{figure}
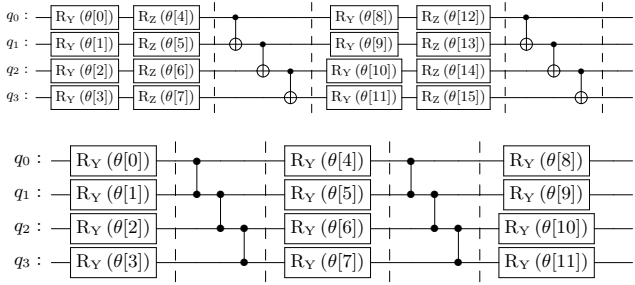

The Hardware Efficient Ansatz (HEA) is based on the idea to create short, expressible circuits by using only the native gates of the hardware and entangling schemes respecting the hardware topology, i.e.\ two-qubit gates are only placed acting on connected qubits. If only single qubit gates are parameterized, its construction follows the following steps:
\begin{enumerate}
    \item Create a rotation layer \(U^{\mathrm{rot}}(\theta_i)\) by adding parameterized single qubit rotations (\Rx, \Ry, \Rz) to every qubit (two of these three are sufficient) \label{HEA_start}
    \item Create an entangling layer \(U^{\mathrm{ent}}\) by entangling all (neighboring) qubits (nearest neighbor, all-to-all, ...) 
    \item Repeat from \ref{HEA_start} until the requested number of layers is achieved
    \item If final layer required, add another set of parameterized single qubit rotations to every qubit    
\end{enumerate}
For \(L\) layers this leads mathematically to the state
\begin{align}
    \left|\Psi(\theta)\right> = \left[\prod_{i=1}^L U^\mathrm{ent} \times U^\mathrm{rot}(\theta_i) \right] \left|\Psi_\textrm{init}\right>.
\end{align}
The construction rules provided above can be adapted to the specific available gates, for example if parameterized two-qubit gates are available on the hardware. Two examples based on layers of parameterized single-qubit \Ry- and \Rz-rotations and a nearest-neighbor entangling scheme is shown in Fig.\ \ref{Fig:HEff_ansatz} which is adapted to a hardware which has \Cx- or \Cz, \(\texttt{S}_\texttt{X}\)- and parameterized \Rz-gates available on the hardware. 

While good results can be obtained with the HEA even on NISQ devices due to its relatively short circuits, its large parameter space poses a problem for optimization due to local minima and barren-plateaus. 
It does not conserve any symmetries of the system so that the initialization to a specific particle number or spin easily gets lost. In order to start from a previously prepared initial state like HF, an initialization strategy as presented in Ref.\ \onlinecite{Grant2019} or \onlinecite{Gratsea2024} should be used, turning the circuit into an identity map. However, for the optimization starting from an initial state with limited entanglement (i.e. satisfying an area law), which is the case for unentangled states like HF, Leone et al.\ point out \cite{Leone2024} that the HEA can always be simulated classically. One can, therefore, not expect a quantum advantage with such a combination of initial state and ansatz, however, a classically simulable approximation to the ground state can be obtained.

\subsubsection{Efficient Symmetry Preserving Ansatz} \label{Sec:symmetry_preserving_ansatz}
\begin{figure}
    \centering
    \input{Eff_Sym_N2}
    \input{Eff_Sym_N1}
    \input{A_gate}
    \caption{Schematic representation of symmetry preserving hardware-efficient ansätze. Top: two electrons in four spin-orbitals, consisting of a state-preparation, an initial entangling and one additional layer which preserves the SU(2) spin-symmetry. Bottom left: one electron in four spin-orbitals with an initial layer and only one entangling ladder including time-reversal symmetry. Buttom right: One possible realization of the A-gate.}
    \label{Fig:EffSym_ansatz}
\end{figure}
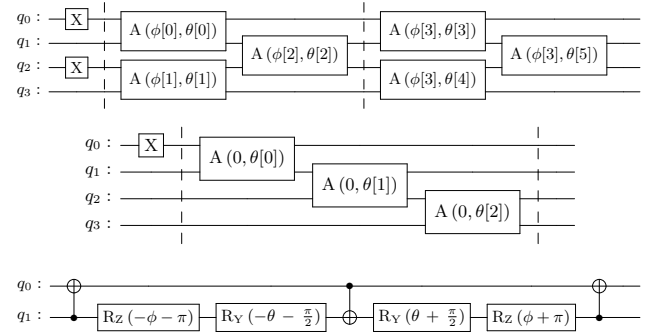
In order to maintain the symmetries of the system, contrary to the HEA, one can construct an ansatz based on two-particle gates which inherently conserve the particle number. These exchange-type gates preserve the states \(\left|00\right>\) and \(\left|11\right>\), while they allow an exchange between and thus a superposition of the states \(\left|01\right>\) and \(\left|10\right>\)~\cite{Roth2017,Barkoutsos2018,Egger2019}. Such gates can be realized in different ways, for example as Givens rotations as introduced in Sec.\ \ref{Sec:Ansatz_Init}. An alternative, generalized realization is given by
\begin{align}
    A(\theta, \phi) = \begin{pmatrix}
                           1 &       0        &      0                 & 0 \\
                           0 & \cos(\theta)   & e^{i\phi} \sin(\theta) & 0 \\
                           0 & e^{-i\phi} \sin(\theta) & -\cos(\theta) & 0 \\
                           0 &       0        &      0                 & 1
                      \end{pmatrix}.
\end{align}
Note that $A(\theta, \phi)$ reduces to the Givens rotation in the case of time-reversal symmetry, which leads to \(\phi=0\) and thus half the number of free parameters (cf.\ Eq.\ \ref{eq:Givens}). This gate can be realized on hardware by three alternating \Cx-gates intertwined by a rotation gate \(\texttt{R}_\texttt{Z}(\phi+\pi)\texttt{R}_\texttt{Z}(\theta+\nicefrac{\pi}{2})\) and its adjoint.

One constructive and general approach to create this kind of ansatz is provided by Gard et al.\ \cite{Gard2020}: %\onlinecite{Gard2020}:
\begin{enumerate}
    \item Map the fermionic orbitals to qubits such that 
    \mbox{(i) the qubits} are split to one set for spin-up and one set for spin-down orbitals,
        and (ii) occupied and empty orbitals alternate within each set.

    \label{SCA}
%    \begin{itemize}
%        \item the qubits are split to one set for spin-up and one set for spin-down orbitals,
%        \item occupied and empty orbitals alternate within each set.
%    \end{itemize} 
    \item Based on this mapping, initialize a previously, classically found state (e.g.\ the HF ground state)
    \item Initial layers: Append an A-Gate to all qubit pairs of which (i) one is empty and the other is occupied, (ii) an A-Gate has previously been applied to only on one of them
    \item Full layers: Append layers of A-Gates on all qubits, start alternating from first and second qubit to create a brick wall pattern
\end{enumerate}
Figure \ref{Fig:EffSym_ansatz} shows two realizations of this pattern. In general, this pattern conserves only the particle number. However, employing the mapping specified in item \ref{SCA} above enables a preservation of \(S_z\) by setting the angles of the entangling gates between the up- and down-qubit sets to zero. In addition,  Gard et al.\ \cite{Gard2020} further present a more involved construction rule which preserves the total spin \(S^2\).

Due to the preservation of symmetries, this ansatz requires less parameters than the HEA, but needs two- to three times more two-qubit gates per layer due to the gate decomposition if a native implementation of exchange-type gates is not available. 

\subsubsection{Unitary Coupled Cluster Ansätze}\label{Sec:UCC}

The coupled cluster (CC) theory is a well established method in quantum chemistry to calculate correlated electronic systems beyond Hartree--Fock. The basic idea is the expansion of a reference wavefunction \(|\Psi_0\rangle\) by \(e^{T}|\Psi_0\rangle\) to include all correlation effects. 
The cluster operator $T$ is given by  
\begin{align}
    T &= \sum_n T_n \qquad \textrm{with} \\
    T_1 &= \sum_{i,j} \theta^j_i a^\dagger_j a_i,  \\
    T_2 &= \sum_{i,j,k,l} \theta^{k,l}_{i,j} a^\dagger_k a^\dagger_l a_i a_j, \\
    \dots\nonumber
\end{align}
and contains all possible n-particle excitations $T_n$. However, due to the strong growth of possible combinations, only a small number of excitations can be considered in practice. Typically, only single (\(T_1\)) and double excitations (\(T_2\)) are considered, which is referred to by the name Coupled Cluster Singles and Doubles (CCSD). In the basic setting only excitations from occupied orbitals  to empty orbitals are considered, where \(a_i\) describes the annihilation and \(a^\dagger_i\) the creation of an electron in orbital \(i\) and $\theta$ is the transition amplitude.

The CC operator \(e^{T}\) is not unitary, so that the energy obtained is not variational and it can not be implemented on QCs. However, the evolution operator \(e^{T-T^\dagger}\) is unitary because \(T-T^\dagger\) is anti-hermitian. The approach 
\begin{align}
    \left|\Psi(\theta)\right> = e^{T(\theta)-T^\dagger(\theta)} \left|\Psi_0\right>
\end{align}
is therefore denoted as Unitary Coupled Cluster (UCC) ansatz, which was first used in the seminal VQE paper \cite{Peruzzo2014} with the transition rates \(\theta\) as variational parameters. The creation of a UCCSD ansatz based on these ideas follows the procedure:   
\begin{enumerate}
    \item Determine an initial state $\left|\Psi_0\right>$ by a classical calculation, e.g. the HF ground state. Build a circuit to initialize the qubit register accordingly.
    \item Generate the excitation operators
    \begin{enumerate}
        \item Choose a set of single excitations \(\mathcal{S}\) and/or double excitations \(\mathcal{D}\) which define the respective operators \(\tilde{T}_1\) and \(\tilde{T}_2\)
        \item Define the hermitian operator \(T_{\textrm{H}}=\tilde{T}_1+\tilde{T}_2-\tilde{T}_1^\dagger-\tilde{T}_2^\dagger\)
        \item Transform \(T_{\textrm{H}}\) to a qubit operator \(T_\textrm{Q}\) by the same transformation as the Hamiltonian
        \item Create the evolution operator \(T_\textrm{e} = e^{T_\textrm{Q}-T^\dagger_\textrm{Q}}\) 
        \item Apply the Suzuki-Trotter expansion
        \begin{align}
            e^{A+B} = \lim_{n\rightarrow \infty} \left(e^{\frac{A}{n}} e^{\frac{B}{n}}\right)^n, \label{Eq:SuzukiTrotter}
        \end{align}
            in first order on \(T_\textrm{e}\)  
        \item Transform each factor \(e^{aO}\), in which \(O\) is a product of qubit operators, to a circuit
        \item Create one layer by combining all operator circuits 
    \end{enumerate}
    \item Append the layer to the initial circuit until the requested number of layers is reached.
\end{enumerate}
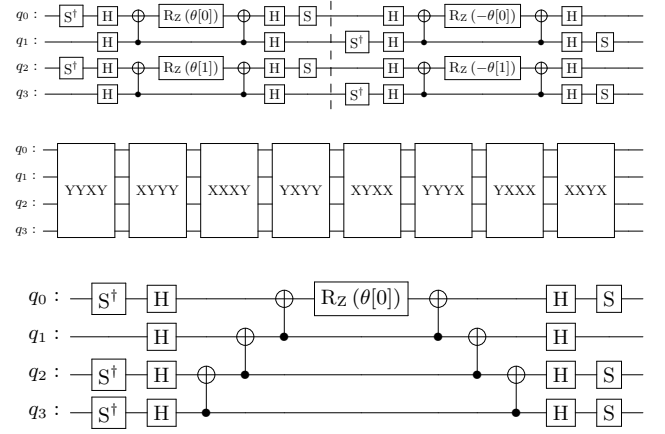
\begin{figure}
\centering
    \input{UCC_s}
    \input{UCC_d}
    \input{UCC_dd}
    \caption{Schematic representation of the circuit representing a single (top) and double (center) excitation contributing to a UCCSD ansatz for two orbitals. The single excitations correspond to \(a^\dagger_0 a_1 - a^\dagger_1 a_0 \rightarrow XY+YX\) for the spin-up part and correspondingly for the spin-down part. The Jordan--Wigner encoding maps the double excitation \(a^\dagger_0 a^\dagger_2 a_1 a_3\) operator to then the eight contributions shown in the center. The contributions written there are exponentiated when realized as gates. The decomposition of the first contribution \(e^{iYYXY}\) in a set of native gates is displayed in the bottom panel, where the \texttt{H} and \texttt{S} gates correspond to the basis transformations between \(Z\) and \(X\) and \(X\) and \(Y\) basis, respectively.}
    \label{Fig:UCC_ansatz}
\end{figure}
We note three aspects for this approach: First, the qubit operators can be grouped by commuting terms in order to reduce the circuit lengths, see Sec.\ \ref{Sec:MeasExpVal}. Second, the operators in general do not form a closed Lie-algebra, as operators resulting from mutual commutation among the pool members are missing \cite{Izmaylov2020, Grimsley2020}. This causes the ordering of the operators to be not arbitrary. Grimsley points out in Ref.\ \onlinecite{Grimsley2020} that having double before single excitations improves the accuracy. Evangelista introduced in Ref.\ \onlinecite{Evangelista2019} a construction scheme to represent any state by a disentangled UCC ansatz (DUCC\footnote{The double UCC is typically also abbreviated by DUCC}), which successively removes all determinants and therefore requires an excitation pool beyond singles and doubles. Alternatively, Ref.\ \onlinecite{Izmaylov2020} proposes to heal the ordering problem by creating a (symmetric) Lie sub-algebra through the addition of missing operators to the pool, which at the same time increases the ansatz length as the Lie algebra grows exponentially in the worst case. Third, the qubit operators are often highly non-local due to the \(Z\)-operators originating from the fermion-to-qubit transformation. When translated to a quantum circuit, this leads to ladders of two-qubit gates with a single qubit rotation on top. Thus, UCC ansätze are typically quite long and have a high two-qubit gate count. 

The Qubit CC \cite{Ryabinkin2018} approach was developed to overcome this non-locality. Rather than using the full linear combinations of Pauli strings that arise from mapping fermionic excitation operators to qubits, QCC treats each Pauli string individually, forming a pool of entanglers. This not only reduces the number of entangling gates but also, owing to the involutory property of Pauli strings (i.e.\ \(P^2 = I\)), causes the Baker–Campbell–Hausdorff expansion of the similarity-transformed Hamiltonian under the circuit unitary to terminate, yielding a closed-form expression in terms of sine and cosine functions of the circuit parameters. The iterative QCC \cite{Ryabinkin2019} exploits this closed form to successively dress the Hamiltonian, thereby shortening the quantum circuit. Ref.\ \onlinecite{Lang2020} extended this idea by using mutually anti-commuting linear combinations of Pauli strings, further simplifying the Hamiltonian dressing. In both approaches, the entangler operators are selected via a pre-screening step that requires additional quantum resources prior to the actual VQE run.

The essential remaining open question for UCC ansätze is the choice of excitation operators. In general, taking more excitation operators into account improves the expressibility of the ansatz but also leads to long circuits. For the task of choosing  excitations, several different approaches have been developed during the last years and we shortly present the ideas of several of them. While the original approach only includes excitations from occupied to unoccupied orbitals, the UCC generalised singles and doubles (UCCGSD) includes all possible excitations \cite{Lee2019}. This leads to higher accuracies of the final states at the cost of significantly longer circuits. Alternatively, in the paired UCCGSD (UpCCGSD) only the single excitation operators are generalized and the two-particle excitations are restricted to those, where a pair of spin-up and spin-down electrons is mutually excited from one orbital to another \cite{Lee2019}. In order to improve the expressivity, this ansatz is repeated \(k\) times (k-UpCCGSD), which still shows high flexibility with an improved depth and parameter scaling in contrast to UCCSD. There exist several other extensions of UCC, like the orbital optimized UCC (OO-UCC), which optimize the one-body part with classical computations \cite{Sokolov2020}.

In order to select the most important excitations, while going to higher excitation orders like triple and quadruple excitations, selective rules and pre-screening methods were recently introduced. The symmetry-UCCSDT (sym-UCCSDT) performs a pre-screening in order to remove all excitations which can be excluded due to spin, orbital and point group symmetries \cite{Haidar2023}. Another simple approach is to neglect all terms whose contribution according to a classical MP2 calculations is expected to be below a chosen threshold \cite{Romero2019}. 

The unitary selective coupled cluster method with triples and quadruples (UsCCSDTQ) \cite{Fedorov2022} introduces selection rules for operators in an initial UCCSD calculation, whose optimization parameters are used for the selection of triple and quadruple excitations. In the commutativity pre-screened automated selection of scatterers (COMPASS) \cite{Mondal2023} approach, VQEs optimizing only one parameter are used to obtain a pre-screened pool of single and two-particle operators based on their potential energy reduction. Their commutators create higher excitations, which are selected based on the results of two parameter VQEs. The operators of both sets are used to create the final ansatz. The non-iterative disentangled unitary coupled cluster (NI-DUCC) \cite{Haidar2025} approach combines several considerations mentioned above. It starts with a similar VQE pre-screening as UsCCD to select the fermionic operators leading to strong correlations, adds their corresponding qubit strings based on selection rules determined by symmetries as in sym-UCCSD and uses the closure of the Lie-algebra to create the NI-DUCC wave-function ansatz. The cognitive optimization via many body perturbation theory toward ansatz construction and tailoring (COMPACT) \cite{Halder2024} method performs a perturbative expansion of the two-body interactions around the Hartree--Fock state and selects the excitations for the ansatz by their classically obtained amplitudes. Especially, the last three methods provide ansätze that lead to very high agreement with full CI solutions, while maintaining a reasonable parameter and CNOT count for small examples that can be simulated classically. However, the circuits are still too long to be implemented on current NISQ hardware.

\subsubsection{Variational Hamiltonian Ansatz}
This approach is inspired from adiabatic time evolution, which describes the evolution of an initial state to the ground state under the action of the Hamiltonian. In order to mimic this evolution it replaces time by variational parameters \(\theta_\alpha\) leading to its name variational Hamiltonian Ansatz (VHA)~\cite{Wecker2015}. Thus, the ansatz is defined by parameterized unitaries consisting of the excitations of the Hamiltonian and resembles a UCC ansatz with an operator pool defined by the Hamiltonian. The construction of the corresponding ansatz is given by the following scheme:
\begin{enumerate}
    \item Initialize a previously, classically found state $\left|\Psi_0\right>$ (e.g.\ the HF ground state)
    \item Generate the excitation operators
    \begin{enumerate}
        \item Transform the fermionic Hamiltonian operator \(H\) to its qubit version \(H_{\textrm{Q}}\)
        \item Group the terms of the Hamiltonian into commuting sets \(h_\alpha\) (cf.\ Sec.\ \ref{Sec:MeasExpVal}), s.t.
        \begin{align}
            H_{\textrm{Q}} = \sum_{\tilde{\lambda}}h_{\tilde{\lambda}} \qquad \textrm{ with } [h_\alpha, h_\beta]\neq 0.
        \end{align}
        \item Create the operator \(H_\textrm{e} = e^{\sum_{\lambda} \theta_\lambda h_\lambda}\) 
        \item Apply the Suzuki-Trotter expansion (cf.\ Eq.\ \eqref{Eq:SuzukiTrotter}) in first order on \(H_{\textrm{e}}\)
        \item Transform each factor \(e^{\theta_\lambda h_\lambda}\) to a quantum circuit 
        \item Create one layer by combining all operator circuits
    \end{enumerate}
    \item Append layers to the initial circuit until the requested number of layers is reached.
\end{enumerate}
The resulting state is given by
\begin{align}
    \left|\Psi(\theta)\right> = \prod_{i=1}^L \left(\prod_{\alpha}e^{-i\theta_{\alpha,i}h_\alpha} \right) \left|\Psi_0\right>,
\end{align}
naturally preserving the symmetries of the Hamiltonian. It was successfully used for VQE simulations of lattice models, as their Hamiltonians usually have a strong locality\cite{Li_2023}. It was shown that local minima can be avoided by an over-parameterization through layer repetition while maintaining a strong resilience to barren plateaus~\cite{Wiersema2020}. This approach was further developed to a Fourier transform VHA \cite{Babbush2018,Choquette2021} for periodic systems and a symmetry breaking VHA \cite{Vogt2020,Choquette2021}, which targets systems where the ground state might be in a symmetry broken state by adding symmetry breaking terms in the operator.

\subsubsection{ADAPT-VQE}\label{Sec:AdaptiveVQE} 

Instead of optimizing a fixed ansatz the ADAPT-VQE approach~\cite{Grimsley2019, Tang2021} aims to create compact, problem‑tailored ansatz circuits in an iterative fashion.  
In order to do so, a set~$\mathcal{S}$ of parameterized unitary operators is defined, from which the ansatz is build by iteratively selecting and appending operators to the circuit.  
In order to decide which operators to add, a metric~$M_{C_i}$, ranking the operators relative to the existing ansatz~$C_i$, needs to be chosen.

\begin{figure}
    \centering
    \includegraphics[width=\linewidth]{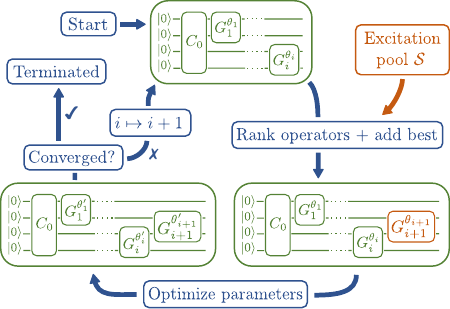}    
    \caption{Schematic of the iterative ADAPT-VQE ansatz construction. In each step, all operators of a pre-defined pool are rated, e.g.\ according to their energy-gradient, and the best one is appended to the ansatz circuit. The full set of parameters is then optimized with a short VQE optimization. This procedure is repeated until a convergence criterion is fulfilled. 
    }
    \label{fig:adapt_vqe}
\end{figure}

In more detail, after preparing an initial state (e.g., the Hartree‑Fock state) resulting in the quantum circuit $C_0$, the following iterative procedure is performed (cf. Figure~\ref{fig:adapt_vqe}):  
\begin{enumerate}
    \item For each operator $G\in \mathcal{S}$, compute the metric~$M_{C_i}(G)$.
    \item Take the operator~$G$ with the best metric and append the quantum circuit of ansatz~$C_{i}$ resulting in the new ansatz~$C_{i+1}$.
    The new ansatz now has an additional parameter~$\theta_{i+1}$.
    \item Optimize the parameters $\theta_1, \dots, \theta_{i+1}$ of the ansatz~$C_{i+1}$ in a standard VQE-run.
    \item Exit if a convergence criterion is met, e.g. the energy improvement is below some threshold.
    Otherwise continue the loop.
\end{enumerate}
The operator pool $\mathcal{S}$ usually consists of parametrized unitary operators $G=e^{i\theta g}$ for Hermitian operators $g$.
The concrete pools used are determined by the Hermitian operators $g$ and vary from more chemically motivated to more hardware‑friendly.
While the original proposal~\cite{Grimsley2019} takes individual single- and double UCCSD excitations as operators, the closely related qubit excitations (QEB‑ADAPT~\cite{Yordanov2021}) remove the $Z$'s in the Pauli strings that enforce the fermionic anti-commutativity, leading to a lower two-qubit gate count.
One can further reduce the resource requirements by combining QEB-operators as done in Ref.\ \onlinecite{Ramoa2025}.
All operators mentioned so far conserve the spin $S_z$ and the particle number of the system, but not the total spin squared $S^2$. 
In order to rectify this, recently circuits for symmetry-adapted operators have been developed~\cite{magoulas2025, magoulas2025symmetry}, albeit the operators require a large number of gates to implement.
On the side of the more hardware-inspired pools, individual Pauli‑strings may be used as the operators $g$, leading to a variant called Qubit‑ADAPT~\cite{Tang2021}.  
It was shown~\cite{Shkolnikov2023} that for an $n$-qubit system a pool consisting of $2n-2$ excitations are required to map any two arbitrary real states onto each other.
Efficiently constructing, e.g. via Lie algebra techniques, such minimal complete pools that also have good convergence properties is still an active area of research \cite{viswanathan2025, Sapova2022} with a crucial aspect being the way the symmetries of the Hamiltonian are taken into account~\cite{Shkolnikov2023}.

For the ranking metric $M_{C_i}$ two variants are used.  
Given $G\in \mathcal{S}$ both use the function $f(\theta)$ which describes the 1-D energy surface created by appending the unitary operator $G=e^{i\theta g}$ to the ansatz~$C_i$, and fixing all of the ansatz parameters except $\theta$. 
The metric employed in the original proposal~\cite{Grimsley2019} and many works since is the (absolute) value of the derivative $f'(0)$, which may be computed as the expectation value of the commutator~$\langle C_i | [H, G] | C_i \rangle$.
The second approach~\cite{Feniou2023, Jaeger2025} instead takes the minimum value along the full 1-D energy surface $f(\theta)$ as the metric~$M_{C_i}(G)$.  
In order to efficiently find this minimum one exploits the fact that~$f(\theta)$ is a trigonometric function of known type and thus can be reconstructed from evaluating $f(\theta)$ at a small number of well‑chosen parameter values $\theta$.
The function $f(\theta)$ can then be minimized on a classical device.

In order to reduce the overhead of the additional optimization runs and observable calculations compared to standard VQE multiple approaches have been proposed~\cite{Liu2021, Nykänen2023, Anastasiou2024, Fitzpatrick2024}. 
Moreover, the question of the existence of small or vanishing gradients has been investigated \cite{Grimsley2023, stadelmann2025}.

\begin{tcolorbox}[colback=blue!5!white, colframe=blue!75!black, title=Choosing Ansatz Circuits]
\begin{itemize}[leftmargin=*]
\item Balance expressivity, symmetry preservation, trainability, and circuit depth. 
\item Prefer problem-inspired and symmetry-preserving ansätze. 
\item Use shallow circuits on noisy hardware; avoid unnecessarily large parameter spaces. 
\item ADAPT-VQE and symmetry-preserving hardware-efficient ansätze are promising choices on current hardware. \end{itemize}
\end{tcolorbox}

\subsection{Transpilation: Adapting Circuits to Hardware Constraints}\label{Sec:Transpilation}

While the initial state preparation and the ansatz are mainly motivated by the physical setting, each quantum computer is restricted to its own set of native gates and may have limited connectivity. The translation of the circuit to this native gate set and the qubit layout is called transpilation. Besides, the process of adapting quantum circuits to a chosen error mitigation technique (Section \ref{Sec:Error_NISQ}) is often also considered part of the transpilation process.

\subsubsection{Native Gates}

Quantum processors are controlled by physical (e.g. microwave) pulses that implement a specific set of elementary operations, known as the \textit{native gate set}. While quantum computation theory often assumes a discrete universal gate set, such as Clifford + \texttt{T} \cite{nielsen2010quantum}, actual hardware typically supports a continuous family of single-qubit rotations alongside specific entangling gates. For example, superconducting transmon qubits natively support arbitrary virtual $Z$-rotations ($R_Z$), which are implemented via frame changes in the control electronics with effectively zero duration and high fidelity \cite{mckay_efficient_2017,vezvaee_virtual_2025}. Rotations in the $XY$-plane require actual microwave pulses. Superconducting processors typically offer fixed-angle $\texttt{X}$ and $\texttt{SX}$ ($\sqrt{\texttt{X}}$) gates, while newer systems support arbitrary-angle rotations in the $XY$-plane (often denoted $R_{XY}$), parameterized by both the rotation angle and the axis orientation within the plane \cite{hyyppa_reducing_2024}. In contrast, trapped-ion systems support arbitrary single-qubit rotations on the full Bloch sphere via laser-driven or microwave-driven transitions \cite{bruzewicz_trapped-ion_2019}. For entanglement, they leverage collective motional modes to implement all-to-all connected gates, such as the Mølmer-Sørensen (\texttt{MS}) gate natively \cite{bruzewicz_trapped-ion_2019}. 

Abstract algorithmic circuits must be translated into this hardware-specific language before execution. The first stage of transpilation involves \textit{decomposition}, where high-level logical gates are synthesized into sequences of native operations \cite{developers_qiskit_2023}. If the hardware supports continuous rotations, arbitrary single-qubit unitaries can be implemented exactly with minimal overhead. However, two-qubit gates often require basis changes if the target entangling interaction differs from the native one (e.g., decomposing a $\texttt{CX}$ into $\texttt{MS}$ and single-qubit rotations). Furthermore, multi-qubit interactions common in electronic structure ansätze, such as fermionic swap networks or multi-excitation operators, must be decomposed into pairwise native entangling gates \cite{sivarajah_tket_2020}. This decomposition inevitably increases the circuit depth and gate count, introducing a baseline overhead before any optimization is applied. 

Following decomposition, standard transpilation pipelines apply \textit{exact optimization} passes to reduce the resource cost without altering the unitary operation \cite{developers_qiskit_2023}. These \textit{lossless} techniques rely on algebraic identities and commutation relations to simplify the circuit. Common examples include the cancellation of inverse gates ($U U^\dagger = I$), the merging of consecutive rotations about the same axis ($R_Z(\alpha) R_Z(\beta) \rightarrow R_Z(\alpha+\beta)$), and the commutation of single-qubit gates through CNOTs to enable further cancellations \cite{sivarajah_tket_2020}. It is crucial to distinguish these steps from the approximate optimizations discussed below in Sec.\ \ref{Sec:circuit_compression_optimization}: the procedures described here preserve the mathematical equivalence of the original circuit exactly. Nevertheless, even with aggressive exact optimization, the depth resulting from decomposition can remain prohibitive for noisy hardware, motivating the need for hardware-aware ansatz design, such as ordering excitations in UCCSD to minimize non-local interactions \cite{Grimsley2019}.

\subsubsection{Limited Connectivity}

Modern quantum processors exhibit distinct connectivity graphs that constrain which qubit pairs can interact directly. Superconducting devices typically feature fixed, nearest-neighbor architectures, such as square lattices or the \textit{heavy-hex} topology. In contrast, trapped-ion and neutral-atom systems offer all-to-all connectivity in principle, as qubits can be coupled via shared motional modes or optical tweezers. However, even in these systems, interaction strength may decay with distance, or shuttling operations may introduce overhead, effectively creating a weighted connectivity graph \cite{bruzewicz_trapped-ion_2019}.

When a logical circuit requires a two-qubit gate between non-adjacent physical qubits, the transpiler must insert \texttt{SWAP} gates to route quantum information across the hardware graph. Since a logical \texttt{SWAP} typically decomposes into three native entangling gates (e.g.\ \texttt{CX}), each routing step significantly inflates circuit depth. 

The accumulation of \texttt{SWAP} gates directly impacts algorithm fidelity. In the NISQ regime, gate errors and decoherence accumulate exponentially with circuit depth \cite{preskill_quantum_2018}. To mitigate this, specific strategies for quantum chemistry have been developed, such as fermionic SWAP networks \cite{kivlichan_quantum_2018}. Unlike standard \texttt{SWAP}s, fermionic SWAPs exchange both the qubit state and the orbital index, preserving the antisymmetry of the wavefunction without introducing additional phase gates. This allows for efficient sorting of orbitals to minimize non-local interactions during Trotterization or UCCSD execution.

Finally, minimizing routing overhead requires solving the qubit mapping problem: assigning logical qubits to physical qubits to minimize the total number of \texttt{SWAP}s. This is an NP-hard optimization problem, but efficient heuristic algorithms exist, such as SABRE (Swapping Based Routing), which is implemented in frameworks like Qiskit \cite{li_tackling_2019}. While the transpiler can automatically search for an optimal layout, providing a hardware-aware initial layout (for example, ordering logical qubits to match the linear chain of a heavy-hex lattice) can significantly reduce compilation time and final circuit depth. Thus, co-designing the ansatz with the hardware connectivity graph is essential for scaling VQE to large molecular systems.

\subsubsection{Circuit Compression and Optimization}\label{Sec:circuit_compression_optimization}
The transpilation steps described until now focus on mapping the logical circuit to the physical hardware without altering the underlying unitary operation of the quantum circuit. However, on NISQ devices, an exact but deep circuit may suffer from such severe decoherence and gate infidelity that the signal-to-noise ratio becomes negligible, rendering the computation useless. In this regime, \textit{lossy} or approximate optimization strategies may become preferable. Techniques such as ignoring rotation angles below a specific threshold (treating them as identity operations) or approximating complex gate sequences with shorter, near-equivalent sequences can significantly reduce circuit depth \cite{nishio_extracting_2020}.  While these methods introduce coherent errors by deviating from the target unitary, they mitigate the stochastic noise accumulation associated with longer execution times \cite{escofet_quantum_2026}.

Since VQE relies on a classical optimizer to minimize an energy expectation value, the algorithm exhibits a degree of resilience to small systematic errors introduced during transpilation. The variational parameters can effectively absorb minor perturbations in the ansatz, allowing the optimizer to converge to a solution close to the ground state despite the approximate nature of the compiled circuit \cite{de_keijzer_optimization_2022}. Consequently, modern transpilation pipelines for chemistry applications often include optimization levels that trade strict mathematical equivalence for reduced depth, prioritizing the feasibility of sampling a meaningful signal over the exactness of the unitary corresponding to the quantum circuit \cite{sivarajah_tket_2020}.

\subsubsection{Noise Reduction and Tailoring}
During transpilation, the unavoidable presence of noise should already be taken into account by applying techniques that suppress or modify its effect. While a qubit is idle, interactions with its environment lead to decoherence and dephasing. Since these noise processes are often slowly varying, sequences of single-qubit pulses can be applied during idle periods to periodically invert the system evolution with respect to the dominant noise terms. The resulting error accumulation is partially canceled, thereby suppressing decoherence and extending coherence times. This approach is known as dynamical decoupling (DD) and numerous DD sequences have been proposed, each targeting specific noise sources\cite{Souza2012, Ezzel2023}. Consequently, DD is commonly applied as a final transpilation step, after circuit optimization and scheduling, by filling idle time intervals with suitable pulse sequences.

Another approach to tailor the effective noise channel is \textit{Pauli twirling}. In this method, randomized circuit instances are generated by surrounding gates, in particular two-qubit gates, with suitably chosen single-qubit Pauli operations that leave the logical circuit action unchanged. Averaging over these randomized implementations transforms coherent errors into a more stochastic error channel, which is often easier to model and mitigate. This approach is discussed in more detail in Sec.~\ref{Sec:Twirling}.

\begin{tcolorbox}[colback=blue!5!white, colframe=blue!75!black, title=Transpilation]
Transpilation is a hard optimization problem; results depend on the chosen strategy. 
\begin{itemize}[leftmargin=*]
    \item Optimization of pre-transpiled circuits (e.g.\ efficient Trotter ordering) and manual gate-translation can significantly reduce gate-count.
    \item Provide hardware-aware initial qubit layout avoiding worst performing qubits.
    \item Use highest suitable optimization level.
    \item Employ transpilation-time noise-reduction and tailoring techniques (dynamical decoupling, Pauli twirling).
\end{itemize}
\end{tcolorbox}

\subsection{Expectation Values}\label{Sec:MeasExpVal} 
\begin{figure*}
\includegraphics[width = \textwidth]{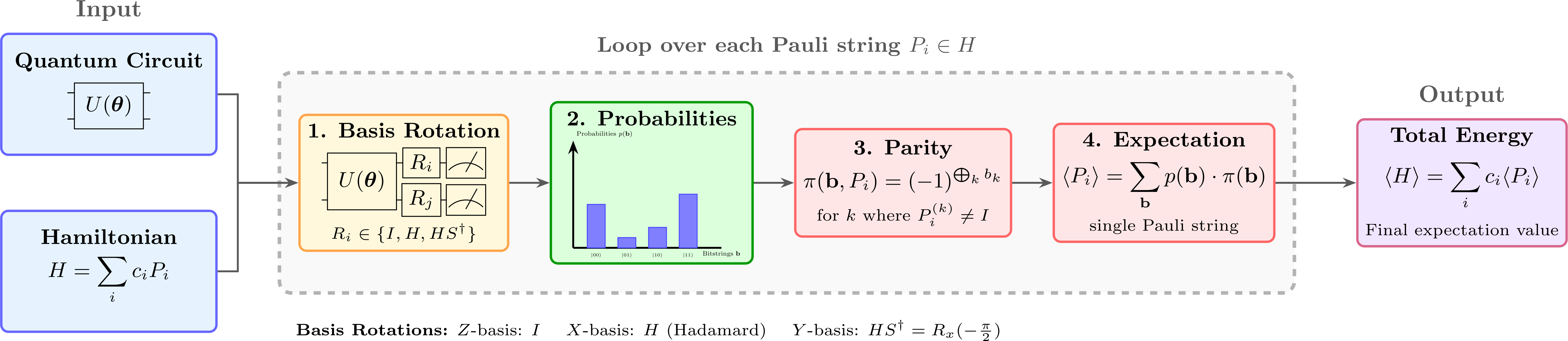}
\caption{Flowchart for computing the expectation value $\langle H \rangle = \langle \psi(\boldsymbol{\theta}) | H | \psi(\boldsymbol{\theta}) \rangle$ 
via Pauli string decomposition and measurement-based evaluation.}
\label{Fig:expectation_flowchart}
\end{figure*}
In the previous sections, we have discussed how to generate a parametrized ansatz circuit for the problem of interest and transpile it to the quantum hardware at hand. After running the circuit on the hardware and measuring the qubits, we can proceed by computing expectation values of the Hamiltonian (and other operators), which serve as cost function of the optimizer. In order to illustrate the calculation of an expectation value, let us first consider the simplest case of only one qubit where the operator is any of the spin-operators. As a quantum computer always measures the projection to the \texttt{Z} basis, the \(Z\) operator can be evaluated directly. Its expectation value then is the sum of the probability of measuring zero (\(p_{0}\)) and one (\(p_1\)) times its eigenvalue \(+1\) and \(-1\), respectively, that is
\begin{align}\label{Eq:1QEW}
\left<S_z\right> = p_0 - p_1 .
\end{align}
Due to the incompatibility of measurements in other spin-directions, a basis transformation before the measurement is required. A Hadamard gate provides the required basis rotation to the \(X\)-basis, while a combination of Hadamard and an \texttt{S}-gate is required for the rotation to the \(Y\)-basis. In their corresponding basis the eigenvalues of the operators are \(+1\) and \(-1\) for the \(\left|0\right>\) and \(\left|1\right>\) states, respectively, so that the expectation value can be calculated analogously to Eq.\ \eqref{Eq:1QEW}. The identity operator can be measured in any basis and has eigenvalues \(+1\) for both states, such that its expectation value is always \(1\).

This procedure is displayed in Fig.\ \ref{Fig:expectation_flowchart} for the full Hamiltonian mapped to qubits (cf.\ Sec.\ \ref{Sec:Mapping}). The full Hamiltonian is represented by a sum of Pauli strings \(\mathbf{P}_i\), a direct product of spin-operators each acting on one qubit, 
\begin{align}
 H = \sum_i c_i  \mathbf{P}_i = \sum_i c_i \bigotimes_{j=0}^{N-1} S_j^i %S^j_{\sigma_{i,j}}
\end{align}
weighted by \(c_i\). The spin-operator \(S^j_{i}\) acts on qubit \(j\) and refers to the spin-direction in the \(i\)th Pauli string for this qubit according to the mapping, i.e.\ \(S_i\in\{I, X,Y,Z\}\). Due to the direct product of the operators, the approach of basis rotation and calculation of expectation values introduced above has to be performed for each qubit. The expectation value for each Pauli string is then calculated by 
\begin{align}
    \left<\mathcal{P}_i\right> = \sum_{b} p_{b} \prod_{j=0}^{N-1} P(b_j, \mathbf{P}_j),
\end{align}
with the probability \(p_{b}\) of obtaining the bitstring \(b\) as returned from the backend after circuit execution and the parity function \(P(b_j, \mathbf{P}_j) \) which returns \(-1\) only if the \(j\)th bit is \(1\) and the \(j\)th operator is not the identity and \(+1\) in all other cases. 

This primitive approach requires one separate circuit for almost all Pauli-strings of the Hamiltonian, which scale as \(O(N^4)\) with the considered number of orbitals \(N\) in molecular models, and it scales with the number of edges for mappings tailored to lattice models\cite{Tilly2022} (e.g. the number of edges scales as \(O(nD)\) for a regular square lattice Hubbard model of dimension \(D\) and \(n\) sites).

\subsubsection{Shot Reduction Approaches}
As the expectation values are based on a stochastic sampling process, the standard error is \(\varepsilon = \frac{\sigma}{\sqrt{S}}\) with the standard deviation \(\sigma\) and the shot number \(S\). Thus, to achieve an error \(\varepsilon\) the total number of shots is given by
\begin{align}
 S = \frac{\sum_i c_i^2 \ \text{Var}[\mathbf{P}_i]}{\varepsilon^2}
\end{align}
for each energy expectation value estimation \cite{Wecker2015b, Rubin2018, Arrasmith2020}.

As the Pauli terms contribute with different weights, to the energy, we can reduce the number of shots by measuring those with high weights more precisely than those with a small weight instead of uniformly distributing the available shots. It was shown \cite{Rubin2018} that \(S_i\propto |c_i|\sqrt{\text{Var}[\mathbf{P}_i]}\) is optimal. As the variation of the Pauli string may not be easily accessible, choosing \(S_i\propto |c_i|\) is often already outperforming the uniform distribution~\cite{Arrasmith2020}.

The contribution of every Pauli term is further bounded by its associated weight, so that some terms may have contributions below the error tolerance threshold. Leaving out corresponding terms has shown to have a negligible impact on the accuracy of the energy estimation~\cite{McClean2014}. However, as several terms can add up to have a relevant contribution, it is suggested~\cite{McClean2016} to choose a constant \(C\in[0,1[\) and consider only the \(k\) terms with highest weight, such that \(\sum_i^k|c_i|<C\varepsilon\).

Another alternative to reduce the number of shots is to measure the probability distributions of several Pauli strings simultaneously, which is possible only if they commute, i.e.\ 
\begin{align}\label{Eq:Commutation}
    [P_i, P_j] = P_i P_j - P_j P_i = 0.
\end{align}
With these commutativity-groupings, the expectation values for several Pauli-strings are based on the same measuring result, which introduces a covariance between them, increasing the shot-noise. Therefore, while an efficient grouping significantly decreases the total number of shots, the required number of shots for each performed measurement should be increased compared to individual measurements.

The partitioning of the Pauli strings into commuting groups itself is an NP-hard problem which can lead to significant classical costs. In the following we shortly present two different approaches and recommend the interested reader to the detailed and instructive papers \onlinecite{Gokhale2019} and \onlinecite{Yen2020}. In addition to the two presented here, we want to point out the grouping of anti-commuting terms introduced in \cite{Izmaylov2019} which we will not consider here.

\subsubsection{Qubit Wise Commutativity and Tensor Product Basis}
As the Pauli strings are a tensor product of spin-operators on different qubits, one can analyse the commutativity qubit-wise, leading to the name Qubit Wise Commutativity (QWC). This means that two Pauli strings for which the operators on every qubit commute are sorted into the same group. As every one-qubit operator commutes with itself and with the identity, the necessary basis transformation of a group is determined by its Pauli-string with the highest number of non-identity operators in the same way as for the primitive approach described above. This grouping is ambiguous, at least due to the commutativity of the identity operator. The task of Pauli string grouping can be mapped to the minimum clique cover (MCC) or graph coloring problem (GCP) for which the determination of the overall best solution is an NP-hard problem. However, efficient classical heuristics have been developed and employed to find a good partitioning \cite{Gokhale2019}. \\
We remark that this approach is equivalently called Tensor Product Basis (TPB), as the measured Pauli strings are based on the tensor product of Pauli operators for each qubit without any additional entanglement. An appealing feature of this method is its easy implementation (at least for a suboptimal partitioning). A numerical analysis \cite{Gokhale2019, Yen2020} further showed a reduction of the Pauli strings to be measured by a factor of three. However, in the asymptotic limit of infinitely long Pauli strings the reduction of measured Pauli terms vanishes~\cite{Gokhale2019}.

\subsubsection{General Commutativity}
As the spin-operators form a cyclic ordering, i.e.
\begin{align}
    [S_x, S_y] &= i S_z \\ 
    [S_y, S_z] &= iS_x \\
    [S_z, S_x] &= iS_y
\end{align}
they anti-commute with each other, s.t.\ \(S_iS_j = -S_jS_i\) for \(i\neq j\). For two Pauli-strings \(\mathbf{P}_a\) and \(\mathbf{P}_b\) this leads to
\begin{align}
    \mathbf{P}_a\mathbf{P}_b &= \bigotimes_{j=0}^{N-1} S_j^a S_j^b = \bigotimes_{j=0}^{N-1}\begin{cases}
        S_j^b S_j^a \quad \textrm{ if } [S_j^a, S_j^b] = 0 \\
        -S_j^b S_j^a \; \textrm{ if } [S_j^a, S_j^b] \neq 0 
    \end{cases} \\ \nonumber
    & =(-1)^k \mathbf{P}_b\mathbf{P}_a
\end{align}
with \(k\) the number of indices where the single-qubit operators are not commuting. Thus, both Pauli strings commute, if and only if they anti-commute on an even number of indices \cite{Gokhale2019}. As the commutativity of the full Pauli string is taken into consideration, this approach is referred to as General Commutativity (GC). As in the QWC case, the partitioning of the Pauli strings into general commuting groups is equivalent to the MCC or GCP graph problems~\cite{Gokhale2019}, for which only heuristic solution algorithms are available. 

In addition, it is necessary to create the rotation to the common basis for the measurement for all Pauli strings in one group~\cite{Hamamura2020}. One example for the circuit synthesis procedure based on stabilizer matrices is provided in Ref.\ \onlinecite{Gokhale2019}, which includes the two-qubit gates \texttt{CX}, \texttt{SWAP}, \texttt{CZ} and the single qubit gates \texttt{S} and \texttt{H}. In general, this requires at maximum \(O(N^2)\) gates. If this is a significant elongation of the ansatz circuit, this may lead to a relevant increase in errors for NISQ devices. However, it was shown that the number of terms to be measured separately can be reduced by one order of magnitude to \(O(N^3)\)~\cite{Gokhale2019b}.

As QWC and GC as well as the anti-commuting grouping are mapped to graph problems, classical heuristic algorithms for their solution are required. A list of those algorithms is provided in Sec.\ 5.2.5 of Ref.\ \onlinecite{Tilly2022} and a more detailed review from a mathematical viewpoint can be found in Ref.\ \onlinecite{Gokhale2019}.

\begin{tcolorbox}[colback=blue!5!white, colframe=blue!75!black, title=Measuring Observables]
Expectation value estimation from finite-shot measurements is often the dominant runtime cost of VQE.
\begin{itemize}[leftmargin=*]
\item Reduce sampling costs by truncating small Hamiltonian terms when justified. 
\item Measure commuting Pauli strings in a common basis (use at least qubit-wise commutativity).
\item Allocate shots according to the importance, coefficient magnitude, or variance of individual measurements. 
\end{itemize}
\end{tcolorbox}

\subsection{Choosing an Optimizer}\label{Sec:Optimizer}
After measuring and computing expectation values of the ansatz circuit on the quantum computer, the remaining bit that closes the VQE loop is to classically optimize the ansatz parameters.
The problem of finding the optimal parameter set to minimize a scalar function is a well known but also difficult task in optimization theory. Several efficient numerical algorithms have been developed, which scan the cost function landscape starting from an initial parameter set by varying them, in order to find the global minimum. In the following we introduce some of the most widely used concepts. 

\subsubsection{Gradient-Based Optimizers}
Many optimizers rely on the gradient, as it directly provides information on the direction, in which a variation of the parameters $\theta$ would decrease the cost function $C(\theta)$. However, as an analytical form of the cost function and its gradient with respect to the parameters is typically unknown, the gradient can be approximated by finite differences, i.e.\
\begin{align}\label{Eq:finite_difference}
\partial_i C(\theta) \approx \frac{C(\theta + e_i \varepsilon)-C(\theta - e_i \varepsilon)}{2\varepsilon}.
\end{align}
This gradient requires two function evaluations for each parameter which is very costly for actual application. Therefore, stochastic approximation methods were developed. The finite difference stochastic approximation (FDSA) \cite{Kiefer1952} uses Eq.\ \eqref{Eq:finite_difference} to update one randomly chosen parameter in each iteration. As this becomes inefficient for a large number of parameters, the simultaneous perturbation stochastic approximation (SPSA) \cite{Durrani2003, Finck2012, Wang2018} calculates the gradient in a random direction \(\Delta\) in parameter space instead of the single parameter dimension \(e_i\). 

An alternative to these finite difference methods are methods, which exploit the analytical structure of quantum circuits as sets of unitary operations. The partial derivative of the expectation value of an observable is
\begin{align}
\frac{\partial \langle O(\theta)\rangle}{\partial \theta_j} = 2\text{Im}\left(\langle\phi_0
| (\partial_j U^\dagger(\theta)) O U(\theta) |\phi_0\rangle \right).
\label{Eq:partial_diff}
\end{align}
Since the unitary is given by a product of exponentials of parameterized Pauli-strings, the derivative can be evaluated analytically as
\begin{align}
\partial_j U(\theta) = ie^{i\theta_N P_N}\dots e^{i\theta_j P_j} P_j e^{i\theta_{j-1}P_{j-1}}\dots e^{i\theta_1P_1}.
\end{align}
As the operator in the expectation value in Eq.~\eqref{Eq:partial_diff} is non-Hermitian, the direct evaluation as described in Sec.\ \ref{Sec:MeasExpVal} can not be applied. Instead, one can either perform a direct measurement via Hadamard test\footnote{The Hadamard test\cite{Romero2019} is a quantum circuit technique that estimates the real (or imaginary) part of an expectation value $\langle \psi | U | \psi \rangle$ by using an ancilla qubit prepared in superposition via a Hadamard gate, applying a controlled-$U$ operation, and measuring the ancilla to extract the desired quantity from the resulting interference pattern.}  or reformulate the problem such that the parameter shift rule~\cite{Schuld2019} can be applied\footnote{The parameter shift rule allows the exact analytic gradient of a variational quantum circuit's expectation value with respect to a gate parameter to be computed by evaluating the circuit at two shifted parameter values (typically $\pm\pi/2$) and taking their difference, avoiding the need for finite-difference approximations.}, so that
\begin{align}
\frac{\partial \langle O(\theta)\rangle}{\partial \theta_j}  = \langle O(\theta+\frac{\pi}{2}\ e_j)\rangle - \langle O(\theta-\frac{\pi}{2}\ e_j)\rangle
\end{align}
which, again, are Hermitian expectation values. By default, classical optimizers like those from \texttt{scipy} calculate the gradients by two-point finite differences, unless a function providing a rule for its calculation is provided. 

The simple gradient descent (GD) algorithm \cite{Lemarechel2012, Courant1943} directly optimizes the parameters in the direction of the calculated gradient with a fixed learning rate. Instead of directly optimizing in the direction of the gradient, the conjugate gradient (CG) optimizer \cite{Fletcher1964} conjugates the gradient with a contribution from the previous optimization direction and chooses a step size based on a line search in this direction. This approach is further refined by the adaptive momentum optimizer (Adam) \cite{Kingma2017}, which determines the optimization direction and the learning rate from a weighted combination of the recent gradient and previous direction on the one hand and recent square of gradients and previous learning rate, on the other hand. As moving averages are used, this optimizer is also applicable in noisy surroundings, which are typical for quantum hardware.

The Broyden-Fletcher-Goldfarb-Shanno (BFGS) algorithm \cite{Broyden1970, Fletcher1970, Goldfarb1970, Shanno1970} determines the descent direction using information of the curvature. This requires the Hessian matrix \(H\) and, thus, the calculation of second derivatives. More explicitly, the descent direction is obtained as vector solving the equation \(H(\theta_k)\mathbf{d}_k = - \nabla C(\theta_k)\) and the step width is obtained by searching the minimum of the cost function along this direction. As the Hessian is difficult to calculate, different variants of BFGS approximate it like the limited-memory BFGS (L-BFGS) \cite{Byrd1995}. Similar to BFGS, the sequential least square programming (SLSQP) optimizer \cite{Kraft1988} obtains the optimization direction from the solution of a quadratic programming subproblem, which is \(\min_s(\nicefrac{1}{2} s^T H s +(\nabla f(x))^T\cdot s) \). This solver can also handle linearized constraints and bounds.

\subsubsection{Gradient-Free Optimizers}
The constrained optimization by linear approximation (COBYLA) optimizer~\cite{Powell1994} does what its name says: the cost function and constraints are approximated linearly at the actual point and this linear optimization problem is solved within a trust region. This result is the new evaluation point at which the next iteration is performed. In this process, the trust region is successively reduced, until the optimal point is found. While this optimizer can handle stochastic noise well, it tends to fail with too noisy data as obtained from NISQ devices. 

The Nelder-Mead algorithm \cite{Nelder1965} searches the optimal set of \(n\) parameters by updating the \(n+1\) edges of a simplex iteratively. Starting from a random simplex, in each iteration the vertices are first ordered according to their corresponding expectation value, then the most expensive point is replaced by the reflection, expansion or contraction point if they have lower cost, and finally the simplex is shrunk towards the best point. This procedure iteratively encircles the best solution. It was shown in Refs.\ \onlinecite{Pellow_Jarman2021, Lavrijsen2020} that this optimizer works only for low-noise environments. 

The Powells conjugate direction algorithm \cite{Powell1964} starts from an initial point and two non-parallel vectors. The algorithm performs a bidirectional line search which takes all of these search vectors into consideration. The minimum defines the new position vector while the new displacement vector is added to the search vector list, the most relevant search vector is removed. This procedure is repeated until convergence is reached.

The optimization method suggested by Vidal \cite{Vidal2018}, the Nakanishi-Fuji-Todo (NFT) optimizer \cite{Nakanishi2020} and the Rotosolve optimizer \cite{Ostaszewski2021} exploit the analytical structure of quantum circuits as a product of unitary operations. Each of those unitaries can be expressed as an exponential function of Pauli strings, which can be expanded to a sum of sine and cosine functions. Thus, if we allow variation in only a single parameter, the cost function becomes a sinusoidal function in this one, so that three measurements are sufficient to fit the parameters and find the minimum in this direction. This procedure relies on the condition that all parameterized gates have a \(2\pi\)-periodicity, which is not the case for some ansätze. However, there exists an extension \cite{Wierichs2022} which allows quantum gates with arbitrary frequencies. \\

\subsubsection{General Remarks}
Overall, the main criteria for a comparison between different optimizers are the quality of the solution and the total number of shots performed on a quantum computer until convergence is achieved. As the spectrum of the Hamiltonian might have excited states close to the ground state, which appear as local minima in the energy landscape, another quality criterion than solely the energy is required. A direct comparison of the optimal parameters as such is difficult, due to the periodicity with unknown frequency and as their exact result is unknown. However, the fidelity of the state realized by these parameters in the ansatz circuit with the ground state from a classical solution of the interacting problem solves this issue, although this approach is limited to small, classically solvable system sizes.

For the total measurement cost, it is important to note that optimizer performance depends strongly on the ansatz, problem instance, noise level, and hyperparameter choices, which directly influence the number of iterations required for convergence. In noiseless simulations, gradient-based optimizers such as CG, BFGS, and SLSQP often converge efficiently. As shot and device noise increase, however, their performance can deteriorate, while optimizers designed for noisy objective functions, such as COBYLA, SPSA, NFT, or Rotosolve, frequently become more competitive. Consequently, no single optimizer can be recommended universally, and empirical benchmarking remains an important part of VQE studies. Although several comparisons of optimization methods have been reported in the literature \cite{Lavrijsen2020}, a comprehensive benchmark spanning different ans\"atze, Hamiltonians, noise models, and hardware platforms is still lacking.

There are various approaches in order to increase the performance of the different optimizers and to reduce their computational costs. While there is a large variation of expectation values at the beginning of the optimization, it becomes smaller close to the convergence. Therefore, the required accuracy becomes more important towards the end, so that the shots per iteration can be reduced at the beginning~\cite{Kubler2020, Arrasmith2020}. A layerwise learning procedure was suggested as another approach to reduce the optimization costs, where only the parameters of one ansatz layer are optimized successively with a few iterations until convergence is reached\cite{Liu2022, Gard2020}. Especially for gradient-based optimizers this reduces the number of function evaluations significantly as the optimal solution in a smaller subspace is searched and, therefore, also helps to avoid barren plateaus.

\begin{tcolorbox}[colback=blue!5!white, colframe=blue!75!black, title=Choosing Optimizer]
\begin{itemize}[leftmargin=*]
    \item Optimizer performance can differ substantially between noiseless and noisy settings.
    \item Exploit optimizer-specific features: trust regions, adaptive learning rates, shot adaptation, and natural-gradient preconditioning (when affordable).
    \item Gradient-free optimizers are often more robust to noise, while gradient-based methods can converge faster in low-noise regimes.
    \item Monitor convergence and consider restarting from different initial parameters if optimization stagnates.
\end{itemize}
\end{tcolorbox}

\subsection{Barren Plateaus} \label{Sec:barren_plateaus}

Despite the rapid development of variational quantum algorithms in recent years, scaling the hybrid quantum-classical workflow to larger systems remains challenging. One important obstacle is the flattening of the optimization landscape with increasing system size, which can impair the performance of the optimization techniques discussed in Sec.\ \ref{Sec:Optimizer}. When the energy landscape becomes nearly flat, gradients vanish and provide little information about promising search directions, particularly in the presence of noise. This phenomenon is known as a barren plateau (BP) \cite{larocca_barren_2025} and has been identified as one of the factors limiting the successful application of VQE to larger problem instances.

%Despite the rapid development of variational quantum algorithms in recent years, the scalability of the hybrid quantum-classical workflow is fundamentally limited by the exponential flattening of the parameter optimization landscape as the problem size increases. This is a fundamental bottleneck faced by the optimization techniques in Sec.\ \ref{Sec:Optimizer} and arises when the energy landscape becomes flat, such that there is no hint for a search direction, especially in the presence of noise. This phenomenon of vanishing gradients is called barren plateau (BP) \cite{larocca_barren_2025} and is one of the reasons for the limited size of problems investigated sucessfully so far with VQE. 

\subsubsection{Origins and Analytical Bounds}

In recent years, many theoretical studies have been conducted to understand the origin of BPs, as well as the development of various heuristic techniques to mitigate or avoid BPs as much as possible. In quantum computing, the exponential growth of the Hilbert space with the problem size (the curse of dimensionality) has been signaled as the ultimate culprit for the emergence of BPs. In general, BPs can be induced by the expressiveness of the circuit, the entanglement of the input data or the locality of the observables needed to compute the loss function \cite{ragone_lie_2024}. We emphasize that BPs already emerge in the ideal case of noiseless quantum computers and do not necessarily stem from hardware noise. Incoherent noise in quantum processors further reduces the magnitude of gradients, giving rise to noise-induced barren plateaus that are conceptually different from noise-free BPs \cite{wang_noise-induced_2021}. Structural BPs are tied to the algorithmic design of the ansatz, while noise-induced BPs are dictated by the physical constraints of the quantum hardware. 

The emergence of barren plateaus in a quantum loss landscape is a statement about vanishing gradients on average, and much effort has been devoted to providing guarantees for the existence of small subspaces with substantial gradients \cite{mhiri_unifying_2025} by providing lower bounds on loss variances. A mathematical analysis of BPs has been developed based on the study of the Lie group and the associated Lie algebra generated by a particular parametrized quantum circuit to precisely compute the variance of the loss function for a known input state \cite{ragone_lie_2024}.  In fact, this formalism has been employed to provide evidence that, for certain models, the absence of BPs in the variational optimization landscape may be compatible with classical simulability, at least when some classical data can be obtained from quantum devices during an initial data-acquisition phase\cite{cerezo_does_2025}.

Fundamental limitations of variational quantum algorithms, particularly the role and inevitability of barren plateaus, remain an active and open area of research \cite{zimboras_myths_2025}. Although it is still unclear whether BPs constitute a fundamental, unavoidable bottleneck of variational optimization workflows, we discuss next (Sec.\ \ref{sec:strategies_to_avoid_BPs}) approaches demonstrating that the underlying structure of atomic and molecular Hamiltonians may be exploited to avoid or partially mitigate BPs.

\subsubsection{Strategies to Avoid or Mitigate BPs} \label{sec:strategies_to_avoid_BPs}

While barren plateaus represent a general challenge for variational quantum algorithms, their manifestation and potential mitigation within electronic structure calculations are profoundly influenced by the unique properties of molecular Hamiltonians \cite{sarkar_quantum_2026}. The Hamiltonians describing atoms and molecules are typically sums of local operators, namely one-body and two-body fermionic interaction terms \cite{mukherjee_comparative_2023}. Theoretical analysis demonstrates that the use of local observables is a key factor in mitigating the severity of BPs compared to scenarios involving global observables \cite{ragone_lie_2024}.

Beyond Hamiltonian locality, the conservation laws governing molecular systems provide another powerful avenue for avoiding BPs. Molecular Hamiltonians are invariant under certain symmetry operations, such as the conservation of total particle number (associated with $U(1)$ symmetry) and the conservation of total spin (associated with $SU(2)$ symmetry) \cite{gibbs_exploiting_2025}. Hardware-efficient ansätze (Sec.\ \ref{sec:HEA}), which are easy to implement on quantum processors, in general do not conserve these properties, lack chemical intuition and quickly suffer from BPs at modest depths due to the high number of parameters. A critical mitigation strategy is therefore to construct parametrized quantum circuits, or ansätze, that explicitly preserve these symmetries (such as the symmetry-preserving ansatz in Sec.\ \ref{Sec:symmetry_preserving_ansatz}) or the chemically motivated UCCSD-Ansatz or HVA. For these ansatz-types the quantum state remains confined to the correct symmetry sector throughout the optimization process. This restriction effectively reduces the dimensionality of the search space from an exponentially large Hilbert space to a much smaller, polynomially-sized subspace corresponding to the irreducible representation of the symmetry group \cite{cerezo_does_2025}. 
% In contrast, chemically-motivated ansätze such as the Symmetry-Preserving Ansatz (Sec.\ \ref{Sec:symmetry_preserving_ansatz}) or the Unitary Coupled Cluster with Singles and Doubles (UCCSD) ansatz (Sec.\ \ref{Sec:UCC}), for instance, is based on the coupled cluster method from classical computational chemistry and is designed to generate correlated wavefunctions. 
Nonetheless, UCCSD can still face trainability challenges depending on the molecule and basis-set \cite{boy_energy_2025}. A promising class of ansätze in this respect is the family of adaptive algorithms, most notably ADAPT-VQE (Sec.\ \ref{Sec:AdaptiveVQE}), which constructs an ansatz by iteratively adding those elements which provide the largest gradient.

This strategy provides short, efficiently optimizable circuits for electronic structure problems \cite{Grimsley2023}.
Further refinements, such as pruning redundant operators after optimization (Pruned-ADAPT-VQE) \cite{vaquero-sabater_pruned-adapt-vqe_2025} continue to push the boundaries of what is achievable in this domain.

Another powerful empirically-driven strategy is the use of more sophisticated initial states than the standard Hartree--Fock state, of which some ideas were already discussed in Sec.\ \ref{Sec:Ansatz_Init}. 
Going beyond mean-field approximations, the Generative Flow-based Warm Start (Flow-VQE) approach employs a classical generative model to learn a mapping from Hamiltonian parameters to a high-quality initial wavefunction \cite{zou_generative_2025}. This model is trained on a dataset of previously solved problems to directly generate initial states with a high overlap with the true ground state of new, unseen Hamiltonians \cite{lee_improving_2026}. By starting the VQE optimization from such a refined state, the search distance to the optimal solution is drastically reduced, placing the optimizer firmly within a region of the landscape with large gradients and improving the overall success rate of the algorithm \cite{lee_improving_2026}. 

Beyond ansatz design and initialization, improvements in the classical optimization loop itself also contribute to navigating barren landscapes. Techniques such as the Quantum Natural Gradient (QNG) and its more practical approximations aim to optimize using the geometry of the quantum state manifold, which can help navigate sharp, curved regions of the landscape more effectively than standard gradient descent methods \cite{shi_weighted_2026}. Other empirical work focuses on balancing the training process, for instance, by optimizing the contributions from different terms of the Hamiltonian to ensure stable convergence \cite{bhowmick_enhancing_2025}.

These empirically validated strategies form a practical toolkit that, when combined with theoretically sound principles, offers a promising path toward solving challenging electronic structure problems with current and near-term quantum hardware. Barren plateaus severely hinder the successful implementation of VQE algorithms, underscoring the need for a thorough analytical understanding and the development of effective mitigation strategies, which we overview in the next Section\ \ref{Sec:Error_NISQ}.

\begin{tcolorbox}[colback=blue!5!white, colframe=blue!75!black, title=Barren Plateaus]
Mitigation strategies of barren plateaus include: 
    \begin{itemize}[leftmargin=*]
        \item Adaptive or layer-wise ansätze that minimize unnecessary circuit depth. 
        \item Symmetry-preserving and problem-inspired circuits that restrict the search space.
        \item Warm-start initialization from classical approximations like multi-configuration post-Hartree--Fock methods. 
        \item Local cost functions whenever possible.
    \end{itemize} 
\end{tcolorbox}

\subsection{Error Mitigation} \label{Sec:Error_NISQ}

    The main obstacle preventing quantum algorithms with provable performance guarantees, like quantum phase estimation, to be directly executed on current devices is the noise inherently present in all realizations of quantum computers. This leads to errors during the execution of a quantum circuit and limits the size of circuits that may be executed on a quantum hardware platform.
    In order to overcome this hurdle, quantum error correction (QEC)~\cite{Terhal2015} encodes logical qubits as a larger ensemble of physical qubits, allowing for real time detection and correction of errors.
    When the physical error rate of the quantum hardware lies below the threshold of a QEC-code family, logical errors can be exponentially suppressed to arbitrarily low levels~\cite{Kitaev1997, Aharonov1997} by scaling up the number of physical qubits used for the code, allowing for the execution of arbitrary deep circuits.
    However, even though progress in QEC-experiments are made at a fast pace~\cite{Bluvstein2024, Eickbusch2025, DaSilva2024, yamamoto2025, SalesRodriguez2025, rosenfeld2025, Lacroix2025}, error correction that allows for large-scale fault-tolerant computation is not expected to be feasible in the near future.
    
    In the meantime, to make use of today's noisy quantum hardware that does not allow for deep circuits, NISQ-algorithms~\cite{Cerezo2021}, like VQE, usually work by gathering a large statistical ensemble of measurement results gained by executing a family~$\mathcal{C}$ of short quantum circuits.
    From this ensemble the algorithms output is calculated.
    With the inaccessibility of error correction, a broad array of alternative techniques for mitigating errors in the algorithms output, coming from noise during circuit execution, have been developed.
    These techniques, termed quantum error mitigation (QEM)~\cite{Cai2023}, generally are comprised of running a family of circuits~$\tilde{\mathcal{C}}$ built by adding and/or replacing the circuits in~$\mathcal{C}$ together with classical pre- and post-processing.
    Note that usually $\tilde{\mathcal{C}}$ is larger than $\mathcal{C}$.
    In contrast to QEC, this approach does not change the effective level of noise during the execution of an individual circuit, but can only mitigate the effect of noise on the algorithms output.
    
    As in many cases, like VQE, this output is an estimate of some observables expectation value with respect to a prepared quantum state. QEM-methods aim to decrease the noise-induced bias of this estimate.
    Generally, any such method leads to an increase in variance and therefore the need to increase the number of samples for estimating the expectation value.
    This sampling overhead, together with guarantees for the remaining bias, are characteristics for the error mitigation scheme giving rise to bias-variance trade-off.
    Completely mitigating the effect of noise for estimating an observable has an exponential overhead~\cite{Quek2024, Takagi2022}, which sets limits on the applicability of QEM-techniques. However, these methods have become a key ingredient for stretching the scope of applications of variational algorithms on current and near-term devices until fault-tolerant hardware platforms become available.
    Realistically, particularly in the era of early-fault-tolerant machines, QEM-methods will complement QEC~\cite{zhou2025, Luthra_2025, piveteau} and might be useful far outside the NISQ-era they were originally developed for.

    In the following we describe the QEM-methods most widely used on today's hardware, introducing the necessary vocabulary for describing quantum noise first.
    The state-of-the-art methods often revolve around the idea of gaining an effective model of the noise acting on a repeatedly executed circuit, shaping the effect of this model on the measured observable by adapting the circuits executed and then combining the collective measurement results in a post-processing step to give the mitigated result.

    \subsubsection{Quantum Noise}
    \label{Sec:QuantumNoise}
        A quantum computer, at its core, is a sufficiently stable quantum mechanical system over which we have an extremely high degree of control enabling precise manipulations like unitary operations and measurements.
        Like any quantum system it can not be considered in isolation but coupled to the environment, making it necessary to adopt the language of density matrices and quantum operations for an accurate description~\cite{nielsen2010quantum}.
        
        In addition to decoherence due to environment coupling, the control operations needed for implementing gates and measurements are not perfect and can lead to unwanted processes like crosstalk (where an operation on one pair of qubits affects other qubits) or over/under-rotations.
        On many platforms those unitary errors are further amplified by drifts in the system over time, requiring periodical recalibrations of the control pulses.
        
        In order to describe the processes mentioned above we write the computational state of the $n$-qubit system as a density matrix~$\rho$, i.e.\ a positive, trace-one matrix of dimension $2^n$, and the process of transforming the state~$\rho$ into another density matrix~$\rho'$ by a quantum channel~$\xi$, i.e.\ a linear, completely positive and trace-preserving map with $\rho' = \xi(\rho)$.
        For a pure state~$|\rho \rangle$ the corresponding density matrix is given by $\rho = |\rho \rangle \langle \rho |$ and the channel~$\mathcal{U}$ induced by a unitary operation~$U$ by $\mathcal{U} (\rho) = U\rho U^{\dagger}$.\footnote{We denote quantum operations by calligraphic letters.}
        
        It can be shown that any quantum channel~$\xi$ may be written in an operator sum representation 
        \begin{equation}
        \xi(\rho) = \sum_k E_k \rho E_k^{\dagger}
        \label{eq:kraus}
        \end{equation}
        \noindent with operators~$E_k$ such that $\sum_k E_k^\dagger E_k = I$.
        The operators~$E_k$ are known as Kraus-operators.
        Example are one-qubit amplitude damping $\mathcal{A}(\rho) \coloneqq A_0 \rho A_0^\dagger + A_1 \rho A_1^\dagger$ with 
        \begin{equation}
            A_0 \coloneqq \begin{pmatrix}
                1 & 0 \\
                0 & \sqrt{1-\gamma} \\
            \end{pmatrix}, \;\;\;\;\;\;\;\;\; A_1 \coloneqq \begin{pmatrix}
                0 & \sqrt\gamma \\
                0 & 0 \\
            \end{pmatrix}
        \end{equation}
        for $1 > \gamma > 0$ or the $n$-qubit Pauli channel
        \begin{equation}
            \mathcal{N}(\rho) \coloneqq \sum_{P\in \mathcal{P}_n} \lambda_P P\rho P^\dagger
        \end{equation}
        for the group of $n$-qubit Pauli strings~$\mathcal{P}_n$ with $\lambda_P\geq 0$ satisfying $\sum_P \lambda_P = 1$. 
        The latter may be interpreted as randomly applying an $n$-qubit Pauli operator~$P$ with probability~$\lambda_P$.
        A special case of a Pauli channel is the $n$-qubit depolarization channel 
        \begin{equation}
            \mathcal{D}(\rho) \coloneqq (1-\lambda)\rho + \frac{\lambda}{2^n}I = (1-\lambda')\rho + \frac{\lambda'}{2^n-1}\sum_{P\in \mathcal{P}_n\setminus \{I\}} P\rho P^\dagger
        \end{equation}
        with \(\lambda' \coloneqq \frac{2^n -1}{2^n}\lambda\).
        It describes the process of replacing $\rho$ by the completely mixed state with probability~$\lambda$ and leaving it untouched with probability~$1-\lambda$.
        One speaks about global depolarization noise for the $n$-qubit channel and local depolarization noise in the case of independent $1$-qubit depolarization channels on all $n$~qubits.
        
        A very helpful representation of a quantum operation~$\xi$ is given by its Pauli transfer matrix (PTM). 
        It is a matrix~$M(\xi)$ of size~$4^n$ lexicographical indexed by elements from $\mathcal{P}_n$ with entries~$M_{P}^{Q} \coloneqq \frac{1}{2^n} \text{tr}(Q\xi (P))$.
        The entry~$M_{P}^{Q}$ may be understood as the coefficient of $Q$ in the Pauli-basis decomposition of the density matrix~$\xi(P)$.
        For $n=1$ the examples from above take the form
        \begin{equation}
            M(\mathcal{A}) = \begin{pmatrix}
                1 & 0 & 0 & 0 \\
                0 & \sqrt{1-\gamma} & 0 & 0 \\
                0 & 0 & \sqrt{1-\gamma} & 0 \\
                \gamma & 0 & 0 & 1-\gamma \\
            \end{pmatrix}
        \end{equation}
        \begin{equation}
        \adjustbox{max width=\columnwidth}{$
            M(\mathcal{N}) = \begin{pmatrix}
                1 & 0 & 0 & 0 \\
                0 & \lambda_X - \lambda_Y - \lambda_Z & 0 & 0 \\
                0 & 0 & \lambda_Y - \lambda_X - \lambda_Z & 0 \\
                0 & 0 & 0 & \lambda_Z - \lambda_Y - \lambda_X \\
            \end{pmatrix}$}
            \label{eq:PTM_PauliChannel}
        \end{equation}
        and the unitary channel $\mathcal{U}$ induced by $U = e^{-i\frac{\theta}{2}Z}$ is represented by
        \begin{equation}
            M(\mathcal{U}) = \begin{pmatrix}
                  1 & 0 & 0 & 0\\
                  0 & \cos(\theta) & -\sin(\theta) & 0\\
                  0 & \sin(\theta) & \cos(\theta) & 0\\
                  0 & 0 & 0 & 1 \\
            \end{pmatrix}.
            \label{eq:PTM_UnitaryChannel}
        \end{equation}
        One can see that the Pauli channel is diagonal in the PTM-representation with the diagonal entries given by the Walsh-Hadamard transform of the Pauli coefficients $\lambda_P$, cf.\ Eq.\ \ref{eq:PTM_PauliChannel}
        In particular, any Pauli operator $Q$ is an eigenvector of any Pauli channel with the eigenvalue given by the entry $M_Q^Q$ in the PTM.
        
        From the Pauli transfer matrix representation one can easily check if a channel $\xi$ is unital, i.e.\ if $\xi(I) = I$, by checking whether there is any non-zero entry in the first column except of the $1$ on the top left.
        Moreover, the channel is incoherent (stochastic) if and only if there are no off-diagonal elements outside the first column.

        In the context of QEM one usually works with a noise model in which the quantum operation $\tilde{\mathcal{U}}\coloneqq \tilde{\mathcal{U}}_m \dots \tilde{\mathcal{U}}_1$ of the noisy circuit $U$ is decomposed into smaller pieces $\tilde{\mathcal{U}}_i$ (e.g.\ noisy gates) which may further be written $\tilde{\mathcal{U}}_i \coloneqq \mathcal{N}_i \mathcal{U}_i$ as the noiseless unitary operation $\mathcal{U}_i$ followed by noise $\mathcal{N}_i$.
        It is moreover assumed that the noise $\mathcal{N}_i$ does not change throughout repeated circuit runs.

    \subsubsection{Noise Tailoring}\label{Sec:Twirling}
        Having seen that the sources and types of errors that may occur during the execution of a quantum circuit are vast and moreover difficult to characterize \cite{Hashim2024}, it is desirable to have a method transforming generic noise into a known noise channel.
        For this reason, the technique of \textbf{Pauli twirling} via \textbf{randomized compiling} has become a standard tool in QEM\footnote{Both terms are sometimes used equivalently in the literature.}.
        
        Twirling a noise process $\xi$ on $n$ qubits over the Pauli group $\mathcal{P}_n$ is the mathematical process of taking the average 
        \begin{equation}
            T(\xi) \coloneqq \frac{1}{4^n} \sum_{Q\in \mathcal{P}_n} \mathcal{Q}\xi \mathcal{Q}
        \end{equation}
        over $\xi$ conjugated by the unitary channels $\mathcal{Q}$ induced by the Pauli elements $Q \in \mathcal{P}_n$.
        It can be shown that the Pauli transfer matrix of the twirled channel $T(\xi)$ is given by the PTM of $\xi$ with all non-diagonal entries removed, meaning that $T(\xi)$ is now a Pauli channel~\cite{Cai2019}.
        
        As an example we consider an arbitrary noise process $\xi$ on a single qubit and compute the off diagonal entry $M(T\xi))_{X}^Z$ of the twirled channels PTM $M(T(\xi))$.
        Using the commutation relation of the Pauli matrices one has
        \begin{multline}
            M(T(\xi)))_{XZ}= \frac{1}{8} (\text{tr}(X\xi Z) + \text{tr}(XXX\xi XZX) \\
            + \text{tr}(YXY\xi YZY)+ \text{tr}(ZXZ\xi ZZZ))\\
            = \frac{\text{tr}(X\xi Z)}{8}(1-1+1-1) = 0
        \end{multline}
        which generalizes to all off diagonal entries and the $n$-qubit case.
        Even though there are still $4^n$ parameters needed to characterize an $n$-qubit Pauli noise channel, all coherent and non-unital noise contributions have been removed by the twirling.

\begin{figure}[t]
            \centering
            \includegraphics[width=0.35\linewidth]{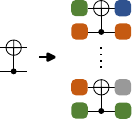}    
            \caption{Diagrammatic description of Pauli twirling via randomized compiling. The two-qubit {\Cx} gate is randomly dressed with one-qubit Pauli gates such that the overall circuit is equivalent to the original {\Cx}.}
            \label{fig:pauli_twirling}
        \end{figure}

        In order to implement the formal twirling construction in practice, an approach called randomized compiling \cite{Wallman2016} is used.
        For this, consider a noisy circuit $\tilde{\mathcal{U}} = \mathcal{N} \mathcal{U}$ which consists of a single $n$-qubit Clifford gate $U$, i.e.\ a gate that maps all $n$-qubit Pauli operators into another Pauli operator.
        In practice one usually twirls native 2-qubit gates like {\Cx}, as shown in Fig.\ \ref{fig:pauli_twirling}.
        In order to twirl the noise $\mathcal{N}$ we replace the circuit $U$ by a family of circuits $(V_P)_{P\in \mathcal{P}_n}$ in which the gate $U$ is replaced by $V_P \coloneqq PU Q$ with $Q \coloneqq U^{\dagger}PU$.
        Since $U$ is a Clifford gate, $Q$ is, by definition, also a Pauli operator and $PU Q=U$.
        Assuming that we can neglect the noise of the one-qubit Pauli rotations, which in practice usually is weak compared to the noise of multi-qubit gates, we thus end up with a circuit $V_P$ that is logical equivalent to $U$, but with the noise $\mathcal{N}$ being conjugated by $\mathcal{P}$ due to
        \begin{equation}
            \tilde{\mathcal{V}}_{P} = \mathcal{P}\mathcal{N} \mathcal{U}\mathcal{Q} = \mathcal{P}\mathcal{N}\mathcal{P} \mathcal{U}.
        \end{equation}
        
        If one now executes the circuits $(V_P)_{P\in \mathcal{P}_n}$ instead of repeatedly sampling results from $U$, the output data is as if one had sampled from the circuit $U$ with the twirled noise channel $T(\mathcal{N})$, i.e.\ {\Cx}. The effective noise channel of the circuit family $(V_P)_{P\in \mathcal{P}_n}$ is the Pauli channel $T(\mathcal{N})$.
        
        For general circuits with multiple multi-qubit Clifford gates one simply twirls each of them individually.
        One should note that in practice not all combinations of Pauli conjugations are applied to all multi-qubit gates in the circuit.
        Instead it has been experimentally observed that it suffices to select a reasonably small number of circuits chosen at random to get a good twirling result \cite{Hashim2021, Ware2021}.

        With the knowledge that the effective noise channels of the multi-qubit Clifford gates are given by Pauli channels the second step is to measure those Pauli channels.
        In order to do this, a protocol called cycle benchmarking \cite{Erhard2019, Flammia2020} has proven useful, which is insensitive to state preparation and measurement (SPAM) errors.
        
        To apply this, we consider the family of twirled circuits $(V_P)_{P\in \mathcal{P}_n}$ from above, whose effective channel is given by $\tilde{\mathcal{V}} \coloneqq T(\mathcal{N}) \mathcal{U}$. Note that if a circuit contains multiple different multi-qubit Clifford gates the noise channel of each such gate needs to be measured individually.
        To measure a single entry of \(T(\mathcal{N})\), we choose, for simplicity, a Pauli $Q\in \mathcal{P}_n$ that is an eigenvector of the unitary channel $\mathcal{U}$ with eigenvalue $1$. Repeated application of \(\tilde{\mathcal{V}}(Q)\) gives $\tilde{\mathcal{V}}^n(Q) = \lambda_Q^n Q$ where $\lambda_Q \coloneqq M(T(\mathcal{N}))_Q^Q$ is the eigenvalue of $Q$ with respect to $T(\mathcal{N})$.
        This eigenvalue can be extracted experimentally by preparing the state $Q$, applying $\mathcal{V}^n$ for varying depths $n$, and measuring in the $Q$-basis. The resulting data points then correspond to $f_n = S \lambda_Q^n$ where $S$ is a factor modeling the SPAM errors independent of $n$. A fit to \(f_n\) then yields the desired PTM entry $M(T(\mathcal{N}))_Q^Q$.

        In practice $Q$ is generally not an eigenvector of $\mathcal{U}$, making it necessary to apply correction gates and assume some symmetries in computing the PTM diagonal entries \cite{vandenBerg2023, vandenBerg2024, Chen2023}.
        More importantly, it is not feasible to measure each PTM entry of an $n$-qubit system.
        Therefore one has to assume that the Pauli channel of interest is given by some sparse model that is efficiently learnable \cite{vandenBerg2023, Flammia2020}, e.g.\ 2-local Pauli errors on neighboring qubits.

    \subsubsection{(Probabilistic) Error Cancellation}
        A natural idea for mitigating errors occurring during the execution of a quantum circuit is to first characterize the noise processes present and applying the inverse of those processes on the hardware or in post processing to get an unbiased estimator.
        Since the inverse of a noise channel is generally not physical, the idea of probabilistic error cancellation (PEC) \cite{Temme2017, Endo2018} is to implement this inverse as the probabilistic mixture of physical circuits.
        
        Following \cite{Ferracin2024, vandenBerg2023} we consider an $n$-qubit circuit $U = R_m C_m R_{m-1}\cdots R_1 C_1$ consisting of Clifford gates $C_i$ interspersed by non-Clifford one-qubit gates $R_i$ that are assumed to be noiseless.
        Using Pauli twirling via randomized compiling and cycle benchmarking with respect to some sparse Pauli model, the effective noise channel of the Clifford gates may be assumed to be known as Pauli channels $\mathcal{N}_i$ with the diagonal Pauli transfer matrix $M(\mathcal{N}_i) = \text{diag}(d^i_Q)_{Q \in \mathcal{P}_n}$.
        We denote by $\mathcal{U}_i$ the unitary channel induced by $R_i C_i$.

        The inverse $\mathcal{N}_i^{-1}(\rho) = \sum_{P_i\in \mathcal{P}_n} \lambda_{P_i} P_i\rho P_i^\dagger$ of the channel $\mathcal{N}_i$, whose coefficients 
        \begin{equation}
            \lambda_{P_i} = \frac{1}{4^n} \sum_{Q\in \mathcal{P}_n} (-1)^{\langle Q, P_i\rangle} \frac{1}{d^i_Q}
        \end{equation} 
        may be calculated via the Walsh-Hadamard transform with $\langle Q, P_i\rangle = 0$ if $Q$ and $P_i$ commute and $1$ otherwise, looks like a Pauli channel, however, the coefficients $\lambda_{P_i}$ can also be negative and therefore do not define a probability distribution.
        In order to mitigate the errors caused by $\mathcal{N}_i$ while computing the expectation value of an observable $\mathcal{O}$ with respect to the state prepared by $U$ one therefore rewrites
        \begin{equation}
        \label{Eq::InverseChannel}
            \mathcal{N}_i^{-1}(\rho) = \mu_i \sum_{P_i\in \mathcal{P}_n} \text{sgn}(\lambda_{P_i}) \frac{|\lambda_{P_i}|}{\mu_i} P_i\rho P_i^\dagger
        \end{equation}
        for $\mu_i \coloneqq \sum_{P_i\in \mathcal{P}_n} |\lambda_{P_i}| \geq 1$ and runs the family of circuits defined by applying the Pauli unitary $P_i\in \mathcal{P}_n$ after $C_i$ with probability $\frac{|\lambda_{P_i}|}{\mu_i}$.
        Doing this antinoise-insertion in parallel for all $i$, with the measured expectation value for the circuit with inserted Paulis $(P_1, \dots, P_m)\in \mathcal{P}_n^m$ denoted by $\langle \mathcal{O} \rangle_{(P_1, \dots, P_m)}$, one gets an unbiased estimator $\langle \mathcal{O} \rangle'$ for the sought after expectation value in
        \begin{equation}
        \label{Eq::MonteCarlo}
            \langle \mathcal{O} \rangle'= \mu \sum_{(P_1, \dots, P_m)\in \mathcal{P}_n^m} \text{sgn}(\lambda_{P_1}\cdots \lambda_{P_m}) \langle \mathcal{O} \rangle_{(P_1, \dots, P_m)}.
        \end{equation}
        While the bias of this estimator is zero, the sampling overhead is given by $\mu^2$ for the pre-factor $\mu \coloneqq \mu_1 \cdots \mu_m$. 
        It can be shown, that for common noise models this sampling overhead scales exponentially with the circuit depth \cite{vandenBerg2023, Takagi2022}.

        Even though probabilistic error cancellation has been applied successfully in practice \cite{vandenBerg2023}, the amount of different circuits to run and the exponential overhead is a bottleneck of the approach \cite{Kim2023}.
        As described in Refs.\ \onlinecite{Tran2023, Eddins2024} this overhead can be reduced by restricting the noise inversion to noise that actually changes the measured expectation value, i.e., the noise in the lightcone of the observable\footnote{The lightcone of an observable is the set of gates whose noise can causally influence its expectation value, analogous to the causal light cone in special relativity.}.

        Another way to reduce the overhead of error cancellation is by using an alternative way of noise inversion.
        In tensor-network error mitigation (TEM) \cite{Filippov2023}, instead of locally implementing the inverse as the probabilistic mixture of Pauli gates in the circuit, matrix product state representations of the noise channels $\mathcal{N}_i$ and unitary channels $\mathcal{U}_i$ are used to get a compact approximation of the total inverse error channel 
        \begin{equation}
            \mathcal{N}^{-1} = (\mathcal{N}_m \mathcal{U}_m \mathcal{N}_1 \mathcal{U}_1)^{-1} \mathcal{U}_1 \cdots \mathcal{U}_m
        \end{equation} 
        by contracting the tensors iteratively starting from the middle and compressing the bond dimension if needed.
        Using informationally complete Positive Operator-Valued Measurements\footnote{A positive operator-valued measurement (POVM) is a generalized quantum measurement described by positive operators that sum to the identity. Informationally complete POVMs provide sufficient information to reconstruct expectation values of arbitrary observables from a single measurement scheme.} (POVMs) optimized for the observable $\mathcal{O}$ on the noisy circuit $\tilde{\mathcal{U}} = \mathcal{N} \mathcal{U}_m \cdots \mathcal{U}_1$ allows the relevant information of the prepared state for the observable to be captured.
        The measurement results together with the duals of the POVM can then be used to calculate the mitigated value in postprocessing by applying the inverse map $\mathcal{N}^{-1}$. 
        It can be shown that for realistic noise models this classical post-processing inversion is quadratically more efficient than probabilistic error cancellation \cite{Filippov2023, Filippov2024}.
        
    \subsubsection{Zero Noise Extrapolation}
\begin{figure}
            \centering
            \includegraphics[width=0.75\linewidth]{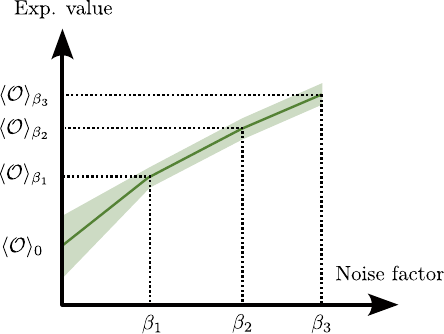}
            \caption{Schematic of the ZNE. Expectation values at different noise levels are measured and extrapolated to the zero noise level.}
            \label{fig:zne}
        \end{figure}

        An alternative \cite{Temme2017, Endo2018} to canceling the noise during circuit execution is to controllably increase the noise strength $\beta$ for different circuit runs and extrapolate the measured expectation values $\langle \mathcal{O} \rangle_{\beta_1}, \dots, \langle \mathcal{O} \rangle_{\beta_k}$ at different noise levels $\beta_1, \dots \beta_k$ to the zero noise limit $\langle \mathcal{O} \rangle_{0}$ as shown in Figure~\ref{fig:zne}.

        For this zero noise extrapolation (ZNE) either a low-order Taylor approximation \cite{Temme2017} or a (multi-) exponential ansatz \cite{Cai2021} are being used.
        While the polynomial ansatz in form of Richardson extrapolation is valid for small $\beta$ and has been analyzed in terms of bias and overhead \cite{Giurgica2020}, recent experiments \cite{Kim2023} often work with an exponential ansatz which can be shown to be accurate, e.g.\ in the case of Pauli noise in Clifford circuits. 
        Generally, for realistic noise models and circuits, ZNE has significantly less sampling overhead compared to PEC at the cost of not being fully unbiased \cite{Filippov2024}.
       
        Early approaches of ZNE used the stretching of pulses \cite{Kandala2019, Kim2023b} or the replacement \cite{He2020, Giurgica2020} of a noisy gate (e.g.\ \Cx) by a logical equivalent sequence of gates (e.g.\ three \Cx) for boosting the error rates.
        While the latter approach is valid for depolarization noise, it is not faithfully amplifying more complex noise processes like general Pauli noise.
        Hence, in most recent implementations of ZNE all effective Pauli noise channels $\mathcal{N}_i$ of twirled Clifford gates $C_i$ are learned and then amplified by the probabilistic insertion of unitary Pauli operators \cite{Ferracin2024, Kim2023}.

        In more detail, after having found $M(\mathcal{N}_i)$ we can compute $\mathcal{N}_i^\alpha$ exploiting $M(\mathcal{N}_i^\alpha) = M(\mathcal{N}_i)^\alpha$ and the Walsh-Hadamard transform yielding
        \(\mathcal{N}_i^\alpha = \sum_{P_i\in \mathcal{P}_n} \lambda_{P_i,\alpha} P_i\rho P_i^\dagger\).
        Now, if $\alpha > 0$, the $\lambda_{P_i,\alpha}$ define a probability distribution and we can implement the effective noise amplification from \(\mathcal{N}_i\) to \(\mathcal{N}_i^{(1+\alpha)}\) by randomly implementing the Pauli $P_i$ after $C_i$ according to this distribution.
        If $\alpha < 0$, as seen in PEC, the $\lambda_{P_i,\alpha}$ do not define a probability distribution and hence one has to introduce the sampling overhead factor $\mu_i$ (cf. Eq.\ \eqref{Eq::InverseChannel}) for physically implementing the channel $\mathcal{N}_i^\alpha$.
        By choosing $\alpha \in (-1, 0)$, however, this approach can be used for reducing the noise level \cite{McDonough2022} and obtaining samples $\langle \mathcal{O} \rangle_{\beta}$ for $\beta \in [0, \beta_0]$ albeit incurring overhead due to $\mu_i$.

        Apart from the schemes explained above, other variants of ZNE have been developed, to learn a realistic noise-rate from quantum circuit executions~\cite{zhang2025, Koenig}.
        
        \subsubsection{Readout Error Mitigation}
        One distinguished class of errors we have not talked about yet are errors occurring during the measurement stage of a quantum circuit.
        The sources of those readout errors are diverse and depend highly on the hardware technology used.
        However, in order to understand the mitigation techniques for this class of errors, it suffices to work with a simple error model. 
        Noise processes like coherent errors that can not be included in this model are usually considered as part of the pre-measurement circuit noise.
        
        For the model we assume to measure $n$ qubits in the computational $Z$-basis and write down a stochastic transition matrix $A$ with rows and columns indexed by the set of binary output strings of length $n$ which we denote by $[n]$.
        For two such strings $a$ and $b$ the matrix entry $M_{ab}$ describes the probability of the measurement noise channel to change the readout result from $a$ to $b$.
        If there is no noise channel $M$ is the identity matrix.
        Measuring in a different basis can be done by applying a base change unitary before the measurement (see Sec.\ \ref{Sec:MeasExpVal}).
        The matrix element $M_{ab}$ may be obtained by repeatedly preparing the quantum states $|a\rangle$, carrying out the noisy measurement and using the results for calculating the ratio of how often $b$ was obtained.
        For a single qubit this is 
        \begin{equation}
            M = \begin{pmatrix}
                M_{0 \rightarrow 0} & M_{1 \rightarrow 0} \\
                M_{0 \rightarrow 1} & M_{1 \rightarrow 1} \\
            \end{pmatrix}.
        \end{equation} 
        
        Given $M$ we consider the vector $\tilde{v}$ whose entries $\tilde{v}_a$ are the probabilities of obtaining the result $a$ when applying the noisy measurement process to a state $|\rho \rangle$.
        Denoting by $v$ the result vector a noiseless measurement would produce, one has $\tilde{v} = Mv$.
        Now, in case that $M$ is known and invertible, this allows to get mitigated results $v$ from noisy measurements via $v = M^{-1} \tilde{v}$.
        If $v$ is computed in that way, it might have negative values due to the sampling noise. Therefore, it is not a probability vector in general.
        Hence, in practice one often finds $v$ by minimizing $||Mv - \tilde{v}||^2$.

        As the cost of obtaining and handling the square matrix $M$ of dimension $2^n$ grows exponentially in $n$, working with the full matrix $M$ is only feasible for a small number of qubits.
        In order to efficiently treat larger instances one can make the assumption that the noise on different qubits are not correlated which makes $M$ the tensor product of the one-qubit transition matrix of the individual qubits.
        Another approach was proposed in Ref.\ \onlinecite{Nation2021} where it is assumed that $M$ is sparse and the non-zero entries $M_{ab}$ restricted to $a$ and $b$ that are present in the noisy measurement vector $\tilde{v}$.

        The techniques mentioned so far aim at reconstructing the noiseless probability vector $v$ from the noisy result vector $\tilde{v}$ of measuring a state $|\phi \rangle$ in the $Z$-basis.
        If we want to get the mitigated  expectation value of the Pauli observable $\mathcal{O} = \otimes_{i=1}^n Z^{u_i}$ with $u_i \in \{0, 1\}$ with respect to $|\phi \rangle$, we simply compute
        \begin{equation}
           \langle \phi | \mathcal{O}  |\phi \rangle = \sum_{a \in [n]} (-1)^{\sum_{i=1}^n u_i a_i} v_a.
        \end{equation}
        For measurements of general Pauli strings one needs to apply one-qubit rotations before the $Z$-measurements.
        
        If one is only interested in computing expectation values, it is not necessary to find the noiseless probability vector $v$.
        Instead, one can apply a technique called readout twirling \cite{VanDenBerg2022}. This gives an unbiased estimator of the expectation value by creating a probabilistic mixture of error models, which is unbiased after taking the average.
        
        In order to do so, instead of creating $\tilde{v}$ from a single circuit that is repeatedly sampled, for each shot and for each qubit $i$ we decide based on a uniform distribution if we measure the qubit as it is or apply an $X$-gate right before the measurement, and undo this change by inverting the binary result string of the measurement at position $i$.
        Denoting by $M'$ the transition matrix that describes this twirled measurement process, one may check that the expectation value with respect to the noisy probability vector $\tilde{v} = M' v$ is given as
        \begin{equation}
        \label{Eq::readout_twirling}
            \langle \phi | \mathcal{O}  |\phi \rangle' = (-1)^{\sum_{i=1}^n u_i b_i} (M' v)_b = \langle 0 | \mathcal{O}  |0 \rangle' \langle \phi | \mathcal{O}  |\phi \rangle.
        \end{equation}
        For a single qubit this is easily seen by noting
        \begin{equation}
            M' = \frac{1}{2} \begin{pmatrix}
                M_{0\rightarrow 0} & M_{1\rightarrow 0} \\
                M_{0\rightarrow 1} & M_{1\rightarrow 1} \\
            \end{pmatrix} + \frac{1}{2}
            \begin{pmatrix}
                M_{1\rightarrow 1} & M_{0\rightarrow 1} \\
                M_{1\rightarrow 0} & M_{0\rightarrow 0} \\
            \end{pmatrix}
        \end{equation} 
        and 
        \begin{multline}
            \langle \phi | Z  |\phi \rangle' = (M' v)_0 ´- (M' v)_1 =\\ \frac{1}{2} ( M_{0\rightarrow 0} - M_{1\rightarrow 0} + M_{1\rightarrow 1} - M_{0\rightarrow 1}) (v_0 - v_1).
        \end{multline}
        The mitigation strategy is therefore to compute the noisy expectation values $\langle \phi | \mathcal{O}  |\phi \rangle'$ and $\langle 0 | \mathcal{O}  |0 \rangle'$ which then may be used to get the mitigated value $\langle \phi | \mathcal{O}  |\phi \rangle$ using \eqref{Eq::readout_twirling}.
        Once $\langle 0 | \mathcal{O}  |0 \rangle'$ is computed the sampling overhead is given by $\langle 0 | \mathcal{O}  |0 \rangle'^{-2}$.

    \subsubsection{Further Error Mitigation Methods}
        Apart from the methods mentioned above, there are multiple other approaches that have been used to increase the accuracy of VQE-calculations.
        Deciding to give a short introduction to quantum noise and a more detailed treatment of the most used techniques of QEM we just briefly mention some of those other approaches and refer to the recent review \cite{Cai2023} for a detailed exposition.

        Learning based methods try to learn the relationship between exact and noisy expectation values of circuits given such training pairs and then apply the learned model to circuits for which no exact result can be classically computed.
        The classically simulatable circuits used are often Clifford- or matchgate circuits.

        The approach of virtual distillation tries to mitigate incoherent noise by distilling the mixed state $\rho$ of the system to a purer state which is then used to measure expectation values.
        The virtual distillation process usually requires multiple copies of $\rho$.

        Knowing a symmetry, i.e.\ an operator commuting with the systems Hamiltonian, of the ansatz state to be measured, can be used to filter out states that  due to noise have been evolved out of the correct symmetry sector, e.g.\ states with the wrong number of particles.
        The filtering can be done using measurement on ancillary qubits or through the measurement of additional Pauli observables together with post-processing. %\cite{Bonet-Monroig_2018}.

        Finally, an approach called subspace expansion may be used to increase the accuracy of a VQE calculation by considering a subspace generated by the application of expansion operators to the noisy prepared state $\rho$, measuring the Hamiltonian matrix elements and overlaps between the subspace states and then solving a generalized eigenvalue problem in order to get the groundstate energy within the subspace. %\cite{McClean2017}.

\begin{tcolorbox}[colback=blue!5!white, colframe=blue!75!black, title=Error Mitigation]
\begin{itemize}[leftmargin=*]
    \item Exploit zero- and low cost methods (e.g.\ Pauli twirling).
    \item Further use (combination of) methods that fit the circuit structure, hardware and runtime constraints, taking into account the bias/variance trade-off.
    \item Exploit problem-specific knowledge, e.g. symmetries, conservation laws or classical reference data, of the problem at hand.
\end{itemize}
\end{tcolorbox}

%% file: 2D_lattice.tex
\begin{tikzpicture}[
        scale=0.7,  % Scale down to fit column
        transform shape,
		site/.style={
			circle,
			draw=black,
			fill=white,
			line width=0.8pt,
			minimum size=9mm,
			inner sep=0pt,
			font=\normalsize
		},
		bond/.style={
			line width=1pt,
			black
		}
		]
		% ===== PARAMETERS =====
		\def\Nrows{4}
		\def\Ncols{6}
		\def\spacing{1.5}  % Distance between nodes
		
		% Calculate max indices
		\pgfmathtruncatemacro{\maxrow}{\Nrows-1}
		\pgfmathtruncatemacro{\maxcol}{\Ncols-1}
		\pgfmathtruncatemacro{\maxcolm}{\Ncols-2}
		\pgfmathtruncatemacro{\maxrowm}{\Nrows-2}
		
		% ===== DRAW EDGES FIRST (behind nodes) =====
		% Horizontal bonds
		\foreach \row in {0,...,\maxrow} {
			\foreach \col in {0,...,\maxcolm} {
				\draw[bond] (\col*\spacing, -\row*\spacing) -- (\col*\spacing+\spacing, -\row*\spacing);
			}
		}
		% Vertical bonds
		\foreach \row in {0,...,\maxrowm} {
			\foreach \col in {0,...,\maxcol} {
				\draw[bond] (\col*\spacing, -\row*\spacing) -- (\col*\spacing, -\row*\spacing-\spacing);
			}
		}
		
		% ===== DRAW NODES ON TOP =====
		\foreach \row in {0,...,\maxrow} {
			\foreach \col in {0,...,\maxcol} {
				% Snake enumeration: even rows L→R, odd rows R→L
				\pgfmathtruncatemacro{\nodenum}{%
					mod(\row,2)==0 ? 
					(\row*\Ncols + \col + 1) : 
					(\row*\Ncols + \Ncols - \col)%
				}
				\node[site] at (\col*\spacing, -\row*\spacing) {\nodenum};
			}
		}
	\end{tikzpicture}

%% file: Givens_prep.tex
\resizebox{\columnwidth}{!}{
\Qcircuit @C=1.0em @R=0.2em @!R {
\nghost{{q}_{0} :  } & \lstick{q_0} & \gate{X} & \qw & \qw              & \multigate{1}{G} & \qw              & \multigate{1}{G} & \qw              & \qw \\
\nghost{{q}_{1} :  } & \lstick{q_1} & \gate{X} & \qw & \multigate{1}{G} & \ghost{G}        & \multigate{1}{G} & \ghost{G}        & \multigate{1}{G} & \qw \\
\nghost{{q}_{2} :  } & \lstick{q_2} & \qw      & \qw & \ghost{G}        & \multigate{1}{G} & \ghost{G}        & \multigate{1}{G} & \ghost{G}        & \qw \\
\nghost{{q}_{3} :  } & \lstick{q_3} & \qw      & \qw & \qw              & \ghost{G}        & \qw              & \ghost{G}        & \qw              & \qw \\
}}

%% file: EffSU2_CX.tex
\resizebox{\columnwidth}{!}{
\Qcircuit @C=1.0em @R=0.2em @!R { \\
	 	\nghost{{q}_{0} :  } & \lstick{{q}_{0} :  } & \gate{\mathrm{R_Y}\,(\mathrm{{\ensuremath{\theta}}[0]})} & \gate{\mathrm{R_Z}\,(\mathrm{{\ensuremath{\theta}}[4]})} \barrier[0em]{3} & \qw & \ctrl{1} & \qw & \qw \barrier[0em]{3} & \qw & \gate{\mathrm{R_Y}\,(\mathrm{{\ensuremath{\theta}}[8]})} & \gate{\mathrm{R_Z}\,(\mathrm{{\ensuremath{\theta}}[12]})} \barrier[0em]{3} & \qw & \ctrl{1} & \qw & \qw \barrier[0em]{3} & \qw & \qw & \qw\\
	 	\nghost{{q}_{1} :  } & \lstick{{q}_{1} :  } & \gate{\mathrm{R_Y}\,(\mathrm{{\ensuremath{\theta}}[1]})} & \gate{\mathrm{R_Z}\,(\mathrm{{\ensuremath{\theta}}[5]})} & \qw & \targ & \ctrl{1} & \qw & \qw & \gate{\mathrm{R_Y}\,(\mathrm{{\ensuremath{\theta}}[9]})} & \gate{\mathrm{R_Z}\,(\mathrm{{\ensuremath{\theta}}[13]})} & \qw & \targ & \ctrl{1} & \qw & \qw & \qw & \qw\\
	 	\nghost{{q}_{2} :  } & \lstick{{q}_{2} :  } & \gate{\mathrm{R_Y}\,(\mathrm{{\ensuremath{\theta}}[2]})} & \gate{\mathrm{R_Z}\,(\mathrm{{\ensuremath{\theta}}[6]})} & \qw & \qw & \targ & \ctrl{1} & \qw & \gate{\mathrm{R_Y}\,(\mathrm{{\ensuremath{\theta}}[10]})} & \gate{\mathrm{R_Z}\,(\mathrm{{\ensuremath{\theta}}[14]})} & \qw & \qw & \targ & \ctrl{1} & \qw & \qw & \qw\\
	 	\nghost{{q}_{3} :  } & \lstick{{q}_{3} :  } & \gate{\mathrm{R_Y}\,(\mathrm{{\ensuremath{\theta}}[3]})} & \gate{\mathrm{R_Z}\,(\mathrm{{\ensuremath{\theta}}[7]})} & \qw & \qw & \qw & \targ & \qw & \gate{\mathrm{R_Y}\,(\mathrm{{\ensuremath{\theta}}[11]})} & \gate{\mathrm{R_Z}\,(\mathrm{{\ensuremath{\theta}}[15]})} & \qw & \qw & \qw & \targ & \qw & \qw & \qw\\
\\ }}

%% file: EffSU2_CZ.tex
\resizebox{\columnwidth}{!}{
\Qcircuit @C=1.0em @R=0.2em @!R { \\
	 	\nghost{{q}_{0} :  } & \lstick{{q}_{0} :  } & \gate{\mathrm{R_Y}\,(\mathrm{{\ensuremath{\theta}}[0]})} \barrier[0em]{3} & \qw & \ctrl{1} & \qw & \qw \barrier[0em]{3} & \qw & \gate{\mathrm{R_Y}\,(\mathrm{{\ensuremath{\theta}}[4]})} \barrier[0em]{3} & \qw & \ctrl{1} & \qw & \qw \barrier[0em]{3} & \qw & \gate{\mathrm{R_Y}\,(\mathrm{{\ensuremath{\theta}}[8]})} & \qw & \qw\\
	 	\nghost{{q}_{1} :  } & \lstick{{q}_{1} :  } & \gate{\mathrm{R_Y}\,(\mathrm{{\ensuremath{\theta}}[1]})} & \qw & \control\qw & \ctrl{1} & \qw & \qw & \gate{\mathrm{R_Y}\,(\mathrm{{\ensuremath{\theta}}[5]})} & \qw & \control\qw & \ctrl{1} & \qw & \qw & \gate{\mathrm{R_Y}\,(\mathrm{{\ensuremath{\theta}}[9]})} & \qw & \qw\\
	 	\nghost{{q}_{2} :  } & \lstick{{q}_{2} :  } & \gate{\mathrm{R_Y}\,(\mathrm{{\ensuremath{\theta}}[2]})} & \qw & \qw & \control\qw & \ctrl{1} & \qw & \gate{\mathrm{R_Y}\,(\mathrm{{\ensuremath{\theta}}[6]})} & \qw & \qw & \control\qw & \ctrl{1} & \qw & \gate{\mathrm{R_Y}\,(\mathrm{{\ensuremath{\theta}}[10]})} & \qw & \qw\\
	 	\nghost{{q}_{3} :  } & \lstick{{q}_{3} :  } & \gate{\mathrm{R_Y}\,(\mathrm{{\ensuremath{\theta}}[3]})} & \qw & \qw & \qw & \control\qw & \qw & \gate{\mathrm{R_Y}\,(\mathrm{{\ensuremath{\theta}}[7]})} & \qw & \qw & \qw & \control\qw & \qw & \gate{\mathrm{R_Y}\,(\mathrm{{\ensuremath{\theta}}[11]})} & \qw & \qw\\
\\ }}

%% file: Eff_Sym_N2.tex
\resizebox{\columnwidth}{!}{
\Qcircuit @C=1.0em @R=0.2em @!R { \\
	 	\nghost{{q}_{0} :  } & \lstick{{q}_{0} :  } & \gate{\mathrm{X}} \barrier[0em]{3} & \qw & \multigate{1}{\mathrm{A}\,(\mathrm{\phi[0],\theta[0]})}_<<<{} & \qw \barrier[0em]{3} & \qw & \multigate{1}{\mathrm{A}\,(\mathrm{\phi[3],\theta[3]})}_<<<{} & \qw & \qw & \qw\\
	 	\nghost{{q}_{1} :  } & \lstick{{q}_{1} :  } & \qw & \qw & \ghost{\mathrm{A}\,(\mathrm{\phi[0],\theta[0]})}_<<<{} & \multigate{1}{\mathrm{A}\,(\mathrm{\phi[2],\theta[2]})}_<<<{} & \qw & \ghost{\mathrm{A}\,(\mathrm{\phi[3],\theta[3]})}_<<<{} & \multigate{1}{\mathrm{A}\,(\mathrm{\phi[3],\theta[5]})}_<<<{} & \qw & \qw\\
	 	\nghost{{q}_{2} :  } & \lstick{{q}_{2} :  } & \gate{\mathrm{X}} & \qw & \multigate{1}{\mathrm{A}\,(\mathrm{\phi[1],\theta[1]})}_<<<{} & \ghost{\mathrm{A}\,(\mathrm{\phi[2],\theta[2]})}_<<<{} & \qw & \multigate{1}{\mathrm{A}\,(\mathrm{\phi[3],\theta[4]})}_<<<{} & \ghost{\mathrm{A}\,(\mathrm{\phi[3],\theta[5]})}_<<<{} & \qw & \qw\\
	 	\nghost{{q}_{3} :  } & \lstick{{q}_{3} :  } & \qw & \qw & \ghost{\mathrm{A}\,(\mathrm{\phi[1],\theta[1]})}_<<<{} & \qw & \qw & \ghost{\mathrm{A}\,(\mathrm{\phi[3],\theta[4]})}_<<<{} & \qw & \qw & \qw\\
\\ }}

%% file: Eff_Sym_N1.tex
\resizebox{0.8\columnwidth}{!}{
\Qcircuit @C=1.0em @R=0.2em @!R { \\
	 	\nghost{{q}_{0} :  } & \lstick{{q}_{0} :  } & \gate{\mathrm{X}} \barrier[0em]{3} & \qw & \multigate{1}{\mathrm{A}\,(\mathrm{0,\theta[0]})}_<<<{} & \qw & \qw \barrier[0em]{3} & \qw & \qw & \qw\\
	 	\nghost{{q}_{1} :  } & \lstick{{q}_{1} :  } & \qw & \qw & \ghost{\mathrm{A}\,(\mathrm{0,\theta[0]})}_<<<{} & \multigate{1}{\mathrm{A}\,(\mathrm{0,\theta[1]})}_<<<{} & \qw & \qw & \qw & \qw\\
	 	\nghost{{q}_{2} :  } & \lstick{{q}_{2} :  } & \qw & \qw & \qw & \ghost{\mathrm{A}\,(\mathrm{0,\theta[1]})}_<<<{} & \multigate{1}{\mathrm{A}\,(\mathrm{0,\theta[2]})}_<<<{} & \qw & \qw & \qw\\
	 	\nghost{{q}_{3} :  } & \lstick{{q}_{3} :  } & \qw & \qw & \qw & \qw & \ghost{\mathrm{A}\,(\mathrm{0,\theta[2]})}_<<<{} & \qw & \qw & \qw\\
\\ }}

%% file: A_gate.tex
\resizebox{\columnwidth}{!}{
\Qcircuit @C=1.0em @R=0.2em @!R { \\
	 	\nghost{{q}_{0} :  } & \lstick{{q}_{0} :  } & \targ & \qw & \qw & \ctrl{1} & \qw & \qw & \targ & \qw & \qw\\
	 	\nghost{{q}_{1} :  } & \lstick{{q}_{1} :  } & \ctrl{-1} & \gate{\mathrm{R_Z}\,(\mathrm{-\phi-\pi})} & \gate{\mathrm{R_Y}\,(\mathrm{-\theta\,-\,\frac{\pi}{2}})} & \targ & \gate{\mathrm{R_Y}\,(\mathrm{\theta\,+\,\frac{\pi}{2}})} & \gate{\mathrm{R_Z}\,(\mathrm{\phi+\pi})} & \ctrl{-1} & \qw & \qw\\
\\ }}

%% file: UCC_s.tex
\resizebox{\columnwidth}{!}{
\Qcircuit @C=1.0em @R=0.2em @!R { \\
	 	\nghost{{q}_{0} :  } & \lstick{{q}_{0} :  } & \gate{\mathrm{S^\dagger}} & \gate{\mathrm{H}} & \targ & \gate{\mathrm{R_Z}\,(\mathrm{\theta[0]})} & \targ & \gate{\mathrm{H}} & \gate{\mathrm{S}} \barrier[0em]{3} &\qw & \qw & \gate{\mathrm{H}} & \targ & \gate{\mathrm{R_Z}\,(\mathrm{-\theta[0]})} & \targ & \gate{\mathrm{H}} & \qw & \qw & \qw\\
	 	\nghost{{q}_{1} :  } & \lstick{{q}_{1} :  } & \qw & \gate{\mathrm{H}}  & \ctrl{-1} & \qw & \ctrl{-1} & \gate{\mathrm{H}} & \qw & \qw &  \gate{\mathrm{S^\dagger}} & \gate{\mathrm{H}} & \ctrl{-1} & \qw & \ctrl{-1} & \gate{\mathrm{H}} & \gate{\mathrm{S}} & \qw & \qw\\
	 	\nghost{{q}_{2} :  } & \lstick{{q}_{2} :  } & \gate{\mathrm{S^\dagger}} & \gate{\mathrm{H}} & \targ & \gate{\mathrm{R_Z}\,(\mathrm{\theta[1]})} & \targ & \gate{\mathrm{H}} & \gate{\mathrm{S}} & \qw & \qw & \gate{\mathrm{H}} & \targ & \gate{\mathrm{R_Z}\,(\mathrm{-\theta[1]})} & \targ & \gate{\mathrm{H}} & \qw & \qw & \qw\\
	 	\nghost{{q}_{3} :  } & \lstick{{q}_{3} :  } & \qw & \gate{\mathrm{H}}  & \ctrl{-1} & \qw & \ctrl{-1} & \gate{\mathrm{H}} & \qw & \qw &  \gate{\mathrm{S^\dagger}} & \gate{\mathrm{H}} & \ctrl{-1} & \qw & \ctrl{-1} & \gate{\mathrm{H}} & \gate{\mathrm{S}} & \qw & \qw\\
\\ }}

%% file: UCC_d.tex
\resizebox{\columnwidth}{!}{
\Qcircuit @C=1.0em @R=1.0em @!R { \\
	 	\nghost{{q}_{0} :  } & \lstick{{q}_{0} :  } & \multigate{3}{\mathrm{ YYXY}} & \multigate{3}{\mathrm{ XYYY}} & \multigate{3}{\mathrm{ XXXY}} & \multigate{3}{\mathrm{ YXYY}} & \multigate{3}{\mathrm{ XYXX}} & \multigate{3}{\mathrm{ YYYX}} & \multigate{3}{\mathrm{ YXXX}} & \multigate{3}{\mathrm{ XXYX}} & \qw & \qw\\
	 	\nghost{{q}_{1} :  } & \lstick{{q}_{1} :  } & \ghost{\mathrm{ YYXY}} & \ghost{\mathrm{ XYYY}} & \ghost{\mathrm{ XXXY}} & \ghost{\mathrm{ YXYY}} & \ghost{\mathrm{ XYXX}} & \ghost{\mathrm{ YYYX}} & \ghost{\mathrm{ YXXX}} & \ghost{\mathrm{ XXYX}} & \qw & \qw\\
	 	\nghost{{q}_{2} :  } & \lstick{{q}_{2} :  } & \ghost{\mathrm{ YYXY}} & \ghost{\mathrm{ XYYY}} & \ghost{\mathrm{ XXXY}} & \ghost{\mathrm{ YXYY}} & \ghost{\mathrm{ XYXX}} & \ghost{\mathrm{ YYYX}} & \ghost{\mathrm{ YXXX}} & \ghost{\mathrm{ XXYX}} & \qw & \qw\\
	 	\nghost{{q}_{3} :  } & \lstick{{q}_{3} :  } & \ghost{\mathrm{ YYXY}} & \ghost{\mathrm{ XYYY}} & \ghost{\mathrm{ XXXY}} & \ghost{\mathrm{ YXYY}} & \ghost{\mathrm{ XYXX}} & \ghost{\mathrm{ YYYX}} & \ghost{\mathrm{ YXXX}} & \ghost{\mathrm{ XXYX}} & \qw & \qw\\
\\ }}

%% file: UCC_dd.tex
\resizebox{\columnwidth}{!}{
\Qcircuit @C=1.0em @R=0.2em @!R { \\
	 	\nghost{{q}_{0} :  } & \lstick{{q}_{0} :  } & \gate{\mathrm{S^\dagger}} & \gate{\mathrm{H}} & \qw & \qw & \targ & \gate{\mathrm{R_Z}\,(\mathrm{\theta[0]})} & \targ & \qw & \qw &\gate{\mathrm{H}} & \gate{\mathrm{S}} &\qw\\
	 	\nghost{{q}_{1} :  } & \lstick{{q}_{1} :  } & \qw & \gate{\mathrm{H}} & \qw & \targ & \ctrl{-1} & \qw & \ctrl{-1} & \targ & \qw & \gate{\mathrm{H}} & \qw & \qw\\
	 	\nghost{{q}_{2} :  } & \lstick{{q}_{2} :  } & \gate{\mathrm{S^\dagger}} & \gate{\mathrm{H}} & \targ & \ctrl{-1} & \qw & \qw & \qw & \ctrl{-1} & \targ & \gate{\mathrm{H}} & \gate{\mathrm{S}} & \qw\\
	 	\nghost{{q}_{3} :  } & \lstick{{q}_{3} :  } & \gate{\mathrm{S^\dagger}} & \gate{\mathrm{H}} & \ctrl{-1} & \qw & \qw & \qw & \qw & \qw & \ctrl{-1} & \gate{\mathrm{H}} & \gate{\mathrm{S}} & \qw\\
\\ }}

%% file: simulating.tex
\section{Performing Quantum Simulations} \label{Sec:simulating_qcircuits}

As quantum computing is a rapidly evolving field, there are numerous software tools to prepare the necessary circuits, group observables into a batch of measurement protocols, and implement error mitigation methods. A compact description of the VQE workflow introduced in Fig.\ \ref{Fig:schematic} is shown step by step in the following pseudocode snippet.

\begin{algorithm}[H]
\caption{VQE Hybrid Workflow}
\begin{algorithmic}  % <-- no [1], so no line numbers
  \Require Electronic or model Hamiltonian $H$, basis/model definition, threshold $\varepsilon$
  \Ensure Estimate ground state energy $E_0$, parameters $\boldsymbol{\theta}^\star$

  \State \textbf{Model construction:} define the relevant active/correlated subspace and construct the effective Hamiltonian $H_{\mathrm{eff}}$.
   \State \textbf{Reference state:} prepare a suitable initial state $\ket{\Psi_{\mathrm{ref}}}$ (e.g.\ HF state, product state, or classically approximated ground state).
  \State \textbf{Ansatz:} choose parametrized $U(\boldsymbol{\theta})$ with
         $\ket{\Psi(\boldsymbol{\theta})} = U(\boldsymbol{\theta})\ket{\Psi_{\mathrm{HF}}}$.
  \State \textbf{Optimizer:} choose classical optimizer $\mathcal{O}$.
  \State \textbf{Observables:} map $H_{\mathcal{A}} = \sum_j c_j P_j$ and group commuting $\{P_j\}$.

  \State Initialize parameters $\boldsymbol{\theta}^{(0)}$; set $t \gets 0$.
  \Repeat
    \State Build circuits for $\ket{\Psi(\boldsymbol{\theta}^{(t)})}$ and each observable group.
    \State Transpile with hardware constraints and error suppression techniques.
    \State Execute circuits on quantum hardware (or emulator), collect measurement outcomes.
    \State Post-process measurement outcomes with error mitigation techniques.
    \State Estimate $\langle P_j \rangle_{\boldsymbol{\theta}^{(t)}}$ and compute
           $E(\boldsymbol{\theta}^{(t)}) = \sum_j c_j \langle P_j \rangle_{\boldsymbol{\theta}^{(t)}}$.
    \State Update parameters
           $\boldsymbol{\theta}^{(t+1)} \gets
            \mathcal{O}(\boldsymbol{\theta}^{(t)}, E(\boldsymbol{\theta}^{(t)}))$.
    \State $t \gets t + 1$
  \Until{$|E(\boldsymbol{\theta}^{(t)}) - E(\boldsymbol{\theta}^{(t-1)})| < \varepsilon$}

  \State \Return $E_0 = E(\boldsymbol{\theta}^{(t)})$, $\boldsymbol{\theta}^\star = \boldsymbol{\theta}^{(t)}$.
\end{algorithmic}
\end{algorithm}

In Sec.\ \ref{sec:building_fermHam}, we introduce the initial stage of building the fermionic Hamiltonian for the target system. 
In Sec.\ \ref{sec:sims_on_QCs}, we describe the general procedure of submitting circuits to a hardware provider when performing VQE simulations on real quantum computers. The large number of required shots makes simulations on real quantum computers challenging in the current time of limited accessibility. Therefore, simulators that mimic the execution of quantum circuits on QPUs are an important tool for the development and investigation of quantum algorithms like VQE, as discussed in Sec.\ \ref{Sec:numerically_exact_sims}.

\subsection{Building the Fermionic Hamiltonian}\label{sec:building_fermHam}

The first step in a VQE calculation is the construction and parametrization of the fermionic Hamiltonian (see Sec.\ \ref{Sec:Hamiltonian}) in Eq.~\eqref{eq:H_second_q}. In quantum chemistry, this is usually performed using a quantum chemistry package, while in condensed-matter physics the Hamiltonian may originate from a correlated-subspace construction or an effective model. This step provides both the Hamiltonian whose expectation value is evaluated and, in many cases, a suitable trial state for the VQE optimization loop.
%The first step in VQE calculations of electronic structure is the parametrization of the fermionic Hamiltonian (see Sec.\ \ref{Sec:Hamiltonian}) in Eq.~\eqref{eq:H_second_q} using a quantum chemistry package. This will provide on the one hand the trial state for the VQE optimization loop and, on the other hand, it will determine the required measurements to estimate the energy.

A variety of classical electronic structure codes can be used for this purpose, with PySCF \cite{sun_pyscf_2018,sun_recent_2020} being a popular choice.
PySCF computes molecular orbitals, occupations, and electron integrals using a mean field approximation, and effectively translating complex chemical systems into forms suitable for quantum simulation. By interfacing with quantum computing frameworks like Qiskit \cite{developers_qiskit_2023} and OpenFermion \cite{mcclean_openfermion_2019}, quantum chemistry drivers like PySCF integrate classical computational chemistry results into quantum algorithms.

\subsection{Calculations on Quantum Computers} \label{sec:sims_on_QCs}

In most cases, the submission of quantum circuits to hardware providers takes place via a job management system analogous to those found in high-performance computing (HPC) clusters. These systems handle queue management, resource allocation, compilation, and execution of quantum jobs.  The optimization of parametrized quantum circuits in variational algorithms like VQE involves evaluating a large number of circuit instances. Due to the iterative nature of these algorithms, it is strategically advantageous to bundle all independent measurements into a single batch at each optimization step to minimize queue wait times and maximize hardware utilization.

To address the challenge of long delays between submission and execution, major hardware providers offer both fair-sharing queue systems and priority session systems.  Some providers may offer dedicated sessions during which the user's jobs are prioritized in the queue, significantly reducing wait times. This prioritization becomes particularly critical when circuits have been pre-optimized according to the calibration data of a specific quantum processor at a particular point in time. Quantum hardware characteristics such as gate fidelities and coherence rates can drift significantly over hours or even minutes, making the temporal proximity between calibration-based optimization and circuit execution crucial for obtaining meaningful results.

\subsection{Emulators of Quantum Computers} \label{Sec:numerically_exact_sims}

Classical emulation of quantum circuits is indispensable for the development and validation of VQE workflows. Emulators serve two primary purposes: algorithmic verification and hardware benchmarking. First, in a noiseless environment, simulators allow researchers to verify that the circuit logic, ansatz structure, and classical optimizer function as intended, isolating algorithmic errors from hardware noise. Second, noisy simulators provide a critical reference point for interpreting experimental results. By comparing hardware outputs against simulations incorporating calibrated noise models, one can quantify the impact of decoherence and gate errors. Furthermore, for system sizes beyond exact simulation limits, approximate tensor-network methods offer a classical baseline to assess the quality of quantum results.

Numerically-exact simulators perform linear algebra calculations to evolve the state of qubits in the register in an exact manner, including both unitary operations and errors introduced by noise channels.
In Section \ref{Sec:QuantumNoise} we discussed how quantum computers are subject to various types of noise. These include errors from state preparation and measurement (SPAM), coherent noise due to imperfections in the realization of parametrized quantum gates, and, finally, sources of incoherent noise like amplitude damping, pure dephasing or depolarizing noise. These non-unitary processes can be probed and characterized experimentally on the quantum hardware via a set of standardized measurements that are later taken into account in simulations by employing the formalism of quantum operations (Eq.\ \eqref{eq:kraus}) and noisy channels introduced in Section \ref{Sec:QuantumNoise}. The resulting model of the quantum computer can then serve as an input to a classical simulator that emulates the execution of arbitrary quantum circuits on that particular hardware. In the absence of noise, the evolution is entirely governed by unitary processes that maintain the purity of the quantum state in the register ($\Tr \rho^2 =1$). In that case, a statevector picture can be employed and the density matrix formalism is no longer necessary. 

As the Hilbert space size increases exponentially with the number of qubits, this approach is feasible only for small systems. In order to perform classical simulations of quantum circuits involving a large number of qubits, tensor-network techniques can be employed to represent the quantum state in a memory-efficient manner \cite{provazza_fast_2024}. The maximum amount of entanglement encoded in the tensor representation is determined by the so-called "bond dimension" parameter. This allows for a rigorous control of the error introduced in the simulation by the compressed tensor representation because arbitrary accuracy can always be achieved by increasing the bond dimension.  

Emulators of quantum computers can reproduce the shot-based, stochastic nature of measurements in quantum computers. The outcome of each circuit evaluation is a single bit-string consisting of ``0''s and ``1''s, obtained by measuring the qubits in the computational basis. Thus, each quantum circuit is executed a given number of times \(N_\textrm{shots}\), to recover in a statistical manner the underlying probability distribution corresponding to the exact evolution of the density matrix. These probability distributions are then used to compute the expectation values of the desired observables (see. Sec.\ \ref{Sec:MeasExpVal}). For a number of shot with \(N_\textrm{shots}\), the resulting precision in the measurement is given by \(\epsilon\propto\nicefrac{1}{\sqrt{N_\textrm{shots}}}\). 

%% file: exampleO2.tex
\section{Example: The Oxygen Molecule}\label{Sec:O2}

Dioxygen ($\mathrm{O_2}$, or simply oxygen) has long served as a benchmark for electronic-structure theory. An early theoretical treatment of the molecule was presented in 1929 by Lennard-Jones \cite{lennard-jones_electronic_1929} and the molecule remains a core member of modern benchmark sets such as W4-17 \cite{karton_w417_2017}. In this example we demonstrate a bond dissociation simulation of the oxygen molecule using VQE. We present idealized noiseless simulations with all-to-all connectivity, representing an upper bound on achievable accuracy for current quantum hardware. We particularly emphasize important pre- and post-processing steps often neglected in other literature. We show how our noiseless VQE calculations compare to the CASCI method. Furthermore, we demonstrate that an N-Electron Valence State Perturbation Theory (NEVPT2) calculation built on the CASCI or VQE reference yields a dissociation energy in agreement with high-accuracy experimental data.

\subsection{Choosing an Appropriate Basis Set}
The ground state of molecular oxygen is weakly bound and exhibits numerous low-lying excited states \cite{bauschlicher_full_1987, partridge_theoretical_1991}. We require a relatively large basis set to correctly distinguish the nearly degenerate states. The weak binding energy furthermore necessitates diffuse functions in the basis set. Therefore we use the aug-cc-pVTZ basis set.

\subsection{Construction of the Active Space}
Unlike other abundant molecules, oxygen has a triplet ground state. The near-degenerate highest occupied and low-lying excited states give rise to significant static correlation \cite{zen_properties_2014}. Special care has to be taken to choose an appropriate set of active molecular orbitals. Our active space selection scheme largely follows the autoCAS \cite{stein_automated_2016} and Active Space Finder \cite{shirazi_performance_2025} software packages. We use PySCF \cite{sun_pyscf_2018, sun_recent_2020} to perform the classical simulations.

First, we start by calculating CCSD(T) from an unrestricted HF (UHF) reference. We use the stability analysis to ensure we have converged UHF to a true minimum. UHF allows us to already include some static correlation which is complemented by CCSD(T). We continue by calculating the spin-traced natural orbitals and occupation numbers from the unrelaxed one-electron reduced density matrix (1-RDM) of the CCSD calculation.

To select a first active space, we freeze orbitals with a natural occupation larger than 1.99 using the HF-embedding. Any orbital with a natural occupation smaller than 0.01 is excluded from the active space and becomes a virtual orbital.

Using this first reduced active space, we calculate CASCI to obtain the one-orbital entropies. We can do this because the largest active space after the truncation has only 11 spatial orbitals. Larger active spaces can be solved approximately, e.g.\ with DMRG, to obtain the entropies.

We collect orbitals with one-orbital entropy larger than 0.07 for all bond distances examined. To obtain a smooth dissociation curve, we include all orbital indices found for any bond distance. This results in a final active space of (8,8), i.e.\ eight electrons in eight spatial orbitals.

Because we want to target a triplet state, we assume that five of the eight electrons are in the alpha-sector, i.e., they have spin-up, and three electrons are in the beta-sector, i.e., they have spin-down.

\subsection{The Ansatz}

We perform the VQE calculation using the UsCCSDTQ Ansatz \cite{Fedorov2022} and a Hartree-Fock initial state.

\paragraph*{Mapping and Symmetry Reduction}

Using Qiskit \cite{developers_qiskit_2023}, we first map the active space Hamiltonian using the Jordan-Wigner mapping and then apply a symmetry reduction as described in Sec.~\ref{Sec:symmetry_reduction} \cite{Romero2019}. We are able to reduce the system size by two to four qubits for the different bond distances investigated. This means, in some cases we are able to reduce to the optimal count of 12 qubits, while in some cases we require one or two additional qubits. We use the Hartree-Fock state to identify the target symmetry sector. The fermionic excitation operators are mapped in the same way.

\paragraph*{Selection of excitation operators and optimization}

In UsCCSDTQ excitation operators are selected iteratively. We perform a first screening of single and double excitations based on the magnitude of the elements of the corresponding electronic integrals. 

Then for each iteration, the magnitudes of the optimized single and double excitation parameters and the elements of the integrals are used to select additional excitations. Each iteration only includes excitations up to a certain cutoff, which shrinks progressively. We use the values \numlist{0.04; 0.02; 0.01; 0.005; 0.002; 0.001; 5e-4; 2e-4; 1e-4; 1e-5}. 

To simulate the circuit, we apply the Suzuki-Trotter approximation in second order for the excitation operators. Since all Pauli strings for an individual fermionic excitation do commute (see Appendix of Ref.\ \onlinecite{Romero2019}) the exponential of the sum of Pauli strings takes a product form $\exp(\sum_j c_jP_j) = \prod_j\exp(c_jP_j)$. For each iteration, we apply double excitations first, then single, then triple and then quadruple excitations. We order new excitations within each group starting with the largest coefficients. Using the QuEST simulator \cite{jones_quest_2019}, the circuit is executed and the expectation value of the Hamiltonian is evaluated. We use the classically efficient algorithm \cite{jones_efficient_2020} to evaluate the gradient and perform the optimization with the scipy implementation of SLSQP.

\subsection{Results and Discussion}

\paragraph*{VQE Energy Quality}

\begin{figure}[!htb]
    \centering
    \includegraphics[width=\linewidth]{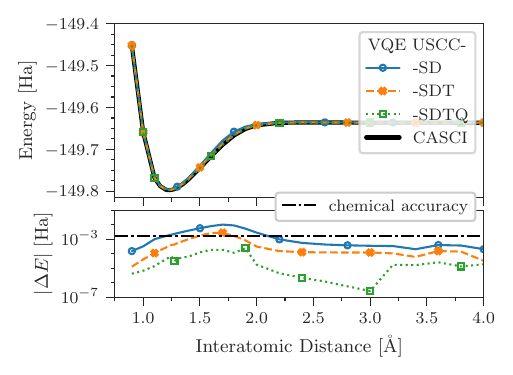}
    \caption{
        VQE Ansatz Convergence: Energy convergence with increasing excitation 
        levels (USCC-SD, SDT, SDTQ) relative to CASCI reference. The error panel 
        shows the energy difference $|\Delta E| = |E_{\text{VQE}} - E_{\text{CASCI}}|$ as a 
        function of bond distance. The dashed horizontal line indicates the 
        threshold for chemical accuracy (\SI{1.6e-3}{Ha}).
    }
    \label{fig:o2_ansatz}
\end{figure}

Figure~\ref{fig:o2_ansatz} illustrates the convergence of the VQE energy with respect to the excitation level of the unitary selected coupled cluster (USCC) ansatz. Moving from USCC-SD to USCC-SDTQ improves the description of the wavefunction, bringing the VQE results closer to the CASCI reference. The energy differences between VQE and CASCI at each bond distance are shown in the error panel of Fig.~\ref{fig:o2_ansatz}, demonstrating that USCC-SDTQ achieves convergence within the chemical accuracy threshold.

\paragraph*{NEVPT2 Correction Performance}

\begin{figure}[!htb]
    \centering
    \includegraphics[width=\linewidth]{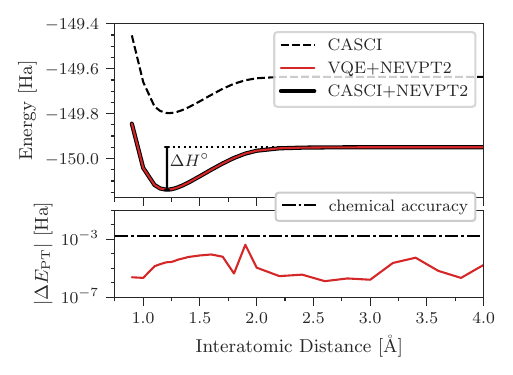}
    \caption{
        VQE+NEVPT2 vs CASCI+NEVPT2: Comparison of energy profiles and absolute 
        errors. The error panel shows the energy difference 
        $|\Delta E_\text{PT}| = |E_{\text{VQE+NEVPT2}} - E_{\text{CASCI+NEVPT2}}|$ as a function of bond 
        distance. The dashed horizontal line indicates the threshold for 
        chemical accuracy (\SI{1.6e-3}{Ha}).
    }
    \label{fig:o2_nevpt2}
\end{figure}

Figure~\ref{fig:o2_nevpt2} presents the potential energy curves as a function of interatomic distance from \SI{1.0}{\angstrom} to \SI{4.0}{\angstrom}. The NEVPT2 correction applied on top of the VQE reference state effectively recovers the dynamic correlation, minimizing the dependence on the specific ansatz truncation for the final dissociation energy.

The energy difference between VQE+NEVPT2 and CASCI+NEVPT2 across the dissociation curve is shown in the error panel of Fig.~\ref{fig:o2_nevpt2}. This comparison demonstrates that NEVPT2 performs equally well whether built on a quantum or classical reference wavefunction, with differences remaining well below the chemical accuracy threshold.

\paragraph*{Comparison with Experiment}

The experimental dissociation energy is taken as \SI{0.188035636}{Ha}, sourced from the Active Thermochemical Tables (ATcT) database (version 1.122) \cite{ruscic_active_2015}. Table~\ref{tab:errors} summarizes the absolute errors in the dissociation energy relative to this experimental reference.

Both VQE+NEVPT2 and CASCI+NEVPT2 achieve errors well within the regime of chemical accuracy (\SI{1}{kcal/mol} \cite{helgaker_molecular_2014}). 

In contrast, the single-reference CCSD(T) method exhibits a significantly larger error, highlighting the importance of multi-reference methods for systems with strong static correlation \cite{cao_quantum_2019, raghavachari_fifth-order_1989}.

\begin{table}[H]
    \centering
    \caption{
        Absolute errors in dissociation energy for $\mathrm{O_2}$ relative to 
        the experimental value from ATcT (version 1.122, \SI{0.188035636}{Ha}) 
        \cite{ruscic_active_2015}. Chemical accuracy is defined as \SI{1}{kcal/mol} 
        (\SI{1.6e-3}{Ha}) \cite{helgaker_molecular_2014}.
    }
    \label{tab:errors}
    \begin{tabular}{l|r|r}
        \toprule
        Method & Error (Ha) & Error (kcal/mol) \\ \hline
        \midrule
        CASCI+NEVPT2 & \SI{1.16e-3}{} & \num{0.73} \\
        VQE+NEVPT2 & \SI{1.18e-3}{} & \num{0.74} \\
        CCSD(T) & \SI{1.88e-2}{} & \num{11.8} \\
        \bottomrule
    \end{tabular}
\end{table}

In summary, our analysis demonstrates that the VQE wavefunction converges to CASCI quality with the USCC-SDTQ ansatz, that NEVPT2 corrections are equally effective when built on quantum or classical reference states, and that both hybrid methods achieve chemical accuracy relative to experimental benchmarks.

The agreement between VQE+NEVPT2 and CASCI+NEVPT2 suggests that near-term quantum devices, when combined with classical perturbation theory, can achieve accuracy competitive with state-of-the-art classical multi-reference methods for systems with strong static correlation. 

%% file: exampleHubbard.tex
\section{Example: The Hubbard Model}\label{Sec:Hubbard}
The Hubbard model \cite{Hubbard1963} is the simplest model for interacting electrons on a lattice that captures the interplay between itineracy (driven by the kinetic energy and the hopping $t$ between neighboring sites) and locality (driven by the on-site repulsion $U$ between electrons of different spin). The Hubbard Hamiltonian reads
\begin{align}
 H = U \sum_{i} \hat{n}_{i\uparrow} \hat{n}_{i\downarrow} + t \sum_{\langle i, j\rangle,\sigma} \left(
 c^\dagger_{i\sigma} c_{j\sigma} + c^\dagger_{j\sigma} c_{i\sigma}
 \right) \label{Eq:Hubbard_model}
\end{align}
where \(c^\dagger_i\) (\(c_{i\sigma}\)) creates (annihilates) an electron on site \(i\) with spin \(\sigma\), for an illustration see Fig.\ \ref{Fig:Hub_mapping}.
Despite its simplicity, the Hubbard model can only be solved exactly in specific cases as one or infinite dimensions, such that numerical solutions within some approximations are required to solve it in two- and three dimensions. Their solutions suggest a rich competition between different phases like spin-density waves or superconductivity. Due to the limitations of classical solvers, the investigation of this model is a highly interesting application of quantum computers and has been the subject of several publications \cite{Wecker2015b, kivlichan_quantum_2018, Cade2020, Hartnett2026, Chowdhury2026, Jafarizadeh2025}. In the following, we will use the one-dimensional Hubbard model in order to demonstrate the different steps for a QC-simulation and point to difficulties arising from higher dimensionality. 

\subsection{Qubit-Mapping and Hardware Topology}
\begin{table*}
\renewcommand{\arraystretch}{1.5}
\begin{tabular}{c|c|c|c|c|}
Map & \(U n_{0\uparrow} n_{0\downarrow}\) & \(U n_{1\uparrow} n_{1\downarrow}\) &  \(t\left(c_{0\uparrow}^\dagger c_{1\uparrow} + c_{1\uparrow}^\dagger c_{0\uparrow}\right)\) & \(t\left(c_{0\downarrow}^\dagger c_{1\downarrow} + c_{1\downarrow}^\dagger c_{0\downarrow}\right)\) \\ \hline 
(a) JW & \(\frac{U}{4}\left(I - Z_0 - Z_2 + Z_0 Z_2 \right)\) & \(\frac{U}{4}\left(I - Z_1 - Z_3 + Z_1 Z_3 \right)\) & \(\frac{t}{2}\left(Y_0 Y_1 + X_0 X_1\right)\) & \(\frac{t}{2}\left(Y_2 Y_3 + X_2 X_3\right)\) \\ 
(b) JW & \(\frac{U}{4}\left(I - Z_0 - Z_1 + Z_0 Z_1 \right)\) & \(\frac{U}{4}\left(I - Z_2 - Z_3 + Z_2 Z_3 \right)\) & \(\frac{t}{2}\left(Y_0 Z_1 Y_2 + X_0 Z_1 X_2\right)\) & \(\frac{t}{2}\left(Y_1 Z_2 Y_3 + X_1 Z_2 X_3\right)\) \\ 
(a) P & \(\frac{U}{4}\left(I - Z_0 - Z_1 Z_2 + Z_0 Z_1 Z_2 \right)\) & \(\frac{U}{4}\left(I - Z_0 Z_1 - Z_2 Z_3 + Z_0 Z_1 Z_2 Z_3 \right)\) & \(\frac{t}{2}\left(X_0 - X_0 Z_1\right)\) & \(\frac{t}{2}\left(X_2 - Z_1 X_2 Z_3\right)\) \\ 
(b) P & \(\frac{U}{4}\left(I - Z_0 - Z_0 Z_1 + Z_1 \right)\) & \(\frac{U}{4}\left(I - Z_1 Z_2 - Z_2 Z_3 + Z_1 Z_3 \right)\) & \(-\frac{t}{2}\left(Y_0 Y_1  + X_0 X_1 Z_2\right)\) & \(-\frac{t}{2}\left(Y_1 Y_2 + Z_0 X_1 X_2 Z_3\right)\) \\
(a) PR & \(\frac{U}{4}\left( I - Z_0 + Z_1 - Z_0Z_1\right) \) & \(\frac{U}{4}\left(I + Z_0 - Z_1 - Z_0 Z_1\right)\) & \(t X_0\) & \(t X_1\) \\
%(b) PR & & & & \\
\end{tabular}
\caption{Qubit operators for the terms of the two-site Hubbard Hamiltonian (Eq.\ \eqref{Eq:Hubbard_model}) in orbital ordering (a) and (b) (cf.\ Eq.\ \eqref{Eq:TwoSiteOrdering}), respectively, after mapping with the Jordan-Wigner mapper (JW), and the Parity mapper with (PR) and without (P) two-qubit tapering. For the tapering we chose one electron in each spin-sector allowing a reduction by two qubits. Here, we assume the identity operator acting on all not-explicitly mentioned qubits and the operator \(I\) refers to the identity operator acting on all qubits.}\label{Tab:HubbardMapping}
\end{table*}
First, we consider the Fermion-to-qubit mapping for the Hamiltonian and its implication on preferable QC topology and the Pauli-weight of its Pauli-strings. Within such mapping, we have the freedom to order the orbitals in different ways, for example, first all spin-up and then all spin-down orbitals or by alternating between both spin-states for each orbital successively, i.e.\ for \(N\)-sites
%\begin{align}
\begin{equation}
\begin{split}
(a)\quad& \{0\uparrow, 1\uparrow, \dots, N\uparrow, 0\downarrow, 1\downarrow , \dots, N\downarrow\} \quad \textrm{or }\\
 (b)\quad&\{0\uparrow, 0\downarrow, 1\uparrow, 1\downarrow,\dots,N\uparrow, N\downarrow\}.
\end{split}
\label{Eq:TwoSiteOrdering}
\end{equation}
%\end{align}
Table \ref{Tab:HubbardMapping} shows the result of mapping the Hamiltonian of a two-site Hubbard model with the Jordan-Wigner (cf.\ Sec.\ \ref{Sec:JWMapping}) and Parity mapping (cf.\ Sec.\ \ref{Sec:PMapping}) to qubits for these two orderings. The combination of JW-mapping and ordering (a) has a maximum Pauli weight of 2 as the parity strings in the hopping operator cancel for operators acting on the same and neighbouring qubits. However, for JW-mapping in combination with ordering (b) the hopping terms have to bridge the qubit referring to the other spin-order, which leads to a Pauli-weight of 3. Using the parity mapper for two operators acting on the same qubit, as it is the case for density operators, the \(X\)-chains cancel but the the occupancy check \((I-Z)_{p-1}\) still appears. For the interaction term, this results in a Pauli weight of 4 for ordering (a), whereas the weight reduces to 2 because of the closer placement of the interacting modes in ordering (b) which leads to the cancellation of some contributions. For the kinetic terms, the situation reverses due to the hopping to either neighboring or second neighboring qubits, so that ordering (a) yields a maximum Pauli weight of 3, while ordering (b) produces weight 4. These observations hold for larger systems than the minimal one considered here, so that the Parity mapper appears less favorable regarding the Pauli-weight. However, under the assumption of one electron in each spin-sector, \(Z_1\) and \(Z_3\) are fixed to \(-1\) and \(+1\), respectively, allowing qubits 1 and 3 to be removed by qubit reduction. This reduces the problem to two qubits, as shown in the last line of Table \ref{Tab:HubbardMapping}. 

The topology of the quantum computer becomes especially relevant, when we consider a variational Hamiltonian ansatz or consider physical excitations in the ansatz like those of UCC-type. However, it also affects how easily correlations can spread within the system even for other ansatz types. To determine an efficient topology if an all-to-all connectivity is not available, we investigate which qubits are present in the different terms of the Hamiltonian. For the JW-mapping and ordering (a) the hopping terms always act on terms with neighboring indices, while the interaction terms act on qubits with labels differing by the number of sites. Thus, two lines of qubits which are connected with each other would be a matching topology for this mapping, as displayed in the lower part of Fig.\ \ref{Fig:Hub_mapping}. For ordering (b), the interaction and the hopping always act on two and three qubits with neighboring indices, so the same topology but with another indexing of qubits appears suitable. Regarding the parity-mapper, these arguments also hold, such that this square-lattice topology appears to be best suited for the one-dimensional Hubbard-chain. In the following, we will continue the discussion with the JW-mapping for ordering (a).

We remark, that the discussion of mapping becomes more challenging for the two-dimensional Hubbard-lattice (for all possible lattice geometries), as it requires a hopping to more than two-neighbors, such that the parity or spin-flip chain for all of them except of two will not cancel. Therefore, efficient mappings for this problem have been subject of research as discussed in Sec.\ \ref{Sec:lattice_mappings}.

\subsection{Initial State and Ansatz}
Having established how different fermion-to-qubit mappings transform the Hubbard Hamiltonian into Pauli-strings, we now turn to the choice of initial state and ansatz circuit as second ingredient of a VQE simulation. As discussed in Section \ref{Sec:Ansatz}, a good ansatz is a balance between expressive power and circuit compactness with the aim to reach the ground state with as few parameters and gate operations as possible.

\begin{figure}
    \centering
    \resizebox{\columnwidth}{!}{
    \input{Figures_Hubbard/Hubbard_mapping}
    }
    \caption{Representation of the 6-site Hubbard model and its mapping to two adjacent chains of qubits.}
    \label{Fig:Hub_mapping}
\end{figure}
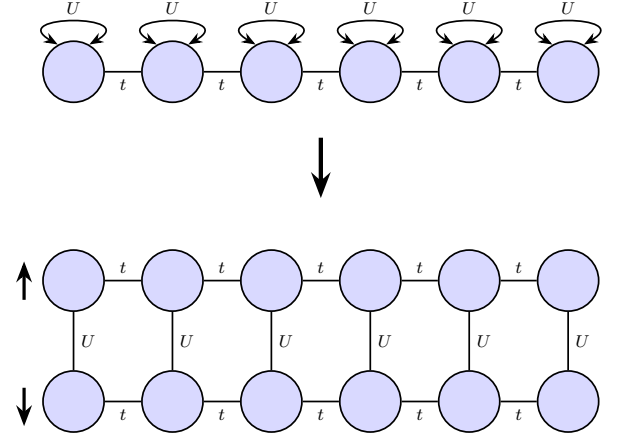
\begin{figure}
\centering
\input{Figures_Hubbard/AFM_init}
\input{Figures_Hubbard/pair_init}
\input{Figures_Hubbard/pair_spin}

\caption{Circuits representing the different initial states investigated in this section: AFM initialization (top left), pair initialization (top right) and pair-spin initialization (bottom). The Hartree-Fock circuit corresponds to the Givens rotation circuit represented in Section \ref{Sec:Ansatz_Init}} %and HF circuit initialization.}
\label{Fig:Hub_init_circs}
\end{figure}
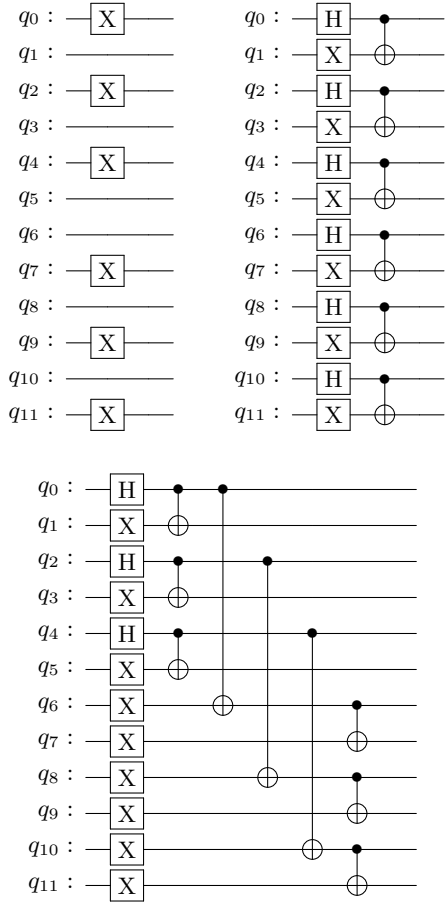
To illustrate how these choices influence VQE performance, we consider the 6-site Hubbard chain at half-filling, mapped to 12 qubits using the Jordan-Wigner transformation. This system is still small enough to allow reference energies and states from exact diagonalization. At half-filling, the 6 electrons are equally distributed over both spin-sectors, which must be respected when constructing particle number-preserving ansätze. This requires an appropriate initial state for which we discuss four options in the following:
For all cases we start with the initial  \(|0\rangle\)-state for all qubits, which corresponds to empty orbitals in the JW-encoding. Thus, applying an X-gate on a qubit corresponds to occupying the associated state. As a first initialization strategy, the even numbered spin-up qubits and the uneven numbered spin-down qubits are occupied, which results in an antiferromagnetically (AFM) ordered state, see Fig.\ \ref{Fig:Hub_init_circs} (top left). Second, we use the circuits shown in Fig.\ \ref{Fig:Hub_init_circs} (top right) to prepare an equal-weighted superposition for one electron over pairs of neighboring spin-up sites \(\frac{1}{\sqrt{2}}\left(|\uparrow\rangle_{2i} + |\uparrow\rangle_{2i+1}\right)\) for \(i\in{0,1,2}\), and analogously for spin-down sites, which we refer to as pair initialization. It is tempting to apply this strategy also to spin-up and spin-down pairs on the same site, but this would not necessarily lead to the correct \(S_z\)-value. Therefore, we employ the circuit in Fig.\ \ref{Fig:Hub_init_circs} (bottom), which ensures that neighboring pairs always have the opposite spin, i.e.\ \(\frac{1}{\sqrt{2}}\left(|\uparrow\rangle_{2i} |\downarrow\rangle_{2i+1}+|\downarrow\rangle_{2i} |\uparrow\rangle_{2i+1}\right)\) for \(i\in{0,1,2}\), which we refer to as pair-spin initialization. Finally, we set \(U=0\), solve the resulting non-interacting Hamiltonian classically, and use its ground state for our fourth initialization strategy. As HF solves the non-interacting model exactly, this is equal to a HF initialization. As the ground-state is a Slater-determinant built from cosine-like orbitals, this state can be realized either by a Quantum Fourier Transformation (QFT) or by an approach based on Givens rotations \cite{Jiang2018}, as we do in the following.

\begin{figure}
    \centering
    \includegraphics[width = 0.9\columnwidth]{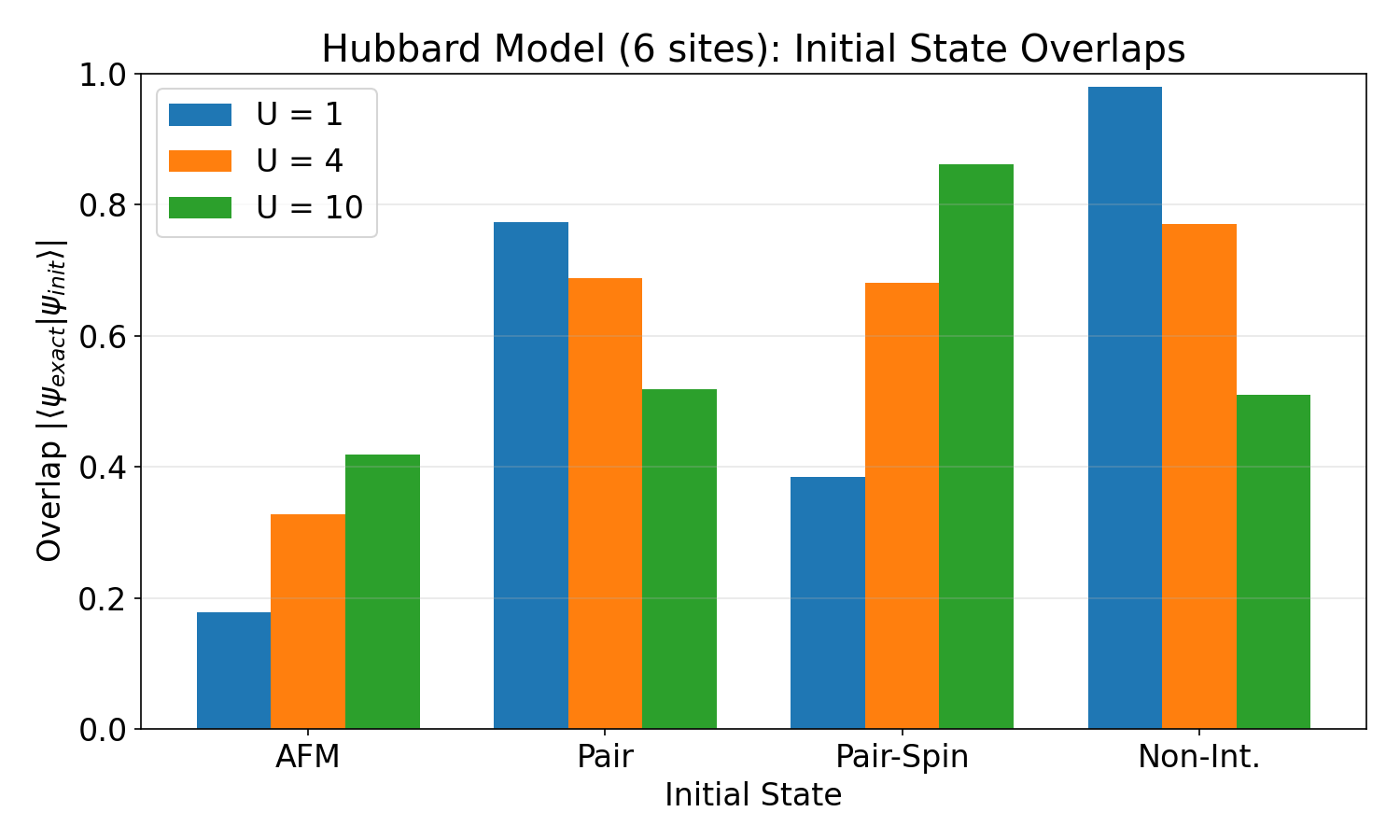}
    \caption{Overlap between four different initial states and the exact solution of the 6-site Hubbard model for different interaction strengths.}
    \label{Fig:Init_state_fid}
\end{figure}
Fig.\ \ref{Fig:Init_state_fid} shows the overlap of these four initial states with the final ground state obtained from exact diagonalization for the weakly coupled (\(U=1\,t\)), intermediate coupling (\(U=4\,t\)) and strong coupling (\(U=10\,t\)) regime. As the weakly coupled system is close to the non-interacting one, the non-interacting initial state has an overlap close to \(1\) with the exact result, significantly larger than the one of the other initial states. However, the overlap reduces significantly with increasing interaction strength to \(0.5\) in the strong-coupling regime. The pair initialization exhibits a similar trend from \(0.78\) for \(U=1\,t\) to \(0.52\) for \(U=10\,t\). The trend is reversed for the AFM and pair-spin initialization, as they avoid double occupancies, which contribute to the weak interacting state but are severely penalized by the strong-interacting case. As the AFM initialization only realizes one particular spin-ordered contribution, while the pair-spin initialization contains a superposition of many different spin-orderings, the latter has a larger overlap for all interactions. Especially in the strong-interaction limit it reaches an overlap above \(0.8\), while it is only slightly less favorable than the non-interacting initialization for the \(U=4\,t\) case. Thus, the non-interacting state is a good initial state for the weakly interacting Hubbard model, while the pair-spin state is a good initial state in the strong limit. In the intermediate regime both appear as a good option. In the following, we consider the intermediate coupling regime at \(U=4\,t\) for an analysis of possible ansatz circuits and use the non-interacting initial state.

\begin{figure}
    \centering
    \includegraphics[width=\columnwidth]{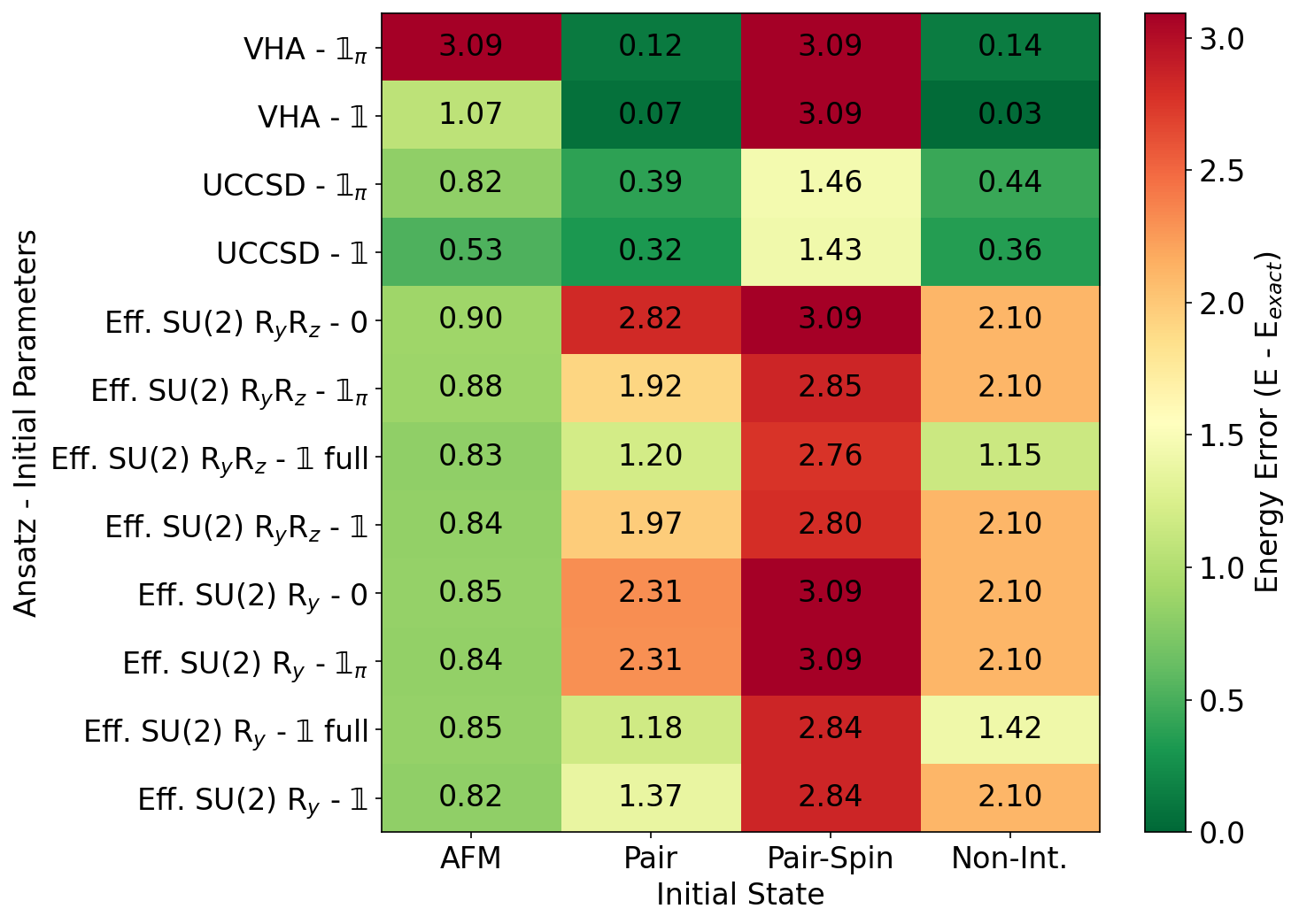}
    \includegraphics[width=\columnwidth]{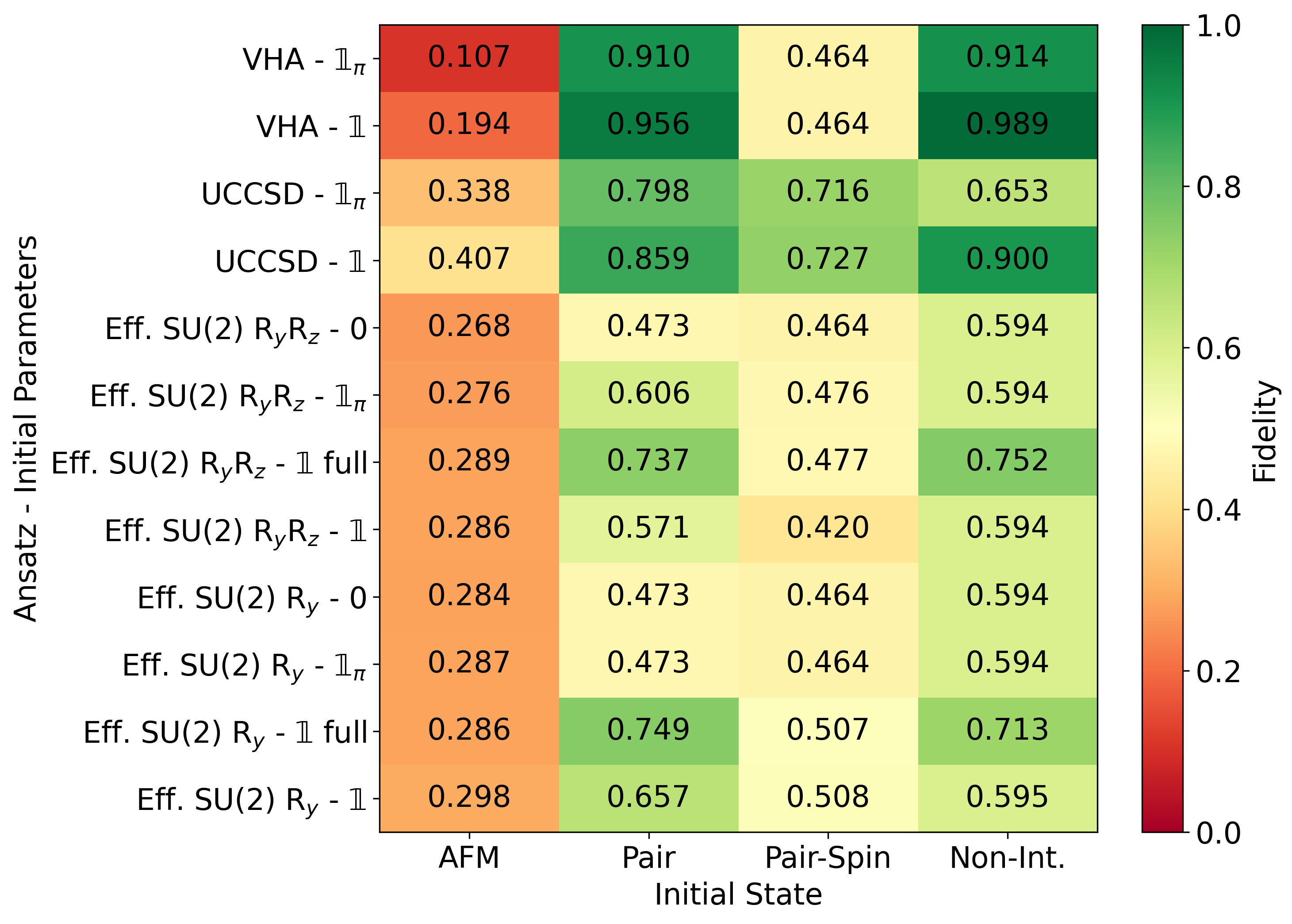}
    \caption{Comparison of the final VQE energy and state in terms of energy-error (top) and fidelity (bottom) with respect to the exact ground state from exact diagonalization with six ansatz layers.}
    \label{Fig:init_compare}
\end{figure}
In order to compare the different initial states in VQE calculations, we performed VQE simulations for a representative selection of initial parameter choices and ansatz settings. We use a statevector simulator, which provides access to the exact quantum state and thus yields a deterministic, (stochastic) noise-free energy landscape in conjunction with the COBYLA optimizer, which is well suited for noise-free optimization. Figure \ref{Fig:init_compare} summarizes the resulting energy errors and fidelities obtained at the optimal parameters for the Eff-SU(2) ansatz with either only \Ry-gates, or with a combination of \Ry- and \Rz-gates, the UCCSD ansatz and the variational Hamiltonian ansatz (cf.\ Sec.\ \ref{Sec:Ansatz}), all evaluated at a fixed depth of six layers. 

For the VHA and the UCCSD ansatz, the choice of initial parameters is particularly transparent. By construction, setting all variational parameters to zero reduces these circuits to the identity operation, such that the prepared quantum state coincides with the chosen initial reference state, referred to by \(\mathbb{1}\). An alternative identity realization can be obtained by setting all parameters to \(\pi\), exploiting the periodicity of the underlying rotation gates, denoted by  \(\mathbb{1}_\pi\). Although both choices correspond to the same unitary operation, they define different local parameter neighborhoods in the variational landscape. In practice, we find that the all‑zero initialization leads to slightly more favorable results in terms of energy and fidelity. For the Eff-SU(2) ansätze we tested several parameter initialization strategies, including zero-initialization (0), \(\pi\)-periodic (\(\mathbb{1}_\pi\))and structured parameter patterns with random compensating rotation pairs on all (\(\mathbb{1}_\mathrm{full}\)) or only on the outermost layers (\(\mathbb{1}\)), which all lead to an identity mapping. We find that the full structured parameter initialization \(\mathbb{1}_\mathrm{full}\) performs best for the Eff‑SU(2) ansatz. All these parameterizations are different realizations of identity mappings, which indicates that not only the initial state itself, but also the local geometry of the parameter space influences the optimization.
\begin{figure}
\centering
\includegraphics[width = \columnwidth]{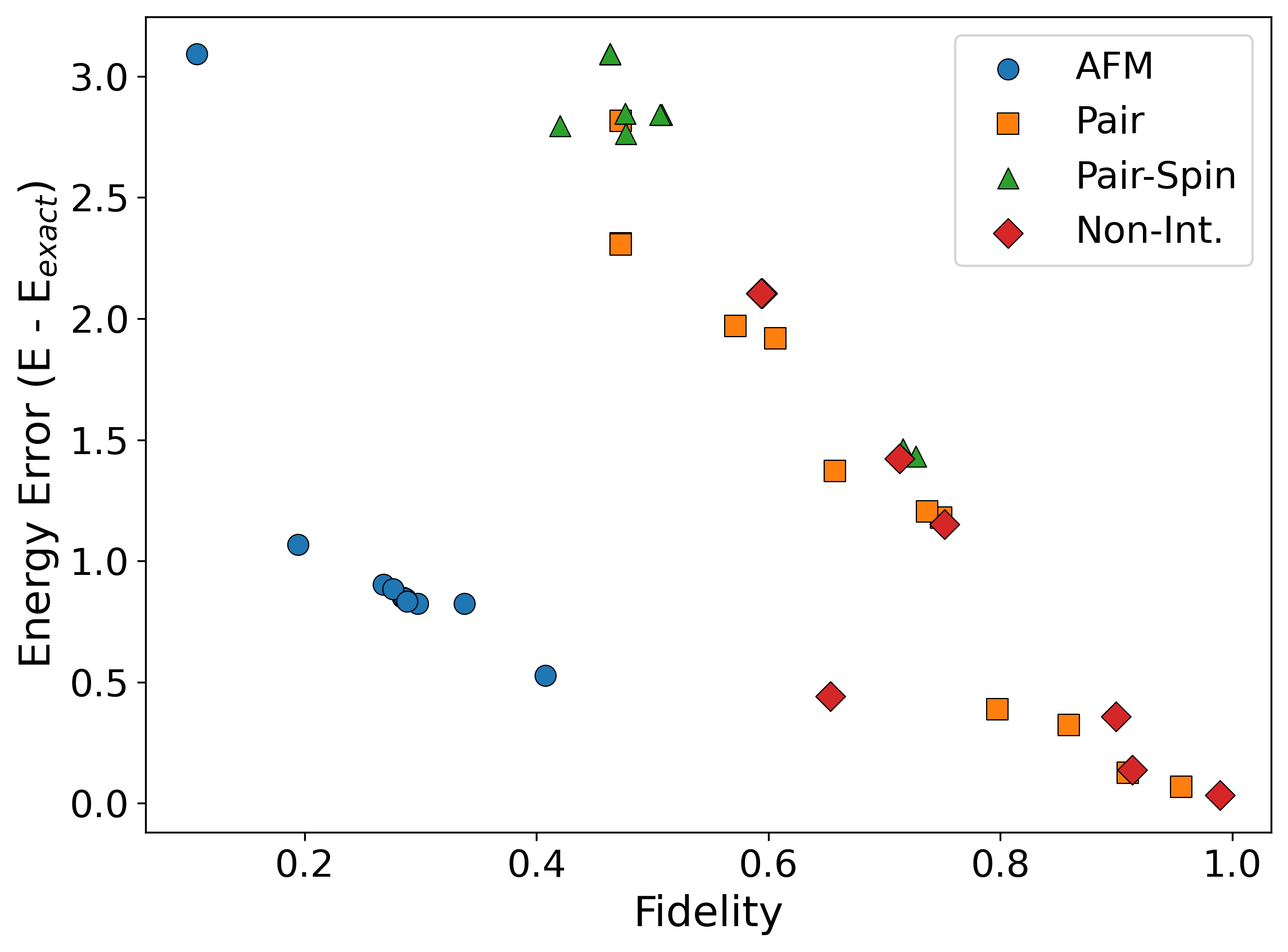}
\caption{Comparison of energy error and reached fidelity at the end of VQE-optimization for the different initial states as in Fig.\ \ref{Fig:init_compare}.}
\label{Fig:EvsFid}
\end{figure}
The initial quantum state strongly affects the result obtained. In all cases, the non‑interacting and pair initial states consistently provide the most reliable results with respect to energy error and fidelities. These states already capture essential features of the target ground state, where delocalization is important, and thus reduce the variational effort required during optimization. In contrast, the antiferromagnetic (AFM) initial state often leads to comparatively or even lower energy errors but with very low fidelities, as shown in Fig.\ \ref{Fig:EvsFid}. This is especially evident in the case of the highly-paremeterized hardware efficient ansätze, which reach a higher fidelity but lower energy error with the other initial states. This suggests that we might have ended in a local minimum from which the optimizer could not escape and that minimizing the energy alone does not guarantee that the correct many‑body wavefunction is obtained. In contrast, the pair-spin state performs worst for all ansätze in terms of energy error, but reaches better fidelities, especially in combination with the UCCSD ansatz. As the overlap with the exact ground state is larger, we start with a higher fidelity but, again, fail to escape the local minimum.

\begin{figure}
    \centering
    \includegraphics[width=\columnwidth]{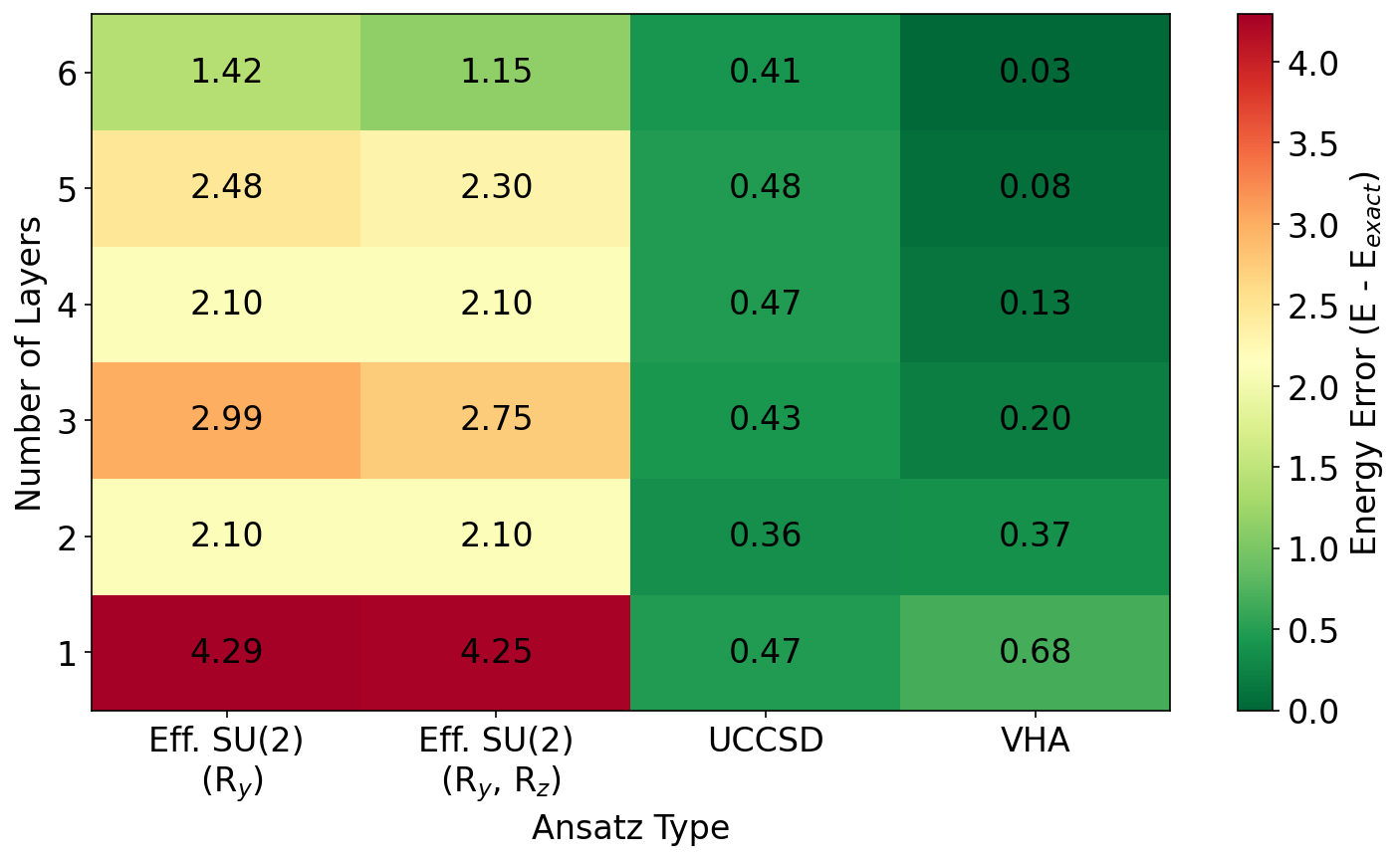}
    \includegraphics[width=\columnwidth]{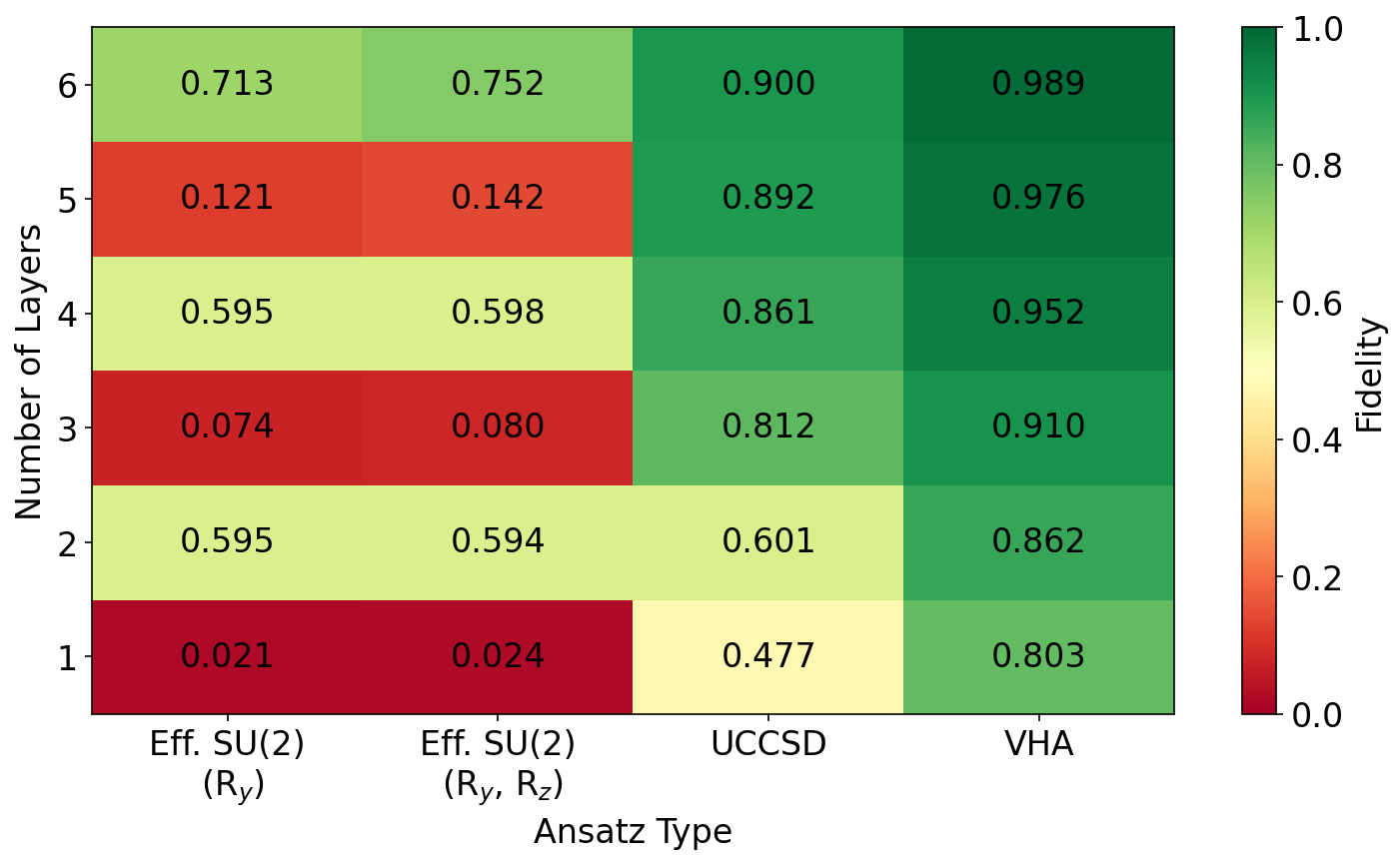}
    \caption{Comparison of the final VQE energy and state in terms of energy-error (top) and fidelity (bottom) with respect to the exact ground state from exact diagonalization for different ansatz types and number of ansatz layers.}
    \label{Fig:Ansatz_compare}
\end{figure}
In order to increase the fidelity of the ansätze we can add layers, as is shown in Fig.\ \ref{Fig:Ansatz_compare} based on VQE calculations using the optimal initialization strategy in combination with the non-interacting initial state as just discussed. Overall, an increase in circuit depth increases the fidelity and reduces the energy error except for the case of UCCSD. Here it appears that some relevant contributions might be missing in the ansatz, which would allow an energy reduction. One peculiarity in this dataset are the significant bad results for the efficient SU(2) ansätze with an odd number of layers. This can be traced back directly to the initialization mechanism, where pairs of entangling blocks cancel for an even number of layers, allowing the circuit to approximate the identity and thus preserve the intended initial state. For an odd number of layers, however, an uncompensated entangling block remains, effectively modifying the initial state by introducing global entanglement. This leads to a substantially more difficult optimization problem and explains the degraded performance observed in these cases.

Concluding, we compare the different ansätze among each other. The two Eff‑SU(2) variants, with and without additional {\Rz} rotations have no significant difference in performance. As the half-filled Hubbard model has time‑reversal symmetry, the ground state can be chosen to be purely real. Consequently, rotations around the y-axis are sufficient to represent the target state, and the additional variational freedom introduced by {\Rz} gates does not lead to a noticeable advantage in this setting. Overall, the VHA and UCCSD ansatz perform significantly better, as they are optimized in the part of the Hilbert space with correct particle number and \(S_z\)-quantum number. Therefore, they require significantly less parameters which helps the optimizer to find the minimum. However, one layer of these ansätze consists of a relevant number of two-qubit gates, so that a realization on real noisy quantum hardware becomes challenging. This highlights the importance of balancing between realizability on hardware and desired precision.

%% file: Figures_Hubbard/Hubbard_mapping.tex
\usetikzlibrary{arrows.meta}

\begin{tikzpicture}[
    scale=0.9,
    site/.style={circle, draw, thick, minimum size=10mm, fill=blue!15},
    lbl/.style={font=\footnotesize},
    >=Stealth
]

% ===== TOP: Hubbard Model with 6 sites =====
%\node[anchor=east] at (0.8, 0) {\textbf{Hubbard Model}};

\foreach \i in {1,...,6} {
    \node[site] (H\i) at ({1.8*\i}, 0) {};
    % U interaction loop on each site
    \draw[thick, <->] (H\i.120) to[out=150, in=30, looseness=5] 
        node[above, lbl] {$U$} (H\i.60);
}

% Hopping connections between neighbors
\foreach \i [evaluate=\i as \j using int(\i+1)] in {1,...,5} {
    \draw[thick] (H\i) -- node[below, lbl] {$t$} (H\j);
}

% ===== MAPPING ARROW =====
\draw[->, line width=2pt] (6.3, -1.2) -- (6.3, -2.3);

% ===== BOTTOM: Two qubit chains =====
%\node[anchor=east] at (0.8, -4.9) {\textbf{Qubit Chains}};

% Spin-up chain
\foreach \i in {1,...,6} {
    \node[site] (U\i) at ({1.8*\i}, -3.8) {};
}

% Spin-down chain
\foreach \i in {1,...,6} {
    \node[site] (D\i) at ({1.8*\i}, -6) {};
}

% Horizontal connections (t)
\foreach \i [evaluate=\i as \j using int(\i+1)] in {1,...,5} {
    \draw[thick] (U\i) -- node[above, lbl] {$t$} (U\j);
    \draw[thick] (D\i) -- node[below, lbl] {$t$} (D\j);
}

% Vertical connections (U)
\foreach \i in {1,...,6} {
    \draw[thick] (U\i) -- node[right, lbl] {$U$} (D\i);
}

% Spin arrows in front of rows
\draw[->, thick, line width=1.5pt] (0.9, -4.15) -- (0.9, -3.45);

\draw[->, thick, line width=1.5pt] (0.9, -5.75) -- (0.9, -6.45);

\end{tikzpicture}

%% file: Figures_Hubbard/AFM_init.tex
\scalebox{1.0}{
\Qcircuit @C=1.0em @R=0.2em @!R { \\
	 	\nghost{{q}_{0} :  } & \lstick{{q}_{0} :  } & \gate{\mathrm{X}} & \qw & \qw\\
	 	\nghost{{q}_{1} :  } & \lstick{{q}_{1} :  } & \qw & \qw & \qw\\
	 	\nghost{{q}_{2} :  } & \lstick{{q}_{2} :  } & \gate{\mathrm{X}} & \qw & \qw\\
	 	\nghost{{q}_{3} :  } & \lstick{{q}_{3} :  } & \qw & \qw & \qw\\
	 	\nghost{{q}_{4} :  } & \lstick{{q}_{4} :  } & \gate{\mathrm{X}} & \qw & \qw\\
	 	\nghost{{q}_{5} :  } & \lstick{{q}_{5} :  } & \qw & \qw & \qw\\
	 	\nghost{{q}_{6} :  } & \lstick{{q}_{6} :  } & \qw & \qw & \qw\\
	 	\nghost{{q}_{7} :  } & \lstick{{q}_{7} :  } & \gate{\mathrm{X}} & \qw & \qw\\
	 	\nghost{{q}_{8} :  } & \lstick{{q}_{8} :  } & \qw & \qw & \qw\\
	 	\nghost{{q}_{9} :  } & \lstick{{q}_{9} :  } & \gate{\mathrm{X}} & \qw & \qw\\
	 	\nghost{{q}_{10} :  } & \lstick{{q}_{10} :  } & \qw & \qw & \qw\\
	 	\nghost{{q}_{11} :  } & \lstick{{q}_{11} :  } & \gate{\mathrm{X}} & \qw & \qw\\
\\ }}

%% file: Figures_Hubbard/pair_init.tex
\scalebox{1.0}{
\Qcircuit @C=1.0em @R=0.2em @!R { \\
	 	\nghost{{q}_{0} :  } & \lstick{{q}_{0} :  } & \gate{\mathrm{H}} & \ctrl{1} & \qw & \qw\\
	 	\nghost{{q}_{1} :  } & \lstick{{q}_{1} :  } & \gate{\mathrm{X}} & \targ & \qw & \qw\\
	 	\nghost{{q}_{2} :  } & \lstick{{q}_{2} :  } & \gate{\mathrm{H}} & \ctrl{1} & \qw & \qw\\
	 	\nghost{{q}_{3} :  } & \lstick{{q}_{3} :  } & \gate{\mathrm{X}} & \targ & \qw & \qw\\
	 	\nghost{{q}_{4} :  } & \lstick{{q}_{4} :  } & \gate{\mathrm{H}} & \ctrl{1} & \qw & \qw\\
	 	\nghost{{q}_{5} :  } & \lstick{{q}_{5} :  } & \gate{\mathrm{X}} & \targ & \qw & \qw\\
	 	\nghost{{q}_{6} :  } & \lstick{{q}_{6} :  } & \gate{\mathrm{H}} & \ctrl{1} & \qw & \qw\\
	 	\nghost{{q}_{7} :  } & \lstick{{q}_{7} :  } & \gate{\mathrm{X}} & \targ & \qw & \qw\\
	 	\nghost{{q}_{8} :  } & \lstick{{q}_{8} :  } & \gate{\mathrm{H}} & \ctrl{1} & \qw & \qw\\
	 	\nghost{{q}_{9} :  } & \lstick{{q}_{9} :  } & \gate{\mathrm{X}} & \targ & \qw & \qw\\
	 	\nghost{{q}_{10} :  } & \lstick{{q}_{10} :  } & \gate{\mathrm{H}} & \ctrl{1} & \qw & \qw\\
	 	\nghost{{q}_{11} :  } & \lstick{{q}_{11} :  } & \gate{\mathrm{X}} & \targ & \qw & \qw\\
\\ }}

%% file: Figures_Hubbard/pair_spin.tex
\scalebox{1.0}{
\Qcircuit @C=1.0em @R=0.2em @!R { \\
	 	\nghost{{q}_{0} :  } & \lstick{{q}_{0} :  } & \gate{\mathrm{H}} & \ctrl{1} & \ctrl{6} & \qw & \qw & \qw & \qw & \qw\\
	 	\nghost{{q}_{1} :  } & \lstick{{q}_{1} :  } & \gate{\mathrm{X}} & \targ & \qw & \qw & \qw & \qw & \qw & \qw\\
	 	\nghost{{q}_{2} :  } & \lstick{{q}_{2} :  } & \gate{\mathrm{H}} & \ctrl{1} & \qw & \ctrl{6} & \qw & \qw & \qw & \qw\\
	 	\nghost{{q}_{3} :  } & \lstick{{q}_{3} :  } & \gate{\mathrm{X}} & \targ & \qw & \qw & \qw & \qw & \qw & \qw\\
	 	\nghost{{q}_{4} :  } & \lstick{{q}_{4} :  } & \gate{\mathrm{H}} & \ctrl{1} & \qw & \qw & \ctrl{6} & \qw & \qw & \qw\\
	 	\nghost{{q}_{5} :  } & \lstick{{q}_{5} :  } & \gate{\mathrm{X}} & \targ & \qw & \qw & \qw & \qw & \qw & \qw\\
	 	\nghost{{q}_{6} :  } & \lstick{{q}_{6} :  } & \gate{\mathrm{X}} & \qw & \targ & \qw & \qw & \ctrl{1} & \qw & \qw\\
	 	\nghost{{q}_{7} :  } & \lstick{{q}_{7} :  } & \gate{\mathrm{X}} & \qw & \qw & \qw & \qw & \targ & \qw & \qw\\
	 	\nghost{{q}_{8} :  } & \lstick{{q}_{8} :  } & \gate{\mathrm{X}} & \qw & \qw & \targ & \qw & \ctrl{1} & \qw & \qw\\
	 	\nghost{{q}_{9} :  } & \lstick{{q}_{9} :  } & \gate{\mathrm{X}} & \qw & \qw & \qw & \qw & \targ & \qw & \qw\\
	 	\nghost{{q}_{10} :  } & \lstick{{q}_{10} :  } & \gate{\mathrm{X}} & \qw & \qw & \qw & \targ & \ctrl{1} & \qw & \qw\\
	 	\nghost{{q}_{11} :  } & \lstick{{q}_{11} :  } & \gate{\mathrm{X}} & \qw & \qw & \qw & \qw & \targ & \qw & \qw\\
\\ }}

%% file: outlook.tex
\section{State-of-the-art, Current Trends and Outlook}

The Variational Quantum Eigensolver has matured from proof-of-principle demonstrations on minimal systems to small-scale electronic structure calculations on current quantum hardware.
The largest experimental demonstrations of standard VQE using chemistry-inspired ansätze have reached approximately 12 qubits, successfully capturing molecular properties such as bond dissociation curves and reaction energies \cite{guo_experimental_2024}. These experiments employed optimized unitary coupled cluster (UCC) ansätze with rigorous error mitigation on superconducting processors, demonstrating that VQE can achieve chemical accuracy in the minimal basis (STO-3G) for small molecules like $\text{H}_2$, $\text{LiH}$, and $\text{BeH}_2$ despite hardware noise. To extend this reach without abandoning the variational framework, hybrid strategies like Contextual Subspace VQE have enabled calculations on larger active spaces (8–14 orbitals) by solving the classically tractable parts of the Hamiltonian on conventional computers, reserving the quantum processor only for the strongly correlated components within a reduced variational subspace \cite{weaving_contextual_2025}. Similarly, overlap-guided adaptive methods (Overlap-ADAPT-VQE) lead to compact circuits by leveraging classical selected configuration interaction (SCI) references to construct the ansatz, demonstrating scalability on systems like $\text{N}_2$ and linear hydrogen chains ($\text{H}_6$) while maintaining the core VQE optimization loop \cite{feniou_overlap-adapt-vqe_2023}. 

Despite this progress, scaling VQE to industrially relevant molecules requires overcoming three fundamental bottlenecks. Measurement overhead can become substantial due to the \(O(N^4)\) Pauli terms in molecular Hamiltonians, although the number of circuits to be measured can be reduced to \(O(N^3)\) by partitioning the terms into commuting families. Further optimizations can be applied by exploiting the locality of interactions (see Sec.~\ref{Sec:MeasExpVal}). 
Techniques inspired by classical shadows and covariance-based optimization offer a path to drastically reduce the shot budget by reconstructing multiple observables from randomized measurements \cite{boyd_training_2022,patel_quantum_2025}. Sparse-Pauli dynamics and Pauli propagation frameworks further enable rapid energy estimation with reduced measurement cycles \cite{shrikhande_rapid_2025,rudolph_pauli_2025}.
Trainability remains threatened by barren plateaus. However, recent advances in initialization strategies (Hartree-Fock warm starts), symmetry-preserving ansätze, and noise-aware optimizers provide robust safeguards against vanishing gradients. Circuit depth is constrained by coherence times, but effective transpilation (including noise-aware compilation and fermionic SWAP networks) alongside hardware improvements in $T_1$/$T_2$ times and gate fidelities, will extend the maximum trainable depth. Critically, reducing error rates linearly extends the reachable circuit depth before noise-induced barren plateaus dominate \cite{wang_noise-induced_2021}.

Recent analysis by Weaving et al. \cite{weaving_JCTC_2026} documents that pure VQE demonstrations on NISQ hardware were effectively capped at approximately 12 qubits through 2023. This practical boundary arose from the confluence of statistical noise, hardware errors, barren plateaus, and prohibitive measurement overhead. By late 2023, it became apparent that a paradigm shift was necessary to scale beyond this limitation. 
To tackle larger molecular systems beyond the reach of pure VQE, quantum-selected approaches have emerged that combine classical pre-selection with quantum evaluation. The most ambitious demonstrations to date have employed quantum-centric supercomputing frameworks, achieving effective system sizes of 50–77 qubits by integrating classical selected configuration interaction (SCI) with quantum matrix element evaluation \cite{javier_robledo_SciAdv_2025}. In these quantum-selected CI (QSCI) methods, the quantum computer identifies the most important determinants from an exponentially large space, while the Hamiltonian matrix elements within this selected subspace are evaluated on a classical computer.

This hybrid strategy effectively circumvents the circuit depth limitations of pure VQE while still leveraging quantum resources for the classically intractable components of strongly correlated systems \cite{sugisaki_hamiltonian_2025}. Hamiltonian time evolution provides a parameter-free alternative to variational ansätze for preparing QSCI input states, avoiding optimization difficulties while maintaining sampling quality \cite{mikkelsen_quantum-selected_2025}. As the preparation of an approximate ground state is still required, a protocol called HI-VQE was developed, which combines the variational ground state optimization with the QSCI basis-selection and diagonalization\cite{Jarman2025}.

Looking beyond standard ground-state VQE, several extensions promise to enhance accuracy and broaden applicability. Quantum Subspace Expansion (QSE) and Krylov methods allow for post-VQE refinement by projecting the variational state into a small subspace of excited determinants, improving ground-state energies and enabling excited-state calculations without deep circuits \cite{yoshioka_generalized_2022,McClean2017,urbanek_chemistry_2020,takeshita_increasing_2020}. 

Time-evolved states prepared via the target Hamiltonian may circumvent optimization challenges in state preparation \cite{mikkelsen_quantum-selected_2025}. 
Dissipative Quantum Eigensolvers represent a paradigm shift, utilizing engineered dissipation to guide the system toward the ground state, potentially offering robustness against noise compared to unitary ansätze \cite{cubitt_dissipative_2023}. Furthermore, VQE serves as a critical precursor to fault-tolerant algorithms \cite{wang_review_2026}. A high-quality VQE state can bootstrap Quantum Phase Estimation (QPE), significantly reducing the required Trotter steps and overall resource cost. Finally, extensions to excited states via Variational Quantum Deflation (VQD) \cite{Higgott2019}, Multi-Configuration VQE (MC-VQE) \cite{parrish_quantum_2019,greene-diniz_quantum_2026}, and Subspace-Search VQE (SSVQE) \cite{Nakanishi2019} enable computation of full spectral properties essential for photochemistry and materials science \cite{li_efficient_2026}.

In parallel to variational strategies, adiabatic quantum computing (or quantum annealing) offers a complementary paradigm for eigenstate preparation \cite{yip_variational_2025}. Adiabatic paths can be hybridized with VQE: for example, VQE may initialize a state at a point in the adiabatic path, reducing required annealing time, or vice versa \cite{hegade_shortcuts_2021}. As hardware progresses toward coherence times capable of supporting adiabatic sweeps, hybrid adiabatic‑variational protocols may become practical for certain fixed Hamiltonians.

In the near term, VQE remains a foundational algorithm for harnessing quantum computational power, serving both as a practical tool for current hardware and retaining significant utility within fault-tolerant architectures. However, achieving quantum advantage will hinge on integrating algorithmic innovations with continued hardware scaling. The synergy between classical and quantum computation (exemplified by contextual subspaces, QSCI, and overlap-guided ansätze) suggests that hybrid workflows will dominate the intermediate future. As coherence times improve and error rates decline, the transition from variational to phase-estimation-based methods will become increasingly viable. Ultimately, VQE's legacy may lie not only in its standalone applications but also in its role as a bootstrap for fault-tolerant quantum chemistry, bridging the gap between today's noisy processors and tomorrow's error-corrected machines.

%% file: literature.bib
@article{verstraeteDMRG2023,
	author = {Verstraete, Frank and Nishino, Tomotoshi and Schollwöck, Ulrich and Bañuls, Mari Carmen and Chan, Garnet K. and Stoudenmire, Miles E.},
	journal = {Nat. Rev. Phys.},
	title = {Density matrix renormalization group, 30 years on},
	year = {2023},	
	issn = {2522-5820},
	month = May,
	number = {5},
	pages = {273--276},
	volume = {5},
	doi = {10.1038/s42254-023-00572-5},
	publisher = {Nature Publishing Group},
}

@article{schollwockDMRG2011,
	author = {Schollwöck, Ulrich},
	journal = {Annals of Physics},
	title = {The density-matrix renormalization group in the age of matrix product states},
	year = {2011},
	issn = {0003-4916},	
	month = jan,
	number = {1},
	pages = {96--192},
	volume = {326},
  doi = {10.1016/j.aop.2010.09.012},
	publisher = {Elsevier}
}

@Article{weaving_JCTC_2026,
  author    = {Weaving, Tim and Mingare, Angus and Ralli, Alexis and V. Coveney, Peter},
  journal   = {J. Chem. Theory Comput.},
  title     = {Selected Configuration Interaction Using Time-Evolved Population Statistics},
  year      = {2026},
  issn      = {1549-9626},
  month     = Apr,
  number    = {9},
  pages     = {4315--4328},
  volume    = {22},
  doi       = {10.1021/acs.jctc.5c01994},
  publisher = {American Chemical Society (ACS)},
}

@Article{wang_review_2026,
  author     = {Wang, Zhirao and Huang, Junxiang and Ye, Runyu and Li, Qingyu and Ding, Qi-Ming and Huang, Yiming and Zhang, Ting and Zeng, Yumeng and Gao, Jianshuo and Yuan, Xiao and Yao, Yuan},
  title      = {A {Review} of {Variational} {Quantum} {Algorithms}: {Insights} into {Fault}-{Tolerant} {Quantum} {Computing}},
  year       = {2026},
  month      = apr,
  doi        = {10.48550/arXiv.2604.07909},
  eprint     = {2604.07909},
  eprinttype = {arxiv},
  publisher  = {arXiv},
  shorttitle = {A {Review} of {Variational} {Quantum} {Algorithms}},
}

@article{Jarman2025,
      title={HIVQE: Handover Iterative Variational Quantum Eigensolver for Efficient Quantum Chemistry Calculations}, 
      author={Aidan Pellow-Jarman and Shane McFarthing and Doo Hyung Kang and Pilsun Yoo and Eyuel Eshetu Elala and Rowan Pellow-Jarman and P. Mai Nakliang and Jaewan Kim and June-Koo Kevin Rhee},
      year={2025},
      month={mar},
      doi = {10.48550/arXiv.2503.06292},
      eprint={2503.06292},
      eprinttype = {arxiv},
      publisher  = {arXiv},
}

@Article{lee_approaching_2021,
  author     = {Lee, Joonho and Feng, Xintian and Cunha, Leonardo A. and Gonthier, Jérôme F. and Epifanovsky, Evgeny and Head-Gordon, Martin},
  journal    = {J. Chem. Phys.},
  title      = {Approaching the basis set limit in {Gaussian}-orbital-based periodic calculations with transferability: {Performance} of pure density functionals for simple semiconductors},
  year       = {2021},
  issn       = {0021-9606, 1089-7690},
  month      = oct,
  number     = {16},
  pages      = {164102},
  volume     = {155},
  doi        = {10.1063/5.0069177},
  language   = {en},
  shorttitle = {Approaching the basis set limit in {Gaussian}-orbital-based periodic calculations with transferability},
  url        = {https://pubs.aip.org/jcp/article/155/16/164102/199787/Approaching-the-basis-set-limit-in-Gaussian},
  urldate    = {2026-04-24},
}

@Article{papajak_what_2012,
  author   = {Papajak, Ewa and Truhlar, Donald G.},
  journal  = {J. Chem. Phys.},
  title    = {What are the most efficient basis set strategies for correlated wave function calculations of reaction energies and barrier heights?},
  year     = {2012},
  issn     = {0021-9606, 1089-7690},
  month    = aug,
  number   = {6},
  pages    = {064110},
  volume   = {137},
  doi      = {10.1063/1.4738980},
  language = {en},
  url      = {https://pubs.aip.org/jcp/article/137/6/064110/72533/What-are-the-most-efficient-basis-set-strategies},
  urldate  = {2026-04-24},
}

@Article{mester_basis_2023,
  author     = {Mester, Dávid and Kállay, Mihály},
  journal    = {J. Chem. Theory Comput.},
  title      = {Basis {Set} {Limit} of {CCSD}({T}) {Energies}: {Explicit} {Correlation} {Versus} {Density}-{Based} {Basis}-{Set} {Correction}},
  year       = {2023},
  issn       = {1549-9618, 1549-9626},
  month      = nov,
  number     = {22},
  pages      = {8210--8222},
  volume     = {19},
  copyright  = {https://creativecommons.org/licenses/by/4.0/},
  doi        = {10.1021/acs.jctc.3c00979},
  language   = {en},
  shorttitle = {Basis {Set} {Limit} of {CCSD}({T}) {Energies}},
  urldate    = {2026-04-24},
}

@InCollection{parrill_basis_2017,
  author    = {Nagy, Balazs and Jensen, Frank},
  booktitle = {Reviews in {Computational} {Chemistry}},
  publisher = {Wiley},
  title     = {Basis {Sets} in {Quantum} {Chemistry}},
  year      = {2017},
  edition   = {1},
  editor    = {Parrill, Abby L. and Lipkowitz, Kenny B.},
  isbn      = {978-1-119-35543-4 978-1-119-35605-9},
  month     = apr,
  pages     = {93--149},
  volume    = {30},
  doi       = {10.1002/9781119356059.ch3},
  language  = {en},
  urldate   = {2026-04-24},
}

@Article{greene-diniz_quantum_2026,
  author   = {Greene-Diniz, Gabriel and Prokopiou, Georgia and Zsolt Manrique, David and Ramo, David Muñoz},
  journal  = {Digit. Discov.},
  title    = {Quantum state preparation of multiconfigurational states for quantum chemistry},
  year     = {2026},
  number   = {1},
  pages    = {134--152},
  volume   = {5},
  doi      = {10.1039/D5DD00350D},
  language = {en},
  url      = {https://pubs.rsc.org/en/content/articlelanding/2026/dd/d5dd00350d},
  urldate  = {2026-03-31},
}

@Article{parrish_quantum_2019,
  author  = {Parrish, Robert M. and Hohenstein, Edward G. and McMahon, Peter L. and Martínez, Todd J.},
  journal = {Phys. Rev. Lett.},
  title   = {Quantum {Computation} of {Electronic} {Transitions} {Using} a {Variational} {Quantum} {Eigensolver}},
  year    = {2019},
  month   = jun,
  number  = {23},
  pages   = {230401},
  volume  = {122},
  doi     = {10.1103/PhysRevLett.122.230401},
  urldate = {2026-03-31},
}

@Article{li_efficient_2026,
  author   = {Li, Tianyi and Zeng, Yumeng and Ding, Qiming and Huo, Zixuan and Xu, Xiaosi and Ren, Jiajun and Tang, Diandong and Cai, Xiaoxia and Yuan, Xiao},
  journal  = {Digit. Discov.},
  title    = {Efficient quantum simulation of non-adiabatic molecular dynamics with precise electronic structure},
  year     = {2026},
  issn     = {2635-098X},
  month    = feb,
  number   = {2},
  pages    = {548--570},
  volume   = {5},
  doi      = {10.1039/D5DD00433K},
  language = {en},
  url      = {https://pubs.rsc.org/en/content/articlelanding/2026/dd/d5dd00433k},
  urldate  = {2026-03-31},
}

@Article{ruscic_active_2015,
  author     = {Ruscic, Branko},
  journal    = {J. Phys. Chem. A.},
  title      = {Active {Thermochemical} {Tables}: {Sequential} {Bond} {Dissociation} {Enthalpies} of {Methane}, {Ethane}, and {Methanol} and the {Related} {Thermochemistry}},
  year       = {2015},
  issn       = {1089-5639, 1520-5215},
  month      = jul,
  number     = {28},
  pages      = {7810--7837},
  volume     = {119},
  copyright  = {http://pubs.acs.org/page/policy/authorchoice\_ccby\_termsofuse.html},
  doi        = {10.1021/acs.jpca.5b01346},
  language   = {en},
  shorttitle = {Active {Thermochemical} {Tables}},
  urldate    = {2026-01-14},
}

@book{helgaker_molecular_2014,
	address = {Hoboken},
	title = {Molecular {Electronic}-{Structure} {Theory}},
	isbn = {978-1-118-53147-1},
	language = {eng},
	publisher = {Wiley},
	author = {Helgaker, Trygve and Jorgensen, Poul and Olsen, Jeppe},
	year = {2014},
}

@Article{raghavachari_fifth-order_1989,
  author   = {Raghavachari, Krishnan and Trucks, Gary W. and Pople, John A. and Head-Gordon, Martin},
  journal  = {Chem. Phys. Lett.},
  title    = {A fifth-order perturbation comparison of electron correlation theories},
  year     = {1989},
  issn     = {0009-2614},
  month    = may,
  number   = {6},
  pages    = {479--483},
  volume   = {157},
  doi      = {10.1016/S0009-2614(89)87395-6},
  language = {en},
  url      = {https://linkinghub.elsevier.com/retrieve/pii/S0009261489873956},
  urldate  = {2026-03-17},
}

@Article{cao_quantum_2019,
  author   = {Cao, Yudong and Romero, Jonathan and Olson, Jonathan P. and Degroote, Matthias and Johnson, Peter D. and Kieferová, Mária and Kivlichan, Ian D. and Menke, Tim and Peropadre, Borja and Sawaya, Nicolas P. D. and Sim, Sukin and Veis, Libor and Aspuru-Guzik, Alán},
  journal  = {Chem. Rev.},
  title    = {Quantum {Chemistry} in the {Age} of {Quantum} {Computing}},
  year     = {2019},
  issn     = {0009-2665, 1520-6890},
  month    = oct,
  number   = {19},
  pages    = {10856--10915},
  volume   = {119},
  doi      = {10.1021/acs.chemrev.8b00803},
  language = {en},
  urldate  = {2023-09-13},
}

@Article{provazza_fast_2024,
  author  = {Provazza, Justin and Gunst, Klaas and Zhai, Huanchen and Chan, Garnet K.-L. and Shiozaki, Toru and Rubin, Nicholas C. and White, Alec F.},
  journal = {J. Chem. Theory Comput.},
  title   = {Fast {Emulation} of {Fermionic} {Circuits} with {Matrix} {Product} {States}},
  year    = {2024},
  issn    = {1549-9618},
  month   = may,
  number  = {9},
  pages   = {3719--3728},
  volume  = {20},
  doi     = {10.1021/acs.jctc.4c00200},
  urldate = {2026-03-10},
}

@InProceedings{li_tackling_2019,
  author    = {Li, Gushu and Ding, Yufei and Xie, Yuan},
  booktitle = {Proceedings of the {Twenty}-{Fourth} {International} {Conference} on {Architectural} {Support} for {Programming} {Languages} and {Operating} {Systems}},
  title     = {Tackling the {Qubit} {Mapping} {Problem} for {NISQ}-{Era} {Quantum} {Devices}},
  year      = {2019},
  address   = {New York, NY, USA},
  month     = apr,
  pages     = {1001--1014},
  publisher = {Association for Computing Machinery},
  series    = {{ASPLOS} '19},
  doi       = {10.1145/3297858.3304023},
  isbn      = {978-1-4503-6240-5},
  urldate   = {2026-03-09},
}

@Article{preskill_quantum_2018,
  author  = {Preskill, John},
  journal = {Quantum},
  title   = {Quantum computing in the {NISQ} era and beyond},
  year    = {2018},
  issn    = {2521-327X},
  number  = {July},
  pages   = {1--20},
  volume  = {2},
  doi     = {10.22331/q-2018-08-06-79},
}

@Article{kivlichan_quantum_2018,
  author  = {Kivlichan, Ian D. and McClean, Jarrod and Wiebe, Nathan and Gidney, Craig and Aspuru-Guzik, Alán and Chan, Garnet Kin-Lic and Babbush, Ryan},
  journal = {Phys. Rev. Lett.},
  title   = {Quantum {Simulation} of {Electronic} {Structure} with {Linear} {Depth} and {Connectivity}},
  year    = {2018},
  month   = mar,
  number  = {11},
  pages   = {110501},
  volume  = {120},
  doi     = {10.1103/PhysRevLett.120.110501},
  urldate = {2024-09-27},
}

@Article{bruzewicz_trapped-ion_2019,
  author     = {Bruzewicz, Colin D. and Chiaverini, John and McConnell, Robert and Sage, Jeremy M.},
  journal    = {Appl. Phys. Rev.},
  title      = {Trapped-ion quantum computing: {Progress} and challenges},
  year       = {2019},
  issn       = {1931-9401},
  month      = may,
  number     = {2},
  pages      = {021314},
  volume     = {6},
  doi        = {10.1063/1.5088164},
  shorttitle = {Trapped-ion quantum computing},
  urldate    = {2026-03-09},
}

@Article{hyyppa_reducing_2024,
  author  = {Hyyppä, Eric and Vepsäläinen, Antti and Papič, Miha and Chan, Chun Fai and Inel, Sinan and Landra, Alessandro and Liu, Wei and Luus, Jürgen and Marxer, Fabian and Ockeloen-Korppi, Caspar and Orbell, Sebastian and Tarasinski, Brian and Heinsoo, Johannes},
  journal = {PRX Quantum},
  title   = {Reducing {Leakage} of {Single}-{Qubit} {Gates} for {Superconducting} {Quantum} {Processors} {Using} {Analytical} {Control} {Pulse} {Envelopes}},
  year    = {2024},
  month   = sep,
  number  = {3},
  pages   = {030353},
  volume  = {5},
  doi     = {10.1103/PRXQuantum.5.030353},
  urldate = {2025-05-13},
}

@Article{vezvaee_virtual_2025,
  author   = {Vezvaee, Arian and Tripathi, Vinay and Kowsari, Daria and Levenson-Falk, Eli and Lidar, Daniel A.},
  journal  = {PRX Quantum},
  title    = {Virtual- {Z} {Gates} and {Symmetric} {Gate} {Compilation}},
  year     = {2025},
  issn     = {2691-3399},
  month    = jun,
  number   = {2},
  pages    = {020348},
  volume   = {6},
  doi      = {10.1103/PRXQuantum.6.020348},
  language = {en},
  urldate  = {2025-07-02},
}

@Article{mckay_efficient_2017,
  author  = {McKay, David C. and Wood, Christopher J. and Sheldon, Sarah and Chow, Jerry M. and Gambetta, Jay M.},
  journal = {Phys. Rev. A},
  title   = {Efficient \${Z}\$ gates for quantum computing},
  year    = {2017},
  month   = aug,
  number  = {2},
  pages   = {022330},
  volume  = {96},
  doi     = {10.1103/PhysRevA.96.022330},
  urldate = {2024-12-10},
}

@Article{de_keijzer_optimization_2022,
  author  = {de Keijzer, R. J. P. T. and Colussi, V. E. and Škorić, B. and Kokkelmans, S. J. J. M. F.},
  journal = {AVS Quantum Sci.},
  title   = {Optimization of the variational quantum eigensolver for quantum chemistry applications},
  year    = {2022},
  issn    = {2639-0213},
  month   = feb,
  number  = {1},
  pages   = {013803},
  volume  = {4},
  doi     = {10.1116/5.0076435},
  urldate = {2026-03-09},
}

@Article{nishio_extracting_2020,
  author  = {Nishio, Shin and Pan, Yulu and Satoh, Takahiko and Amano, Hideharu and Meter, Rodney Van},
  journal = {J. Emerg. Technol. Comput. Syst.},
  title   = {Extracting {Success} from {IBM}’s 20-{Qubit} {Machines} {Using} {Error}-{Aware} {Compilation}},
  year    = {2020},
  issn    = {1550-4832},
  number  = {3},
  pages   = {32:1--32:25},
  volume  = {16},
  doi     = {10.1145/3386162},
  urldate = {2026-03-09},
}

@Article{escofet_quantum_2026,
  author     = {Escofet, Pau and Rodrigo, Santiago and Sarkar, Rohit Sarma and Almudéver, Carmen G. and Alarcón, Eduard and Abadal, Sergi},
  title      = {Quantum {Circuit} {Pruning}: {Improving} {Fidelity} via {Compilation}-{Aware} {Circuit} {Approximation}},
  year       = {2026},
  month      = jan,
  doi        = {10.48550/arXiv.2601.13322},
  eprint     = {2601.13322},
  eprinttype = {arxiv},
  publisher  = {arXiv},
  shorttitle = {Quantum {Circuit} {Pruning}},
}

@Article{sivarajah_tket_2020,
  author     = {Sivarajah, Seyon and Dilkes, Silas and Cowtan, Alexander and Simmons, Will and Edgington, Alec and Duncan, Ross},
  journal    = {Quantum Sci. Technol.},
  title      = {t{\textbar}ket⟩: a retargetable compiler for {NISQ} devices},
  year       = {2020},
  issn       = {2058-9565},
  month      = nov,
  number     = {1},
  pages      = {014003},
  volume     = {6},
  doi        = {10.1088/2058-9565/ab8e92},
  language   = {en},
  shorttitle = {t{\textbar}ket⟩},
  urldate    = {2026-03-09},
}

@Article{bhowmick_enhancing_2025,
  author     = {Bhowmick, Rahul and Wadhwa, Harsh and Singh, Avinash and Sidana, Tania and Tran, Quoc Hoan and Sabapathy, Krishna Kumar},
  title      = {Enhancing variational quantum algorithms by balancing training on classical and quantum hardware},
  year       = {2025},
  month      = mar,
  doi        = {10.48550/arXiv.2503.16361},
  eprint     = {2503.16361},
  eprinttype = {arxiv},
  publisher  = {arXiv},
}

@Article{shi_weighted_2026,
  author   = {Shi, Chenyu and Dunjko, Vedran and Wang, Hao},
  journal  = {Quantum Sci. Technol.},
  title    = {Weighted approximate quantum natural gradient for variational quantum eigensolver},
  year     = {2026},
  issn     = {2058-9565},
  month    = feb,
  number   = {1},
  pages    = {015060},
  volume   = {11},
  doi      = {10.1088/2058-9565/ae3fc8},
  language = {en},
  urldate  = {2026-03-09},
}

@Article{lee_improving_2026,
  author     = {Lee, Jungyun and Park, Daniel K.},
  title      = {Improving {Generalization} and {Trainability} of {Quantum} {Eigensolvers} via {Graph} {Neural} {Encoding}},
  year       = {2026},
  month      = feb,
  doi        = {10.48550/arXiv.2602.19752},
  eprint     = {2602.19752},
  eprinttype = {arxiv},
  publisher  = {arXiv},
}

@Article{zou_generative_2025,
  author    = {Zou, Hang and Rahm, Martin and Kockum, Anton Frisk and Olsson, Simon},
  journal   = {npj Quantum Inf.},
  title     = {Generative flow-based warm start of the variational quantum eigensolver},
  year      = {2025},
  issn      = {2056-6387},
  month     = dec,
  number    = {1},
  pages     = {5},
  volume    = {12},
  copyright = {2025 The Author(s)},
  doi       = {10.1038/s41534-025-01159-x},
  language  = {en},
  url       = {https://www.nature.com/articles/s41534-025-01159-x},
  urldate   = {2026-03-09},
}

@Article{vaquero-sabater_pruned-adapt-vqe_2025,
  author     = {Vaquero-Sabater, Nonia and Carreras, Abel and Casanova, David},
  title      = {Pruned-{ADAPT}-{VQE}: compacting molecular ansätze by removing irrelevant operators},
  year       = {2025},
  month      = aug,
  doi        = {10.48550/arXiv.2504.04652},
  eprint     = {2504.04652},
  eprinttype = {arxiv},
  publisher  = {arXiv},
  shorttitle = {Pruned-{ADAPT}-{VQE}},
}

@Article{boy_energy_2025,
  author  = {Boy, Choy and Filip, Maria-Andreea and Wales, David J.},
  journal = {J. Chem. Theory Comput.},
  title   = {Energy {Landscapes} for the {Unitary} {Coupled} {Cluster} {Ansatz}},
  year    = {2025},
  issn    = {1549-9618},
  month   = feb,
  number  = {4},
  pages   = {1739--1751},
  volume  = {21},
  doi     = {10.1021/acs.jctc.4c01667},
  urldate = {2026-03-09},
}

@Article{gibbs_exploiting_2025,
  author   = {Gibbs, Joe and Holmes, Zoë and Stevenson, Paul},
  journal  = {Quantum Mach. Intell.},
  title    = {Exploiting symmetries in nuclear {Hamiltonians} for ground state preparation},
  year     = {2025},
  issn     = {2524-4914},
  month    = jan,
  number   = {1},
  pages    = {14},
  volume   = {7},
  doi      = {10.1007/s42484-025-00242-y},
  language = {en},
  urldate  = {2026-03-09},
}

@Article{mukherjee_comparative_2023,
  author    = {Mukherjee, Anirban and Berthusen, Noah F. and Getelina, João C. and Orth, Peter P. and Yao, Yong-Xin},
  journal   = {Commun. Phys.},
  title     = {Comparative study of adaptive variational quantum eigensolvers for multi-orbital impurity models},
  year      = {2023},
  issn      = {2399-3650},
  month     = jan,
  number    = {1},
  pages     = {4},
  volume    = {6},
  copyright = {2023 The Author(s)},
  doi       = {10.1038/s42005-022-01089-6},
  language  = {en},
  url       = {https://www.nature.com/articles/s42005-022-01089-6},
  urldate   = {2026-03-09},
}

@Article{sarkar_quantum_2026,
  author    = {Sarkar, Maitreyee and Roy, Lisa and Gutal, Akash Popat and Kumar, Atul and Paranjothy, Manikandan},
  journal   = {J. Phys. Chem. A},
  title     = {Symmetry-Match Fraction for Active-Space Selection: A Simple Route to Accurate Reaction Energies in VQE/UCCSD},
  year      = {2026},
  issn      = {1520-5215},
  month     = June,
  number    = {28},
  pages     = {5594--5601},
  volume    = {130},
  doi       = {10.1021/acs.jpca.6c00363},
  publisher = {American Chemical Society (ACS)},
}

@Article{wang_noise-induced_2021,
  author    = {Wang, Samson and Fontana, Enrico and Cerezo, M. and Sharma, Kunal and Sone, Akira and Cincio, Lukasz and Coles, Patrick J.},
  journal   = {Nat. Commun.},
  title     = {Noise-induced barren plateaus in variational quantum algorithms},
  year      = {2021},
  issn      = {2041-1723},
  month     = nov,
  number    = {1},
  pages     = {6961},
  volume    = {12},
  copyright = {2021 The Author(s)},
  doi       = {10.1038/s41467-021-27045-6},
  language  = {en},
  url       = {https://www.nature.com/articles/s41467-021-27045-6},
  urldate   = {2026-03-05},
}

@Article{harville_recent_2026,
  author     = {Harville, Taylor and Khurana, Rishu and Grizzi, Vitor F. and Liu, Cong},
  title      = {Recent {Developments} in {VQE}: {Survey} and {Benchmarking}},
  year       = {2026},
  month      = feb,
  doi        = {10.48550/arXiv.2602.11384},
  eprint     = {2602.11384},
  eprinttype = {arxiv},
  publisher  = {arXiv},
  shorttitle = {Recent {Developments} in {VQE}},
}

@Article{zimboras_myths_2025,
  author     = {Zimborás, Zoltán and Koczor, Bálint and Holmes, Zoë and Borrelli, Elsi-Mari and Gilyén, András and Huang, Hsin-Yuan and Cai, Zhenyu and Acín, Antonio and Aolita, Leandro and Banchi, Leonardo and Brandão, Fernando G. S. L. and Cavalcanti, Daniel and Cubitt, Toby and Filippov, Sergey N. and García-Pérez, Guillermo and Goold, John and Kálmán, Orsolya and Kyoseva, Elica and Rossi, Matteo A. C. and Sokolov, Boris and Tavernelli, Ivano and Maniscalco, Sabrina},
  title      = {Myths around quantum computation before full fault tolerance: {What} no-go theorems rule out and what they don't},
  year       = {2025},
  month      = jan,
  doi        = {10.48550/arXiv.2501.05694},
  eprint     = {2501.05694},
  eprinttype = {arxiv},
  publisher  = {arXiv},
  shorttitle = {Myths around quantum computation before full fault tolerance},
}

@Article{cheng_optimal_2025,
  author   = {Cheng, M. H. and Chen, Yu-Cheng and Wang, Qian and Bartsch, V. and Medina, A. C. and Kim, M. S. and Hu, Alice and Hsieh, Min-Hsiu},
  journal  = {Phys. Rev. Research},
  title    = {Optimal number-conserved linear encoding for practical fermionic simulation},
  year     = {2025},
  issn     = {2643-1564},
  month    = sep,
  number   = {3},
  pages    = {033280},
  volume   = {7},
  doi      = {10.1103/lzf5-x6hc},
  language = {en},
  urldate  = {2026-01-27},
}

@Article{jones_efficient_2020,
  author     = {Jones, Tyson and Gacon, Julien},
  title      = {Efficient calculation of gradients in classical simulations of variational quantum algorithms},
  year       = {2020},
  month      = sep,
  doi        = {10.48550/arXiv.2009.02823},
  eprint     = {2009.02823},
  eprinttype = {arxiv},
  publisher  = {arXiv},
}

@Article{shee_qubit-efficient_2022,
  author   = {Shee, Yu and Tsai, Pei-Kai and Hong, Cheng-Lin and Cheng, Hao-Chung and Goan, Hsi-Sheng},
  journal  = {Phys. Rev. Research},
  title    = {Qubit-efficient encoding scheme for quantum simulations of electronic structure},
  year     = {2022},
  issn     = {2643-1564},
  month    = may,
  number   = {2},
  pages    = {023154},
  volume   = {4},
  doi      = {10.1103/PhysRevResearch.4.023154},
  language = {en},
  urldate  = {2026-01-27},
}

@Article{gandon_quantum_2025,
  author   = {Gandon, Anthony and Baiardi, Alberto and Rossmannek, Max and Dobrautz, Werner and Tavernelli, Ivano},
  journal  = {PRX Quantum},
  title    = {Quantum {Computing} in {Spin}-{Adapted} {Representations} for {Efficient} {Simulations} of {Spin} {Systems}},
  year     = {2025},
  issn     = {2691-3399},
  month    = jul,
  number   = {3},
  pages    = {030306},
  volume   = {6},
  doi      = {10.1103/dbnd-sl4j},
  language = {en},
  urldate  = {2026-01-27},
}

@Article{rudolph_pauli_2025,
  author     = {Rudolph, Manuel S. and Jones, Tyson and Teng, Yanting and Angrisani, Armando and Holmes, Zoë},
  title      = {Pauli {Propagation}: {A} {Computational} {Framework} for {Simulating} {Quantum} {Systems}},
  year       = {2025},
  month      = may,
  doi        = {10.48550/arXiv.2505.21606},
  eprint     = {2505.21606},
  eprinttype = {arxiv},
  publisher  = {arXiv},
  shorttitle = {Pauli {Propagation}},
}

@Article{lennard-jones_electronic_1929,
  author   = {Lennard-Jones, J. E.},
  journal  = {Trans. Faraday Soc.},
  title    = {The electronic structure of some diatomic molecules},
  year     = {1929},
  issn     = {0014-7672},
  pages    = {668},
  volume   = {25},
  doi      = {10.1039/tf9292500668},
  language = {en},
  url      = {https://xlink.rsc.org/?DOI=tf9292500668},
  urldate  = {2026-01-13},
}

@Article{bauschlicher_full_1987,
  author     = {Bauschlicher, Charles W. and Langhoff, Stephen R.},
  journal    = {J. Chem. Phys.},
  title      = {Full {CI} benchmark calculations on {N2}, {NO}, and {O2}: {A} comparison of methods for describing multiple bonds},
  year       = {1987},
  issn       = {0021-9606, 1089-7690},
  month      = may,
  number     = {10},
  pages      = {5595--5599},
  volume     = {86},
  doi        = {10.1063/1.452534},
  language   = {en},
  shorttitle = {Full {CI} benchmark calculations on {N2}, {NO}, and {O2}},
  url        = {https://pubs.aip.org/jcp/article/86/10/5595/93182/Full-CI-benchmark-calculations-on-N2-NO-and-O2-A},
  urldate    = {2026-01-13},
}

@Article{moll_optimizing_2016,
  author  = {Moll, Nikolaj and Fuhrer, Andreas and Staar, Peter and Tavernelli, Ivano},
  journal = {J. Phys. A: Math. Theor.},
  title   = {Optimizing qubit resources for quantum chemistry simulations in second quantization on a quantum computer},
  year    = {2016},
  issn    = {1751-8113, 1751-8121},
  month   = jul,
  number  = {29},
  pages   = {295301},
  volume  = {49},
  doi     = {10.1088/1751-8113/49/29/295301},
  urldate = {2026-01-17},
}

@Article{bravyi_tapering_2017,
  author     = {Bravyi, Sergey and Gambetta, Jay M. and Mezzacapo, Antonio and Temme, Kristan},
  title      = {Tapering off qubits to simulate fermionic {Hamiltonians}},
  year       = {2017},
  copyright  = {arXiv.org perpetual, non-exclusive license},
  doi        = {10.48550/ARXIV.1701.08213},
  eprint     = {1701.08213},
  eprinttype = {arxiv},
  publisher  = {arXiv},
}

@Article{zen_properties_2014,
  author   = {Zen, Andrea and Trout, Bernhardt L. and Guidoni, Leonardo},
  journal  = {J. Chem. Phys.},
  title    = {Properties of reactive oxygen species by quantum {Monte} {Carlo}},
  year     = {2014},
  issn     = {0021-9606, 1089-7690},
  month    = jul,
  number   = {1},
  pages    = {014305},
  volume   = {141},
  doi      = {10.1063/1.4885144},
  language = {en},
  url      = {https://pubs.aip.org/jcp/article/141/1/014305/352636/Properties-of-reactive-oxygen-species-by-quantum},
  urldate  = {2026-01-14},
}

@Article{stein_automated_2016,
  author   = {Stein, Christopher J. and Reiher, Markus},
  journal  = {J. Chem. Theory Comput.},
  title    = {Automated {Selection} of {Active} {Orbital} {Spaces}},
  year     = {2016},
  issn     = {1549-9618, 1549-9626},
  month    = apr,
  number   = {4},
  pages    = {1760--1771},
  volume   = {12},
  doi      = {10.1021/acs.jctc.6b00156},
  language = {en},
  urldate  = {2026-01-14},
}

@Article{shirazi_performance_2025,
  author     = {Shirazi, Reza G. and Zech, Alexander and Pinski, Peter and Rybkin, Vladimir V.},
  title      = {Performance of {Automatic} {Active} {Space} {Selection} for {Electronic} {Excitation} {Energies}},
  year       = {2025},
  copyright  = {Creative Commons Attribution 4.0 International},
  doi        = {10.48550/ARXIV.2511.05732},
  eprint     = {2511.05732},
  eprinttype = {arxiv},
  publisher  = {arXiv},
}

@Article{jones_quest_2019,
  author   = {Jones, Tyson and Brown, Anna and Bush, Ian and Benjamin, Simon C.},
  journal  = {Sci. Rep.},
  title    = {{QuEST} and {High} {Performance} {Simulation} of {Quantum} {Computers}},
  year     = {2019},
  issn     = {2045-2322},
  month    = jul,
  number   = {1},
  pages    = {10736},
  volume   = {9},
  doi      = {10.1038/s41598-019-47174-9},
}

@Article{partridge_theoretical_1991,
  author   = {Partridge, Harry and Bauschlicher, Charles W. and Langhoff, Stephen R. and Taylor, Peter R.},
  journal  = {J. Chem. Phys.},
  title    = {Theoretical study of the low-lying bound states of {O2}},
  year     = {1991},
  issn     = {0021-9606, 1089-7690},
  month    = dec,
  number   = {11},
  pages    = {8292--8300},
  volume   = {95},
  doi      = {10.1063/1.461309},
  language = {en},
  url      = {https://pubs.aip.org/jcp/article/95/11/8292/443622/Theoretical-study-of-the-low-lying-bound-states-of},
  urldate  = {2026-01-13},
}

@Article{karton_w417_2017,
  author     = {Karton, Amir and Sylvetsky, Nitai and Martin, Jan M. L.},
  journal    = {J. Comput. Chem.},
  title      = {W4‐17: {A} diverse and high‐confidence dataset of atomization energies for benchmarking high‐level electronic structure methods},
  year       = {2017},
  issn       = {0192-8651, 1096-987X},
  month      = sep,
  number     = {24},
  pages      = {2063--2075},
  volume     = {38},
  copyright  = {http://onlinelibrary.wiley.com/termsAndConditions\#am},
  doi        = {10.1002/jcc.24854},
  language   = {en},
  shorttitle = {W4‐17},
  urldate    = {2026-01-13},
}

@Article{shrikhande_rapid_2025,
  author     = {Shrikhande, Chinmay and Bachhar, Arnab and Jimenez, Aaron Rodriguez and Mayhall, Nicholas J.},
  title      = {Rapid ground state energy estimation with a {Sparse} {Pauli} {Dynamics}-enabled {Variational} {Double} {Bracket} {Flow}},
  year       = {2025},
  month      = nov,
  doi        = {10.48550/arXiv.2511.21651},
  eprint     = {2511.21651},
  eprinttype = {arxiv},
  publisher  = {arXiv},
}

@Article{Grimsley2023,
  author  = {Grimsley, Harper R. and Barron, George S. and Barnes, Edwin and Economou, Sophia E. and Mayhall, Nicholas J.},
  journal = {npj Quantum Inf.},
  title   = {Adaptive, problem-tailored variational quantum eigensolver mitigates rough parameter landscapes and barren plateaus},
  year    = {2023},
  month   = {Mar},
  number  = {1},
  pages   = {19},
  volume  = {9},
  day     = {01},
  url     = {https://www.nature.com/articles/s41534-023-00681-0}
}

@Article{Ramoa2025,
  author  = {Ram{\^o}a, Mafalda and Anastasiou, Panagiotis G. and Santos, Luis Paulo and Mayhall, Nicholas J. and Barnes, Edwin and Economou, Sophia E.},
  journal = {npj Quantum Inf.},
  title   = {Reducing the resources required by ADAPT-VQE using coupled exchange operators and improved subroutines},
  year    = {2025},
  month   = {May},
  number  = {1},
  pages   = {86},
  volume  = {11},
  day     = {26},
  url     = {https://www.nature.com/articles/s41534-025-01039-4}
}

@Article{stadelmann2025,
  author        = {Jonas Stadelmann and Julian Übelher and Mafalda Ramôa and Bharath Sambasivam and Edwin Barnes and Sophia E. Economou},
  title         = {Strategies for Overcoming Gradient Troughs in the ADAPT-VQE Algorithm},
  year          = {2025},
  archiveprefix = {arXiv},
  eprint        = {2512.25004},
  eprinttype    = {arxiv},
  primaryclass  = {quant-ph},
}

@Article{viswanathan2025,
  author        = {Yaromir Viswanathan and Olivier Adjoua and César Feniou and Siwar Badreddine and Jean-Philip Piquemal},
  title         = {An Optimal Framework for Constructing Lie-Algebra Generator Pools: Application to Variational Quantum Eigensolvers for Chemistry},
  year          = {2025},
  archiveprefix = {arXiv},
  eprint        = {2511.22593},
  eprinttype    = {arxiv},
  primaryclass  = {quant-ph},
}

@Article{magoulas2025symmetry,
  author    = {Magoulas, Ilias and Zhang, Muhan and Evangelista, Francesco A.},
  journal   = {J. Chem. Phys.},
  title     = {Symmetry dilemmas in quantum computing for chemistry: A comprehensive analysis},
  year      = {2026},
  issn      = {1089-7690},
  month     = Apr,
  number    = {14},
  pages     = {144113},
  volume    = {164},
  doi       = {10.1063/5.0316482},
  publisher = {AIP Publishing},
}

@Article{magoulas2025,
  author        = {Ilias Magoulas and Francesco A. Evangelista},
  title         = {Spin-Adapted Fermionic Unitaries: From Lie Algebras to Compact Quantum Circuits},
  year          = {2025},
  archiveprefix = {arXiv},
  eprint        = {2511.13485},
  eprinttype    = {arxiv},
  primaryclass  = {quant-ph},
}

@Article{Sapova2022,
  author  = {Sapova, Mariia D. and Fedorov, Aleksey K.},
  journal = {Commun. Phys.},
  title   = {Variational quantum eigensolver techniques for simulating carbon monoxide oxidation},
  year    = {2022},
  month   = {Aug},
  number  = {1},
  pages   = {199},
  volume  = {5},
  day     = {06},
  url     = {https://www.nature.com/articles/s42005-022-00982-4}
}

@Article{Jaeger2025,
  author  = {J{\"a}ger, Jonas and Kaldenbach, Thierry N. and Haas, Max and Schultheis, Erik},
  journal = {Commun. Phys.},
  title   = {Fast gradient-free optimization of excitations in variational quantum eigensolvers},
  year    = {2025},
  month   = {Oct},
  number  = {1},
  pages   = {418},
  volume  = {8},
  day     = {30},
  url     = {https://www.nature.com/articles/s42005-025-02375-9}
}

@Article{ragone_lie_2024,
  author    = {Ragone, Michael and Bakalov, Bojko N. and Sauvage, Frédéric and Kemper, Alexander F. and Ortiz Marrero, Carlos and Larocca, Martín and Cerezo, M.},
  journal   = {Nat. Commun.},
  title     = {A {Lie} algebraic theory of barren plateaus for deep parameterized quantum circuits},
  year      = {2024},
  issn      = {2041-1723},
  month     = aug,
  number    = {1},
  pages     = {7172},
  volume    = {15},
  copyright = {2024 The Author(s)},
  doi       = {10.1038/s41467-024-49909-3},
  url       = {https://www.nature.com/articles/s41467-024-49909-3},
  urldate   = {2025-11-21},
}

@Article{cerezo_does_2025,
  author    = {Cerezo, M. and Larocca, Martin and García-Martín, Diego and Diaz, N. L. and Braccia, Paolo and Fontana, Enrico and Rudolph, Manuel S. and Bermejo, Pablo and Ijaz, Aroosa and Thanasilp, Supanut and Anschuetz, Eric R. and Holmes, Zoë},
  journal   = {Nat. Commun.},
  title     = {Does provable absence of barren plateaus imply classical simulability?},
  year      = {2025},
  issn      = {2041-1723},
  month     = aug,
  number    = {1},
  pages     = {7907},
  volume    = {16},
  copyright = {2025 The Author(s)},
  doi       = {10.1038/s41467-025-63099-6},
  url       = {https://www.nature.com/articles/s41467-025-63099-6},
  urldate   = {2025-11-13},
}

@Article{mhiri_unifying_2025,
  author     = {Mhiri, Hela and Puig, Ricard and Lerch, Sacha and Rudolph, Manuel S. and Chotibut, Thiparat and Thanasilp, Supanut and Holmes, Zoë},
  title      = {A unifying account of warm start guarantees for patches of quantum landscapes},
  year       = {2025},
  month      = feb,
  doi        = {10.48550/arXiv.2502.07889},
  eprint     = {2502.07889},
  eprinttype = {arxiv},
  publisher  = {arXiv},
}

@Article{larocca_barren_2025,
  author    = {Larocca, Martín and Thanasilp, Supanut and Wang, Samson and Sharma, Kunal and Biamonte, Jacob and Coles, Patrick J. and Cincio, Lukasz and McClean, Jarrod R. and Holmes, Zoë and Cerezo, M.},
  journal   = {Nat. Rev. Phys.},
  title     = {Barren plateaus in variational quantum computing},
  year      = {2025},
  issn      = {2522-5820},
  month     = apr,
  number    = {4},
  pages     = {174--189},
  volume    = {7},
  copyright = {2025 Springer Nature Limited},
  doi       = {10.1038/s42254-025-00813-9},
  url       = {https://www.nature.com/articles/s42254-025-00813-9},
  urldate   = {2025-11-21},
}

@Article{feniou_overlap-adapt-vqe_2023,
  author     = {Feniou, César and Hassan, Muhammad and Traoré, Diata and Giner, Emmanuel and Maday, Yvon and Piquemal, Jean-Philip},
  journal    = {Commun. Phys.},
  title      = {Overlap-{ADAPT}-{VQE}: practical quantum chemistry on quantum computers via overlap-guided compact {Ansätze}},
  year       = {2023},
  issn       = {2399-3650},
  month      = jul,
  number     = {1},
  pages      = {192},
  volume     = {6},
  copyright  = {2023 The Author(s)},
  doi        = {10.1038/s42005-023-01312-y},
  shorttitle = {Overlap-{ADAPT}-{VQE}},
  url        = {https://www.nature.com/articles/s42005-023-01312-y},
  urldate    = {2025-10-28},
}

@Article{guo_experimental_2024,
  author    = {Guo, Shaojun and Sun, Jinzhao and Qian, Haoran and Gong, Ming and Zhang, Yukun and Chen, Fusheng and Ye, Yangsen and Wu, Yulin and Cao, Sirui and Liu, Kun and Zha, Chen and Ying, Chong and Zhu, Qingling and Huang, He-Liang and Zhao, Youwei and Li, Shaowei and Wang, Shiyu and Yu, Jiale and Fan, Daojin and Wu, Dachao and Su, Hong and Deng, Hui and Rong, Hao and Li, Yuan and Zhang, Kaili and Chung, Tung-Hsun and Liang, Futian and Lin, Jin and Xu, Yu and Sun, Lihua and Guo, Cheng and Li, Na and Huo, Yong-Heng and Peng, Cheng-Zhi and Lu, Chao-Yang and Yuan, Xiao and Zhu, Xiaobo and Pan, Jian-Wei},
  journal   = {Nat. Phys.},
  title     = {Experimental quantum computational chemistry with optimized unitary coupled cluster ansatz},
  year      = {2024},
  issn      = {1745-2481},
  month     = aug,
  number    = {8},
  pages     = {1240--1246},
  volume    = {20},
  copyright = {2024 The Author(s), under exclusive licence to Springer Nature Limited},
  doi       = {10.1038/s41567-024-02530-z},
  url       = {https://www.nature.com/articles/s41567-024-02530-z},
  urldate   = {2025-10-28},
}

@Article{weaving_contextual_2025,
  author    = {Weaving, Tim and Ralli, Alexis and Love, Peter J. and Succi, Sauro and Coveney, Peter V.},
  journal   = {npj Quantum Inf.},
  title     = {Contextual subspace variational quantum eigensolver calculation of the dissociation curve of molecular nitrogen on a superconducting quantum computer},
  year      = {2025},
  issn      = {2056-6387},
  month     = feb,
  number    = {1},
  pages     = {25},
  volume    = {11},
  copyright = {2025 The Author(s)},
  doi       = {10.1038/s41534-024-00952-4},
  url       = {https://www.nature.com/articles/s41534-024-00952-4},
  urldate   = {2025-10-28},
}

@Article{yip_variational_2025,
  author     = {Yip, Ka-Wa and Yeter-Aydeniz, Kübra and Dong, Sijia S.},
  title      = {Variational {Quantum} {Annealing} for {Quantum} {Chemistry}},
  year       = {2025},
  month      = mar,
  doi        = {10.48550/arXiv.2503.15473},
  eprint     = {2503.15473},
  eprinttype = {arxiv},
  publisher  = {arXiv},
}

@Article{hegade_shortcuts_2021,
  author  = {Hegade, Narendra N. and Paul, Koushik and Ding, Yongcheng and Sanz, Mikel and Albarrán-Arriagada, F. and Solano, Enrique and Chen, Xi},
  journal = {Phys. Rev. Appl.},
  title   = {Shortcuts to {Adiabaticity} in {Digitized} {Adiabatic} {Quantum} {Computing}},
  year    = {2021},
  month   = feb,
  number  = {2},
  pages   = {024038},
  volume  = {15},
  doi     = {10.1103/PhysRevApplied.15.024038},
  urldate = {2024-02-26},
}

@Article{mikkelsen_quantum-selected_2025,
  author  = {Mikkelsen, Mathias and Nakagawa, Yuya O.},
  journal = {Phys. Rev. Research},
  title   = {Quantum-selected configuration interaction with time-evolved state},
  year    = {2025},
  month   = oct,
  number  = {4},
  pages   = {043043},
  volume  = {7},
  doi     = {10.1103/75pv-hbrx},
  urldate = {2025-10-28},
}

@Article{javier_robledo_SciAdv_2025,
  author   = {Javier Robledo-Moreno and Mario Motta and Holger Haas and Ali Javadi-Abhari and Petar Jurcevic and William Kirby and Simon Martiel and Kunal Sharma and Sandeep Sharma and Tomonori Shirakawa and Iskandar Sitdikov and Rong-Yang Sun and Kevin J. Sung and Maika Takita and Minh C. Tran and Seiji Yunoki and Antonio Mezzacapo},
  journal  = {Sci. Adv.},
  title    = {Chemistry beyond the scale of exact diagonalization on a quantum-centric supercomputer},
  year     = {2025},
  number   = {25},
  pages    = {eadu9991},
  volume   = {11},
  doi      = {10.1126/sciadv.adu9991},
}

@Article{sugisaki_hamiltonian_2025,
  author    = {Sugisaki, Kenji and Kanno, Shu and Itoko, Toshinari and Sakuma, Rei and Yamamoto, Naoki},
  journal   = {Phys. Chem. Chem. Phys.},
  title     = {Hamiltonian simulation-based quantum-selected configuration interaction for large-scale electronic structure calculations with a quantum computer},
  year      = {2025},
  issn      = {1463-9084},
  month     = sep,
  number    = {38},
  pages     = {20869--20884},
  volume    = {27},
  doi       = {10.1039/D5CP02202A},
  publisher = {Royal Society of Chemistry (RSC)},
  url       = {https://pubs.rsc.org/en/content/articlelanding/2025/cp/d5cp02202a},
  urldate   = {2025-09-16},
}

@Article{takeshita_increasing_2020,
  author   = {Takeshita, Tyler and Rubin, Nicholas C. and Jiang, Zhang and Lee, Eunseok and Babbush, Ryan and McClean, Jarrod R.},
  journal  = {Phys. Rev. X},
  title    = {Increasing the {Representation} {Accuracy} of {Quantum} {Simulations} of {Chemistry} without {Extra} {Quantum} {Resources}},
  year     = {2020},
  issn     = {2160-3308},
  month    = jan,
  number   = {1},
  pages    = {011004},
  volume   = {10},
  doi      = {10.1103/PhysRevX.10.011004},
  language = {en},
  urldate  = {2023-10-30},
}

@Article{yoshioka_generalized_2022,
  author  = {Yoshioka, Nobuyuki and Hakoshima, Hideaki and Matsuzaki, Yuichiro and Tokunaga, Yuuki and Suzuki, Yasunari and Endo, Suguru},
  journal = {Phys. Rev. Lett.},
  title   = {Generalized {Quantum} {Subspace} {Expansion}},
  year    = {2022},
  month   = jul,
  number  = {2},
  pages   = {020502},
  volume  = {129},
  doi     = {10.1103/PhysRevLett.129.020502},
  urldate = {2025-09-16},
}

@Article{urbanek_chemistry_2020,
  author  = {Urbanek, Miroslav and Camps, Daan and Van Beeumen, Roel and De Jong, Wibe A.},
  journal = {J. Chem. Theory Comput.},
  title   = {Chemistry on {Quantum} {Computers} with {Virtual} {Quantum} {Subspace} {Expansion}},
  year    = {2020},
  issn    = {1549-9618, 1549-9626},
  month   = sep,
  number  = {9},
  pages   = {5425--5431},
  volume  = {16},
  doi     = {10.1021/acs.jctc.0c00447},
  urldate = {2023-10-30},
}

@Article{patel_quantum_2025,
  author     = {Patel, Smik and Jayakumar, Praveen and Yen, Tzu-Ching and Izmaylov, Artur F.},
  title      = {Quantum {Measurement} for {Quantum} {Chemistry} on a {Quantum} {Computer}},
  year       = {2025},
  month      = may,
  doi        = {10.48550/arXiv.2501.14968},
  eprint     = {2501.14968},
  eprinttype = {arxiv},
  publisher  = {arXiv},
}

@Article{tranter_comparison_2018,
  author       = {Tranter, Andrew and Love, Peter J. and Mintert, Florian and Coveney, Peter V.},
  title        = {A Comparison of the Bravyi–Kitaev and Jordan–Wigner Transformations for the Quantum Simulation of Quantum Chemistry},
  year         = {2018},
  issn         = {1549-9618},
  number       = {11},
  pages        = {5617--5630},
  volume       = {14},
  date         = {2018-11-13},
  doi          = {10.1021/acs.jctc.8b00450},
  journaltitle = {J. Chem. Theory Comput.},
}

@Article{Jiang2020optimalfermionto,
  author    = {Jiang, Zhang and Kalev, Amir and Mruczkiewicz, Wojciech and Neven, Hartmut},
  journal   = {{Quantum}},
  title     = {Optimal fermion-to-qubit mapping via ternary trees with applications to reduced quantum states learning},
  year      = {2020},
  issn      = {2521-327X},
  month     = jun,
  pages     = {276},
  volume    = {4},
  doi       = {10.22331/q-2020-06-04-276},
  publisher = {{Verein zur F{\"{o}}rderung des Open Access Publizierens in den Quantenwissenschaften}},
}

@Article{iqbal2022preentanglingquantumalgorithms,
  author        = {Mohsin Iqbal and David Muñoz Ramo and Henrik Dreyer},
  title         = {Preentangling Quantum Algorithms -- the Density Matrix Renormalization Group-assisted Quantum Canonical Transformation},
  year          = {2022},
  archiveprefix = {arXiv},
  eprint        = {2209.07106},
  eprinttype    = {arxiv},
  primaryclass  = {quant-ph},
}

@Article{PhysRevA.111.062413,
  author    = {Cheng, M. H. and Khosla, K. E. and Self, C. N. and Lin, M. and Li, B. X. and Medina, A. C. and Kim, M. S.},
  journal   = {Phys. Rev. A},
  title     = {Clifford circuit initialization for variational quantum algorithms},
  year      = {2025},
  month     = {Jun},
  pages     = {062413},
  volume    = {111},
  doi       = {10.1103/PhysRevA.111.062413},
  issue     = {6},
  numpages  = {12},
  publisher = {American Physical Society},
}

@Article{jaderberg2025variationalpreparationnormalmatrix,
  author        = {Ben Jaderberg and George Pennington and Kate V. Marshall and Lewis W. Anderson and Abhishek Agarwal and Lachlan P. Lindoy and Ivan Rungger and Stefano Mensa and Jason Crain},
  title         = {Variational preparation of normal matrix product states on quantum computers},
  year          = {2025},
  archiveprefix = {arXiv},
  eprint        = {2503.09683},
  eprinttype    = {arxiv},
  primaryclass  = {quant-ph},
}

@Article{cubitt_dissipative_2023,
  author     = {Cubitt, Toby S.},
  title      = {Dissipative ground state preparation and the {Dissipative} {Quantum} {Eigensolver}},
  year       = {2023},
  month      = sep,
  doi        = {10.48550/arXiv.2303.11962},
  eprint     = {2303.11962},
  eprinttype = {arxiv},
  publisher  = {arXiv},
}

@Article{Kitaev1995,
  author        = {Kitaev, A. Yu.},
  title         = {{Quantum measurements and the Abelian Stabilizer Problem}},
  year          = {1995},
  pages         = {1--22},
  archiveprefix = {arXiv},
  arxivid       = {quant-ph/9511026},
  eprint        = {quant-ph/9511026},
  eprintclass   = {quant-ph},
  eprinttype    = {arxiv},
  primaryclass  = {quant-ph},
}

@Article{Peruzzo2014,
  author   = {Peruzzo, Alberto and McClean, Jarrod and Shadbolt, Peter and Yung, Man Hong and Zhou, Xiao Qi and Love, Peter J. and Aspuru-Guzik, Al{\'{a}}n and O'Brien, Jeremy L.},
  journal  = {Nat. Commun.},
  title    = {{A variational eigenvalue solver on a photonic quantum processor}},
  year     = {2014},
  issn     = {2041-1723},
  month    = may,
  pages    = {4213},
  volume   = {5},
  doi      = {10.1038/ncomms5213},
}

@Article{Ostaszewski2021,
  author    = {Ostaszewski, Mateusz and Grant, Edward and Benedetti, Marcello},
  journal   = {{Quantum}},
  title     = {Structure optimization for parameterized quantum circuits},
  year      = {2021},
  issn      = {2521-327X},
  month     = jan,
  pages     = {391},
  volume    = {5},
  doi       = {10.22331/q-2021-01-28-391},
  publisher = {{Verein zur F{\"{o}}rderung des Open Access Publizierens in den Quantenwissenschaften}},
}

@Article{Nakanishi2020,
  author    = {Nakanishi, Ken M. and Fujii, Keisuke and Todo, Synge},
  journal   = {Phys. Rev. Res.},
  title     = {Sequential minimal optimization for quantum-classical hybrid algorithms},
  year      = {2020},
  month     = {Oct},
  pages     = {043158},
  volume    = {2},
  doi       = {10.1103/PhysRevResearch.2.043158},
  issue     = {4},
  numpages  = {10},
  publisher = {American Physical Society},
}

@Article{Ollitrault2020,
  author    = {Ollitrault, Pauline J. and Kandala, Abhinav and Chen, Chun-Fu and Barkoutsos, Panagiotis Kl. and Mezzacapo, Antonio and Pistoia, Marco and Sheldon, Sarah and Woerner, Stefan and Gambetta, Jay M. and Tavernelli, Ivano},
  journal   = {Phys. Rev. Res.},
  title     = {Quantum equation of motion for computing molecular excitation energies on a noisy quantum processor},
  year      = {2020},
  month     = {Oct},
  pages     = {043140},
  volume    = {2},
  doi       = {10.1103/PhysRevResearch.2.043140},
  issue     = {4},
  numpages  = {13},
  publisher = {American Physical Society},
}

@Article{Tilly2022,
  author   = {Jules Tilly and Hongxiang Chen and Shuxiang Cao and Dario Picozzi and Kanav Setia and Ying Li and Edward Grant and Leonard Wossnig and Ivan Rungger and George H. Booth and Jonathan Tennyson},
  journal  = {Phys. Rep.},
  title    = {The Variational Quantum Eigensolver: A review of methods and best practices},
  year     = {2022},
  issn     = {0370-1573},
  pages    = {1-128},
  volume   = {986},
  doi      = {10.1016/j.physrep.2022.08.003},
  url      = {https://www.sciencedirect.com/science/article/pii/S0370157322003118},
}

@Article{derby_compact_2021,
  author   = {Derby, Charles and Klassen, Joel and Bausch, Johannes and Cubitt, Toby},
  journal  = {Phys. Rev. B},
  title    = {Compact fermion to qubit mappings},
  year     = {2021},
  month    = jul,
  number   = {3},
  pages    = {035118},
  volume   = {104},
  doi      = {10.1103/PhysRevB.104.035118},
  urldate  = {2024-03-13},
}

@Article{Derby2021b,
  author        = {Charles Derby and Joel Klassen},
  title         = {A Compact Fermion to Qubit Mapping Part 2: Alternative Lattice Geometries},
  year          = {2021},
  archiveprefix = {arXiv},
  eprint        = {2101.10735},
  eprinttype    = {arxiv},
  primaryclass  = {quant-ph},
}

@Article{Grimsley2020,
  author    = {Grimsley, Harper R. and Claudino, Daniel and Economou, Sophia E. and Barnes, Edwin and Mayhall, Nicholas J.},
  journal   = {J. Chem. Theory Comput.},
  title     = {Is the Trotterized UCCSD Ansatz Chemically Well-Defined?},
  year      = {2020},
  issn      = {1549-9618},
  month     = {Jan},
  number    = {1},
  pages     = {1-6},
  volume    = {16},
  day       = {14},
  doi       = {10.1021/acs.jctc.9b01083},
  publisher = {American Chemical Society},
}

@Article{Ryabinkin2018,
  author    = {Ryabinkin, Ilya G. and Yen, Tzu-Ching and Genin, Scott N. and Izmaylov, Artur F.},
  journal   = {J. Chem. Theory Comput.},
  title     = {Qubit Coupled Cluster Method: A Systematic Approach to Quantum Chemistry on a Quantum Computer},
  year      = {2018},
  issn      = {1549-9618},
  month     = {Dec},
  number    = {12},
  pages     = {6317-6326},
  volume    = {14},
  day       = {11},
  doi       = {10.1021/acs.jctc.8b00932},
  publisher = {American Chemical Society},
}

@Article{Evangelista2019,
  author   = {Evangelista, Francesco A. and Chan, Garnet Kin-Lic and Scuseria, Gustavo E.},
  journal  = {J. Chem. Phys.},
  title    = {Exact parameterization of fermionic wave functions via unitary coupled cluster theory},
  year     = {2019},
  number   = {24},
  pages    = {244112},
  volume   = {151},
  doi      = {10.1063/1.5133059},
}

@Article{Haidar2023,
  author     = {Haidar, Mohammad and Ran{\v{c}}i{\'c}, Marko J. and Maday, Yvon and Piquemal, Jean-Philip},
  journal    = {J. Phys. Chem. A.},
  title      = {Extension of the Trotterized Unitary Coupled Cluster to Triple Excitations},
  year       = {2023},
  issn       = {1089-5639},
  number     = {15},
  pages      = {3543--3550},
  volume     = {127},
  doi        = {10.1021/acs.jpca.3c01753},
  eprint     = {2212.12462v3},
  eprinttype = {arxiv},
}

@Article{Haidar2025,
  author   = {Haidar, Mohammad and Adjoua, Olivier and Badreddine, Siwar and Peruzzo, Alberto and Piquemal, Jean-Philip},
  journal  = {Quantum Sci. Technol.},
  title    = {Non-iterative disentangled unitary coupled-cluster based on lie-algebraic structure},
  year     = {2025},
  number   = {2},
  pages    = {025031},
  volume   = {10},
  doi      = {10.1088/2058-9565/adb3c5},
}

@Article{Halder2024,
  author   = {Halder, Dipanjali and Mondal, Dibyendu and Maitra, Rahul},
  journal  = {J. Chem. Phys.},
  title    = {Noise-independent route toward the genesis of a COMPACT ansatz for molecular energetics: A dynamic approach},
  year     = {2024},
  number   = {12},
  pages    = {124104},
  volume   = {160},
  doi      = {10.1063/5.0198277},
}

@Article{Izmaylov2020,
  author     = {Izmaylov, Artur F. and D{\'i}az-Tinoco, Manuel and Lang, Robert A.},
  journal    = {Phys. Chem. Chem. Phys.},
  title      = {On the order problem in construction of unitary operators for the Variational Quantum Eigensolver},
  year       = {2020},
  issn       = {1463-9076},
  number     = {23},
  pages      = {12980--12986},
  volume     = {22},
  doi        = {10.1039/D0CP01707H},
  eprint     = {2003.07351v3},
  eprinttype = {arxiv},
}

@Article{Lang2020,
  author     = {Lang, Robert A. and Ryabinkin, Ilya G. and Izmaylov, Artur F.},
  title      = {Unitary transformation of the electronic Hamiltonian with an exact quadratic truncation of the Baker-Campbell-Hausdorff expansion},
  year       = {2020},
  date       = {13.02.2020},
  eprint     = {2002.05701v2},
  eprinttype = {arxiv},
}

@Article{Mondal2023,
  author   = {Mondal, Dibyendu and Halder, Dipanjali and Halder, Sonaldeep and Maitra, Rahul},
  journal  = {J. Chem. Phys.},
  title    = {Development of a compact Ansatz via operator commutativity screening: Digital quantum simulation of molecular systems},
  year     = {2023},
  number   = {1},
  pages    = {014105},
  volume   = {159},
  doi      = {10.1063/5.0153182},
}

@Article{Romero2019,
  author   = {Romero, Jonathan and Babbush, Ryan and McClean, Jarrod R. and Hempel, Cornelius and Love, Peter J. and Aspuru-Guzik, Al{\'a}n},
  journal  = {Quantum Sci. Technol.},
  title    = {Strategies for quantum computing molecular energies using the unitary coupled cluster ansatz},
  year     = {2019},
  number   = {1},
  pages    = {014008},
  volume   = {4},
  doi      = {10.1088/2058-9565/aad3e4},
}

@Article{Babbush2018,
  author    = {Babbush, Ryan and Wiebe, Nathan and McClean, Jarrod and McClain, James and Neven, Hartmut and Chan, Garnet Kin-Lic},
  journal   = {Phys. Rev. X},
  title     = {Low-Depth Quantum Simulation of Materials},
  year      = {2018},
  issn      = {2160-3308},
  month     = Mar,
  number    = {1},
  pages     = {011044},
  volume    = {8},
  doi       = {10.1103/PhysRevX.8.011044},
  publisher = {American Physical Society (APS)},
}

@Article{Shkolnikov2023,
  author     = {Shkolnikov, V. O. and Mayhall, Nicholas J. and Economou, Sophia E. and Barnes, Edwin},
  journal    = {Quantum},
  title      = {Avoiding symmetry roadblocks and minimizing the measurement overhead of adaptive variational quantum eigensolvers},
  year       = {2023},
  issn       = {2521-327X},
  pages      = {1040},
  volume     = {7},
  doi        = {10.22331/q-2023-06-12-1040},
  eprint     = {2109.05340v2},
  eprinttype = {arxiv},
}

@Article{Vogt2020,
  author     = {Vogt, Nicolas and Zanker, Sebastian and Reiner, Jan-Michael and Eckl, Thomas and Marusczyk, Anika and Marthaler, Michael},
  title      = {Preparing symmetry broken ground states with variational quantum algorithms},
  year       = {2020},
  date       = {03.07.2020},
  eprint     = {2007.01582v2},
  eprinttype = {arxiv},
}

@Article{Wiersema2020,
  author    = {Wiersema, Roeland and Zhou, Cunlu and de Sereville, Yvette and Carrasquilla, Juan Felipe and Kim, Yong Baek and Yuen, Henry},
  journal   = {PRX Quantum},
  title     = {Exploring Entanglement and Optimization within the Hamiltonian Variational Ansatz},
  year      = {2020},
  issn      = {2691-3399},
  month     = Dec,
  number    = {2},
  pages     = {020319},
  volume    = {1},
  doi       = {10.1103/PRXQuantum.1.020319},
  publisher = {American Physical Society (APS)},
}

@Article{Lee2019,
  author    = {Lee, Joonho and Huggins, William J. and Head-Gordon, Martin and Whaley, K. Birgitta},
  journal   = {J. Chem. Theory Comput.},
  title     = {Generalized Unitary Coupled Cluster Wave functions for Quantum Computation},
  year      = {2019},
  issn      = {1549-9618},
  month     = {Jan},
  number    = {1},
  pages     = {311-324},
  volume    = {15},
  day       = {08},
  doi       = {10.1021/acs.jctc.8b01004},
  publisher = {American Chemical Society},
}

@Article{Roth2017,
  author    = {Roth, Marco and Ganzhorn, Marc and Moll, Nikolaj and Filipp, Stefan and Salis, Gian and Schmidt, Sebastian},
  journal   = {Phys. Rev. A},
  title     = {Analysis of a parametrically driven exchange-type gate and a two-photon excitation gate between superconducting qubits},
  year      = {2017},
  month     = {Dec},
  pages     = {062323},
  volume    = {96},
  doi       = {10.1103/PhysRevA.96.062323},
  issue     = {6},
  numpages  = {9},
  publisher = {American Physical Society},
}

@Article{Barkoutsos2018,
  author    = {Barkoutsos, Panagiotis Kl. and Gonthier, Jerome F. and Sokolov, Igor and Moll, Nikolaj and Salis, Gian and Fuhrer, Andreas and Ganzhorn, Marc and Egger, Daniel J. and Troyer, Matthias and Mezzacapo, Antonio and Filipp, Stefan and Tavernelli, Ivano},
  journal   = {Phys. Rev. A},
  title     = {Quantum algorithms for electronic structure calculations: Particle-hole Hamiltonian and optimized wave-function expansions},
  year      = {2018},
  month     = {Aug},
  pages     = {022322},
  volume    = {98},
  doi       = {10.1103/PhysRevA.98.022322},
  issue     = {2},
  numpages  = {13},
  publisher = {American Physical Society},
}

@Article{Egger2019,
  author    = {Egger, D.J. and Ganzhorn, M. and Salis, G. and Fuhrer, A. and M\"uller, P. and Barkoutsos, P.Kl. and Moll, N. and Tavernelli, I. and Filipp, S.},
  journal   = {Phys. Rev. Appl.},
  title     = {Entanglement Generation in Superconducting Qubits Using Holonomic Operations},
  year      = {2019},
  month     = {Jan},
  pages     = {014017},
  volume    = {11},
  doi       = {10.1103/PhysRevApplied.11.014017},
  issue     = {1},
  numpages  = {11},
  publisher = {American Physical Society},
}

@Article{Gard2020,
  author    = {Gard, Bryan T. and Zhu, Linghua and Barron, George S. and Mayhall, Nicholas J. and Economou, Sophia E. and Barnes, Edwin},
  journal   = {npj Quantum Inf.},
  title     = {{Efficient symmetry-preserving state preparation circuits for the variational quantum eigensolver algorithm}},
  year      = {2020},
  issn      = {2056-6387},
  number    = {1},
  pages     = {10},
  volume    = {6},
  doi       = {10.1038/s41534-019-0240-1},
  publisher = {Springer US},
}

@Article{Grimsley2019,
  author  = {Grimsley, Harper R. and Economou, Sophia E. and Barnes, Edwin and Mayhall, Nicholas J.},
  journal = {Nat. Commun.},
  title   = {An adaptive variational algorithm for exact molecular simulations on a quantum computer},
  year    = {2019},
  month   = {Jul},
  number  = {1},
  pages   = {3007},
  volume  = {10},
  day     = {08},
  doi     = {10.1038/s41467-019-10988-2},
}

@Misc{developers_qiskit_2023,
  author    = {{The Qiskit Nature developers and contributors}},
  month     = apr,
  title     = {Qiskit {Nature} 0.6.0},
  year      = {2023},
  doi       = {10.5281/zenodo.7828768},
  publisher = {Zenodo},
}

@Article{sun_pyscf_2018,
  author     = {Sun, Qiming and Berkelbach, Timothy C. and Blunt, Nick S. and Booth, George H. and Guo, Sheng and Li, Zhendong and Liu, Junzi and McClain, James D. and Sayfutyarova, Elvira R. and Sharma, Sandeep and Wouters, Sebastian and Chan, Garnet Kin-Lic},
  journal    = {Wiley Interdiscip. Rev. Comput. Mol. Sci.},
  title      = {{PySCF}: the {Python}-based simulations of chemistry framework},
  year       = {2018},
  issn       = {1759-0884},
  number     = {1},
  pages      = {e1340},
  volume     = {8},
  copyright  = {© 2017 Wiley Periodicals, Inc.},
  doi        = {10.1002/wcms.1340},
  language   = {en},
  shorttitle = {{PySCF}},
  urldate    = {2024-11-07},
}

@Article{mcclean_openfermion_2019,
  author     = {McClean, Jarrod R. and Sung, Kevin J. and Kivlichan, Ian D. and Cao, Yudong and Dai, Chengyu and Fried, E. Schuyler and Gidney, Craig and Gimby, Brendan and Gokhale, Pranav and Häner, Thomas and Hardikar, Tarini and Havlíček, Vojtěch and Higgott, Oscar and Huang, Cupjin and Izaac, Josh and Jiang, Zhang and Liu, Xinle and McArdle, Sam and Neeley, Matthew and O'Brien, Thomas and O'Gorman, Bryan and Ozfidan, Isil and Radin, Maxwell D. and Romero, Jhonathan and Rubin, Nicholas and Sawaya, Nicolas P. D. and Setia, Kanav and Sim, Sukin and Steiger, Damian S. and Steudtner, Mark and Sun, Qiming and Sun, Wei and Wang, Daochen and Zhang, Fang and Babbush, Ryan},
  title      = {{OpenFermion}: {The} {Electronic} {Structure} {Package} for {Quantum} {Computers}},
  year       = {2019},
  month      = feb,
  doi        = {10.48550/arXiv.1710.07629},
  eprint     = {1710.07629},
  eprinttype = {arxiv},
  publisher  = {arXiv},
  shorttitle = {{OpenFermion}},
}

@Article{sun_recent_2020,
  author   = {Sun, Qiming and Zhang, Xing and Banerjee, Samragni and Bao, Peng and Barbry, Marc and Blunt, Nick S. and Bogdanov, Nikolay A. and Booth, George H. and Chen, Jia and Cui, Zhi-Hao and Eriksen, Janus J. and Gao, Yang and Guo, Sheng and Hermann, Jan and Hermes, Matthew R. and Koh, Kevin and Koval, Peter and Lehtola, Susi and Li, Zhendong and Liu, Junzi and Mardirossian, Narbe and McClain, James D. and Motta, Mario and Mussard, Bastien and Pham, Hung Q. and Pulkin, Artem and Purwanto, Wirawan and Robinson, Paul J. and Ronca, Enrico and Sayfutyarova, Elvira R. and Scheurer, Maximilian and Schurkus, Henry F. and Smith, James E. T. and Sun, Chong and Sun, Shi-Ning and Upadhyay, Shiv and Wagner, Lucas K. and Wang, Xiao and White, Alec and Whitfield, James Daniel and Williamson, Mark J. and Wouters, Sebastian and Yang, Jun and Yu, Jason M. and Zhu, Tianyu and Berkelbach, Timothy C. and Sharma, Sandeep and Sokolov, Alexander Yu. and Chan, Garnet Kin-Lic},
  journal  = {J. Chem. Phys.},
  title    = {Recent developments in the {PySCF} program package},
  year     = {2020},
  issn     = {0021-9606},
  month    = jul,
  number   = {2},
  pages    = {024109},
  volume   = {153},
  doi      = {10.1063/5.0006074},
  urldate  = {2024-11-07},
}

@Article{Filippov2023,
  author        = {Sergei Filippov and Matea Leahy and Matteo A. C. Rossi and Guillermo García-Pérez},
  title         = {Scalable tensor-network error mitigation for near-term quantum computing},
  year          = {2023},
  archiveprefix = {arXiv},
  eprint        = {2307.11740},
  eprinttype    = {arxiv},
  primaryclass  = {quant-ph},
}

@Article{Filippov2024,
  author        = {Sergey N. Filippov and Sabrina Maniscalco and Guillermo García-Pérez},
  title         = {Scalability of quantum error mitigation techniques: from utility to advantage},
  year          = {2024},
  archiveprefix = {arXiv},
  eprint        = {2403.13542},
  eprinttype    = {arxiv},
  primaryclass  = {quant-ph},
}

@Article{Takagi2022,
  author  = {Takagi, Ryuji and Endo, Suguru and Minagawa, Shintaro and Gu, Mile},
  journal = {npj Quantum Inf.},
  title   = {Fundamental limits of quantum error mitigation},
  year    = {2022},
  month   = {Sep},
  number  = {1},
  pages   = {114},
  volume  = {8},
  day     = {22},
  doi     = {10.1038/s41534-022-00618-z},
}

@Article{Hashim2024,
  author    = {Hashim, Akel and Nguyen, Long B. and Goss, Noah and Marinelli, Brian and Naik, Ravi K. and Chistolini, Trevor and Hines, Jordan and Marceaux, J.P. and Kim, Yosep and Gokhale, Pranav and Tomesh, Teague and Chen, Senrui and Jiang, Liang and Ferracin, Samuele and Rudinger, Kenneth and Proctor, Timothy and Young, Kevin C. and Siddiqi, Irfan and Blume-Kohout, Robin},
  journal   = {PRX Quantum},
  title     = {Practical Introduction to Benchmarking and Characterization of Quantum Computers},
  year      = {2025},
  month     = {Aug},
  pages     = {030202},
  volume    = {6},
  doi       = {10.1103/PRXQuantum.6.030202},
  issue     = {3},
  numpages  = {132},
  publisher = {American Physical Society},
}

@Article{piveteau,
  author    = {Piveteau, Christophe and Sutter, David and Bravyi, Sergey and Gambetta, Jay M. and Temme, Kristan},
  journal   = {Phys. Rev. Lett.},
  title     = {Error Mitigation for Universal Gates on Encoded Qubits},
  year      = {2021},
  month     = {Nov},
  pages     = {200505},
  volume    = {127},
  doi       = {10.1103/PhysRevLett.127.200505},
  issue     = {20},
  numpages  = {6},
  publisher = {American Physical Society},
}

@Article{Luthra_2025,
  author    = {Luthra, Surabhi and Moylett, Alexandra E and Browne, Dan E and Campbell, Earl T},
  journal   = {Quantum Sci. Technol.},
  title     = {Unlocking early fault-tolerant quantum computing with mitigated magic dilution},
  year      = {2025},
  month     = {oct},
  number    = {4},
  pages     = {045066},
  volume    = {10},
  doi       = {10.1088/2058-9565/ae0aef},
  publisher = {IOP Publishing},
}

@Article{zhou2025,
  author        = {Zeyuan Zhou and Shaun Pexton and Aleksander Kubica and Yongshan Ding},
  title         = {Error Mitigation of Fault-Tolerant Quantum Circuits with Soft Information},
  year          = {2025},
  archiveprefix = {arXiv},
  eprint        = {2512.09863},
  eprinttype    = {arxiv},
  primaryclass  = {quant-ph},
}

@Article{Lacroix2025,
  author   = {Lacroix, N. and Bourassa, A. and Heras, F. J. H. and Zhang, L. M. and Bausch, J. and Senior, A. W. and Edlich, T. and Shutty, N. and Sivak, V. and Bengtsson, A. and McEwen, M. and Higgott, O. and Kafri, D. and Claes, J. and Morvan, A. and Chen, Z. and Zalcman, A. and Madhuk, S. and Acharya, R. and Aghababaie Beni, L. and Aigeldinger, G. and Alcaraz, R. and Andersen, T. I. and Ansmann, M. and Arute, F. and Arya, K. and Asfaw, A. and Atalaya, J. and Babbush, R. and Ballard, B. and Bardin, J. C. and Bilmes, A. and Blackwell, S. and Bovaird, J. and Bowers, D. and Brill, L. and Broughton, M. and Browne, D. A. and Buchea, B. and Buckley, B. B. and Burger, T. and Burkett, B. and Bushnell, N. and Cabrera, A. and Campero, J. and Chang, H.-S. and Chiaro, B. and Chih, L.-Y. and Cleland, A. Y. and Cogan, J. and Collins, R. and Conner, P. and Courtney, W. and Crook, A. L. and Curtin, B. and Das, S. and Demura, S. and De Lorenzo, L. and Di Paolo, A. and Donohoe, P. and Drozdov, I. and Dunsworth, A. and Eickbusch, A. and Elbag, A. Moshe and Elzouka, M. and Erickson, C. and Ferreira, V. S. and Flores Burgos, L. and Forati, E. and Fowler, A. G. and Foxen, B. and Ganjam, S. and Garcia, G. and Gasca, R. and Genois, {\'E} and Giang, W. and Gilboa, D. and Gosula, R. and Grajales Dau, A. and Graumann, D. and Greene, A. and Gross, J. A. and Ha, T. and Habegger, S. and Hansen, M. and Harrigan, M. P. and Harrington, S. D. and Heslin, S. and Heu, P. and Hiltermann, R. and Hilton, J. and Hong, S. and Huang, H.-Y. and Huff, A. and Huggins, W. J. and Jeffrey, E. and Jiang, Z. and Jin, X. and Joshi, C. and Juhas, P. and Kabel, A. and Kang, H. and Karamlou, A. H. and Kechedzhi, K. and Khaire, T. and Khattar, T. and Khezri, M. and Kim, S. and Klimov, P. V. and Kobrin, B. and Korotkov, A. N. and Kostritsa, F. and Kreikebaum, J. Mark and Kurilovich, V. D. and Landhuis, D. and Lange-Dei, T. and Langley, B. W. and Laptev, P. and Lau, K.-M. and Ledford, J. and Lee, K. and Lester, B. J. and Le Guevel, L. and Li, W. Yan and Li, Y. and Lill, A. T. and Livingston, W. P. and Locharla, A. and Lucero, E. and Lundahl, D. and Lunt, A. and Maloney, A. and Mandr{\`a}, S. and Martin, L. S. and Martin, O. and Maxfield, C. and McClean, J. R. and Meeks, S. and Megrant, A. and Miao, K. C. and Molavi, R. and Molina, S. and Montazeri, S. and Movassagh, R. and Neill, C. and Newman, M. and Nguyen, A. and Nguyen, M. and Ni, C.-H. and Niu, M. Y. and Oas, L. and Oliver, W. D. and Orosco, R. and Ottosson, K. and Pizzuto, A. and Potter, R. and Pritchard, O. and Quintana, C. and Ramachandran, G. and Reagor, M. J. and Resnick, R. and Rhodes, D. M. and Roberts, G. and Rosenberg, E. and Rosenfeld, E. and Rossi, E. and Roushan, P. and Sankaragomathi, K. and Schurkus, H. F. and Shearn, M. J. and Shorter, A. and Shvarts, V. and Small, S. and Smith, W. Clarke and Springer, S. and Sterling, G. and Suchard, J. and Szasz, A. and Sztein, A. and Thor, D. and Tomita, E. and Torres, A. and Torunbalci, M. Mert and Vaishnav, A. and Vargas, J. and Vdovichev, S. and Vidal, G. and Vollgraff Heidweiller, C. and Waltman, S. and Waltz, J. and Wang, S. X. and Ware, B. and Weidel, T. and White, T. and Wong, K. and Woo, B. W. K. and Woodson, M. and Xing, C. and Yao, Z. Jamie and Yeh, P. and Ying, B. and Yoo, J. and Yosri, N. and Young, G. and Zhang, Y. and Zhu, N. and Zobrist, N. and Neven, H. and Kohli, P. and Davies, A. and Boixo, S. and Kelly, J. and Jones, C. and Gidney, C. and Satzinger, K. J.},
  journal  = {Nature},
  title    = {Scaling and logic in the colour code on a superconducting quantum processor},
  year     = {2025},
  issn     = {1476-4687},
  month    = {Sep},
  number   = {8081},
  pages    = {614-619},
  volume   = {645},
  day      = {01},
  doi      = {10.1038/s41586-025-09061-4},
}

@Article{rosenfeld2025,
  author        = {Emma Rosenfeld and Craig Gidney and Gabrielle Roberts and Alexis Morvan and Nathan Lacroix and Dvir Kafri and Jeffrey Marshall and Ming Li and Volodymyr Sivak and Dmitry Abanin and Amira Abbas and Rajeev Acharya and Laleh Aghababaie Beni and Georg Aigeldinger and Ross Alcaraz and Sayra Alcaraz and Trond I. Andersen and Markus Ansmann and Frank Arute and Kunal Arya and Walt Askew and Nikita Astrakhantsev and Juan Atalaya and Ryan Babbush and Brian Ballard and Joseph C. Bardin and Hector Bates and Andreas Bengtsson and Majid Bigdeli Karimi and Alexander Bilmes and Simon Bilodeau and Felix Borjans and Jenna Bovaird and Dylan Bowers and Leon Brill and Peter Brooks and Michael Broughton and David A. Browne and Brett Buchea and Bob B. Buckley and Tim Burger and Brian Burkett and Nicholas Bushnell and Jamal Busnaina and Anthony Cabrera and Juan Campero and Hung-Shen Chang and Silas Chen and Zijun Chen and Ben Chiaro and Liang-Ying Chih and Agnetta Y. Cleland and Bryan Cochrane and Matt Cockrell and Josh Cogan and Paul Conner and Harold Cook and Rodrigo G. Cortiñas and William Courtney and Alexander L. Crook and Ben Curtin and Martin Damyanov and Sayan Das and Dripto M. Debroy and Sean Demura and Paul Donohoe and Ilya Drozdov and Andrew Dunsworth and Valerie Ehimhen and Alec Eickbusch and Aviv Moshe Elbag and Lior Ella and Mahmoud Elzouka and David Enriquez and Catherine Erickson and Lara Faoro and Vinicius S. Ferreira and Marcos Flores and Leslie Flores Burgos and Sam Fontes and Ebrahim Forati and Jeremiah Ford and Brooks Foxen and Masaya Fukami and Alan Wing Lun Fung and Lenny Fuste and Suhas Ganjam and Gonzalo Garcia and Christopher Garrick and Robert Gasca and Helge Gehring and Robert Geiger and Élie Genois and William Giang and Dar Gilboa and James E. Goeders and Edward C. Gonzales and Raja Gosula and Stijn J. de Graaf and Alejandro Grajales Dau and Dietrich Graumann and Joel Grebel and Alex Greene and Jonathan A. Gross and Jose Guerrero and Loïck Le Guevel and Tan Ha and Steve Habegger and Tanner Hadick and Ali Hadjikhani and Michael C. Hamilton and Monica Hansen and Matthew P. Harrigan and Sean D. Harrington and Jeanne Hartshorn and Stephen Heslin and Paula Heu and Oscar Higgott and Reno Hiltermann and Jeremy Hilton and Hsin-Yuan Huang and Mike Hucka and Christopher Hudspeth and Ashley Huff and William J. Huggins and Lev B. Ioffe and Evan Jeffrey and Shaun Jevons and Zhang Jiang and Xiaoxuan Jin and Chaitali Joshi and Pavol Juhas and Andreas Kabel and Hui Kang and Kiseo Kang and Amir H. Karamlou and Ryan Kaufman and Kostyantyn Kechedzhi and Tanuj Khattar and Mostafa Khezri and Seon Kim and Paul V. Klimov and Can M. Knaut and Bryce Kobrin and Alexander N. Korotkov and Fedor Kostritsa and John Mark Kreikebaum and Ryuho Kudo and Ben Kueffler and Arun Kumar and Vladislav D. Kurilovich and Vitali Kutsko and Tiano Lange-Dei and Brandon W. Langley and Pavel Laptev and Kim-Ming Lau and Emma Leavell and Justin Ledford and Joy Lee and Kenny Lee and Brian J. Lester and Wendy Leung and Lily Li and Wing Yan Li and Alexander T. Lill and William P. Livingston and Matthew T. Lloyd and Aditya Locharla and Laura De Lorenzo and Erik Lucero and Daniel Lundahl and Aaron Lunt and Sid Madhuk and Aniket Maiti and Ashley Maloney and Salvatore Mandrà and Leigh S. Martin and Orion Martin and Eric Mascot and Paul Masih Das and Dmitri Maslov and Melvin Mathews and Cameron Maxfield and Jarrod R. McClean and Matt McEwen and Seneca Meeks and Anthony Megrant and Kevin C. Miao and Zlatko K. Minev and Reza Molavi and Sebastian Molina and Shirin Montazeri and Charles Neill and Michael Newman and Anthony Nguyen and Murray Nguyen and Chia-Hung Ni and Murphy Yuezhen Niu and Nicholas Noll and Logan Oas and William D. Oliver and Raymond Orosco and Kristoffer Ottosson and Alice Pagano and Agustin Di Paolo and Sherman Peek and David Peterson and Alex Pizzuto and Elias Portoles and Rebecca Potter and Orion Pritchard and Michael Qian and Chris Quintana and Ganesh Ramachandran and Arpit Ranadive and Matthew J. Reagor and Rachel Resnick and David M. Rhodes and Daniel Riley and Roberto Rodriguez and Emma Ropes and Lucia B. De Rose and Eliott Rosenberg and Dario Rosenstock and Elizabeth Rossi and Pedram Roushan and David A. Rower and Robert Salazar and Kannan Sankaragomathi and Murat Can Sarihan and Max Schaefer and Sebastian Schroeder and Henry F. Schurkus and Aria Shahingohar and Michael J. Shearn and Aaron Shorter and Noah Shutty and Vladimir Shvarts and Spencer Small and W. Clarke Smith and David A. Sobel and Barrett Spells and Sofia Springer and George Sterling and Jordan Suchard and Aaron Szasz and Alexander Sztein and Madeline Taylor and Jothi Priyanka Thiruraman and Douglas Thor and Dogan Timucin and Eifu Tomita and Alfredo Torres and M. Mert Torunbalci and Hao Tran and Abeer Vaishnav and Justin Vargas and Sergey Vdovichev and Guifre Vidal and Benjamin Villalonga and Catherine Vollgraff Heidweiller and Meghan Voorhees and Steven Waltman and Jonathan Waltz and Shannon X. Wang and Danni Wang and Brayden Ware and James D. Watson and Yonghua Wei and Travis Weidel and Theodore White and Kristi Wong and Bryan W. K. Woo and Christopher J. Wood and Maddy Woodson and Cheng Xing and Z. Jamie Yao and Ping Yeh and Bicheng Ying and Juhwan Yoo and Noureldin Yosri and Elliot Young and Grayson Young and Adam Zalcman and Ran Zhang and Yaxing Zhang and Ningfeng Zhu and Nicholas Zobrist and Zhenjie Zou and Hartmut Neven and Sergio Boixo and Cody Jones and Julian Kelly and Alexandre Bourassa and Kevin J. Satzinger},
  title         = {Magic state cultivation on a superconducting quantum processor},
  year          = {2025},
  archiveprefix = {arXiv},
  eprint        = {2512.13908},
  eprinttype    = {arxiv},
  primaryclass  = {quant-ph},
}

@book{nielsen2010quantum,
  title={Quantum Computation and Quantum Information: 10th Anniversary Edition},
  author={Nielsen, M.A. and Chuang, I.L.},
  isbn={9781139495486},
  url={https://books.google.de/books?id=-s4DEy7o-a0C},
  year={2010},
  publisher={Cambridge University Press}
}

@Article{SalesRodriguez2025,
  author  = {Sales Rodriguez, Pedro and Robinson, John M. and Jepsen, Paul Niklas and He, Zhiyang and Duckering, Casey and Zhao, Chen and Wu, Kai-Hsin and Campo, Joseph and Bagnall, Kevin and Kwon, Minho and Karolyshyn, Thomas and Weinberg, Phillip and Cain, Madelyn and Evered, Simon J. and Geim, Alexandra A. and Kalinowski, Marcin and Li, Sophie H. and Manovitz, Tom and Amato-Grill, Jesse and Basham, James I. and Bernstein, Liane and Braverman, Boris and Bylinskii, Alexei and Choukri, Adam and DeAngelo, Robert J. and Fang, Fang and Fieweger, Connor and Frederick, Paige and Haines, David and Hamdan, Majd and Hammett, Julian and Hsu, Ning and Hu, Ming-Guang and Huber, Florian and Jia, Ningyuan and Kedar, Dhruv and Kornja{\v{c}}a, Milan and Liu, Fangli and Long, John and Lopatin, Jonathan and Lopes, Pedro L. S. and Luo, Xiu-Zhe and Macr{\`i}, Tommaso and Markovi{\'{c}}, Ognjen and Mart{\'i}nez-Mart{\'i}nez, Luis A. and Meng, Xianmei and Ostermann, Stefan and Ostroumov, Evgeny and Paquette, David and Qiang, Zexuan and Shofman, Vadim and Singh, Anshuman and Singh, Manuj and Sinha, Nandan and Thoreen, Henry and Wan, Noel and Wang, Yiping and Waxman-Lenz, Daniel and Wong, Tak and Wurtz, Jonathan and Zhdanov, Andrii and Zheng, Laurent and Greiner, Markus and Keesling, Alexander and Gemelke, Nathan and Vuleti{\'{c}}, Vladan and Kitagawa, Takuya and Wang, Sheng-Tao and Bluvstein, Dolev and Lukin, Mikhail D. and Lukin, Alexander and Zhou, Hengyun and Cant{\'u}, Sergio H.},
  journal = {Nature},
  title   = {Experimental demonstration of logical magic state distillation},
  year    = {2025},
  issn    = {1476-4687},
  month   = {Sep},
  number  = {8081},
  pages   = {620-625},
  volume  = {645},
  day     = {01},
  doi     = {10.1038/s41586-025-09367-3},
}

@Article{yamamoto2025,
  author    = {Kentaro Yamamoto and Yuta Kikuchi and David Amaro and Ben Criger and Silas Dilkes and Ciarán Ryan-Anderson and Andrew Tranter and Joan M. Dreiling and Dan Gresh and Cameron Foltz and Michael Mills and Steven A. Moses and Peter E. Siegfried and Maxwell D. Urmey and Justin J. Burau and Aaron Hankin and Dominic Lucchetti and John P. Gaebler and Natalie C. Brown and Brian Neyenhuis and David Muñoz Ramo},
  journal   = {PRX Quantum},
  title     = {Quantum error-corrected computation of molecular energies},
  year      = {2026},
  issn      = {2691-3399},
  month     = Mar,
  pages     = {020319},
  volume    = {7},
  doi       = {10.1103/m7j3-5sk6},
  publisher = {American Physical Society (APS)},
}

@Article{Eddins2024,
  author        = {Andrew Eddins and Minh C. Tran and Patrick Rall},
  title         = {Lightcone shading for classically accelerated quantum error mitigation},
  year          = {2024},
  archiveprefix = {arXiv},
  eprint        = {2409.04401},
  eprinttype    = {arxiv},
  primaryclass  = {quant-ph},
}

@Article{Tran2023,
  author        = {Minh C. Tran and Kunal Sharma and Kristan Temme},
  title         = {Locality and Error Mitigation of Quantum Circuits},
  year          = {2023},
  archiveprefix = {arXiv},
  eprint        = {2303.06496},
  eprinttype    = {arxiv},
  primaryclass  = {quant-ph},
}

@Article{Kim2023,
author={Kim, Youngseok
and Eddins, Andrew
and Anand, Sajant
and Wei, Ken Xuan
and van den Berg, Ewout
and Rosenblatt, Sami
and Nayfeh, Hasan
and Wu, Yantao
and Zaletel, Michael
and Temme, Kristan
and Kandala, Abhinav},
title={Evidence for the utility of quantum computing before fault tolerance},
journal={Nature},
year={2023},
month={Jun},
day={01},
volume={618},
number={7965},
pages={500-505},
doi={10.1038/s41586-023-06096-3},
}

@Article{Erhard2019,
  author  = {Erhard, Alexander and Wallman, Joel J. and Postler, Lukas and Meth, Michael and Stricker, Roman and Martinez, Esteban A. and Schindler, Philipp and Monz, Thomas and Emerson, Joseph and Blatt, Rainer},
  journal = {Nat. Commun.},
  title   = {Characterizing large-scale quantum computers via cycle benchmarking},
  year    = {2019},
  month   = {Nov},
  number  = {1},
  pages   = {5347},
  volume  = {10},
  day     = {25},
  doi     = {10.1038/s41467-019-13068-7},
}

@Article{Wallman2016,
  author    = {Wallman, Joel J. and Emerson, Joseph},
  journal   = {Phys. Rev. A},
  title     = {Noise tailoring for scalable quantum computation via randomized compiling},
  year      = {2016},
  month     = {Nov},
  pages     = {052325},
  volume    = {94},
  doi       = {10.1103/PhysRevA.94.052325},
  issue     = {5},
  numpages  = {9},
  publisher = {American Physical Society},
}

@Article{Ware2021,
  author    = {Ware, Matthew and Ribeill, Guilhem and Rist\`e, Diego and Ryan, Colm A. and Johnson, Blake and da Silva, Marcus P.},
  journal   = {Phys. Rev. A},
  title     = {Experimental Pauli-frame randomization on a superconducting qubit},
  year      = {2021},
  month     = {Apr},
  pages     = {042604},
  volume    = {103},
  doi       = {10.1103/PhysRevA.103.042604},
  issue     = {4},
  numpages  = {9},
  publisher = {American Physical Society},
}

@article{Hashim2021,
  title = {Randomized Compiling for Scalable Quantum Computing on a Noisy Superconducting Quantum Processor},
  author = {Hashim, Akel and Naik, Ravi K. and Morvan, Alexis and Ville, Jean-Loup and Mitchell, Bradley and Kreikebaum, John Mark and Davis, Marc and Smith, Ethan and Iancu, Costin and O'Brien, Kevin P. and Hincks, Ian and Wallman, Joel J. and Emerson, Joseph and Siddiqi, Irfan},
  journal = {Phys. Rev. X},
  volume = {11},
  issue = {4},
  pages = {041039},
  numpages = {12},
  year = {2021},
  month = {Nov},
  publisher = {American Physical Society},
  doi = {10.1103/PhysRevX.11.041039},
}

@Article{Cerezo2021,
  author  = {Cerezo, M. and Arrasmith, Andrew and Babbush, Ryan and Benjamin, Simon C. and Endo, Suguru and Fujii, Keisuke and McClean, Jarrod R. and Mitarai, Kosuke and Yuan, Xiao and Cincio, Lukasz and Coles, Patrick J.},
  journal = {Nat. Rev. Phys.},
  title   = {Variational quantum algorithms},
  year    = {2021},
  month   = {Sep},
  number  = {9},
  pages   = {625-644},
  volume  = {3},
  day     = {01},
  doi     = {10.1038/s42254-021-00348-9},
}

@article{Cai2023,
  title = {Quantum error mitigation},
  author = {Cai, Zhenyu and Babbush, Ryan and Benjamin, Simon C. and Endo, Suguru and Huggins, William J. and Li, Ying and McClean, Jarrod R. and O'Brien, Thomas E.},
  journal = {Rev. Mod. Phys.},
  volume = {95},
  issue = {4},
  pages = {045005},
  numpages = {37},
  year = {2023},
  month = {Dec},
  publisher = {American Physical Society},
  doi = {10.1103/RevModPhys.95.045005},
}

@Article{Cai2019,
  author  = {Cai, Zhenyu and Benjamin, Simon C.},
  journal = {Sci. Rep.},
  title   = {Constructing Smaller Pauli Twirling Sets for Arbitrary Error Channels},
  year    = {2019},
  issn    = {2045-2322},
  month   = {Aug},
  number  = {1},
  pages   = {11281},
  volume  = {9},
  day     = {02},
  doi     = {10.1038/s41598-019-46722-7},
}

@Article{Koenig,
  author    = {Koenig, Kathrin F. and Reinecke, Finn and Hahn, Walter and Wellens, Thomas},
  journal   = {Phys. Rev. A},
  title     = {Inverted-circuit zero-noise extrapolation for quantum-gate error mitigation},
  year      = {2024},
  month     = {Oct},
  pages     = {042625},
  volume    = {110},
  doi       = {10.1103/PhysRevA.110.042625},
  issue     = {4},
  numpages  = {14},
  publisher = {American Physical Society},
}

@Article{zhang2025,
  author        = {C. Zhang and R. G. Cortiñas and A. H. Karamlou and N. Noll and J. Provazza and J. Bausch and S. Shirobokov and A. White and M. Claassen and S. H. Kang and A. W. Senior and N. Tomašev and J. Gross and K. Lee and T. Schuster and W. J. Huggins and H. Celik and A. Greene and B. Kozlovskii and F. J. H. Heras and A. Bengtsson and A. Grajales Dau and I. Drozdov and B. Ying and W. Livingstone and V. Sivak and N. Yosri and C. Quintana and D. Abanin and A. Abbas and R. Acharya and L. Aghababaie Beni and G. Aigeldinger and R. Alcaraz and S. Alcaraz and T. I. Andersen and M. Ansmann and F. Arute and K. Arya and W. Askew and N. Astrakhantsev and J. Atalaya and B. Ballard and J. C. Bardin and H. Bates and M. Bigdeli Karimi and A. Bilmes and S. Bilodeau and F. Borjans and A. Bourassa and J. Bovaird and D. Bowers and L. Brill and P. Brooks and M. Broughton and D. A. Browne and B. Buchea and B. B. Buckley and T. Burger and B. Burkett and J. Busnaina and N. Bushnell and A. Cabrera and J. Campero and H. -S. Chang and S. Chen and Z. Chen and B. Chiaro and L. -Y. Chih and A. Y. Cleland and B. Cochrane and M. Cockrell and J. Cogan and R. Collins and P. Conner and H. Cook and W. Courtney and A. L. Crook and B. Curtin and S. Das and M. Damyanov and D. M. Debroy and L. De Lorenzo and S. Demura and L. B. De Rose and A. Di Paolo and P. Donohoe and A. Dunsworth and V. Ehimhen and A. Eickbusch and A. M. Elbag and L. Ella and M. Elzouka and D. Enriquez and C. Erickson and V. S. Ferreira and M. Flores and L. Flores Burgos and E. Forati and J. Ford and A. G. Fowler and B. Foxen and M. Fukami and A. W. L. Fung and L. Fuste and S. Ganjam and G. Garcia and C. Garrick and R. Gasca and H. Gehring and R. Geiger and É. Genois and W. Giang and C. Gidney and D. Gilboa and J. E. Goeders and E. C. Gonzales and R. Gosula and S. J. de Graaf and D. Graumann and J. Grebel and J. Guerrero and J. D. Guimarães and T. Ha and S. Habegger and T. Hadick and A. Hadjikhani and M. P. Harrigan and S. D. Harrington and J. Hartshorn and S. Heslin and P. Heu and O. Higgott and R. Hiltermann and J. Hilton and H. -Y. Huang and M. Hucka and C. Hudspeth and A. Huff and E. Jeffrey and S. Jevons and Z. Jiang and X. Jin and C. Joshi and P. Juhas and A. Kabel and H. Kang and K. Kang and R. Kaufman and K. Kechedzhi and T. Khattar and M. Khezri and S. Kim and R. King and O. Kiss and P. V. Klimov and C. M. Knaut and B. Kobrin and F. Kostritsa and J. M. Kreikebaum and R. Kudo and B. Kueffler and A. Kumar and V. D. Kurilovich and V. Kutsko and N. Lacroix and D. Landhuis and T. Lange-Dei and B. W. Langley and P. Laptev and K. -M. Lau and L. Le Guevel and J. Ledford and J. Lee and B. J. Lester and W. Leung and L. Li and W. Y. Li and M. Li and A. T. Lill and M. T. Lloyd and A. Locharla and D. Lundahl and A. Lunt and S. Madhuk and A. Maiti and A. Maloney and S. Mandra and L. S. Martin and O. Martin and E. Mascot and P. Masih Das and D. Maslov and M. Mathews and C. Maxfield and J. R. McClean and M. McEwen and S. Meeks and K. C. Miao and R. Molavi and S. Molina and S. Montazeri and C. Neill and M. Newman and A. Nguyen and M. Nguyen and C. -H. Ni and M. Y. Niu and L. Oas and R. Orosco and K. Ottosson and A. Pagano and S. Peek and D. Peterson and A. Pizzuto and E. Portoles and R. Potter and O. Pritchard and M. Qian and A. Ranadive and M. J. Reagor and R. Resnick and D. M. Rhodes and D. Riley and G. Roberts and R. Rodriguez and E. Ropes and E. Rosenberg and E. Rosenfeld and D. Rosenstock and E. Rossi and D. A. Rower and M. S. Rudolph and R. Salazar and K. Sankaragomathi and M. C. Sarihan and K. J. Satzinger and M. Schaefer and S. Schroeder and H. F. Schurkus and A. Shahingohar and M. J. Shearn and A. Shorter and N. Shutty and V. Shvarts and S. Small and W. C. Smith and D. A. Sobel and R. D. Somma and B. Spells and S. Springer and G. Sterling and J. Suchard and A. Szasz and A. Sztein and M. Taylor and J. P. Thiruraman and D. Thor and D. Timucin and E. Tomita and A. Torres and M. M. Torunbalci and H. Tran and A. Vaishnav and J. Vargas and S. Vdovichev and G. Vidal and C. Vollgraff Heidweiller and M. Voorhees and S. Waltman and J. Waltz and S. X. Wang and B. Ware and J. D. Watson and Y. Wei and T. Weidel and T. White and K. Wong and B. W. K. Woo and C. J. Wood and M. Woodson and C. Xing and Z. J. Yao and P. Yeh and J. Yoo and E. Young and G. Young and A. Zalcman and R. Zhang and Y. Zhang and N. Zhu and N. Zobrist and Z. Zou and G. Bortoli and S. Boixo and J. Chen and Y. Chen and M. Devoret and M. Hansen and C. Jones and J. Kelly and P. Kohli and A. Korotkov and E. Lucero and J. Manyika and Y. Matias and A. Megrant and H. Neven and W. D. Oliver and G. Ramachandran and R. Babbush and V. Smelyanskiy and P. Roushan and D. Kafri and R. Sarpong and D. W. Berry and C. Ramanathan and X. Mi and C. Bengs and A. Ajoy and Z. K. Minev and N. C. Rubin and T. E. O'Brien},
  title         = {Quantum computation of molecular geometry via many-body nuclear spin echoes},
  year          = {2025},
  archiveprefix = {arXiv},
  eprint        = {2510.19550},
  eprinttype    = {arxiv},
  primaryclass  = {quant-ph},
}

@Article{Tang2021,
  author    = {Tang, Ho Lun and Shkolnikov, V.O. and Barron, George S. and Grimsley, Harper R. and Mayhall, Nicholas J. and Barnes, Edwin and Economou, Sophia E.},
  journal   = {PRX Quantum},
  title     = {Qubit-ADAPT-VQE: An Adaptive Algorithm for Constructing Hardware-Efficient Ans\"atze on a Quantum Processor},
  year      = {2021},
  month     = {Apr},
  pages     = {020310},
  volume    = {2},
  doi       = {10.1103/PRXQuantum.2.020310},
  issue     = {2},
  numpages  = {16},
  publisher = {American Physical Society},
}

@Article{Yordanov2021,
  author    = {Yordanov, Yordan S and Armaos, Vasileios and Barnes, Crispin HW and Arvidsson-Shukur, David RM},
  journal   = {Commun. Phys.},
  title     = {Qubit-excitation-based adaptive variational quantum eigensolver},
  year      = {2021},
  number    = {1},
  pages     = {228},
  volume    = {4},
  doi       = {10.1038/s42005-021-00730-0},
  publisher = {Nature Publishing Group UK London},
}

@Article{Liu2021,
  author  = {Liu, Jie and Li, Zhenyu and Yang, Jinlong},
  journal = {J. Chem. Phys.},
  title   = {{An efficient adaptive variational quantum solver of the Schrödinger equation based on reduced density matrices}},
  year    = {2021},
  month   = {06},
  number  = {24},
  pages   = {244112},
  volume  = {154},
  doi     = {10.1063/5.0054822},
}

@Article{Nykänen2023,
  author        = {Anton Nykänen and Matteo A. C. Rossi and Elsi-Mari Borrelli and Sabrina Maniscalco and Guillermo García-Pérez},
  title         = {Mitigating the measurement overhead of ADAPT-VQE with optimised informationally complete generalised measurements},
  year          = {2023},
  archiveprefix = {arXiv},
  eprint        = {2212.09719},
  eprinttype    = {arxiv},
  primaryclass  = {quant-ph},
}

@Article{Anastasiou2024,
  author    = {Anastasiou, Panagiotis G. and Chen, Yanzhu and Mayhall, Nicholas J. and Barnes, Edwin and Economou, Sophia E.},
  journal   = {Phys. Rev. Res.},
  title     = {TETRIS-ADAPT-VQE: An adaptive algorithm that yields shallower, denser circuit Ans\"atze},
  year      = {2024},
  month     = {Mar},
  pages     = {013254},
  volume    = {6},
  doi       = {10.1103/PhysRevResearch.6.013254},
  issue     = {1},
  numpages  = {10},
  publisher = {American Physical Society},
}

@Article{Fitzpatrick2024,
  author  = {Fitzpatrick, Aaron and Nykänen, Anton and Talarico, N. Walter and Lunghi, Alessandro and Maniscalco, Sabrina and García-Pérez, Guillermo and Knecht, Stefan},
  journal = {J. Phys. Chem. A.},
  title   = {Self-Consistent Field Approach for the Variational Quantum Eigensolver: Orbital Optimization Goes Adaptive},
  year    = {2024},
  number  = {14},
  pages   = {2843-2856},
  volume  = {128},
  doi     = {10.1021/acs.jpca.3c05882},
}

@Article{vandenBerg2023,
  author  = {van den Berg, Ewout and Minev, Zlatko K. and Kandala, Abhinav and Temme, Kristan},
  journal = {Nat. Phys.},
  title   = {Probabilistic error cancellation with sparse Pauli--Lindblad models on noisy quantum processors},
  year    = {2023},
  month   = {Aug},
  number  = {8},
  pages   = {1116-1121},
  volume  = {19},
  day     = {01},
  doi     = {10.1038/s41567-023-02042-2},
}

@Article{vandenBerg2024,
  author    = {van den Berg, Ewout and Wocjan, Pawel},
  journal   = {Quantum},
  title     = {Techniques for learning sparse Pauli-Lindblad noise models},
  year      = {2024},
  issn      = {2521-327X},
  month     = Dec,
  pages     = {1556},
  volume    = {8},
  doi       = {10.22331/q-2024-12-10-1556},
  publisher = {Verein zur Forderung des Open Access Publizierens in den Quantenwissenschaften},
}

@Article{Chen2023,
  author  = {Chen, Senrui and Liu, Yunchao and Otten, Matthew and Seif, Alireza and Fefferman, Bill and Jiang, Liang},
  journal = {Nat. Commun.},
  title   = {The learnability of Pauli noise},
  year    = {2023},
  month   = {Jan},
  number  = {1},
  pages   = {52},
  volume  = {14},
  day     = {04},
  doi     = {10.1038/s41467-022-35759-4},
}

@Article{Flammia2020,
  author     = {Flammia, Steven T. and Wallman, Joel J.},
  journal    = {ACM Trans. Quantum Comput.},
  title      = {Efficient Estimation of Pauli Channels},
  year       = {2020},
  month      = dec,
  number     = {1},
  pages      = {3},
  volume     = {1},
  address    = {New York, NY, USA},
  articleno  = {3},
  doi        = {10.1145/3408039},
  issue_date = {December 2020},
  numpages   = {32},
  publisher  = {Association for Computing Machinery},
}

@Article{Kim2023b,
  author  = {Kim, Youngseok and Wood, Christopher J. and Yoder, Theodore J. and Merkel, Seth T. and Gambetta, Jay M. and Temme, Kristan and Kandala, Abhinav},
  journal = {Nat. Phys.},
  title   = {Scalable error mitigation for noisy quantum circuits produces competitive expectation values},
  year    = {2023},
  month   = {May},
  number  = {5},
  pages   = {752-759},
  volume  = {19},
  day     = {01},
  doi     = {10.1038/s41567-022-01914-3},
}

@Article{Cai2021,
  author  = {Cai, Zhenyu},
  journal = {npj Quantum Inf.},
  title   = {Multi-exponential error extrapolation and combining error mitigation techniques for NISQ applications},
  year    = {2021},
  month   = {May},
  number  = {1},
  pages   = {80},
  volume  = {7},
  day     = {25},
  doi     = {10.1038/s41534-021-00404-3},
}

@InProceedings{Aharonov1997,
  author    = {Aharonov, D. and Ben-Or, M.},
  booktitle = {Proceedings of the Twenty-Ninth Annual ACM Symposium on Theory of Computing},
  title     = {Fault-tolerant quantum computation with constant error},
  year      = {1997},
  address   = {New York, NY, USA},
  pages     = {176–188},
  publisher = {Association for Computing Machinery},
  series    = {STOC '97},
  doi       = {10.1145/258533.258579},
  isbn      = {0897918886},
  location  = {El Paso, Texas, USA},
  numpages  = {13},
}

@Article{Kitaev1997,
  author  = {A Yu Kitaev},
  journal = {Russ. Math. Surv.},
  title   = {Quantum computations: algorithms and error correction},
  year    = {1997},
  month   = {dec},
  number  = {6},
  pages   = {1191},
  volume  = {52},
  doi     = {10.1070/RM1997v052n06ABEH002155},
}

@Article{Ferracin2024,
  author    = {Ferracin, Samuele and Hashim, Akel and Ville, Jean-Loup and Naik, Ravi and Carignan-Dugas, Arnaud and Qassim, Hammam and Morvan, Alexis and Santiago, David I. and Siddiqi, Irfan and Wallman, Joel J.},
  journal   = {{Quantum}},
  title     = {Efficiently improving the performance of noisy quantum computers},
  year      = {2024},
  month     = jul,
  pages     = {1410},
  volume    = {8},
  doi       = {10.22331/q-2024-07-15-1410},
  publisher = {{Verein zur F{\"{o}}rderung des Open Access Publizierens in den Quantenwissenschaften}},
}

@Article{Terhal2015,
  author    = {Terhal, Barbara M.},
  journal   = {Rev. Mod. Phys.},
  title     = {Quantum error correction for quantum memories},
  year      = {2015},
  month     = {Apr},
  pages     = {307--346},
  volume    = {87},
  doi       = {10.1103/RevModPhys.87.307},
  issue     = {2},
  numpages  = {40},
  publisher = {American Physical Society},
}

@Article{Bluvstein2024,
  author  = {Bluvstein, Dolev and Evered, Simon J. and Geim, Alexandra A. and Li, Sophie H. and Zhou, Hengyun and Manovitz, Tom and Ebadi, Sepehr and Cain, Madelyn and Kalinowski, Marcin and Hangleiter, Dominik and Bonilla Ataides, J. Pablo and Maskara, Nishad and Cong, Iris and Gao, Xun and Sales Rodriguez, Pedro and Karolyshyn, Thomas and Semeghini, Giulia and Gullans, Michael J. and Greiner, Markus and Vuleti{\'{c}}, Vladan and Lukin, Mikhail D.},
  journal = {Nature},
  title   = {Logical quantum processor based on reconfigurable atom arrays},
  year    = {2024},
  month   = {Feb},
  number  = {7997},
  pages   = {58-65},
  volume  = {626},
  day     = {01},
  doi     = {10.1038/s41586-023-06927-3},
}

@Article{Eickbusch2025,
  author  = {Eickbusch, Alec and McEwen, Matt and Sivak, Volodymyr and Bourassa, Alexandre and Atalaya, Juan and Claes, Jahan and Kafri, Dvir and Gidney, Craig and Warren, Christopher W. and Gross, Jonathan and Opremcak, Alex and Zobrist, Nicholas and Miao, Kevin C. and Roberts, Gabrielle and Satzinger, Kevin J. and Bengtsson, Andreas and Neeley, Matthew and Livingston, William P. and Greene, Alex and Acharya, Rajeev and Aghababaie Beni, Laleh and Aigeldinger, Georg and Alcaraz, Ross and Andersen, Trond I. and Ansmann, Markus and Arute, Frank and Arya, Kunal and Asfaw, Abraham and Babbush, Ryan and Ballard, Brian and Bardin, Joseph C. and Bilmes, Alexander and Bovaird, Jenna and Bowers, Dylan and Brill, Leon and Broughton, Michael and Browne, David A. and Buchea, Brett and Buckley, Bob B. and Burger, Tim and Burkett, Brian and Bushnell, Nicholas and Cabrera, Anthony and Campero, Juan and Chang, Hung-Shen and Chiaro, Ben and Chih, Liang-Ying and Cleland, Agnetta Y. and Cogan, Josh and Collins, Roberto and Conner, Paul and Courtney, William and Crook, Alexander L. and Curtin, Ben and Das, Sayan and Del Toro Barba, Alexander and Demura, Sean and De Lorenzo, Laura and Di Paolo, Agustin and Donohoe, Paul and Drozdov, Ilya K. and Dunsworth, Andrew and Elbag, Aviv Moshe and Elzouka, Mahmoud and Erickson, Catherine and Ferreira, Vinicius S. and Flores Burgos, Leslie and Forati, Ebrahim and Fowler, Austin G. and Foxen, Brooks and Ganjam, Suhas and Garcia, Gonzalo and Gasca, Robert and Genois, {\'E}lie and Giang, William and Gilboa, Dar and Gosula, Raja and Grajales Dau, Alejandro and Graumann, Dietrich and Ha, Tan and Habegger, Steve and Hamilton, Michael C. and Hansen, Monica and Harrigan, Matthew P. and Harrington, Sean D. and Heslin, Stephen and Heu, Paula and Higgott, Oscar and Hiltermann, Reno and Hilton, Jeremy and Huang, Hsin-Yuan and Huff, Ashley and Huggins, William J. and Jeffrey, Evan and Jiang, Zhang and Jin, Xiaoxuan and Jones, Cody and Joshi, Chaitali and Juhas, Pavol and Kabel, Andreas and Kang, Hui and Karamlou, Amir H. and Kechedzhi, Kostyantyn and Khaire, Trupti and Khattar, Tanuj and Khezri, Mostafa and Kim, Seon and Kobrin, Bryce and Korotkov, Alexander N. and Kostritsa, Fedor and Kreikebaum, John Mark and Kurilovich, Vladislav D. and Landhuis, David and Lange-Dei, Tiano and Langley, Brandon W. and Lau, Kim-Ming and Ledford, Justin and Lee, Kenny and Lester, Brian J. and Le Guevel, Lo{\"i}ck and Li, Wing Yan and Lill, Alexander T. and Locharla, Aditya and Lucero, Erik and Lundahl, Daniel and Lunt, Aaron and Madhuk, Sid and Maloney, Ashley and Mandr{\`a}, Salvatore and Martin, Leigh S. and Martin, Orion and Maxfield, Cameron and McClean, Jarrod R. and Meeks, Seneca and Megrant, Anthony and Molavi, Reza and Molina, Sebastian and Montazeri, Shirin and Movassagh, Ramis and Newman, Michael and Nguyen, Anthony and Nguyen, Murray and Ni, Chia-Hung and Oas, Logan and Orosco, Raymond and Ottosson, Kristoffer and Pizzuto, Alex and Potter, Rebecca and Pritchard, Orion and Quintana, Chris and Ramachandran, Ganesh and Reagor, Matthew J. and Rhodes, David M. and Rosenberg, Eliott and Rossi, Elizabeth and Sankaragomathi, Kannan and Schurkus, Henry F. and Shearn, Michael J. and Shorter, Aaron and Shutty, Noah and Shvarts, Vladimir and Small, Spencer and Smith, W. Clarke and Springer, Sofia and Sterling, George and Suchard, Jordan and Szasz, Aaron and Sztein, Alex and Thor, Douglas and Tomita, Eifu and Torres, Alfredo and Torunbalci, M. Mert and Vaishnav, Abeer and Vargas, Justin and Vdovichev, Sergey and Vidal, Guifre and Vollgraff Heidweiller, Catherine and Waltman, Steven and Waltz, Jonathan and Wang, Shannon X. and Ware, Brayden and Weidel, Travis and White, Theodore and Wong, Kristi and Woo, Bryan W. K. and Woodson, Maddy and Xing, Cheng and Yao, Z. Jamie and Yeh, Ping and Ying, Bicheng and Yoo, Juhwan and Yosri, Noureldin and Young, Grayson and Zalcman, Adam and Zhang, Yaxing and Zhu, Ningfeng and Boixo, Sergio and Kelly, Julian and Smelyanskiy, Vadim and Neven, Hartmut and Bacon, Dave and Chen, Zijun and Klimov, Paul V. and Roushan, Pedram and Neill, Charles and Chen, Yu and Morvan, Alexis},
  journal = {Nat. Phys.},
  title   = {Demonstration of dynamic surface codes},
  year    = {2025},
  issn    = {1745-2481},
  month   = {Dec},
  number  = {12},
  pages   = {1994-2001},
  volume  = {21},
  day     = {01},
  doi     = {10.1038/s41567-025-03070-w},
}

@InProceedings{McDonough2022,
  author    = {McDonough, Benjamin and Mari, Andrea and Shammah, Nathan and Stemen, Nathaniel T. and Wahl, Misty and Zeng, William J. and Orth, Peter P.},
  booktitle = {2022 IEEE/ACM Third International Workshop on Quantum Computing Software (QCS)},
  title     = {{ Automated quantum error mitigation based on probabilistic error reduction }},
  year      = {2022},
  pages     = {83-93},
  doi       = {10.1109/QCS56647.2022.00015},
}

@Article{Kandala2019,
  author  = {Kandala, Abhinav and Temme, Kristan and C{\'o}rcoles, Antonio D. and Mezzacapo, Antonio and Chow, Jerry M. and Gambetta, Jay M.},
  journal = {Nature},
  title   = {Error mitigation extends the computational reach of a noisy quantum processor},
  year    = {2019},
  month   = {Mar},
  number  = {7749},
  pages   = {491-495},
  volume  = {567},
  day     = {01},
  doi     = {10.1038/s41586-019-1040-7},
}

@Article{He2020,
  author    = {He, Andre and Nachman, Benjamin and de Jong, Wibe A. and Bauer, Christian W.},
  journal   = {Phys. Rev. A},
  title     = {Zero-noise extrapolation for quantum-gate error mitigation with identity insertions},
  year      = {2020},
  month     = {Jul},
  pages     = {012426},
  volume    = {102},
  doi       = {10.1103/PhysRevA.102.012426},
  issue     = {1},
  numpages  = {12},
  publisher = {American Physical Society},
}

@Article{Temme2017,
  author    = {Temme, Kristan and Bravyi, Sergey and Gambetta, Jay M.},
  journal   = {Phys. Rev. Lett.},
  title     = {Error Mitigation for Short-Depth Quantum Circuits},
  year      = {2017},
  month     = {Nov},
  pages     = {180509},
  volume    = {119},
  doi       = {10.1103/PhysRevLett.119.180509},
  issue     = {18},
  numpages  = {5},
  publisher = {American Physical Society},
}

@Article{Endo2018,
  author    = {Endo, Suguru and Benjamin, Simon C. and Li, Ying},
  journal   = {Phys. Rev. X},
  title     = {Practical Quantum Error Mitigation for Near-Future Applications},
  year      = {2018},
  month     = {Jul},
  pages     = {031027},
  volume    = {8},
  doi       = {10.1103/PhysRevX.8.031027},
  issue     = {3},
  numpages  = {21},
  publisher = {American Physical Society},
}

@Article{DaSilva2024,
  author        = {A. Paetznick and M. P. da Silva and C. Ryan-Anderson and J. M. Bello-Rivas and J. P. Campora III and A. Chernoguzov and J. M. Dreiling and C. Foltz and F. Frachon and J. P. Gaebler and T. M. Gatterman and L. Grans-Samuelsson and D. Gresh and D. Hayes and N. Hewitt and C. Holliman and C. V. Horst and J. Johansen and D. Lucchetti and Y. Matsuoka and M. Mills and S. A. Moses and B. Neyenhuis and A. Paz and J. Pino and P. Siegfried and A. Sundaram and D. Tom and S. J. Wernli and M. Zanner and R. P. Stutz and K. M. Svore},
  title         = {Demonstration of logical qubits and repeated error correction with better-than-physical error rates},
  year          = {2024},
  archiveprefix = {arXiv},
  eprint        = {2404.02280},
  eprinttype    = {arxiv},
  primaryclass  = {quant-ph},
}

@Article{Nation2021,
  author    = {Nation, Paul D. and Kang, Hwajung and Sundaresan, Neereja and Gambetta, Jay M.},
  journal   = {PRX Quantum},
  title     = {Scalable Mitigation of Measurement Errors on Quantum Computers},
  year      = {2021},
  month     = {Nov},
  number    = {4},
  pages     = {040326},
  volume    = {2},
  doi       = {10.1103/PRXQuantum.2.040326},
  publisher = {American Physical Society},
}

@article{VanDenBerg2022,
  title = {Model-free readout-error mitigation for quantum expectation values},
  author = {van den Berg, Ewout and Minev, Zlatko K. and Temme, Kristan},
  journal = {Phys. Rev. A},
  volume = {105},
  issue = {3},
  pages = {032620},
  numpages = {8},
  year = {2022},
  month = {Mar},
  publisher = {American Physical Society},
  doi = {10.1103/PhysRevA.105.032620},
}

@Article{Quek2024,
  author    = {Quek, Yihui and Stilck Fran{\c{c}}a, Daniel and Khatri, Sumeet and Meyer, Johannes Jakob and Eisert, Jens},
  journal   = {Nat. Phys.},
  title     = {Exponentially tighter bounds on limitations of quantum error mitigation},
  year      = {2024},
  issn      = {1745-2481},
  month     = {Jul},
  number    = {10},
  pages     = {1648--1658},
  volume    = {20},
  day       = {25},
  doi       = {10.1038/s41567-024-02536-7},
  publisher = {Springer Science and Business Media LLC},
}

@Article{mcardle_quantum_2020,
  author       = {{McArdle}, Sam and Endo, Suguru and Aspuru-Guzik, Alán and Benjamin, Simon C. and Yuan, Xiao},
  title        = {Quantum computational chemistry},
  year         = {2020},
  issn         = {0034-6861, 1539-0756},
  number       = {1},
  pages        = {015003},
  volume       = {92},
  date         = {2020-03-30},
  doi          = {10.1103/RevModPhys.92.015003},
  journaltitle = {Rev. Mod. Phys.},
  langid       = {english},
  urldate      = {2023-10-30},
}

@Article{boyd_training_2022,
  author     = {Boyd, Gregory and Koczor, Bálint},
  journal    = {Phys. Rev. X},
  title      = {Training {Variational} {Quantum} {Circuits} with {CoVaR}: {Covariance} {Root} {Finding} with {Classical} {Shadows}},
  year       = {2022},
  month      = nov,
  number     = {4},
  pages      = {041022},
  volume     = {12},
  doi        = {10.1103/PhysRevX.12.041022},
  shorttitle = {Training {Variational} {Quantum} {Circuits} with {CoVaR}},
  urldate    = {2024-12-09},
}

@Article{Feynman1982,
  author  = {Feynman, Richard P.},
  journal = {Int. J. Theor. Phys.},
  title   = {{Simulating physics with computers}},
  year    = {1982},
  issn    = {0020-7748},
  number  = {6-7},
  pages   = {467--488},
  volume  = {21},
  doi     = {10.1007/BF02650179},
}

@Article{Arrasmith2020,
  author        = {Andrew Arrasmith and Lukasz Cincio and Rolando D. Somma and Patrick J. Coles},
  title         = {Operator Sampling for Shot-frugal Optimization in Variational Algorithms},
  year          = {2020},
  archiveprefix = {arXiv},
  eprint        = {2004.06252},
  eprinttype    = {arxiv},
  primaryclass  = {quant-ph},
}

@Article{Rubin2018,
  author    = {Rubin, Nicholas C and Babbush, Ryan and McClean, Jarrod},
  journal   = {New J. Phys.},
  title     = {Application of fermionic marginal constraints to hybrid quantum algorithms},
  year      = {2018},
  month     = {may},
  number    = {5},
  pages     = {053020},
  volume    = {20},
  doi       = {10.1088/1367-2630/aab919},
  publisher = {IOP Publishing},
}

@Article{Gokhale2019,
  author        = {Pranav Gokhale and Olivia Angiuli and Yongshan Ding and Kaiwen Gui and Teague Tomesh and Martin Suchara and Margaret Martonosi and Frederic T. Chong},
  title         = {Minimizing State Preparations in Variational Quantum Eigensolver by Partitioning into Commuting Families},
  year          = {2019},
  archiveprefix = {arXiv},
  eprint        = {1907.13623},
  eprinttype    = {arxiv},
  primaryclass  = {quant-ph},
}

@Article{Kandala2017,
  author   = {Kandala, Abhinav and Mezzacapo, Antonio and Temme, Kristan and Takita, Maika and Brink, Markus and Chow, Jerry M. and Gambetta, Jay M.},
  journal  = {Nature},
  title    = {Hardware-efficient variational quantum eigensolver for small molecules and quantum magnets},
  year     = {2017},
  issn     = {1476-4687},
  month    = {Sep},
  number   = {7671},
  pages    = {242-246},
  volume   = {549},
  day      = {01},
  doi      = {10.1038/nature23879},
}

@Article{Reiner2019,
  author    = {Reiner, Jan-Michael and Wilhelm-Mauch, Frank and Schön, Gerd and Marthaler, Michael},
  journal   = {Quantum Sci. Technol.},
  title     = {Finding the ground state of the Hubbard model by variational methods on a quantum computer with gate errors},
  year      = {2019},
  month     = {may},
  number    = {3},
  pages     = {035005},
  volume    = {4},
  doi       = {10.1088/2058-9565/ab1e85},
  publisher = {IOP Publishing},
}

@Article{Cade2020,
  author    = {Cade, Chris and Mineh, Lana and Montanaro, Ashley and Stanisic, Stasja},
  journal   = {Phys. Rev. B},
  title     = {Strategies for solving the Fermi-Hubbard model on near-term quantum computers},
  year      = {2020},
  month     = {Dec},
  pages     = {235122},
  volume    = {102},
  doi       = {10.1103/PhysRevB.102.235122},
  issue     = {23},
  numpages  = {25},
  publisher = {American Physical Society},
}

@Article{Hartnett2026,
  author        = {Gavin S. Hartnett and Khadijeh Sona Najafi and Aleksei Khindanov and Haoran Liao and Michael Schutzman and Michael R. Hush and Michael J. Biercuk and Yuval Baum},
  title         = {Fast, accurate, high-resolution simulation of large-scale Fermi-Hubbard models on a digital quantum processor},
  year          = {2026},
  archiveprefix = {arXiv},
  eprint        = {2605.04025},
  eprinttype    = {arxiv},
  primaryclass  = {quant-ph},
}

@Article{Jafarizadeh2025,
  author    = {Jafarizadeh, Arash and Pollmann, Frank and Gammon-Smith, Adam},
  journal   = {Phys. Rev. Res.},
  title     = {Recipe for local simulation of strongly correlated fermionic matter on quantum computers: The two-dimensional Fermi-Hubbard model},
  year      = {2025},
  month     = {Oct},
  pages     = {043018},
  volume    = {7},
  doi       = {10.1103/PhysRevResearch.7.043018},
  issue     = {4},
  numpages  = {24},
  publisher = {American Physical Society},
}

@Article{Chowdhury2026,
  author  = {Chowdhury, Talal Ahmed and Korepin, Vladimir and Pascuzzi, Vincent R. and Yu, Kwangmin},
  journal = {Appl. Phys. Rev.},
  title   = {Quantum utility in simulating the real-time dynamics of the Fermi–Hubbard model using superconducting quantum computers},
  year    = {2026},
  issn    = {1931-9401},
  month   = {03},
  number  = {1},
  pages   = {011434},
  volume  = {13},
  doi     = {10.1063/5.0306069},
}

@Article{Rungger2019,
  author        = {Rungger, I. and Fitzpatrick, N. and Chen, H. and Alderete, C. H. and Apel, H. and Cowtan, A. and Patterson, A. and Ramo, D. Munoz and Zhu, Y. and Nguyen, N. H. and Grant, E. and Chretien, S. and Wossnig, L. and Linke, N. M. and Duncan, R.},
  title         = {{Dynamical mean field theory algorithm and experiment on quantum computers}},
  year          = {2019},
  archiveprefix = {arXiv},
  arxivid       = {1910.04735},
  eprint        = {1910.04735},
  eprinttype    = {arxiv},
}

@Article{Ehrlich2025,
  author    = {Ehrlich, Jannis and Urban, Daniel F and Elsässer, Christian},
  journal   = {J. Phys.: Condens. Matter},
  title     = {Variational quantum-algorithm based self-consistent calculations for the two-site DMFT model on noisy quantum computing hardware},
  year      = {2025},
  month     = {may},
  number    = {22},
  pages     = {225901},
  volume    = {37},
  doi       = {10.1088/1361-648X/add4db},
  publisher = {IOP Publishing},
}

@Article{Fedorov2022,
  author   = {Fedorov, Dmitry A. and Peng, Bo and Govind, Niranjan and Alexeev, Yuri},
  journal  = {J Mater. Sci: Mater. Theory},
  title    = {VQE method: a short survey and recent developments},
  year     = {2022},
  number   = {1},
  pages    = {2},
  volume   = {6},
  doi      = {10.1186/s41313-021-00032-6},
}

@Article{Bharti2022,
  author   = {Bharti, Kishor and Cervera-Lierta, Alba and Kyaw, Thi Ha and Haug, Tobias and Alperin-Lea, Sumner and Anand, Abhinav and Degroote, Matthias and Heimonen, Hermanni and Kottmann, Jakob S. and Menke, Tim and Mok, Wai-Keong and Sim, Sukin and Kwek, Leong-Chuan and Aspuru-Guzik, Al{\'a}n},
  journal  = {Rev. Mod. Phys.},
  title    = {Noisy intermediate-scale quantum algorithms},
  year     = {2022},
  issn     = {0034-6861},
  number   = {1},
  pages    = {015004},
  volume   = {94},
  doi      = {10.1103/RevModPhys.94.015004},
}

@Article{Le2025,
  author    = {Le, Duc-Truyen and Nguyen, Vu-Linh and Nguyen, Ha C. and Nguyen, Hung Q. and Nguyen, Van-Duy},
  journal   = {IEEE Trans. Quantum Eng.},
  title     = {Variational Quantum Eigensolver: A Comparative Analysis of Classical and Quantum Optimizer Methods},
  year      = {2026},
  issn      = {2689-1808},
  pages     = {3104115--3104115},
  volume    = {7},
  doi       = {10.1109/tqe.2026.3708537},
  publisher = {Institute of Electrical and Electronics Engineers (IEEE)},
}

@Article{Gratsea2024,
  author    = {Gratsea, Katerina and Selisko, Johannes and Amsler, Maximilian and Wever, Christopher and Eckl, Thomas and Samsonidze, Georgy},
  journal   = {Quantum Sci. Technol.},
  title     = {OnionVQE optimization strategy for ground state preparation on NISQ devices},
  year      = {2024},
  month     = {nov},
  number    = {1},
  pages     = {015024},
  volume    = {10},
  doi       = {10.1088/2058-9565/ad8a85},
  publisher = {IOP Publishing},
}

@Article{McClean2018,
  author   = {McClean, Jarrod R. and Boixo, Sergio and Smelyanskiy, Vadim N. and Babbush, Ryan and Neven, Hartmut},
  journal  = {Nat. Commun.},
  title    = {Barren plateaus in quantum neural network training landscapes},
  year     = {2018},
  issn     = {2041-1723},
  month    = {Nov},
  number   = {1},
  pages    = {4812},
  volume   = {9},
  day      = {16},
  doi      = {10.1038/s41467-018-07090-4},
}

@Article{Grant2019,
  author     = {Grant, Edward and Wossnig, Leonard and Ostaszewski, Mateusz and Benedetti, Marcello},
  journal    = {Quantum},
  title      = {An initialization strategy for addressing barren plateaus in parametrized quantum circuits},
  year       = {2019},
  issn       = {2521-327X},
  pages      = {214},
  volume     = {3},
  doi        = {10.22331/q-2019-12-09-214},
  eprint     = {1903.05076v3},
  eprinttype = {arxiv},
}

@Article{Li_2023,
  author   = {Li, Andy C. Y. and Alam, M. Sohaib and Iadecola, Thomas and Jahin, Ammar and Job, Joshua and Kurkcuoglu, Doga Murat and Li, Richard and Orth, Peter P. and {\"O}zg{\"u}ler, A. Bar{\i}{\c{s}} and Perdue, Gabriel N. and Tubman, Norm M.},
  journal  = {Phys. Rev. Research},
  title    = {Benchmarking variational quantum eigensolvers for the square-octagon-lattice Kitaev model},
  year     = {2023},
  issn     = {2643-1564},
  number   = {3},
  pages    = {033071},
  volume   = {5},
  doi      = {10.1103/PhysRevResearch.5.033071},
}

@Article{Ryabinkin2019,
  author     = {Ryabinkin, Ilya G. and Lang, Robert A. and Genin, Scott N. and Izmaylov, Artur F.},
  title      = {Iterative Qubit Coupled Cluster approach with efficient screening of generators},
  year       = {2019},
  date       = {26.06.2019},
  eprint     = {1906.11192v2},
  eprinttype = {arxiv},
}

@Article{Sokolov2020,
  author   = {Sokolov, Igor O. and Barkoutsos, Panagiotis Kl and Ollitrault, Pauline J. and Greenberg, Donny and Rice, Julia and Pistoia, Marco and Tavernelli, Ivano},
  journal  = {J. Chem. Phys.},
  title    = {Quantum orbital-optimized unitary coupled cluster methods in the strongly correlated regime: Can quantum algorithms outperform their classical equivalents?},
  year     = {2020},
  number   = {12},
  pages    = {124107},
  volume   = {152},
  doi      = {10.1063/1.5141835},
}

@Article{Wecker2015,
  author  = {Wecker, Dave and Hastings, Matthew B. and Wiebe, Nathan and Clark, Bryan K. and Nayak, Chetan and Troyer, Matthias},
  journal = {Phys. Rev. A},
  title   = {Solving strongly correlated electron models on a quantum computer},
  year    = {2015},
  issn    = {2469-9926},
  number  = {6},
  pages   = {062318},
  volume  = {92},
  doi     = {10.1103/PhysRevA.92.062318},
}

@Article{Choquette2021,
  author  = {Choquette, Alexandre and {Di Paolo}, Agustin and Barkoutsos, Panagiotis Kl. and S{\'e}n{\'e}chal, David and Tavernelli, Ivano and Blais, Alexandre},
  journal = {Phys. Rev. Research},
  title   = {Quantum-optimal-control-inspired ansatz for variational quantum algorithms},
  year    = {2021},
  issn    = {2643-1564},
  number  = {2},
  pages   = {023092},
  volume  = {3},
  doi     = {10.1103/PhysRevResearch.3.023092},
}

@article{Leone2024,
 author = {Leone, Lorenzo and Oliviero, Salvatore F.E. and Cincio, Lukasz and Cerezo, M.},
 year = {2024},
 title = {On the practical usefulness of the Hardware Efficient Ansatz},
 pages = {1395},
 volume = {8},
 issn = {2521-327X},
 journal = {Quantum},
 doi = {10.22331/q-2024-07-03-1395},
}

@Article{Gokhale2019b,
  author     = {Gokhale, Pranav and Chong, Frederic T.},
  title      = {{\$}O(N{\^{}}3){\$} Measurement Cost for Variational Quantum Eigensolver on Molecular Hamiltonians},
  year       = {2019},
  date       = {30.08.2019},
  eprint     = {1908.11857v1},
  eprinttype = {arxiv},
}

@Article{Hamamura2020,
  author   = {Hamamura, Ikko and Imamichi, Takashi},
  journal  = {npj Quantum Inf.},
  title    = {Efficient evaluation of quantum observables using entangled measurements},
  year     = {2020},
  number   = {1},
  pages    = {56},
  volume   = {6},
  doi      = {10.1038/s41534-020-0284-2},
}

@Article{Tubman2018,
  author        = {Norm M. Tubman and Carlos Mejuto-Zaera and Jeffrey M. Epstein and Diptarka Hait and Daniel S. Levine and William Huggins and Zhang Jiang and Jarrod R. McClean and Ryan Babbush and Martin Head-Gordon and K. Birgitta Whaley},
  title         = {Postponing the orthogonality catastrophe: efficient state preparation for electronic structure simulations on quantum devices},
  year          = {2018},
  archiveprefix = {arXiv},
  eprint        = {1809.05523},
  eprinttype    = {arxiv},
  primaryclass  = {quant-ph},
}

@Article{Malz2024,
  author    = {Malz, Daniel and Styliaris, Georgios and Wei, Zhi-Yuan and Cirac, J. Ignacio},
  journal   = {Phys. Rev. Lett.},
  title     = {Preparation of Matrix Product States with Log-Depth Quantum Circuits},
  year      = {2024},
  month     = {Jan},
  pages     = {040404},
  volume    = {132},
  doi       = {10.1103/PhysRevLett.132.040404},
  issue     = {4},
  numpages  = {9},
  publisher = {American Physical Society},
}

@Article{Rudolph2022,
  author        = {Manuel S. Rudolph and Jing Chen and Jacob Miller and Atithi Acharya and Alejandro Perdomo-Ortiz},
  title         = {Decomposition of Matrix Product States into Shallow Quantum Circuits},
  year          = {2022},
  archiveprefix = {arXiv},
  eprint        = {2209.00595},
  eprinttype    = {arxiv},
  primaryclass  = {quant-ph},
}

@Article{Dov2024,
  author   = {Ben-Dov, Matan and Shnaiderov, David and Makmal, Adi and Dalla Torre, Emanuele G.},
  journal  = {npj Quantum Inf.},
  title    = {Approximate encoding of quantum states using shallow circuits},
  year     = {2024},
  issn     = {2056-6387},
  month    = {Jul},
  number   = {1},
  pages    = {65},
  volume   = {10},
  day      = {02},
  doi      = {10.1038/s41534-024-00858-1},
}

@Article{Bravyi2002,
  author   = {Sergey B. Bravyi and Alexei Yu. Kitaev},
  journal  = {Ann. Phys.},
  title    = {Fermionic Quantum Computation},
  year     = {2002},
  issn     = {0003-4916},
  number   = {1},
  pages    = {210-226},
  volume   = {298},
  doi      = {10.1006/aphy.2002.6254},
  url      = {https://www.sciencedirect.com/science/article/pii/S0003491602962548},
}

@Article{Seeley2012,
  author   = {Seeley, Jacob T. and Richard, Martin J. and Love, Peter J.},
  journal  = {J. Chem. Phys.},
  title    = {The Bravyi-Kitaev transformation for quantum computation of electronic structure},
  year     = {2012},
  issn     = {0021-9606},
  month    = {12},
  number   = {22},
  pages    = {224109},
  volume   = {137},
  doi      = {10.1063/1.4768229},
}

@Article{Verstraete2005,
  author   = {Verstraete, F and Cirac, J I},
  journal  = {J. Stat. Mech: Theory Exp.},
  title    = {Mapping local Hamiltonians of fermions to local Hamiltonians of spins},
  year     = {2005},
  month    = {sep},
  number   = {09},
  pages    = {P09012},
  volume   = {2005},
  doi      = {10.1088/1742-5468/2005/09/P09012},
}

@Article{Setia2019,
  author    = {Setia, Kanav and Bravyi, Sergey and Mezzacapo, Antonio and Whitfield, James D.},
  journal   = {Phys. Rev. Res.},
  title     = {Superfast encodings for fermionic quantum simulation},
  year      = {2019},
  month     = {Oct},
  pages     = {033033},
  volume    = {1},
  doi       = {10.1103/PhysRevResearch.1.033033},
  issue     = {3},
  numpages  = {8},
  publisher = {American Physical Society},
}

@Article{Setia2018,
  author   = {Setia, Kanav and Whitfield, James D.},
  journal  = {J. Chem. Phys.},
  title    = {Bravyi-Kitaev Superfast simulation of electronic structure on a quantum computer},
  year     = {2018},
  issn     = {0021-9606},
  month    = {04},
  number   = {16},
  pages    = {164104},
  volume   = {148},
  doi      = {10.1063/1.5019371},
}

@Article{Izmaylov2019,
  author  = {Izmaylov, Artur F. and Yen, Tzu-Ching and Lang, Robert A. and Verteletskyi, Vladyslav},
  journal = {J. Chem. Theory Comput.},
  title   = {Unitary Partitioning Approach to the Measurement Problem in the Variational Quantum Eigensolver Method},
  year    = {2020},
  number  = {1},
  pages   = {190-195},
  volume  = {16},
  doi     = {10.1021/acs.jctc.9b00791},
}

@Article{McClean2014,
  author  = {McClean, Jarrod R. and Babbush, Ryan and Love, Peter J. and Aspuru-Guzik, Alán},
  journal = {J. Phys. Chem. Lett.},
  title   = {Exploiting Locality in Quantum Computation for Quantum Chemistry},
  year    = {2014},
  number  = {24},
  pages   = {4368-4380},
  volume  = {5},
  doi     = {10.1021/jz501649m},
}

@Article{McClean2016,
  author    = {McClean, Jarrod R and Romero, Jonathan and Babbush, Ryan and Aspuru-Guzik, Alán},
  journal   = {New J. Phys.},
  title     = {The theory of variational hybrid quantum-classical algorithms},
  year      = {2016},
  month     = {feb},
  number    = {2},
  pages     = {023023},
  volume    = {18},
  doi       = {10.1088/1367-2630/18/2/023023},
  publisher = {IOP Publishing},
}

@article{Ezzel2023,
  title = {Dynamical decoupling for superconducting qubits: A performance survey},
  author = {Ezzell, Nic and Pokharel, Bibek and Tewala, Lina and Quiroz, Gregory and Lidar, Daniel A.},
  journal = {Phys. Rev. Appl.},
  volume = {20},
  issue = {6},
  pages = {064027},
  numpages = {42},
  year = {2023},
  month = {Dec},
  publisher = {American Physical Society},
  doi = {10.1103/PhysRevApplied.20.064027},
  url = {https://link.aps.org/doi/10.1103/PhysRevApplied.20.064027}
}

@article{Souza2012,
    author = {Souza, Alexandre M. and Álvarez, Gonzalo A. and Suter, Dieter},
    title = {Robust dynamical decoupling},
    journal = {Philosophical Transactions of the Royal Society A: Mathematical, Physical and Engineering Sciences},
    volume = {370},
    number = {1976},
    pages = {4748-4769},
    year = {2012},
    month = {10},
    issn = {1364-503X},
    doi = {10.1098/rsta.2011.0355},
    url = {https://doi.org/10.1098/rsta.2011.0355},
    eprint = {https://royalsocietypublishing.org/rsta/article-pdf/370/1976/4748/410848/rsta.2011.0355.pdf},
}

@Article{Wecker2015b,
  author    = {Wecker, Dave and Hastings, Matthew B. and Troyer, Matthias},
  journal   = {Phys. Rev. A},
  title     = {Progress towards practical quantum variational algorithms},
  year      = {2015},
  month     = {Oct},
  pages     = {042303},
  volume    = {92},
  doi       = {10.1103/PhysRevA.92.042303},
  issue     = {4},
  numpages  = {10},
  publisher = {American Physical Society},
}

@Article{Kiefer1952,
  author  = {Kiefer, J. and Wolfowitz, J.},
  journal = {Ann. Math. Statist.},
  title   = {Stochastic Estimation of the Maximum of a Regression Function.},
  year    = {1952},
  pages   = {462-466},
  volume  = {23},
  doi     = {10.1214/aoms/1177729392},
  issue   = {3},
}

@INPROCEEDINGS{Durrani2003,
  author={Durrani, T. and Morison, G.},
  booktitle={2003 IEEE International Conference on Acoustics, Speech, and Signal Processing, 2003. Proceedings. (ICASSP '03).}, 
  title={SPSA for noisy non-stationary blind source separation}, 
  year={2003},
  volume={5},
  number={},
  pages={V-285},
  doi={10.1109/ICASSP.2003.1199925}
}

@Article{Finck2012,
  author   = {Steffen Finck and Hans-Georg Beyer},
  journal  = {Theor. Comput. Sci.},
  title    = {Performance analysis of the simultaneous perturbation stochastic approximation algorithm on the noisy sphere model},
  year     = {2012},
  issn     = {0304-3975},
  pages    = {50-72},
  volume   = {419},
  doi      = {10.1016/j.tcs.2011.11.015},
  url      = {https://www.sciencedirect.com/science/article/pii/S0304397511009340},
}

@INPROCEEDINGS{Wang2018,
  author={Wang, Long and Zhu, Jingyi and Spall, James C.},
  booktitle={2018 Annual American Control Conference (ACC)}, 
  title={Mixed Simultaneous Perturbation Stochastic Approximation for Gradient-Free Optimization with Noisy Measurements}, 
  year={2018},
  volume={},
  number={},
  pages={3774-3779},
  doi={10.23919/ACC.2018.8430974}}

@inproceedings{Lemarechel2012,
author = {Lemarechel, C.},
booktitle = {Optimization Stories},
publisher = {Documenta Mathematica Series},
title = {Cauchy and the gradient method},
year = {2012},
pages = {251-254}, 
url = {https://ems.press/books/dms/251/4938}
}

@article{Courant1943,
author = {Courant, R.},
title = {Variational methods for the solution of problems of equlibrium and vibrations},
journal = {Bull. Amer. Math. Soc.},
volume = {49},
number = {1},
year = {1943},
pages = {1-23},
url = {https://projecteuclid.org/journals/bulletin-of-the-american-mathematical-society-new-series/volume-49/issue-1/Variational-methods-for-the-solution-of-problems-of-equilibrium-and/bams/1183504922.full},
}

@Article{Schuld2019,
  author    = {Schuld, Maria and Bergholm, Ville and Gogolin, Christian and Izaac, Josh and Killoran, Nathan},
  journal   = {Phys. Rev. A},
  title     = {Evaluating analytic gradients on quantum hardware},
  year      = {2019},
  month     = {Mar},
  pages     = {032331},
  volume    = {99},
  doi       = {10.1103/PhysRevA.99.032331},
  issue     = {3},
  numpages  = {7},
  publisher = {American Physical Society},
}

@Article{Fletcher1964,
  author   = {Fletcher, R. and Reeves, C. M.},
  journal  = {Comput. J.},
  title    = {Function minimization by conjugate gradients},
  year     = {1964},
  issn     = {0010-4620},
  month    = {01},
  number   = {2},
  pages    = {149-154},
  volume   = {7},
  doi      = {10.1093/comjnl/7.2.149},
}

@Article{Kingma2017,
  author        = {Diederik P. Kingma and Jimmy Ba},
  title         = {Adam: A Method for Stochastic Optimization},
  year          = {2017},
  archiveprefix = {arXiv},
  eprint        = {1412.6980},
  eprinttype    = {arxiv},
  primaryclass  = {cs.LG},
}

@Article{Broyden1970,
  author   = {Broyden, C. G.},
  journal  = {IMA J. Appl. Math.},
  title    = {The Convergence of a Class of Double-rank Minimization Algorithms 1. General Considerations},
  year     = {1970},
  issn     = {0272-4960},
  month    = {03},
  number   = {1},
  pages    = {76-90},
  volume   = {6},
  doi      = {10.1093/imamat/6.1.76},
}

@Article{Fletcher1970,
  author   = {Fletcher, R.},
  journal  = {Comput. J.},
  title    = {A new approach to variable metric algorithms},
  year     = {1970},
  issn     = {0010-4620},
  month    = {01},
  number   = {3},
  pages    = {317-322},
  volume   = {13},
  doi      = {10.1093/comjnl/13.3.317},
}

@Article{Goldfarb1970,
  author  = {Goldfarb, D.},
  journal = {Math. Comp.},
  title   = {A family of variable-metric methods derived by variational means},
  year    = {1970},
  number  = {109},
  pages   = {23-26},
  volume  = {24},
  doi     = {10.1090/s0025-5718-1970-0258249-6},
  page    = {23},
}

@Article{Shanno1970,
  author  = {Shanno, D. F.},
  journal = {Math. Comp.},
  title   = {Conditioning of quasi-Newton methods for function minimization},
  year    = {1970},
  number  = {111},
  pages   = {647-656},
  volume  = {24},
  doi     = {10.1090/s0025-5718-1970-0274029-x},
  page    = {647},
}

@Article{Byrd1995,
  author   = {Byrd, Richard H. and Lu, Peihuang and Nocedal, Jorge and Zhu, Ciyou},
  journal  = {SIAM J. Sci. Comput.},
  title    = {A Limited Memory Algorithm for Bound Constrained Optimization},
  year     = {1995},
  number   = {5},
  pages    = {1190-1208},
  volume   = {16},
  doi      = {10.1137/0916069},
}

@misc{Kraft1988,
	title = "A software package for sequential quadratic programming",
	author = "Kraft, Dieter",
	volume = "88-28",
	address = "Köln",
	series = "Forschungsbericht, 88-28",
	year = "1988",
}

@InBook{Powell1994,
  author    = {Powell, M. J. D.},
  editor    = {Gomez, Susana and Hennart, Jean-Pierre},
  pages     = {51--67},
  publisher = {Springer Netherlands},
  title     = {A Direct Search Optimization Method That Models the Objective and Constraint Functions by Linear Interpolation},
  year      = {1994},
  address   = {Dordrecht},
  isbn      = {978-94-015-8330-5},
  booktitle = {Advances in Optimization and Numerical Analysis},
  doi       = {10.1007/978-94-015-8330-5_4},
}

@Article{Nelder1965,
  author   = {Nelder, J. A. and Mead, R.},
  journal  = {Comput. J.},
  title    = {A Simplex Method for Function Minimization},
  year     = {1965},
  issn     = {0010-4620},
  month    = {01},
  number   = {4},
  pages    = {308-313},
  volume   = {7},
  doi      = {10.1093/comjnl/7.4.308},
}

@INPROCEEDINGS{Lavrijsen2020,
  author={Lavrijsen, Wim and Tudor, Ana and Müller, Juliane and Iancu, Costin and de Jong, Wibe},
  booktitle={2020 IEEE International Conference on Quantum Computing and Engineering (QCE)}, 
  title={Classical Optimizers for Noisy Intermediate-Scale Quantum Devices}, 
  year={2020},
  volume={},
  number={},
  pages={267-277},
  doi={10.1109/QCE49297.2020.00041}
}

@Article{Powell1964,
  author   = {Powell, M. J. D.},
  journal  = {Comput. J.},
  title    = {An efficient method for finding the minimum of a function of several variables without calculating derivatives},
  year     = {1964},
  issn     = {0010-4620},
  month    = {01},
  number   = {2},
  pages    = {155-162},
  volume   = {7},
  doi      = {10.1093/comjnl/7.2.155},
}

@Article{Vidal2018,
  author        = {Javier Gil Vidal and Dirk Oliver Theis},
  title         = {Calculus on parameterized quantum circuits},
  year          = {2018},
  archiveprefix = {arXiv},
  eprint        = {1812.06323},
  eprinttype    = {arxiv},
  primaryclass  = {quant-ph},
}

@Article{Wierichs2022,
  author    = {Wierichs, David and Izaac, Josh and Wang, Cody and Lin, Cedric Yen-Yu},
  journal   = {{Quantum}},
  title     = {General parameter-shift rules for quantum gradients},
  year      = {2022},
  issn      = {2521-327X},
  month     = mar,
  pages     = {677},
  volume    = {6},
  doi       = {10.22331/q-2022-03-30-677},
  publisher = {{Verein zur F{\"{o}}rderung des Open Access Publizierens in den Quantenwissenschaften}},
}

@Article{Pellow_Jarman2021,
  author  = {Pellow-Jarman, Aidan and Sinayskiy, Ilya and Pillay, Anban and Petruccione, Francesco},
  journal = {Quantum Inf. Process.},
  title   = {A comparison of various classical optimizers for a variational quantum linear solver},
  year    = {2021},
  issn    = {1573-1332},
  month   = {Jun},
  number  = {6},
  pages   = {202},
  volume  = {20},
  day     = {04},
  doi     = {10.1007/s11128-021-03140-x},
}

@Article{Kubler2020,
  author    = {K{\"{u}}bler, Jonas M. and Arrasmith, Andrew and Cincio, Lukasz and Coles, Patrick J.},
  journal   = {{Quantum}},
  title     = {An {A}daptive {O}ptimizer for {M}easurement-{F}rugal {V}ariational {A}lgorithms},
  year      = {2020},
  issn      = {2521-327X},
  month     = may,
  pages     = {263},
  volume    = {4},
  doi       = {10.22331/q-2020-05-11-263},
  publisher = {{Verein zur F{\"{o}}rderung des Open Access Publizierens in den Quantenwissenschaften}},
}

@Article{Liu2022,
  author   = {Liu, Xiaoyuan and Angone, Anthony and Shaydulin, Ruslan and Safro, Ilya and Alexeev, Yuri and Cincio, Lukasz},
  journal  = {IEEE Trans. Quantum Eng.},
  title    = {Layer VQE: A Variational Approach for Combinatorial Optimization on Noisy Quantum Computers},
  year     = {2022},
  pages    = {1-20},
  volume   = {3},
  doi      = {10.1109/TQE.2021.3140190},
}

@Article{Sun2016,
  author    = {Qiming Sun and Garnet Kin-Lic Chan},
  journal   = {Acc. Chem. Res.},
  title     = {Quantum Embedding Theories},
  year      = {2016},
  month     = {nov},
  number    = {12},
  pages     = {2705--2712},
  volume    = {49},
  doi       = {10.1021/acs.accounts.6b00356},
  publisher = {American Chemical Society ({ACS})},
}

@Article{Georges1996,
  author    = {Georges, Antoine and Kotliar, Gabriel and Krauth, Werner and Rozenberg, Marcelo J.},
  journal   = {Rev. Mod. Phys.},
  title     = {Dynamical mean-field theory of strongly correlated fermion systems and the limit of infinite dimensions},
  year      = {1996},
  month     = {Jan},
  pages     = {13--125},
  volume    = {68},
  doi       = {10.1103/RevModPhys.68.13},
  issue     = {1},
  numpages  = {0},
  publisher = {American Physical Society},
}

@Article{Knizia2012,
  author    = {Knizia, Gerald and Chan, Garnet Kin-Lic},
  journal   = {Phys. Rev. Lett.},
  title     = {Density Matrix Embedding: A Simple Alternative to Dynamical Mean-Field Theory},
  year      = {2012},
  month     = {Nov},
  pages     = {186404},
  volume    = {109},
  doi       = {10.1103/PhysRevLett.109.186404},
  issue     = {18},
  numpages  = {5},
  publisher = {American Physical Society},
}

@Article{Wouters2016,
  author    = {Sebastian Wouters and Carlos A. Jim{\'{e}}nez-Hoyos and Qiming Sun and Garnet K.-L. Chan},
  journal   = {J. Chem. Theory Comput.},
  title     = {A Practical Guide to Density Matrix Embedding Theory in Quantum Chemistry},
  year      = {2016},
  month     = {may},
  number    = {6},
  pages     = {2706--2719},
  volume    = {12},
  doi       = {10.1021/acs.jctc.6b00316},
  publisher = {American Chemical Society ({ACS})},
}

@Article{Knizia2013,
  author     = {Knizia, Gerald and Chan, Garnet Kin-Lic},
  journal    = {J. Chem. Theory Comput.},
  title      = {Density {Matrix} {Embedding}: {A} {Strong}-{Coupling} {Quantum} {Embedding} {Theory}},
  year       = {2013},
  issn       = {1549-9618},
  month      = mar,
  number     = {3},
  pages      = {1428--1432},
  volume     = {9},
  doi        = {10.1021/ct301044e},
  publisher  = {American Chemical Society},
  shorttitle = {Density {Matrix} {Embedding}},
  urldate    = {2023-01-31},
}

@InBook{Roos1987,
  author    = {Roos, Björn O.},
  pages     = {399-445},
  publisher = {John Wiley \& Sons, Ltd},
  title     = {The Complete Active Space Self-Consistent Field Method and its Applications in Electronic Structure Calculations},
  year      = {1987},
  isbn      = {9780470142943},
  booktitle = {Advances in Chemical Physics},
  doi       = {10.1002/9780470142943.ch7},
}

@Article{Pulay1988,
  author   = {Pulay, Peter and Hamilton, Tracy P.},
  journal  = {J. Chem. Phys.},
  title    = {{UHF} natural orbitals for defining and starting {MC}‐{SCF} calculations},
  year     = {1988},
  issn     = {0021-9606},
  month    = apr,
  number   = {8},
  pages    = {4926--4933},
  volume   = {88},
  doi      = {10.1063/1.454704},
  urldate  = {2026-01-18},
}

@Article{Ding2023,
  author        = {Lexin Ding and Stefan Knecht and Christian Schilling},
  journal       = {J. Phys. Chem. Lett.},
  title         = {{Quantum Information-Assisted Complete Active Space Optimization (QICAS)}},
  year          = {2023},
  issn          = {1948-7185},
  month         = dec,
  number        = {49},
  pages         = {11022--11029},
  volume        = {14},
  archiveprefix = {arXiv},
  doi           = {10.1021/acs.jpclett.3c02536},
  publisher     = {American Chemical Society (ACS)},
}

@Article{Stein2019,
  author     = {Stein, Christopher J. and Reiher, Markus},
  journal    = {J. Comput. Chem.},
  title      = {{autoCAS}: {A} {Program} for {Fully} {Automated} {Multiconfigurational} {Calculations}},
  year       = {2019},
  issn       = {1096-987X},
  number     = {25},
  pages      = {2216--2226},
  volume     = {40},
  copyright  = {© 2019 Wiley Periodicals, Inc.},
  doi        = {10.1002/jcc.25869},
  shorttitle = {{autoCAS}},
  urldate    = {2024-03-08},
}

@Article{Bao2018,
  author    = {Bao, Jie J. and Dong, Sijia S. and Gagliardi, Laura and Truhlar, Donald G.},
  journal   = {J. Chem. Theory Comput.},
  title     = {Automatic {Selection} of an {Active} {Space} for {Calculating} {Electronic} {Excitation} {Spectra} by {MS}-{CASPT2} or {MC}-{PDFT}},
  year      = {2018},
  issn      = {1549-9618},
  month     = apr,
  number    = {4},
  pages     = {2017--2025},
  volume    = {14},
  doi       = {10.1021/acs.jctc.8b00032},
  publisher = {American Chemical Society},
  urldate   = {2026-01-16},
}

@Article{Khedkar2019,
  author    = {Khedkar, Abhishek and Roemelt, Michael},
  journal   = {J. Chem. Theory Comput.},
  title     = {Active {Space} {Selection} {Based} on {Natural} {Orbital} {Occupation} {Numbers} from n-{Electron} {Valence} {Perturbation} {Theory}},
  year      = {2019},
  issn      = {1549-9618},
  month     = jun,
  number    = {6},
  pages     = {3522--3536},
  volume    = {15},
  doi       = {10.1021/acs.jctc.8b01293},
  publisher = {American Chemical Society},
  urldate   = {2026-01-16},
}

@Article{Sayfutyarova2017,
  author    = {Sayfutyarova, Elvira R. and Sun, Qiming and Chan, Garnet Kin-Lic and Knizia, Gerald},
  journal   = {J. Chem. Theory Comput.},
  title     = {Automated {Construction} of {Molecular} {Active} {Spaces} from {Atomic} {Valence} {Orbitals}},
  year      = {2017},
  issn      = {1549-9618},
  month     = sep,
  number    = {9},
  pages     = {4063--4078},
  volume    = {13},
  doi       = {10.1021/acs.jctc.7b00128},
  publisher = {American Chemical Society},
  urldate   = {2026-01-16},
}

@Article{Claudino2019,
  author    = {Claudino, Daniel and Mayhall, Nicholas J.},
  journal   = {J. Chem. Theory Comput.},
  title     = {Automatic {Partition} of {Orbital} {Spaces} {Based} on {Singular} {Value} {Decomposition} in the {Context} of {Embedding} {Theories}},
  year      = {2019},
  issn      = {1549-9618},
  month     = feb,
  number    = {2},
  pages     = {1053--1064},
  volume    = {15},
  doi       = {10.1021/acs.jctc.8b01112},
  publisher = {American Chemical Society},
  urldate   = {2026-01-16},
}

@Article{Kolodzeiski2023,
  author    = {Kolodzeiski, Elena and Stein, Christopher J.},
  journal   = {J. Chem. Theory Comput.},
  title     = {Automated, {Consistent}, and {Even}-{Handed} {Selection} of {Active} {Orbital} {Spaces} for {Quantum} {Embedding}},
  year      = {2023},
  issn      = {1549-9618},
  month     = oct,
  number    = {19},
  pages     = {6643--6655},
  volume    = {19},
  doi       = {10.1021/acs.jctc.3c00653},
  publisher = {American Chemical Society},
  urldate   = {2026-01-16},
}

@Article{Arute2020,
  author   = {Frank Arute and Kunal Arya and Ryan Babbush and Dave Bacon and Joseph C. Bardin and Rami Barends and Sergio Boixo and Michael Broughton and Bob B. Buckley and David A. Buell and Brian Burkett and Nicholas Bushnell and Yu Chen and Zijun Chen and Benjamin Chiaro and Roberto Collins and William Courtney and Sean Demura and Andrew Dunsworth and Edward Farhi and Austin Fowler and Brooks Foxen and Craig Gidney and Marissa Giustina and Rob Graff and Steve Habegger and Matthew P. Harrigan and Alan Ho and Sabrina Hong and Trent Huang and William J. Huggins and Lev Ioffe and Sergei V. Isakov and Evan Jeffrey and Zhang Jiang and Cody Jones and Dvir Kafri and Kostyantyn Kechedzhi and Julian Kelly and Seon Kim and Paul V. Klimov and Alexander Korotkov and Fedor Kostritsa and David Landhuis and Pavel Laptev and Mike Lindmark and Erik Lucero and Orion Martin and John M. Martinis and Jarrod R. McClean and Matt McEwen and Anthony Megrant and Xiao Mi and Masoud Mohseni and Wojciech Mruczkiewicz and Josh Mutus and Ofer Naaman and Matthew Neeley and Charles Neill and Hartmut Neven and Murphy Yuezhen Niu and Thomas E. O’Brien and Eric Ostby and Andre Petukhov and Harald Putterman and Chris Quintana and Pedram Roushan and Nicholas C. Rubin and Daniel Sank and Kevin J. Satzinger and Vadim Smelyanskiy and Doug Strain and Kevin J. Sung and Marco Szalay and Tyler Y. Takeshita and Amit Vainsencher and Theodore White and Nathan Wiebe and Z. Jamie Yao and Ping Yeh and Adam Zalcman},
  journal  = {Science},
  title    = {Hartree-Fock on a superconducting qubit quantum computer},
  year     = {2020},
  number   = {6507},
  pages    = {1084-1089},
  volume   = {369},
  doi      = {10.1126/science.abb9811},
}

@Article{Shang2023,
  author   = {Shang, Honghui and Fan, Yi and Shen, Li and Guo, Chu and Liu, Jie and Duan, Xiaohui and Li, Fang and Li, Zhenyu},
  journal  = {npj Quantum Inf.},
  title    = {Towards practical and massively parallel quantum computing emulation for quantum chemistry},
  year     = {2023},
  issn     = {2056-6387},
  month    = {Apr},
  number   = {1},
  pages    = {33},
  volume   = {9},
  day      = {07},
  doi      = {10.1038/s41534-023-00696-7},
}

@Article{Hubbard1963,
  author   = {Hubbard, J.},
  journal  = {Proc. R. Soc. A},
  title    = {Electron correlations in narrow energy bands},
  year     = {1963},
  issn     = {0080-4630},
  month    = {11},
  number   = {1365},
  pages    = {238-257},
  volume   = {276},
  doi      = {10.1098/rspa.1963.0204},
}

@Article{Kresse1996,
  author    = {Kresse, G. and Furthm\"uller, J.},
  journal   = {Phys. Rev. B},
  title     = {Efficient iterative schemes for ab initio total-energy calculations using a plane-wave basis set},
  year      = {1996},
  month     = {Oct},
  pages     = {11169--11186},
  volume    = {54},
  doi       = {10.1103/PhysRevB.54.11169},
  issue     = {16},
  numpages  = {0},
  publisher = {American Physical Society},
}

@article{Meyer1998,
  title={FORTRAN 90 program for mixed-basis-pseudopotential calculations for crystals},
  author={Meyer, B and Els{\"a}sser, C and Lechermann, F and F{\"a}hnle, M},
  journal={Max-Planck-Institut f{\"u}r Metallforschung, Stuttgart (unpublished)},
  year={1998}
}

@Article{Elsasser1990,
  author  = {C Elsasser and N Takeuchi and K M Ho and C T Chan and P Braun and M Fahnle},
  journal = {J. Phys.: Condens. Matter},
  title   = {Relativistic effects on ground state properties of 4d and 5d transition metals},
  year    = {1990},
  month   = {may},
  number  = {19},
  pages   = {4371},
  volume  = {2},
  doi     = {10.1088/0953-8984/2/19/006},
}

@Article{Ho1992,
  author  = {K M Ho and C Elsasser and C T Chan and M Fahnle},
  journal = {J. Phys.: Condens. Matter},
  title   = {First-principles pseudopotential calculations for hydrogen in 4d transition metals. I. Mixed-basis method for total energies and forces},
  year    = {1992},
  month   = {jun},
  number  = {22},
  pages   = {5189},
  volume  = {4},
  doi     = {10.1088/0953-8984/4/22/017},
}

@book{Singh2005,
    author = {Singh, David J. and Nordstrom, Lars},
    title = {Planewaves, Pseudopotentials, and the LAPW Method, Springer},
    publisher = {Springer Science and Business Media},
    year = {2005}
}

@Article{Blaha2020,
  author  = {Blaha, Peter and Schwarz, Karlheinz and Tran, Fabien and Laskowski, Robert and Madsen, Georg K. H. and Marks, Laurence D.},
  journal = {J. Chem. Phys.},
  title   = {WIEN2k: An APW+lo program for calculating the properties of solids},
  year    = {2020},
  issn    = {0021-9606},
  month   = {02},
  number  = {7},
  pages   = {074101},
  volume  = {152},
  doi     = {10.1063/1.5143061},
}

@Misc{fleurCode,
  author    = {Wortmann, Daniel and Michalicek, Gregor and Baadji, Nadjib and Betzinger, Markus and Bihlmayer, Gustav and Br\"oder, Jens and Burnus, Tobias and Enkovaara, Jussi and Freimuth, Frank and Friedrich, Christoph and Gerhorst, Christian-Roman and Granberg Cauchi, Sabastian and Grytsiuk, Uliana and Hanke, Andrea and Hanke, Jan-Philipp and Heide, Marcus and Heinze, Stefan and Hilgers, Robin and Janssen, Henning and Kl\"uppelberg, Daniel Aaaron and Kovacik, Roman and Kurz, Philipp and Lezaic, Marjana and Madsen, Georg K. H. and Mokrousov, Yuriy and Neukirchen, Alexander and Redies, Matthias and Rost, Stefan and Schlipf, Martin and Schindlmayr, Arno and Winkelmann, Miriam and Bl\"ugel, Stefan},
  month     = may,
  title     = {{FLEUR}},
  year      = {2023},
  doi       = {10.5281/zenodo.7576163},
  publisher = {Zenodo},
}

@Article{Clark2005,
  author      = {Stewart J. Clark and Matthew D. Segall and Chris J. Pickard and Phil J. Hasnip and Matt I. J. Probert and Keith Refson and Mike C. Payne},
  journal     = {Z. Kristallogr. Cryst. Mater.},
  title       = {First principles methods using CASTEP},
  year        = {2005},
  number      = {5-6},
  pages       = {567--570},
  volume      = {220},
  doi         = {10.1524/zkri.220.5.567.65075},
  lastchecked = {2026-05-18},
}

@Article{Giannozzi2009,
  author  = {Giannozzi, Paolo and Baroni, Stefano and Bonini, Nicola and Calandra, Matteo and Car, Roberto and Cavazzoni, Carlo and Ceresoli, Davide and Chiarotti, Guido L and Cococcioni, Matteo and Dabo, Ismaila and Dal Corso, Andrea and de Gironcoli, Stefano and Fabris, Stefano and Fratesi, Guido and Gebauer, Ralph and Gerstmann, Uwe and Gougoussis, Christos and Kokalj, Anton and Lazzeri, Michele and Martin-Samos, Layla and Marzari, Nicola and Mauri, Francesco and Mazzarello, Riccardo and Paolini, Stefano and Pasquarello, Alfredo and Paulatto, Lorenzo and Sbraccia, Carlo and Scandolo, Sandro and Sclauzero, Gabriele and Seitsonen, Ari P and Smogunov, Alexander and Umari, Paolo and Wentzcovitch, Renata M},
  journal = {J. Phys.: Condens. Matter},
  title   = {QUANTUM ESPRESSO: a modular and open-source software project for quantum simulations of materials},
  year    = {2009},
  month   = {sep},
  number  = {39},
  pages   = {395502},
  volume  = {21},
  doi     = {10.1088/0953-8984/21/39/395502},
}

@article{Lippert1997,
  author    = {Gerald Lippert and Jurg Hutter and Michele Parrinello},
  journal   = {Mol. Phys.},
  title     = {A hybrid Gaussian and plane wave density functional scheme},
  year      = {1997},
  number    = {3},
  pages     = {477--488},
  volume    = {92},
  doi       = {10.1080/002689797170220},
  publisher = {Taylor \& Francis},
  url       = {https://www.tandfonline.com/doi/abs/10.1080/002689797170220},
}

@Article{Kotliar2006,
  author    = {Kotliar, G. and Savrasov, S. Y. and Haule, K. and Oudovenko, V. S. and Parcollet, O. and Marianetti, C. A.},
  journal   = {Rev. Mod. Phys.},
  title     = {Electronic structure calculations with dynamical mean-field theory},
  year      = {2006},
  issn      = {1539-0756},
  month     = Aug,
  number    = {3},
  pages     = {865--951},
  volume    = {78},
  doi       = {10.1103/revmodphys.78.865},
  publisher = {American Physical Society (APS)},
}

@Article{Heisenberg1928,
  author   = {Heisenberg, W.},
  journal  = {Z. Phys.},
  title    = {Zur Theorie des Ferromagnetismus},
  year     = {1928},
  issn     = {0044-3328},
  month    = {Sep},
  number   = {9},
  pages    = {619-636},
  volume   = {49},
  day      = {01},
  doi      = {10.1007/BF01328601},
}

@Article{Aryasetiawan2004,
  author    = {Aryasetiawan, F. and Imada, M. and Georges, A. and Kotliar, G. and Biermann, S. and Lichtenstein, A. I.},
  journal   = {Phys. Rev. B},
  title     = {Frequency-dependent local interactions and low-energy effective models from electronic structure calculations},
  year      = {2004},
  month     = {Nov},
  pages     = {195104},
  volume    = {70},
  doi       = {10.1103/PhysRevB.70.195104},
  issue     = {19},
  numpages  = {8},
  publisher = {American Physical Society},
}

@Article{Kaduk2012,
  author    = {Kaduk, Benjamin and Kowalczyk, Tim and Van Voorhis, Troy},
  journal   = {Chem. Rev.},
  title     = {Constrained Density Functional Theory},
  year      = {2012},
  issn      = {0009-2665},
  month     = {Jan},
  number    = {1},
  pages     = {321-370},
  volume    = {112},
  day       = {11},
  doi       = {10.1021/cr200148b},
  publisher = {American Chemical Society},
}

@Article{Pizzi2020,
  author  = {Pizzi, Giovanni and Vitale, Valerio and Arita, Ryotaro and Bl{\"u}gel, Stefan and Freimuth, Frank and G{\'e}ranton, Guillaume and Gibertini, Marco and Gresch, Dominik and Johnson, Charles and Koretsune, Takashi and Iba{\~n}ez-Azpiroz, Julen and Lee, Hyungjun and Lihm, Jae-Mo and Marchand, Daniel and Marrazzo, Antimo and Mokrousov, Yuriy and Mustafa, Jamal I. and Nohara, Yoshiro and Nomura, Yusuke and Paulatto, Lorenzo and Ponc{\'e}, Samuel and Ponweiser, Thomas and Qiao, Junfeng and Th{\"o}le, Florian and Tsirkin, Stepan S. and Wierzbowska, Ma{\l}gorzata and Marzari, Nicola and Vanderbilt, David and Souza, Ivo and Mostofi, Arash A. and Yates, Jonathan R.},
  journal = {J. Phys.: Condens. Matter},
  title   = {Wannier90 as a community code: new features and applications},
  year    = {2020},
  number  = {16},
  pages   = {165902},
  volume  = {32},
  doi     = {10.1088/1361-648X/ab51ff},
}

@Article{Libisch2014,
  author    = {Libisch, Florian and Huang, Chen and Carter, Emily A.},
  journal   = {Acc. Chem. Res.},
  title     = {Embedded Correlated Wavefunction Schemes: Theory and Applications},
  year      = {2014},
  issn      = {0001-4842},
  month     = {Sep},
  number    = {9},
  pages     = {2768-2775},
  volume    = {47},
  day       = {16},
  doi       = {10.1021/ar500086h},
  publisher = {American Chemical Society},
}

@Article{Wesolowski2015,
  author    = {Wesolowski, Tomasz A. and Shedge, Sapana and Zhou, Xiuwen},
  journal   = {Chem. Rev.},
  title     = {Frozen-Density Embedding Strategy for Multilevel Simulations of Electronic Structure},
  year      = {2015},
  issn      = {0009-2665},
  month     = {Jun},
  number    = {12},
  pages     = {5891-5928},
  volume    = {115},
  day       = {24},
  doi       = {10.1021/cr500502v},
  publisher = {American Chemical Society},
}

@Article{Jacob2014,
  author  = {Jacob, Christoph R. and Neugebauer, Johannes},
  journal = {Wiley Interdiscip. Rev. Comput. Mol. Sci.},
  title   = {Subsystem density-functional theory},
  year    = {2014},
  number  = {4},
  pages   = {325-362},
  volume  = {4},
  doi     = {10.1002/wcms.1175},
}

@Article{Lan2017,
  author    = {Lan, Tran Nguyen and Zgid, Dominika},
  journal   = {J. Phys. Chem. Lett.},
  title     = {Generalized Self-Energy Embedding Theory},
  year      = {2017},
  month     = {May},
  number    = {10},
  pages     = {2200-2205},
  volume    = {8},
  day       = {18},
  doi       = {10.1021/acs.jpclett.7b00689},
  publisher = {American Chemical Society},
}

@Article{Rusakov2019,
  author    = {Rusakov, Alexander A. and Iskakov, Sergei and Tran, Lan Nguyen and Zgid, Dominika},
  journal   = {J. Chem. Theory Comput.},
  title     = {Self-Energy Embedding Theory (SEET) for Periodic Systems},
  year      = {2019},
  issn      = {1549-9618},
  month     = {Jan},
  number    = {1},
  pages     = {229-240},
  volume    = {15},
  day       = {08},
  doi       = {10.1021/acs.jctc.8b00927},
  publisher = {American Chemical Society},
}

@Article{Ma2021,
  author    = {Ma, He and Sheng, Nan and Govoni, Marco and Galli, Giulia},
  journal   = {J. Chem. Theory Comput.},
  title     = {Quantum Embedding Theory for Strongly Correlated States in Materials},
  year      = {2021},
  issn      = {1549-9618},
  month     = {Apr},
  number    = {4},
  pages     = {2116-2125},
  volume    = {17},
  day       = {13},
  doi       = {10.1021/acs.jctc.0c01258},
  publisher = {American Chemical Society},
}

@Article{Vorwerk2022,
  author  = {Vorwerk, Christian and Sheng, Nan and Govoni, Marco and Huang, Benchen and Galli, Giulia},
  journal = {Nat. Comput. Sci.},
  title   = {Quantum embedding theories to simulate condensed systems on quantum computers},
  year    = {2022},
  number  = {7},
  pages   = {424--432},
  volume  = {2},
  doi     = {10.1038/s43588-022-00279-0},
}

@Article{Chiew2023,
  author    = {Chiew, Mitchell and Strelchuk, Sergii},
  journal   = {{Quantum}},
  title     = {Discovering optimal fermion-qubit mappings through algorithmic enumeration},
  year      = {2023},
  issn      = {2521-327X},
  month     = oct,
  pages     = {1145},
  volume    = {7},
  doi       = {10.22331/q-2023-10-18-1145},
  publisher = {{Verein zur F{\"{o}}rderung des Open Access Publizierens in den Quantenwissenschaften}},
}

@Article{Dyrenkova2026,
  author    = {Dyrenkova, Emiliia and Laflamme, Raymond and Vasmer, Michael},
  journal   = {Phys. Rev. A},
  title     = {Scalable simulation of fermionic encoding performance on noisy quantum computers},
  year      = {2026},
  month     = {Feb},
  pages     = {022607},
  volume    = {113},
  doi       = {10.1103/dg5d-m728},
  issue     = {2},
  numpages  = {23},
  publisher = {American Physical Society},
}

@Article{Chiew2026,
  author        = {Mitchell Chiew and Cameron Ibrahim and Ilya Safro and Sergii Strelchuk},
  title         = {Optimal fermion-qubit mappings via quadratic assignment},
  year          = {2026},
  archiveprefix = {arXiv},
  eprint        = {2504.21636},
  eprinttype    = {arxiv},
  primaryclass  = {quant-ph},
}

@Article{Higgott2019,
  author   = {Higgott, Oscar and Wang, Daochen and Brierley, Stephen},
  journal  = {Quantum},
  title    = {{Variational quantum computation of excited states}},
  year     = {2019},
  issn     = {2521-327X},
  month    = jul,
  pages    = {156},
  volume   = {3},
  doi      = {10.22331/q-2019-07-01-156},
  language = {en-GB},
  url      = {https://quantum-journal.org/papers/q-2019-07-01-156/},
  urldate  = {2026-03-31},
}

@Article{Feniou2023,
  author    = {Feniou, César and Hassan, Muhammad and Claudon, Baptiste and Courtat, Axel and Adjoua, Olivier and Maday, Yvon and Piquemal, Jean-Philip},
  journal   = {Sci. Rep.},
  title     = {Greedy gradient-free adaptive variational quantum algorithms on a noisy intermediate scale quantum computer},
  year      = {2025},
  issn      = {2045-2322},
  month     = may,
  number    = {1},
  pages     = {18689},
  volume    = {15},
  copyright = {2025 The Author(s)},
  doi       = {10.1038/s41598-025-99962-1},
  language  = {en},
  url       = {https://www.nature.com/articles/s41598-025-99962-1},
  urldate   = {2026-03-09},
}

@InProceedings{Giurgica2020,
  author    = {Giurgica-Tiron, Tudor and Hindy, Yousef and LaRose, Ryan and Mari, Andrea and Zeng, William J.},
  booktitle = {2020 IEEE International Conference on Quantum Computing and Engineering (QCE)},
  title     = {Digital zero noise extrapolation for quantum error mitigation},
  year      = {2020},
  pages     = {306-316},
  doi       = {10.1109/QCE49297.2020.00045},
}

@Article{Havlicek2017,
  author    = {Havl\'{\i}\ifmmode \check{c}\else \v{c}\fi{}ek, Vojt\ifmmode \check{e}\else \v{e}\fi{}ch and Troyer, Matthias and Whitfield, James D.},
  journal   = {Phys. Rev. A},
  title     = {Operator locality in the quantum simulation of fermionic models},
  year      = {2017},
  month     = {Mar},
  pages     = {032332},
  volume    = {95},
  doi       = {10.1103/PhysRevA.95.032332},
  issue     = {3},
  numpages  = {9},
  publisher = {American Physical Society},
}

@Article{Jiang2018,
  author    = {Jiang, Zhang and Sung, Kevin J. and Kechedzhi, Kostyantyn and Smelyanskiy, Vadim N. and Boixo, Sergio},
  journal   = {Phys. Rev. Appl.},
  title     = {Quantum Algorithms to Simulate Many-Body Physics of Correlated Fermions},
  year      = {2018},
  month     = {Apr},
  number    = {4},
  pages     = {044036},
  volume    = {9},
  doi       = {10.1103/PhysRevApplied.9.044036},
  issue     = {4},
  numpages  = {23},
  publisher = {American Physical Society},
}

@Article{McClean2017,
  author    = {McClean, Jarrod R. and Kimchi-Schwartz, Mollie E. and Carter, Jonathan and de Jong, Wibe A.},
  journal   = {Phys. Rev. A},
  title     = {Hybrid quantum-classical hierarchy for mitigation of decoherence and determination of excited states},
  year      = {2017},
  month     = {Apr},
  number    = {4},
  pages     = {042308},
  volume    = {95},
  doi       = {10.1103/PhysRevA.95.042308},
  issue     = {4},
  numpages  = {10},
  publisher = {American Physical Society},
  urldate   = {2025-09-16},
}

@Article{Nakanishi2019,
  author    = {Nakanishi, Ken M. and Mitarai, Kosuke and Fujii, Keisuke},
  journal   = {Phys. Rev. Research},
  title     = {Subspace-search variational quantum eigensolver for excited states},
  year      = {2019},
  issn      = {2643-1564},
  month     = oct,
  number    = {3},
  pages     = {033062},
  volume    = {1},
  doi       = {10.1103/PhysRevResearch.1.033062},
  publisher = {American Physical Society},
  urldate   = {2026-03-31},
}

@Article{Yen2020,
  author   = {Yen, Tzu-Ching and Verteletskyi, Vladyslav and Izmaylov, Artur F.},
  journal  = {J. Chem. Theory Comput.},
  title    = {Measuring All Compatible Operators in One Series of Single-Qubit Measurements Using Unitary Transformations},
  year     = {2020},
  number   = {4},
  pages    = {2400--2409},
  volume   = {16},
  doi      = {10.1021/acs.jctc.0c00008},
}
